\documentclass[11pt]{article}
\pdfoutput=1
\usepackage{jheppub}
\usepackage{tabu}
\usepackage[vcentermath]{youngtab}
\usepackage[usenames,dvipsnames,table]{xcolor}
\usepackage{graphicx,amsmath,amssymb,amsthm,multirow,array,bm,bbm,esint}
\usepackage[mathscr]{eucal}
\usepackage[bbgreekl]{mathbbol}
\usepackage{epsf,amsfonts}
\usepackage{slashed}
\usepackage[numbers,sort&compress]{natbib}
\usepackage{minibox}
\usepackage{rotating}
\usepackage{pdflscape}
\usepackage{array}

\usepackage{slashed}
\usepackage{subfig}
\usepackage{float}
 \usepackage{tikz}
    \usepackage{pgfplots}
  \usetikzlibrary{backgrounds}
\usetikzlibrary{arrows}
\usetikzlibrary{shapes,shapes.geometric,shapes.misc}
\tikzstyle{tikzfig}=[baseline=-0.25em,scale=0.5]

\usepackage{tikz,tkz-euclide}
\usetkzobj{all}

\usetikzlibrary{fillbetween,backgrounds}

\pgfkeys{/tikz/tikzit fill/.initial=0}
\pgfkeys{/tikz/tikzit draw/.initial=0}
\pgfkeys{/tikz/tikzit shape/.initial=0}
\pgfkeys{/tikz/tikzit category/.initial=0}

\pgfdeclarelayer{edgelayer}
\pgfdeclarelayer{nodelayer}
\pgfsetlayers{background,edgelayer,nodelayer,main}

\tikzstyle{none}=[inner sep=0mm]

\newcommand{\tikzfig}[1]{%
{\tikzstyle{every picture}=[tikzfig]
\IfFileExists{#1.tikz}
  {\input{#1.tikz}}
  {%
    \IfFileExists{./figures/#1.tikz}
      {\input{./figures/#1.tikz}}
      {\tikz[baseline=-0.5em]{\node[draw=red,font=\color{red},fill=red!10!white] {\textit{#1}};}}%
  }}%
}

\tikzstyle{every loop}=[]
\usetikzlibrary{hobby}
\usepackage[T1]{fontenc} 
\usetikzlibrary{decorations.markings,intersections,patterns}
\usepackage{epsfig,amsfonts,amssymb,setspace}
\usepackage{tikz-cd}
\usepackage{relsize}

\definecolor{rust}{rgb}{0.8,0.2,0.2}

\newcommand{\Tr}[1]{\hbox{Tr}\left(#1\right)}

\newcommand{\vev}[1]{\langle #1 \rangle}
\def\ket#1{\mid  \! #1  \rangle}

\def\AdS#1{AdS$_{#1}$}
\def\Sp{{\bf S}}

\def\rA{\mathcal{A}} 
\def\rAc{\mathcal{A}^c} 
\def\rhoA{{\rho_{_{\!\!\rA}}}} 
\def\rhoAh{{\widehat{\rho}_{_{\!\!\rA}}}} 
 
\def\entsurf{\partial \rA} 

\def\rxA{\mathscr{A}} 
\def\rxAc{\mathscr{A}^c} 
\def\rhoAx{\rho_{_{\!\rxA}}}
 
\def\entsurfx{\mathscr{E}}

\def\tsb{\Gamma}

\def\psiSk{|\Psi_{_{\Sp^2}}\rangle}
\def\psiSb{\langle \Psi_{_{\Sp^2}}|}
\def\psiSbn{\langle \widehat{\Psi}_{_{\Sp^2}}|}
\def\psibk{\langle \Psi_{_{\Sp^2}}| \Psi_{_{\Sp^2}}\rangle}
\def\PhiR{\Phi_{_\rco}}

\def\psiTk{|\Psi^i_{_{{\bf T}^2}}\rangle}
\def\psiTb{\langle\Psi^i_{_{{\bf T}^2}}| }
\def\psiTkb{\langle \Psi^i_{_{{\bf T}^2}} |  \Psi^i_{_{{\bf T}^2}}\rangle}

\def\psiSkij{|\Psi^{ij}_{_{\Sp^2}}\rangle}
\def\psiSbij{\langle\Psi^{ij}_{_{\Sp^2}}| }

\def\rAk{\rA_-}
\def\rAb{\rA_+}
\def\rAck{\rAc_-}
\def\rAcb{\rAc_+}
\def\rxAk{\rxA_-}
\def\rxAb{\rxA_+}
\def\rxAck{\rxAc_-}
\def\rxAcb{\rxAc_+}

\def\zcs{\mathcal{Z}_{_\text{CS}}}

\def\zc{\mathcal{Z}_c}
\def\wtimes{\times}

\def\mcs{\mathcal{M}}
\def\mcsT{\mathcal{M}_{_{{\bf T}^2}}}
\def\dco{T^*{\bf S}^3}
\def\ball{\mathbb{B}}
\def\rco{\mathscr{R}}
\def\lagr{\mathscr{L}}
\def\xf{\mathscr{X}}
\def\bket{\mathbb{B}_-}
\def\bbra{\mathbb{B}_+}
\def\disk{\mathbb{D}}

\definecolor{rust}{rgb}{0.8,0.2,0.2}

\pgfplotsset{compat=1.12}
\pgfplotsset{
  torus/.style 2 args={
    surf,
    color=#1!30,faceted color=#1,
    samples=17,
    z buffer=sort,
    domain=0:360, y domain=#2:#2+180
  }
}
\def\m{sin(x)}
\def\n{(2+cos(x))*sin(y)}
\def\p{(2+cos(x))*cos(y)}

\title{Topological string entanglement}

\author{Veronika E.~Hubeny,}
\author{Roji Pius,}
\author{Mukund Rangamani}

\affiliation[]{
Center for Quantum Mathematics and Physics (QMAP),  \\
Department of Physics, University of California, Davis, CA 95616 USA.}

\emailAdd{veronika@physics.ucdavis.edu}
\emailAdd{rpius@ucdavis.edu}
\emailAdd{mukund@physics.ucdavis.edu}

\abstract{ 
We investigate how topological entanglement of Chern-Simons theory is captured in a string theoretic realization. Our explorations are motivated by a desire to understand  how quantum entanglement of low energy open string degrees of freedom is encoded in string theory (beyond the oft discussed classical gravity limit). Concretely, we realize the Chern-Simons theory as the worldvolume dynamics of topological D-branes in the topological A-model string theory on a Calabi-Yau target. Via the open/closed topological string duality one can map this theory onto a pure closed topological A-model string on a different target space, one which is related to the original Calabi-Yau geometry by a geometric/conifold transition.  We demonstrate how to uplift the replica construction of Chern-Simons theory directly onto the closed string and show that it provides a meaningful definition of reduced density matrices in topological string theory. Furthermore, we argue that the replica construction commutes with the geometric transition, thereby providing an explicit closed string dual for computing reduced states, and 
R\'enyi and von Neumann entropies thereof. While most of our  analysis is carried out for Chern-Simons on $\Sp^3$, the emergent picture is rather general. Specifically, we argue that quantum entanglement on the open string side is mapped onto quantum entanglement on the closed string side and briefly comment on the implications of 
our result for physical holographic theories where entanglement has been argued to be crucial ingredient for the emergence of classical geometry.  
}

\begin{document}
\maketitle


\section{Introduction}
\label{sec:intro}

The open/closed topological duality of Gopakumar-Vafa (GV) \cite{Gopakumar:1998ki} between large $N$ Chern-Simons theory on $\Sp^3$ and closed topological string on a resolved conifold through a geometric transition provides a useful context to test the general ideas underlying the gauge/gravity correspondence. As both sides of the duality are topological field theories, one has precision checks. For instance,  \cite{Gopakumar:1998ki} already showed how the   't Hooft expansion of the Chern-Simons partition function matches with the genus expansion of the closed topological string partition function. The match between observables was extended to  Wilson loop expectation values in \cite{Ooguri:1999bv}, and informed the subsequent developments in the subject  such as the all-loop expression for topological string amplitudes \cite{Aganagic:2002qg},  the topological vertex \cite{Aganagic:2003db}, etc.

Whilst the match between conventional observables on the two sides is fascinating, the relative tractability of this topological duality suggests that one ought to be able to do much more. It is instructive to compare the situation with the more familiar examples of gauge/gravity duality. In the physical context, the AdS/CFT correspondence relates large $N$ field theories realized on D-branes, to closed strings propagating on AdS spacetimes \cite{Maldacena:1997re}. Early entries into the holographic dictionary were relations between field theory operators and gravitational fields and prescriptions for computing correlation functions \cite{Gubser:1998bc,Witten:1998qj}. These entries, we now believe, do not altogether capture the complete essence of the holographic duality. Among other things they fail to provide a rationale for how the degrees of freedom of the quantum field theory conspire to build a dynamical spacetime where closed strings propagate. 

While we are yet to fully fathom the story in the physical context, developments in the past decade suggest an intimate connection between the emergence of geometry and the organization of quantum information in the dual field theory. These observations arise from another entry in the holographic dictionary; one relating the computation of von Neumann entropy for a spatial subregion of the field theory to the area of an extremal surface in the dual bulk geometry, viz., the RT/HRT prescriptions of  \cite{Ryu:2006bv,Hubeny:2007xt}. This geometrization of quantum entanglement (to leading order in large $N$), it has been argued, should be interpreted as responsible for the emergence of macroscopic spacetime geometry \cite{VanRaamsdonk:2010pw,Maldacena:2013xja}. An overview of some of the salient developments in this area can be found in \cite{Rangamani:2016dms}.

Given this status quo, we would like to examine the connection between geometry and entanglement in the open/closed topological string duality.  However, we should first convince ourselves that this is a useful exercise which could inform our intuition in the physical setting. Recall that the holographic entanglement entropy prescription is best understood in the limit when the closed string theory truncates to low energy Einstein-Hilbert gravitational dynamics, viz., when $\ell_\text{AdS} \gg \ell_s \gg \ell_{_P}$, which translates to the leading strong coupling, planar limit of the field theory.\footnote{ Concretely, in the familiar duality between $SU(N)$ $\mathcal{N}=4$ Super Yang-Mills (SYM) and string theory on \AdS{5} $\times \Sp^5$, the map between parameters is 
\begin{equation}
g_{YM}^2\,N  \sim \left(\frac{\ell_\text{AdS}}{\ell_s}\right)^4 \,, \qquad  N \sim \left(\frac{\ell_\text{AdS}}{\ell_{_P}}\right)^4 \,.
\label{eq:N4map}
\end{equation}	
\label{fn:N4map}
} 
Stringy corrections are understood perturbatively in $\ell_\text{AdS}/\ell_s$ by encapsulating them into higher derivative gravitational couplings. While this changes the holographic prescription \cite{Dong:2013qoa,Camps:2013zua}, it nevertheless retains the geometric flavour in that one still has to evaluate a geometric functional on a spacetime codimension-2 surface to obtain the field theory entanglement. Quantum gravity effects measured by $\ell_\text{AdS}/\ell_{_P}$ are more subtle -- to first subleading order in the planar expansion they are captured by entanglement of perturbative gravitational degrees of freedom across the extremal surface \cite{Faulkner:2013ana}.  All told this means that we are rather heavily reliant on truncating string theory to classical gravitational dynamics (perhaps with higher derivative interactions) to draw conclusions about how quantum entanglement might conspire to build spacetime geometry. One would hope, at the very least, to have a better understanding of this picture at the level of classical string theory, when $\ell_\text{AdS} \sim \ell_s$, to bypass our current limitations.  

A precise question in this direction is to ask whether the entanglement entropy of a large $N$ gauge theory to leading order in planar perturbation theory can be captured by classical string theory. An affirmative answer would bolster the claim that the closed string description is cognizant of the entanglement patterns of the dual field theory. A-priori, this would appear to require us work with an off-shell classical  closed string field theory.\footnote{ Various authors have attempted to understand 
entanglement of open strings; a clear discussion of issues and subtleties involved is described in \cite{Witten:2018xfj} in the context of the Rindler geometry (cf., also \cite{He:2014gva,Balasubramanian:2018axm} for earlier discussions). } 

To appreciate this issue, let us recall how we currently understand the RT/HRT prescriptions in the classical gravity limit. The RT prescription of was proved in \cite{Lewkowycz:2013nqa} by finding a gravity analog of the replica construction (cf., \cite{Dong:2016hjy} for the covariant HRT version). Here one constructs the field theory density matrix using a Euclidean path integral with suitable boundary conditions. Integer powers of the density matrix are then computed considering the field theory path integral on a suitable branched cover of the background geometry; taking a trace then gives us the  R\'enyi entropies. Finally, analytic continuation of these R\'enyi entropies is used to obtain the von Neumann entropy. The gravity computation  mimics this replica construction -- one identifies the gravitational saddle point solution for each replica branched cover, and extracts the R\'enyi entropy from the on-shell action using the identification between bulk and boundary partition functions, see eg., \cite{Headrick:2010zt,Faulkner:2013yia}.\footnote{ Operationally, there is a significant simplification -- the gravity computation is done by  introducing a \emph{cosmic brane} (a codimension-2 spacelike brane) whose tension is a-priori determined by  the R\'enyi index.  The brane tension can however be freely dialed. One can thus study the problem of cosmic brane back-reacting on geometry independently. The analysis is particularly simple for the von Neumann entropy which involves the leading backreaction of a nearly tensionless cosmic brane (which limits to the extremal surface). } If we were to carry out this exercise when $\ell_\text{AdS} \sim \ell_s$ we would need to be able to solve the off-shell string field equations for the replicated boundary to determine the appropriate target spacetime.

Some aspects of the holographic entanglement entropy prescriptions are greatly clarified by the gravity dual of the replica construction. For one, fluctuations around the saddle point lead to the prescription of \cite{Faulkner:2013ana} for subleading 1-loop corrections (in the planar expansion). This in turn can be used to argue for a semi-classical match between the bulk and boundary relative entropies \cite{Jafferis:2015del} which justifies the relation between the  reduced density matrix of the field theory and the reduced state in the entanglement wedge of the bulk 
geometry  \cite{Headrick:2014cta, Czech:2012bh,Wall:2012uf,Dong:2016eik,Cotler:2017erl}. Specifically, given a field theory reduced density matrix, we can associate a corresponding semi-classical gravitational reduced density matrix which is defined on a (spacetime codimension-1) spatial slice of the bulk geometry that is enclosed between the field theory region prescribed and the bulk extremal surface. The entanglement wedge is the bulk domain of dependence of this spatial region \cite{Headrick:2014cta}. All told, we see that the connection at the end of the day is between (reduced) states of the field theory and (reduced) states of geometry, as we indeed ought to have expected from the basic entries into the holographic dictionary. 

This observation poses a challenge to the general thesis that entanglement and geometry are intimately connected. To be clear, the original discussion of \cite{VanRaamsdonk:2010pw} suggested that a semiclassical geometric picture can only arise when there is sufficient entanglement  in the underlying quantum state, viz., \emph{geometry arises from entanglement}. 
Often this statement is conflated with another,  \emph{entanglement builds geometry}, which a-priori is a lot stronger and seems unlikely to hold in the full quantum string regime. 
Thus, while one could argue that ample EPR is necessary for ER, i.e., building Einstein-Rosen bridges requires sufficient amount of EPR entanglement, it seems unlikely that one can equate  EPR and ER in full string theory.  Moreover, away from the leading semi-classical regime described above, how are we to define reduced density matrices in the dual gravitational or string theory, when the spacetime geometry itself is fluctuating? What covariant prescription can be used to single out some decomposition of the spacetime, or more generally the string field degrees of freedom?\footnote{ For instance the quantum extremal surface proposal of \cite{Engelhardt:2014gca}  (see \cite{Dong:2017xht} for further progress) fails to carefully address these issues.  The prescription, at best, works in the large $N$ perturbation theory, where one can assume a geometric partitioning at each order. }  

A pragmatic approach to addressing these questions is to identify examples where we can make precise statements. This is where the topological open/closed string duality will prove valuable.  While the topological theories are devoid of dynamics, their intrinsic tractability makes them ideal candidates for the exploration we have in mind (absence of dependence on geometry notwithstanding). In particular, we will argue that some of the aforementioned questions can be addressed quite cleanly within this set-up, providing us with some confidence that with further insight one might be able to tackle the physical string theory in due course. Before proceeding, note that \cite{Donnelly:2016jet,Donnelly:2018ppr} have  made some progress in addressing this question in the context of two-dimensional Yang-Mills theory, which is known to have a string dual. The authors define quite generally the notion of an `entanglement brane' in \cite{Donnelly:2016jet} which they have further explored exploiting the axiomatic framework of topological field theories in \cite{Donnelly:2018ppr}. Our discussion will be somewhat complementary, but broadly in keeping with the general idea espoused in these works.

Let us therefore turn to a more careful examination of the open/closed topological string duality \cite{Gopakumar:1998ki}. The basic idea can be traced back to the observation that the open string field theory on the worldvolume of an A-model topological D-brane is given by Chern-Simons theory \cite{Witten:1992fb}. For simplicity we start with Chern-Simons theory on 
$\Sp^3$. This theory is the worldvolume open string theory on topological D-branes in the topological A-model string on the deformed conifold geometry, which is the cotangent bundle ${\bf T^\ast S}^3$. For purposes of the present discussion, the geometry can be simply viewed as a cone over a base which is $\Sp^2 \times \Sp^3$. The $\Sp^2$ shrinks to zero size at the tip of the cone but the $\Sp^3$ remains finite there. The general picture of holography applies here: we dissolve the branes  into flux and correspondingly change the closed string background. In the present case, the resulting closed string background is the resolved conifold, which is still a cone over the same base space, albeit now with the $\Sp^2$ remaining finite while the $\Sp^3$ shrinks to zero at the tip. The two geometries are connected at the singular locus where the entire base shrinks to zero size at the tip which is the singular conifold geometry. The transition between the two pictures  is referred to as the geometric or conifold transition. The statement is that via the geometric transition the open+closed string theory in the deformed conifold side is dual to the closed string theory on the resolved conifold. As mentioned earlier we have a precise match of partition functions \cite{Gopakumar:1998ki} and Wilson loop observables \cite{Ooguri:1999bv,Aganagic:2002qg} (see also \cite{Gomis:2006mv,Gomis:2007kz}); moreover, attempts were also made to give a worldsheet derivation of the duality in \cite{Ooguri:2002gx}.

At zeroth order we would like pose the following question in this setting: \emph{What is the closed topological string quantity which captures the topological entanglement entropy \cite{Kitaev:2005dm,Levin:2006zz} of Chern-Simons theory?} Let us sharpen this question a bit to make closer contact with the story we understand in the physical gauge/gravity context for holographic entanglement entropy  prescriptions.\footnote{ Note in particular, that our considerations are different from those of \cite{Pakman:2008ui} who consider physical gauge/gravity examples and extract topological entanglement contribution from the RT prescription.}  In Chern-Simons theory we imagine that by a suitable partitioning of the wavefunctional on a time-slice we are able to compute spatially-ordered entanglement. This can be done by invoking the seminal analysis of Witten \cite{Witten:1988hf} who provided a picture of the physical Hilbert space of the theory 
in the framework of canonical quantization (see also \cite{Elitzur:1989nr}). Alternately, using surgery techniques for computing Chern-Simons observables \cite{Witten:1988hf}, one can set-up a replica path integral \cite{Dong:2008ft} to obtain spatially ordered entanglement  in Chern-Simons theory. This leads to a flat entanglement spectrum determined by the quantum dimension 
\cite{Kitaev:2005dm} and the number of connected components of the entangling surface.\footnote{ One can extract more interesting information by considering states involving Wilson lines along various knots and links as discussed in \cite{Dong:2008ft, Balasubramanian:2016sro,Balasubramanian:2018por}. For instance, \cite{McGough:2013gka} argues for a relation between topological entanglement and the BTZ black hole entropy. } For example, for Chern-Simons on $\Sp^3$ the entanglement entropy for spatial bipartitioning gives all R\'enyi and von Neumann entropies to be simply $\log \zcs(\Sp^3)$.

Before proceeding, it is worth highlighting one aspect of the replica construction in Chern-Simons theory. In a general topological phase, one expects the entanglement entropy of a subregion to behave as $S_\rA = \frac{L_{\entsurf}}{\epsilon} - S^\text{(top)}_\rA$. However, the replica computation only picks up the finite  topological part of the entanglement: in fact, $S_\rA^\text{(replica)} = +S^\text{(top)}_\rA$.
This is because the replica computation is carried out in a manner that is insensitive to the regulator by staying within the topological sector \cite{Dong:2008ft}. One can expose the divergent terms by inquiring about the boundary conditions (or edge modes) on the entangling surface. We however will frame our discussion strictly within the replica construction in order to be able to employ path integral arguments. 
We will comment on various caveats associated with this restriction at relevant junctures in our analysis.

Given the replica construction of Chern-Simons reduced density matrices and associated entropies, and the fact that they can be realized as the worldvolume theory on topological D-branes, we can ask a more precise version of our question: \emph{How is the replica construction ported across the duality and what is its image on the other side of the geometric transition?} A 
natural way to proceed is to understand the closed string analog of the replica construction. Taking inspiration from \cite{Lewkowycz:2013nqa} we can first imagine 
introducing a bipartition of the closed string theory commensurate with that employed on the D-branes. This is in fact not how the construction is usually phrased: usually one says that  
the replica on the field theory side involves branching over the entangling surface, and mimics this by introducing a source, a codimension-2 cosmic brane, in the bulk.  The cosmic brane however is a spacelike separatrix implementing a partitioning in the closed string theory (which is clear once we account for quantum effects \cite{Faulkner:2013ana,Jafferis:2015del}), so we are justified in making the identification as such. 

As we are imagining constructing the density matrix elements in the closed topological string theory, it is natural to refer to the codimension-2 surface as the entangling surface. From the viewpoint of the duality, this surface is the cosmic brane introduced in the bulk to implement the analog of the spatial bipartition in the open string description. This dual interpretation naturally suggests a portmanteau terminology, \emph{entangling brane}, which we adopt. We should note that \cite{Donnelly:2016jet,Donnelly:2018ppr} prefer to use the phrase `entanglement brane'. We adhere to  our choice as it manifests the dual connection. Nomenclature apart, it should be emphasized that these are not physical objects in the theory, no more than an entangling surface in a QFT. The entangling branes have spacelike worldvolumes,\footnote{ However, note that in contrast to the cosmic brane of \cite{Lewkowycz:2013nqa}, the entangling brane need not select out a definite location in the bulk. In fact, the position of the entangling surface itself in Chern-Simons is immaterial.} have no dynamics/evolution (no temporal extent), and are simply a means to decompose  the closed string Hilbert space (or 
more precisely provide a suitable decomposition of the closed string operator algebra).

In general, one has to worry about this bipartitioning cutting through closed string degrees of freedom, which would involve figuring out what the allowed boundary conditions or superselection sectors are. Fortuitously, we will find that this issue is avoided quite explicitly in the topological string theory (in contrast to the discussion of  \cite{Donnelly:2018ppr} who derive consistency conditions for the decomposition in examples they study). We can give an explicit picture of building a state and bipartitioning it to construct a reduced density matrix directly at the level of the target space topology.  This is broadly facilitated by the following the replica construction in the bulk, which does have the downside, as in the Chern-Simons case, of being insensitive to regulator contributions.

Operationally our construction is implemented as follows: first we identify suitable time-slices in the closed topological string and partition them in accord with that on the D-brane i.e., 
we a-priori pick an ansatz for where the entangling brane resides. Once we have this construct, we can glue together copies of it cyclically and implement the replica construction.
The resulting target space however would have to be such that it corresponds to an admissible closed topological string background. This requires it to admit a Calabi-Yau structure, which 
effectively serves as the dynamical equations, and helps fix the `location' of the entangling brane. Since the A-model closed topological string only cares about K\"ahler data of the target 
\cite{Witten:1988xj}, the constraints we get are quite simple and only involve checking the two-cycles of the replica target. We can compute the closed topological string partition 
function on this replica geometry and check that it reproduces the known Chern-Simons entanglement. We will thereby establish that it is meaningful to examine topological entanglement in 
string theory in a manner consistent with the GV duality. 

Our analysis can be phrased more succinctly in terms of surgery. We essentially uplift three-manifold surgery techniques employed in the computation of Chern-Simons entanglement onto the closed topological string. Consequently, we can argue immediately that we have a result on the closed string side that captures not just the leading planar contribution, but rather the entire 
all-loop expansion for the topological entanglement entropy. This of course immediately follows from the GV duality which maps partition functions across the geometric transition, but it has 
physical implications for thinking about connections between entanglement and geometry/topology.

One natural question we could ask is how the closed string analysis picks out a simple explanation for topological entanglement. For instance, one would like to know
if the resulting  closed topological string partition function computation localizes on some specific submanifold in the target space, which we could identify as a topological string analog of the RT surface. While our target is in a loose sense branched over the entangling brane, it is not entirely straightforward to isolate it as the locus capturing the answer for the partition function. For one the closed string partition function is naturally associated to certain homology 2-cycles (almost by definition) which is insensitive to the location. Topological entanglement on the D-brane open string field theory is just mapped across to topological entanglement of closed strings. In other words, there isn't a simple analog of an RT surface which 
we would identify as capturing any semiclassical piece of entanglement, but rather there is simply consistency between the two sides for prying open the functional integral. We will argue this 
based on the manner in which we map the reduced density matrices on the two sides onto each other. As such, this result should not surprise us; the  leading quantum correction in the physical string duality also implies that one is actually mapping entanglement to entanglement \cite{Jafferis:2015del}. It is just that in certain situations the answer is represented geometrically via the RT/HRT formulae and their generalizations, but these presuppose the dual to be a semiclassical spacetime. The topological open/closed string duality does not appear to offer an analogous simplification; we will return to this point and its implications for the connections between entanglement and geometry at the end.

The outline of the paper is as follows: We begin with a lightning overview of the open/closed topological string duality in \S\ref{sec:topduality} to introduce the basic idea, followed by a 
synopsis of facts relating to quantization of Chern-Simons theory in \S\ref{sec:cscan}. We will then explain how to construct reduced density matrices and compute entanglement  in Chern-Simons theory using surgery techniques in \S\ref{sec:csent}. We will be quite explicit here despite the fact that some of the material has appeared in the literature, to set the stage for 
uplifting the construction to closed topological strings. The remainder of the paper will primarily focus on the closed string:  in \S\ref{sec:entctop} we explain how one can construct states and reduced density matrices for the closed string theory and use it to extract the topological closed string entanglement. The rough idea will be to see how to uplift the Chern-Simons density matrix to open+closed topological string theory, and take the latter through the geometric transition to recover the previously obtained answer.  We discuss generalizations including Wilson lines in \S\ref{sec:genPsi}. Finally, in \S\ref{sec:discuss} we outline some open questions for future exploration. The Appendices contain some useful background material.
 
\section{Topological open/closed string duality}
\label{sec:topduality}

In order to set the stage for our discussion, we briefly review some of the salient facts in the open/closed topological string duality of \cite{Gopakumar:1998ki,Ooguri:1999bv}. Some of the relevant material can be found in reviews \cite{Marino:2004uf,Marino:2005sj}.
    
 Chern-Simons  action is specified by a gauge connection $A$ which transforms in the adjoint of a Lie algebra $\mathfrak{g}$, associated with a gauge group $G$. For definiteness we will focus on $\mathfrak{g} = \mathfrak{su}(N)$. The action which is gauge invariant on a closed three-manifold  $\mcs_3$ is given by the integral of a three-form
\begin{equation}
S_{_\text{CS}} = \frac{k}{4\pi} \, \int _{\mcs_3}  {\rm Tr}\left( A\wedge dA + \frac{2}{3} \, A\wedge A \wedge A \right)\,.
\label{eq:csact}
\end{equation}	
$k$ here is the level of the Chern-Simons theory, and is quantized  to be integral, $k \in {\mathbb Z}$, for the action to be single-valued. We will denote $F = DA$ to be the field strength as usual.  

As defined, the classical theory is topological, in the sense of being independent of the background metric structure, as is manifest from the action \eqref{eq:csact}. The quantum theory,  
first analyzed in \cite{Witten:1988hf}, however does care in a mild manner about the metric structure, which we need to introduce to regulate the theory. This is associated with the `framing ambiguity'. For the most part we will work with a canonical framing choice, which we won't need to specify explicitly. As long as we stick to the topological sector, we are only allowed to consider observables which are similarly independent of the metric structure, i.e., to Wilson loop operators. Among the many results in \cite{Witten:1988hf} it was shown how one can use 
three-manifold surgery techniques to compute partition functions and expectation values of Wilson loop observables defined on knots and links.  In addition, we will also need information regarding the Hilbert space of the theory for our purposes; this was also obtained  in \cite{Witten:1988hf} as we review below.

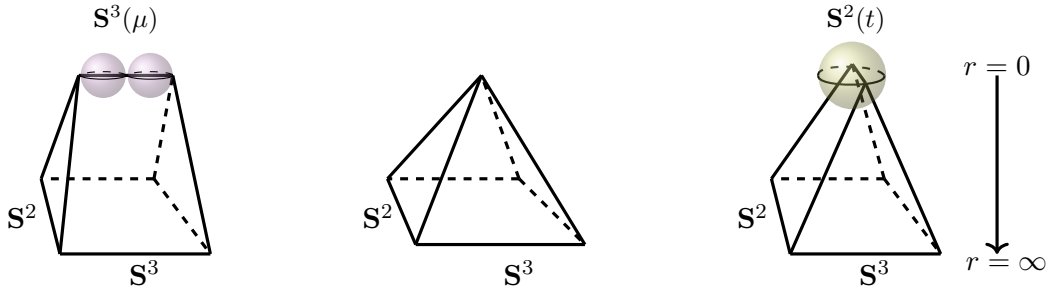
\begin{figure}
\begin{center}
\usetikzlibrary{backgrounds}
\begin{tikzpicture}[scale=.5]
    	\begin{pgfonlayer}{nodelayer}
		\node [style=none] (0) at (-3, -2.75) {};
		\node [style=none] (1) at (-2.5, 0) {};
		\node [style=none] (2) at (-6, -2.75) {};
		\node [style=none] (3) at (-1.5, -4.75) {};
		\node [style=none] (4) at (-5.5, -4.75) {};
		\node [style=none] (5) at (-5, 0) {};
		\node [style=none] (6) at (-3.75, 1.5) {\small{$\Sp^3(\mu)$}};
		\node [style=none] (8) at (-3.25, -5.25) {$\Sp^3$};
		\node [style=none] (10) at (-6.5, -3.75) {$\Sp^2$};
		\node [style=none] (11) at (-1, 0) {};
		\node [style=none] (12) at (-1, -4.75) {};
		\node [style=none] (17) at (-3.5, 0) {};
	\end{pgfonlayer}
	\begin{pgfonlayer}{edgelayer}
		\draw [very thick, style=dashed] (0.center) to (2.center);
		\draw [very thick, style=dashed] (0.center) to (1.center);
		\draw [very thick] (5.center) to (2.center);
		\draw [very thick, in=360, out=-180, looseness=1.25] (1.center) to (5.center);
		\draw [very thick] (5.center) to (4.center);
		\draw [very thick] (1.center) to (3.center);
		\draw [very thick] (2.center) to (4.center);
		\draw [very thick] (3.center) to (4.center);
		\draw [very thick, style=dashed] (0.center) to (3.center);
		\shade[ball color = violet!40, opacity = 0.4] (-4.35, 0) circle (.6cm);
   		\shade[ball color = violet!40, opacity = 0.4]  (-3.1, 0) circle (.6cm);
        \draw   [dashed]  (-3.75, 0) arc (0:180:0.6 and .1);
  		\draw (-3.75,0) arc (360:180:0.6 and .1);
   		\draw   [dashed]  (-2.5,0) arc (0:180:0.6 and .1);
  		\draw (-2.5,0) arc (360:180:0.6 and .1);
	\end{pgfonlayer}
\end{tikzpicture}
\hspace{1.5cm}
\begin{tikzpicture}[scale=.5]
    \begin{pgfonlayer}{nodelayer}
		\node [style=none] (0) at (-3.25, -2.75) {};
		\node [style=none] (1) at (-4.25, 0) {};
		\node [style=none] (2) at (-6.75, -2.75) {};
		\node [style=none] (3) at (-1.5, -4.5) {};
		\node [style=none] (4) at (-6, -4.5) {};
		\node [style=none] (5) at (-4.25, 0) {};
		\node [style=none] (8) at (-3.25, -5.25) {$\Sp^3$};
		\node [style=none] (10) at (-7, -3.75) {$\Sp^2$};
		\node [style=none] (11) at (-1, 0) {};
		\node [style=none] (12) at (-1, -4.5) {};
	\end{pgfonlayer}
	\begin{pgfonlayer}{edgelayer}
		\draw [very thick, style=dashed] (0.center) to (2.center);
		\draw [very thick, style=dashed] (0.center) to (1.center);
		\draw [very thick] (5.center) to (2.center);
		\draw [very thick] (5.center) to (4.center);
		\draw [very thick] (1.center) to (3.center);
		\draw [very thick] (2.center) to (4.center);
		\draw [very thick] (3.center) to (4.center);
		\draw [very thick, style=dashed] (0.center) to (3.center);
	\end{pgfonlayer}
\end{tikzpicture}
\hspace{1.5cm}
\begin{tikzpicture}[scale=.5]
	\begin{pgfonlayer}{nodelayer}
		\node [style=none] (0) at (-3, -2.75) {};
		\node [style=none] (1) at (-3, 0) {};
		\node [style=none] (2) at (-6, -2.75) {};
		\node [style=none] (3) at (-1.5, -4.75) {};
		\node [style=none] (4) at (-5.5, -4.75) {};
		\node [style=none] (5) at (-4.75, 0) {};
		\node [style=none] (6) at (-3.75, 1.5) {\small{$\Sp^2(t)$}};
		\node [style=none] (8) at (-3.25, -5.25) {$\Sp^3$};
		\node [style=none] (10) at (-6.5, -3.75) {$\Sp^2$};
		\node [style=none] (11) at (0, 0) {};
		\node [style=none] (12) at (0, -4.75) {};
		\node [style=none] (14) at (0, 0.25) {$r=0$};
		\node [style=none] (15) at (0.25, -5) {$r=\infty$};
	\end{pgfonlayer}
	\begin{pgfonlayer}{edgelayer}
		\draw [very thick, style=dashed] (-3, -2.75) to (2.center);
		\draw [very thick, style=dashed] (-3, -2.75) to  (-3.85, 0.3);
		\draw [very thick] (-3.85, 0.3) to (2.center);
		\draw [very thick] (-3.5, -0.2) -- (-3.85, 0.3);
		\draw [very thick] (-3.5, -0.2) to (4.center);
		\draw [very thick] (-3.5, -0.2) to (3.center);
		\draw [very thick] (2.center) to (4.center);
		\draw [very thick] (3.center) to (4.center);
		\draw [thick, color=black, bend right=105, looseness=0.50] (5.center) to (1.center);
		\draw [style=dashed, thick, color=black, in=120, out=105, looseness=0.50] (5.center) to (1.center);
		\draw [very thick, ->, in=90, out=-90] (11.center) to (12.center);
		\draw [very thick, style=dashed] (-3, -2.75) to (-1.5, -4.75);
		 \shade[ball color = olive!40, opacity = 0.4] (-3.85,0) circle (.9cm);
	\end{pgfonlayer}
\end{tikzpicture}
\end{center}
\caption{ A illustration of the deformed, singular, and resolved conifold geometries viewed as a cone with a base that is topologically $\Sp^2\times \Sp^3$. We view the base as living far out along the cone $r \to \infty$, and the different topologies of the three geometries are captured by the behaviour near the tip. In the deformed conifold (left),  $\Sp^2$ shrinks to a point and the $\Sp^3$ has radius $\mu$. The singular conifold (middle) instead has both the $\Sp^3$ and the $\Sp^2$ shrinking to a point at the tip. In the  resolved conifold  (right) on the other hand, the $\Sp^3$ shrinks to a point and the $\Sp^2$ has the K\"ahler parameter $t$. We have drawn the three-sphere as two 3-balls, which are to be identified along their boundaries. We will use this representation explicitly below, and also adhere to the color coding differentiating the three-ball $\ball$ from the two-sphere $\Sp^2$ (which will always be in yellowish hue). }
\label{fig:conifold}
\end{figure}

Of primary interest to us is Chern-Simons theory on $\Sp^3$ with the gauge group $SU(N)$. Not only is it an exactly solvable theory, but it also provides an exact effective description of the  A-model open topological string theory with target space  $\dco$ \cite{Witten:1992fb}. In this context, one should view the Chern-Simons theory as the open string field theory of   open string degrees of freedom living on a topological D-brane. The D-branes of the A-model are half-codimension surfaces wrapping a Lagrangian cycle. The topological D-brane of interest wraps the background $\Sp^3$, which is a Lagrangian 3-cycle,  and the topological string target space is $\dco$ which is a Calabi-Yau geometry, the \emph{deformed conifold} (see Fig.~\ref{fig:conifold}). 

The statement of open/closed string duality stems from the observation that the  large $N$ expansion of the $SU(N)$ Chern-Simons partition function on $\Sp^3$ around the classical solution can be interpreted in terms of  a closed A-model topological string theory. More precisely, the closed string dual is the $\mathcal{N}=2$ closed A-model topological string theory whose target space is the \emph{resolved conifold} $\rco \equiv \mathcal{O}(-1)\oplus\mathcal{O}(-1)\to \mathbb{P}^1$.    The resolved conifold is a six dimensional Calabi-Yau manifold with a single  
K\"ahler parameter (which sets the complexified area of the $\Sp^2$).  The string coupling $g_s$ and the K\"ahler parameter $t$ for the closed string target  spacetime are related to rank $N$ and the level $k$ of the Chern-Simons theory as follows:
\begin{equation}
\begin{split}
g_s &= \frac{2\pi}{k+N} \,, \qquad t = i\,  \frac{2\pi\, N}{k+N} = i\, \lambda \,, \\
\end{split}
\label{eq:gstmap}
\end{equation}	
where we have also indicated by $\lambda$ the 't Hooft coupling of the field theory.

Topological A-models define closed string theories on a Calabi-Yau target space and are obtained from the physical Type II string theory by a topological twist of the underlying $(2,2)$ worldsheet CFT \cite{Witten:1988xj}. The A-twist involves shifting the spin current by the vector $R$-current of the superconformal theory, and restricts attention to holomorphic maps from the worldsheet $\Sigma_{\text{ws}}$ to the target Calabi-Yau $\xf_6$. The resulting theory is independent of the complex structure deformations of the target and only depends on the K\"ahler parameters. 

Consider then the deformed conifold $\dco$ which is described by the following hypersurface in $\mathbb{C}^4$ (coordinates 
$\zeta_a = q_a + i \, p_a, \; a = 1,2,3,4$)
\begin{equation}
\sum_{a=1}^4\, \zeta_a^2 =  \mu^2 \;\; \;\; \Longrightarrow \;\; \;\; \sum_{a=1}^4\, \left(|q_a|^2 - |p_a|^2\right)  = \mu^2 \;\;   \& \;\; 
\sum_{a=1}^4\, q_a\, p_a = 0
\label{eq:dcfld}
\end{equation}	
$q_a$ can be thought of as the `coordinates'; so the hypersurface with $p_a = 0$ is indeed an $\Sp^3$ (which is Lagrangian with the canonical symplectic form) with size set by $\mu$. $p_a$ are related to the conjugate momenta.   This geometry can be viewed as a cone with base $\Sp^3 \wtimes \Sp^2$.\footnote{
	The $\Sp^2$ is non-trivially fibered over the $\Sp^3$ but for the most part we will not explicitly need to refer to this fibration structure and continue to use the product notation as is conventional in the literature. Likewise in our illustrations we will simply indicate the $\Sp^2$ and $\Sp^3$  alongside each other. \label{fn:fibration} } As long as $\mu > 0$,  the $\Sp^3$ remains of finite size. The normal bundle is topologically $\mathbb{R}^3$ with a `radial coordinate' along which the $\Sp^2$ shrinks to zero at the tip (where $\Sp^3$ has size $\mu$).

The parameter $\mu$ picks out a complex structure on $\dco$. Since the A-model is insensitive to the  choice of the complex structure, we can vary it at will, and in particular set $\mu \to 0$ whence we get the conifold singularity, where both the $\Sp^3$ and $\Sp^2$ have shrunk to zero size at the tip of the cone.
\begin{equation}
\sum_{a=1}^4\, \zeta_a^2  =0 \,.
\label{eq:scfld}
\end{equation}	
We can parameterize this geometry slightly differently to exhibit its structure. A change of coordinates brings allows us to describe it as the following hypersurface in $\mathbb{C}^4$:
\begin{equation}
x\,y - w\, z =0
\label{eq:scfld2}
\end{equation}	

The singular conifold admits a second desingularization, where we resolve the $\Sp^2 \sim \mathbb{P}^1$. This can be done as follows. Let $\xi$ be an inhomogeneous coordinate on $\mathbb{P}^1$. We can solve \eqref{eq:scfld2} by setting
\begin{equation}
x =  \xi\, z \,, \qquad  w  = \xi\, y \,.
\label{eq:s2res}
\end{equation}	
This parameterization makes manifest the geometry being a $\mathcal{O}(-1) \oplus \mathcal{O}(-1)$ bundle over $\mathbb{P}^1$. This is known as the \emph{resolved conifold} $\rco$.  An alternate way to parameterize the manifold is to use complex coordinates $\xi_a$ such that 
\begin{equation}
|\xi_1|^2 + |\xi_4|^2 - |\xi_2|^2 - |\xi_3|^2 = t
\label{eq:rcfld}
\end{equation}	
The parameter $t$ is complex and captures the K\"ahler modulus of the resolved conifold. Intuitively it is the complexified area of the $\mathbb{P}^1$ which is the locus $\xi_2 = \xi_3 =0$, which identifies $\xi = \xi_1/\xi_4$. 
We explain some more elements of the topology of the conifold in Appendix~\ref{sec:conifold}. Fig.~\ref{fig:conifold} illustrates the basic structure of the topology of the spaces we are interested in. 

The Chern-Simons/topological string correspondence has striking similarities with the more familiar examples of the AdS/CFT correspondence. Consider the duality between $\mathcal{N}=4$ Yang-Mills in $\Sp^4$ and the closed IIB superstring theory in \AdS{5}$\,\times \,\Sp^5$ \cite{Maldacena:1997re}. Working for the moment in Euclidean space,  the boundary of the geometry is $\Sp^4\times \Sp^5$ with the radial direction of \AdS{5} filling in the $\Sp^4$.  As we shall see below, the geometry of the resolved conifold can be understood as a cone with a base which is topologically $\Sp^2\wtimes \Sp^3$, and the space at infinity is $\Sp^2\wtimes \Sp^3$ with the $\Sp^2$ having finite size. One should by way of analogy identify the $\Sp^2$ with the $\Sp^5$ transverse to the D3-branes. Likewise, at a heuristic level, the Chern-Simons theory can be thought of as living on a large $\Sp^3$, far out along the cone, just like the  $\mathcal{N}=4$ Yang-Mills which is living on a large $\Sp^4$ at the conformal infinity of \AdS{5}$\,\times \,\Sp^5$ \cite{Gopakumar:1998ki}.\footnote{ While this perspective provides a useful mnemonic for the duality, there is no real sense in which the Chern-Simons theory lives far out at the base of the cone, just as the $\mathcal{N} =4$ SYM theory doesn't really reside on the boundary of AdS.}

\section{Quantization of Chern-Simons theory}
\label{sec:cscan}

As a prelude to our discussion of entanglement in Chern-Simons theory, we first review some basic facts about the canonical quantization of Chern-Simons gauge theory and the resulting Hilbert space. The essential points we need can be succinctly summarized, see \S\ref{sec:H01summary}. Much of our discussion is drawn from \cite{Witten:1988hf} and \cite{Elitzur:1989nr}.

The action for the theory is given in \eqref{eq:csact} and is invariant under gauge transformation $A_\mu^a\to A_\mu^a-D_\mu\epsilon^a$, for an infinitesimal gauge parameter $\epsilon^a$, where we have made explicit the spacetime and Lie algebra indices. The theory has no local degrees of freedom, but we are free to consider non-local Wilson line observables that can be defined topologically.  Given an oriented knot $C \subset \mcs_3$ and an irreducible representation 
$R$ of the gauge group $G$, the associated Wilson line operator is defined as follows
\begin{equation}
W_{R}(C)=\text{Tr}_R \left( \mathcal{P}\, \exp \left(\int_C A_\mu^a(x) T^a_R \; dx^\mu\right) \right)
\end{equation}
where the path ordering is performed along $C$ and $\{T^a_R\}$ are the generators of $G$ in the representation $R$. More generally one considers Wilson lines defined on oriented links $L$ with $m$ components $\{C_1,\cdots, C_m\}$, where each component  is an oriented knot. Let $R_i$ be the representation associated with the $i^{\text{th}}$ component $C_i$ of the link $L$. Then the physical observables of the theory are expectation values of these gauge invariant Wilson lines on the given link $L$, i.e.,
\begin{equation}
\vev{W(L)} =\frac{1}{\zcs(\mcs_3)} \; \int \left[\mathcal{D}A\right]  e^{i\, S_{_\text{CS}}}\; \prod_{i=1}^m \, W_{R_i}(C_i) \,,
\label{eq:WEV}
\end{equation}
where $\zcs(\mcs_3)$ is the partition function of the theory on the three-manifold $\mcs_3$, viz., 
\begin{equation}
\zcs(\mcs_3 )  \equiv \int_{\mcs_3} \left[\mathcal{D}A\right]  e^{i\, S_{_\text{CS}}}\,.
\label{eq:csZ}
\end{equation}	
%

\subsection{Canonical quantization}
\label{sec:canq}

Consider Chern-Simons theory on a general three-manifold $\mcs_3$ with Wilson lines in it. Let us scissor $\mcs_3$ along a Riemann surface $\Sigma$. The neighborhood of the cut locally looks like $\Sigma\times \mathbb{R}^1$ which will suffice for our purposes. Generically, our cut may also slice through some Wilson lines threading the manifold.  Each of these carries an external charged particle transforming in an associated representation of the gauge group, a description that is natural if we view the coordinate along $\mathbb{R}$ as time.  
Therefore, $\Sigma$ will  have finitely many marked points $P_1,\cdots,P_n$. Each of these points $P_\ell$ are assigned a representation $R_\ell$ of the gauge group $G$. 
What we are after is the physical Hilbert space of the theory  $\mathcal{H}_{\Sigma}$ on such a Riemann surface. This question is quite tractable in our local decomposition of $\mcs_3$ by standard rules of canonical quantization.

Consider first the situation where no Wilson lines are cut when we decompose $\mcs_3$. Denote the coordinate and the gauge field component along $\mathbb{R}$ by $t$
and $A_0$, respectively. On $\Sigma\times \mathbb{R}$, one can naturally choose the gauge 
$A_0=0$ and reduce the Chern-Simons Lagrangian to a quadratic form:
\begin{equation}
S_{_\text{CS}} =\frac{k}{8\pi}\int dt\int_{\Sigma}\quad\epsilon^{ij}\text{Tr}\left( A_i\frac{d}{dt} A_j\right) 
\end{equation}
 where, $i$ and $j$ index coordinates on $\Sigma$. The classical Poisson bracket is given by
 \begin{equation}
 \left\{A_i^a(x),A^b_j(y)\right\}=\frac{4\pi}{k}\epsilon_{ij}\delta^{ab}\delta^2(x-y)
 \label{eq:poisson1}
 \end{equation}
The equation of motion $\frac{\delta}{\delta A_0}\mathcal{L}=0$ for the gauge field component $A_0$ imposes the Gauss law constraint:
\begin{equation}
\epsilon^{ij}\left(\partial_iA_j-\partial_j A_i+\left[A_i,A_j\right] \right)^a=0    
\label{eq:gausslaw}
\end{equation}

As argued in \cite{Witten:1988hf} a useful strategy for this problem is to impose the constraints on the classical phase space and then quantize the gauge fixed theory. Imposing the constraints leads to a finite dimensional phase space immediately. One thus obtains a finite dimensional physical Hilbert space $\mathcal{H}_\Sigma$. The constraints \eqref{eq:poisson1} demand that we focus on flat connections on $\Sigma$, identifying them within equivalence classes as demanded by gauge invariance. The classical phase space is therefore the moduli space of flat connections on $\Sigma$, modulo gauge transformations. This space can be characterized by Wilson lines, or holonomies of the gauge field around non-contractible cycles of the Riemann surface.  As there are $\text{dim}(G)$ gauge field components and $2g$ independent non-contractible cycles on a genus $g$ Riemann surface, the moduli space has a finite dimension 
$2(g-1)\, \text{dim}(G)$. It furthermore admits a natural symplectic structure which can then be quantized.  The Hilbert space  for the quantum problem $\mathcal{H}_\Sigma$ turns out to be given by the space of conformal blocks  of the  $\mathfrak{g}_k$ current algebra on Riemann surface  \cite{Witten:1988hf}.

The story can be repeated when we consider Wilson lines in representations $R_\ell$ piercing through $\Sigma$ at some points $P_\ell$. Quantizing the theory gives a Hilbert space $\mathcal{H}_{\Sigma; \{P_i,R_i\}}$. The main change to account for is a modification of the Gauss law constraint due to the static external charges:
\begin{equation}
\frac{k}{8\pi} \epsilon^{ij}F_{ij}^a(x)=\sum_{\ell=1}^r\delta^2(x-P_\ell)\,T^a_{(\ell)}
\label{eq:gausslawwilson}
\end{equation}
The quantization a-priori appears tricky owing to the fact that the r.h.s.\ of \eqref{eq:gausslawwilson} contains a quantum operator in a particular representation of $G$. 
One makes progress by invoking the Borel-Weil-Bott theorem, which allows us to obtain every irreducible representations $R$ of a compact group $G$ from the quantization of a classical phase space.  The idea is to introduce the manifold $G/T$, with $T$ being a maximal torus in $G$, and for each representation $R$ introduce a symplectic structure $\omega_R$ on $G/T$, such that the quantization of the classical phase space $G/T$, with the symplectic structure $\omega_R$, gives back the representation $R$. Therefore, we first extend the phase space  
$G$-flat connections on $\Sigma$ by including at each marked point $P_i$ a copy of $G/T$, with the symplectic structure appropriate to the $R_i$ representation. This replaces the quantum operator $T^a_{(i)}$ in the right hand side of \eqref{eq:gausslawwilson} by the classical function on $G/T$ whose quantization gives back the $T^a_{(i)}$.  Once again the classical phase space is finite dimensional and can be quantized. The quantum Hilbert space $\mathcal{H}_{\Sigma; \{P_i,R_i\}}$ after picking an arbitrary conformal structure on $\Sigma$ is again identified with the space of conformal blocks in current algebra with primary fields in the $R_i$ representation inserted at the points $P_i$. The main constraint to keep in mind is that in order to get a nontrivial Hilbert space $\mathcal{H}_{\Sigma; \{P_i,R_i\}}$, all the representations $R_i$ must be integrable representations of current algebra for $\mathfrak{g}_k$  \cite{Witten:1988hf}. 
    
\subsection{Physical Hilbert space at genus zero and one}
\label{sec:H01summary}

While the above discussion is quite general, we will primarily be interested in simple three-manifolds like $\Sp^3$, or the solid torus. We need to understand how to decompose these three- manifolds locally into a form that makes them amenable to canonical quantization. The essential idea is to use the so called Heegard splitting of a  3-manifold into handlebodies which have Riemann surfaces as boundaries. What we need is a  corollary of the Dehn-Lickorish theorem \cite{Lickorish:1962rep} which asserts that {\it any arbitrary 3-manifold  $\mcs_3$ can be obtained by cutting out a set of handlebodies whose non-contractible cycles are unknots from $\Sp^3$ and pasting them back in, after applying diffeomorphisms on their boundaries}  \cite{Prasolov:1997kno}.  This is used extensively in the topological decomposition of three-manifolds and allows one to reduce the study of Chern-Simons on a general $\mcs_3$ into suitable combination of elementary building blocks \cite{Witten:1988hf}. 

Consider then $\mcs_3  = \Sp^3$ which we will spend a lot of time with in the sequel. There is a-priori no obvious geometric decomposition of $\Sp^3$ into the desired form of $\Sigma \times \mathbb{R}$. One might be tempted to use the fact that odd spheres can be viewed as Hopf fibrations; eg.,  $\Sp^3$ being anx $\Sp^1$ fibration over $\Sp^2$, but the non-trivial nature of the fibration makes it ill-suited for our purposes. As presaged, we can however use  the Heegard splitting theorem \cite{Prasolov:1997kno} to obtain useful topological decompositions, which will serve our needs. Let us review this in the context of the three-sphere and note  two elementary and inequivalent ways to realize an $\Sp^3$ (we will revisit other decompositions later in our discussion).

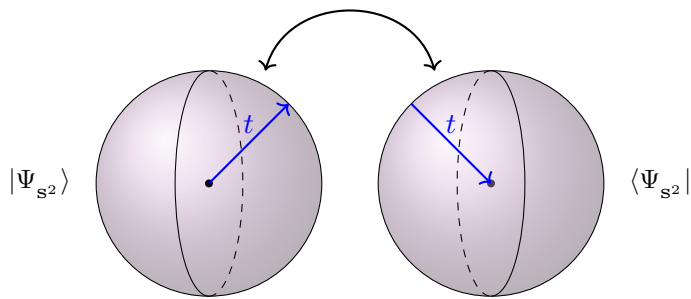
\begin{figure} [h]
\begin{center}
\usetikzlibrary{backgrounds}
\begin{tikzpicture}[scale=.75]
  \shade[ball color = violet!40, opacity = 0.4] (0,0) circle (2cm);
  \draw (0,0) circle (2cm);
  \draw [dashed]  (0,-2) arc (-90:90:0.6 and 2);
  \draw (0,2) arc (90:270:0.6 and 2);
  \fill[fill=black] (0,0) circle (2pt);
  \fill[fill=black] (5,0) circle (2pt);
  \shade[ball color = violet!40, opacity = 0.4] (5,0) circle (2cm);
  \draw (5,0) circle (2cm);
  \draw   (5,-2) arc (-90:90:0.6 and 2);
  \draw [dashed] (5,2) arc (90:270:0.6 and 2);
  \draw[thick, -> , color=blue] (0,0) -- (0.706,0.706) node[above]{$t$} -- (1.414,1.414);
   \draw[thick,-> , color=blue] (3.586,1.414) -- (4.292,0.706) node[above]{$t$} -- (5,0);
  \draw[thick, <->] (1,2)  to [out=75,in=105, looseness=1.25] (4,2);
    \node [thick, font=\fontsize{10}{0}\selectfont] at (-3,0) {$\psiSk$};
    \node [thick, font=\fontsize{10}{0}\selectfont] at (8,0) {$\psiSb$};
\end{tikzpicture}
    \end{center}
\caption{Cutting $\Sp^3$ along an $\Sp^2$ provides two  3-balls $\ball$. The origin of one of the 3-ball corresponds to the origin of $\Sp^3$ which can be viewed as  $t=-\infty$. The origin of the other 3-ball corresponds to $t=\infty$. Canonical quantization should be viewed as radial quantization with time $t$ running from the center of the left $\ball$ to the boundary $\Sp^2$ and then back down through the boundary of the right $\ball$ down to its origin (as indicated by the arrows). This helps prepare the state $\psiSk$ and its conjugate $\psiSb$.}
\label{fig:S3cut}
\end{figure}

One way to realize an $\Sp^3$ is to glue two three-balls $\ball$ together on their bounding $\Sp^2$s as we have attempted to depict in Fig.~\ref{fig:S3cut}. In this decomposition one can view the time direction as the radial direction of each ball, with the time running forward from the center to the boundary of one of the balls, and running back down from the boundary to the center in the other. We will later use this picture to argue that we can view the two balls as a particular slicing of the functional integral to produce a state in the Hilbert space and its conjugate. In this decomposition the relevant local structure for canonical quantization is $\Sp^2 \times \mathbb{R}$. 

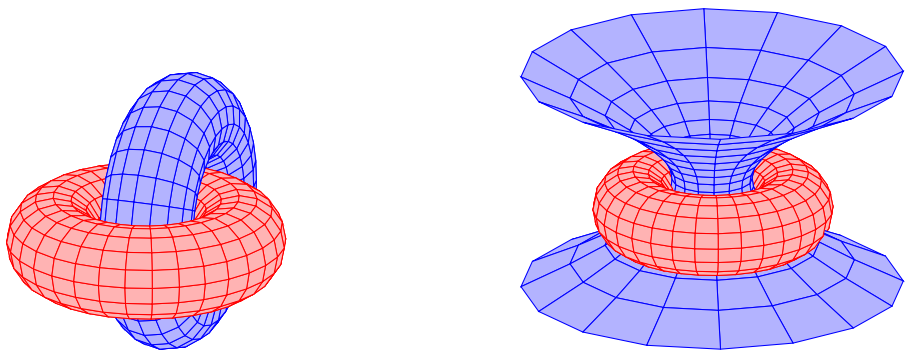
\begin{figure}[h]
\begin{center}
\usetikzlibrary{backgrounds}
 \begin{tikzpicture}
        \begin{axis}[hide axis,axis equal,scale=1.25,view={20}{20}]
        \addplot3[torus={blue}{0}] (\m,\n,\p);
        \addplot3[torus={red}{0}] (\p,\n-2,\m);
        \addplot3[torus={blue}{180}] (\m,\n,\p);
        \addplot3[torus={red}{180}] (\p,\n-2,\m);
      \end{axis}
  \end{tikzpicture}\hspace{3cm}
  \begin{tikzpicture}
        \begin{axis}[hide axis,axis equal,scale=1,view={20}{20}]
         \addplot3[torus={red}{0}] (\p,\n-2,\m);
            \addplot3[torus={blue}{0},domain=0:360,y domain=-1.75:1.75]
                ({-.75+(1+y^2/5)*cosh(y)*cos(x)},{(1+y^2/5)*cosh( y)*sin(x)},{sinh(y)});
                    \addplot3[torus={red}{180}] (\p,\n-2,\m);
        \end{axis}
    \end{tikzpicture}
\end{center}
\caption{$\Sp^3$ can be also obtained by gluing two interlocked solid tori. The gluing is done by identifying the boundaries of them in a way that the cycles homologous to the $a$-cycle of one of the boundary torus are identified to the cycles homologous to the $b$-cycle of the other boundary torus. On the left we have depicted  finite torii to illustrate the interlocking, while the right figure is more true to the spirit of the decomposition (with the point at infinity included).}
\label{fig:S3cuttori}
\end{figure}

A different way to decompose an $\Sp^3$ is to use two solid tori whose boundaries are ${\bf T}^2$. Viewing $\Sp^3$ as $\mathbb{R}^3$ with a point at infinity, we start with a solid torus embedded in $\mathbb{R}^3$. The second solid torus is then simply the complement of the first -- a useful visualization is to imagine the magnetic field lines of a toroid,  see Fig.~\ref{fig:S3cuttori} for an illustration. In this decomposition the two solid torii have different contractible cycles. If the $a$-cycle in the homology basis is contractible for the first solid torus we picked, the second solid torus instead has its $b$-cycle contractible. The local structure attained in this decomposition is ${\bf T}^2\times \mathbb{R}$, except that the swap of contractible cycles should be borne in mind (and can be accounted for by the $S$-transform of the modular $SL(2,\mathbb{Z})$ group on the torus). This particular decomposition is useful to realize that the partition function of Chern-Simons on $\Sp^3$ can be obtained from that on $\Sp^2 \times \Sp^1$.\footnote{ In fact studying Chern-Simons theory on $\Sp^2 \times \Sp^1$ is more intuitive as the angular coordinate along $\Sp^1$ can be viewed as (compactified) Euclidean time. While this makes the analysis of the state space and entanglement properties more straightforward
(see Appendix~\ref{sec:nrdensitym}), it is less well suited to the topological string discussion. }

\begin{figure}
\begin{center}
\usetikzlibrary{backgrounds}
\begin{tikzpicture}[scale=1]
    \begin{axis}[colormap/blackwhite,
       view={30}{60},axis lines=none
       ]
        \addplot3[surf,opacity=0.7,
   samples=25, point meta=x+3*z*z-0.25*y,
       domain=0:2*pi,y domain=0:2*pi,
       z buffer=sort]
       ({(2+cos(deg(x)))*cos(deg(y))}, 
        {(2+cos(deg(x)))*sin(deg(y))}, 
        {sin(deg(x))});

   \end{axis}
   \draw [very  thick,color=black!70] (3.35,2.85) ellipse (1.5cm and .9cm);
   \node [thick, font=\fontsize{14}{0}\selectfont] at (6.75,2.5) {$\psiTk$};
    \node [thick, font=\fontsize{14}{0}\selectfont] at (3.15,2.2) {$R_i$};
 \end{tikzpicture}
\end{center}
\caption{The state $\psiTk \in \mathcal{H}_{{\bf T}^2}$ can be produced by performing  the Chern-Simons theory path integral in a solid torus with a  Wilson line in the representation $R_i$ placed  along the non-contractible cycle of the solid torus.}
\label{fig:torusstate}
\end{figure}
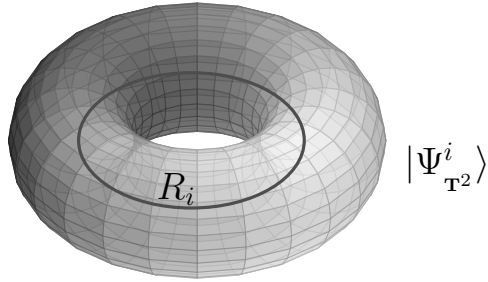

The above discussion makes it clear that knowledge of the physical Hilbert space for the local decomposition into $\Sigma \times \mathbb{R}$ with 
$\Sigma$ being either an $\Sp^2$ at genus $0$, or a ${\bf T}^2$ at genus $1$ will be helpful. Let us therefore collate some salient results on this front for  $\mathcal{H}_{\Sigma}$.
\begin{itemize}
\item For $\Sp^2$ with no Wilson line piercing through it, the physical Hilbert space $\mathcal{H}_{\Sp^2}$ is 1-dimensional. This state which we label $\psiSk$ 
can be produced at the boundary $\Sp^2$ of a 3-ball by performing the Chern-Simons theory  path integral in the 3-ball.
\item For $\Sp^2$ with one marked point in a representation $R_i$, the Hilbert space is  1-dimensional if $R_i$ is trivial; else it is 0-dimensional.
\item For $\Sp^2$ with two marked points with representations $R_i$ and $R_j$, the Hilbert space is one dimensional if $R_j$ is dual of $R_i$, and zero dimensional, otherwise.
\item For $\Sp^2$ with three marked points in representations $R_i, R_j$, and $R_k$, the dimension of the Hilbert space is given by the fusion coefficients $N_{ijk}$.
\item For torus ${\bf T}^2$ with no marked points, the Hilbert space $\mathcal{H}_{{\bf T}^2}$ the Hilbert space is  $m$-dimensional, and they can be associated with the integrable highest weight representations $R_0,R_1,\cdots, R_{m-1}$ of the loop group at level $k$. More precisely, the basis state $\psiTk $ can be produced by performing  the Chern-Simons theory path integral in a solid torus with a  Wilson line in the representation $R_i$ placed  along the non-contractible cycle of the solid torus, as depicted in Fig.~\ref{fig:torusstate}.
\end{itemize}

\section{Entanglement in Chern-Simons theory}
\label{sec:csent}

We have now assembled the necessary machinery to start exploring properties of reduced density matrices in Chern-Simons theory. We can start with a state in the Hilbert space $\mathcal{H}_\Sigma$ constructed above,  consider a bipartitioning of $\Sigma$ into two spatial regions $\Sigma = \rA \cup \rAc$, and ask how these are entangled. Let us see how to extract from the reduced density matrix $\rhoA$ for some subregion $\rA$ the topological entanglement entropy  \cite{Kitaev:2005dm,Levin:2006zz}. At the outset we note that the such a computation has already been carried out in \cite{Dong:2008ft} using the aforementioned logic to define the reduced state, and thence  using the replica method compute $\Tr{\rhoA^{\!q}}$ and finally extract the von Neumann entropy 
\begin{equation}
S_\rA = -\Tr{\rhoA \log \rhoA} = \lim_{q\to 1} \, \frac{1}{1-q} \, \log \Tr{\rhoA^{\!q}}\,.
\label{eq:vNent}
\end{equation}	

The  computation of $\Tr{\rhoA^{\!q}}$ reduces to evaluating the partition function on a new three-manifold $\mcs_3^{(q)}$ which is built as a $q-$fold `branched cover' over $\mcs_3$. The branching occurs along the entangling surface $\entsurf$. The solvability of Chern-Simons theory comes to fore in this computation: one can decompose, via surgery, the branched cover 
$\mcs_3^{(q)}$ into a sequence of topological $\Sp^3$s, see \cite{Witten:1988hf}. This enables one to directly evaluate the partition function on $\mcs_3^{(q)}$ and obtain thereby the R\'enyi and von Neumann entropies. This is explained in some detail in \cite{Dong:2008ft}, and we will provide a slight variant of their argument adapted for our purposes below. Our focus is to follow the construction of the reduced state $\rhoA$ somewhat explicitly in this section, so that we can apply a similar logic on the closed string side in the sequel. Before proceeding, let us also point out other approaches to computing topological entanglement using edge state and CFT methods discussed in \cite{Wen:2016snr} (see also \cite{Das:2015oha}). These ideas can also be extended to computing topological entanglement negativity \cite{Wen:2016bla}. A closely related discussion of entanglement edge modes in Chern-Simons theory can be found in \cite{Wong:2017pdm}.

 \subsection{The reduced density matrix}
 \label{sec:rhoCS}

Let us consider Chern-Simons on $\mcs_3$ decomposed locally into $\Sigma \times \mathbb{R}$ as explained above. $\Sigma$ is a a codimension-1 Cauchy surface in $\mcs_3$ for purposes of our discussion. For starters take it to be a Riemann surface with no punctures on it. Consider the bipartitioning of $\Sigma = \rA \cup \rAc$ with respect to which we wish to define our reduced density matrices.  For simplicity, we can refer to the reduced Hilbert spaces associated with the regions $\rA$ and $\rAc$ as 
$\mathcal{H}_{\rA}$ and $\mathcal{H}_{\rAc}$, respectively. Let us first understand how to go about constructing $\mathcal{H}_\rA$.  The discussion below should 
be familiar to readers acquainted with defining entanglement in gauge theories \cite{Buividovich:2008gq,Casini:2013rba,Donnelly:2014fua,Soni:2015yga,Ghosh:2015iwa}.

The Hilbert space $\mathcal{H}_{\rA}$ is constructed by quantizing the classical phase space obtained by solving the constraint equation \eqref{eq:gausslaw} restricted to region 
$\rA$. However, now since the region $\rA$ has boundary $\entsurf$, we must choose appropriate boundary condition for the fields. Ab-initio, any boundary condition that is consistent 
with the constraint equations is admissible on $\entsurf$. Distinct boundary conditions simply give rise to different {\it superselection sectors} in $\mathcal{H}_\rA$; essentially 
decomposing $\mathcal{H}_\rA = \oplus_{\alpha} \, \mathcal{H}_\rA^{[\alpha]}$ where we are using $\alpha$ to index the different superselection sectors. 
Similarly, we can construct the physical Hilbert space $\mathcal{H}_{\rAc}$ and identify the superselection sectors $\mathcal{H}_{\rAc}^{[\alpha]}$. One can view the choice of boundary conditions and superselection sectors resulting therefrom as including additional charged degrees of freedom on the boundary $\entsurf$.

Since the physical Hilbert space $\mathcal{H}_{\Sigma}$  has to satisfy the Gauss law constraints \eqref{eq:gausslawwilson} (which is now relevant owing to our having introduced charged degrees of freedom on $\entsurf$), we can identify it as the maximal subspace of $\mathcal{H}_\rA\otimes \mathcal{H}_{\rAc}$ that respects the constraints.  We shall  denote this subspace as $\mathcal{H}_\rA\otimes_{\text{inv}} \mathcal{H}_{\rAc}$. The important message of this discussion is that the Hilbert space on $\Sigma$ does not  factorize trivially owing to the underlying gauge invariance, as is well appreciated in the aforementioned works. All told, the essential point for us is that when we compute the entanglement entropy, we must remember to impose the Gauss law constraints.

Fortuitously, there is an efficient method for implementing this. Observe that $\mathcal{H}_\rA\otimes_{\text{inv}} \mathcal{H}_{\rAc}$ can be  decomposed into superselection sectors as follows:
\begin{equation}
\mathcal{H}_\rA \otimes_{\text{inv}} \mathcal{H}_{\rAc}
= \bigoplus_\alpha \mathcal{H}_{\rA}^{[\alpha]} \otimes \mathcal{H}_{\rAc}^{[\alpha]} \,.
\label{eq:superselection}
\end{equation}
 where the sum is over all superselection sectors introduced by the boundary conditions. Using the state-operator correspondence we can describe these superselection sectors quite explicitly.   As explained in \S\ref{sec:H01summary}, any state in $\mathcal{H}_{{\bf T}^2}$ can be obtained by performing path integral over the solid torus $\mcs_{{\bf  T}^2}$ with appropriate Wilson lines insertions along the homologically non-trivial cycle of the solid torus. Similarly, states in $\mathcal{H}_{\Sigma}$ can be obtained by performing path integral over the handlebody $\mcs_{\Sigma}$ bounded by $\Sigma$, with Wilson loops in appropriate representations of the gauge group placed along the non-contractible cycles of $\mcs_{\Sigma}$. 

When we split $\mcs_{\Sigma}$ into two handlebodies $\mcs_\rA$ and $\mcs_{\rAc}$ determined by the bipartitioning of $\Sigma = \rA \cup \rAc$, 
 the resulting handlebodies will have additional boundaries. We denote the new boundary and its decomposition as  
 $\mathcal{D}=\mathcal{D}_{\rA}\cup \mathcal{D}_{\rAc}$, where $\mathcal{D}_{\rA}=\partial \mcs_\rA\setminus \rA$, 
 and similarly for the complement.  We have allowed for multiple components 
 $\mathcal{D}_{\rA}^i$ of $\mathcal{D}_{\rA}$ which may arise from the cut (likewise for $\rAc$).  We should bear in mind that these are two-surfaces obtained from the split.

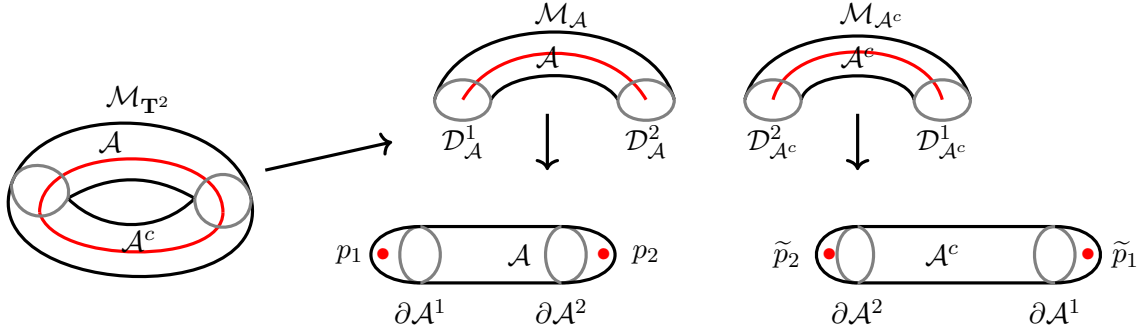
\begin{figure}[h]
\begin{center}
\usetikzlibrary{backgrounds}
	\begin{tikzpicture}[scale=.75]
    \begin{pgfonlayer}{nodelayer}
		\node [style=none] (0) at (-13.25, -6.5) {};
		\node [style=none] (1) at (-9, -6.75) {};
		\node [style=none] (2) at (-12.25, -6.75) {};
		\node [style=none] (3) at (-10, -6.75) {};
		\node [style=none] (4) at (-12.25, -6.75) {};
		\node [style=none] (5) at (-10, -6.75) {};
		\node [style=none] (6) at (-12.75, -7) {};
		\node [style=none] (7) at (-9.5, -7) {};
		\node [style=none] (8) at (-11.5, -5.75) {};
		\node [style=none] (9) at (-11.5, -5.75) {$\rA$};
		\node [style=none] (10) at (-11, -7.5) {$\rAc$};
		\node [style=none] (11) at (-8.75, -6.25) {};
		\node [style=none] (12) at (-6.5, -5.75) {};
		\node [style=none] (13) at (-5.75, -5) {};
		\node [style=none] (14) at (-4.75, -5) {};
		\node [style=none] (15) at (-5.25, -5) {};
		\node [style=none] (16) at (-2.5, -5) {};
		\node [style=none] (17) at (-2, -5) {};
		\node [style=none] (18) at (-1.5, -5) {};
		\node [style=none] (19) at (-5.25, -5.25) {};
		\node [style=none] (20) at (-5.25, -5.75) {$\mathcal{D}_{\rA}^1$};
		\node [style=none] (21) at (-2, -5.75) {$\mathcal{D}_{\rA}^2$};
		\node [style=none] (22) at (-3.75, -4.25) {$\rA$};
		\node [style=none] (23) at (-0.25, -5) {};
		\node [style=none] (24) at (0.25, -5) {};
		\node [style=none] (25) at (0.75, -5) {};
		\node [style=none] (26) at (2.75, -5) {};
		\node [style=none] (27) at (3.25, -5) {};
		\node [style=none] (28) at (3.75, -5) {};
		\node [style=none] (29) at (0.25, -5.75) {$\mathcal{D}_{\rAc}^2$};
		\node [style=none] (30) at (3.25, -5.75) {$\mathcal{D}_{\rAc}^1$};
		\node [style=none] (31) at (1.75, -4.25) {$\rAc$};
		\node [style=none] (32) at (-3.75, -5.25) {};
		\node [style=none] (33) at (-3.75, -6.25) {};
		\node [style=none] (34) at (1.75, -5.25) {};
		\node [style=none] (35) at (1.75, -6.25) {};
		\node [style=none] (36) at (-6, -7.25) {};
		\node [style=none] (37) at (-6, -8.25) {};
		\node [style=none] (38) at (-3.5, -8.25) {};
		\node [style=none] (39) at (-3.5, -7.25) {};
		\node [style=none] (40) at (1.75, -7.25) {};
		\node [style=none] (41) at (1.75, -8.25) {};
		\node [style=none] (42) at (5.25, -7.25) {};
		\node [style=none] (43) at (5.25, -8.25) {};
		\node [style=none] (44) at (-5.75, -7.75) {};
		\node [style=none] (45) at (-6.75, -7.75) {\color{red}{${~\bullet}$}};
		\node [style=none] (46) at (-2.75, -7.75) {\color{red}{$\bullet$}};
		\node [style=none] (47) at (1.25, -7.75) {\color{red}{$\bullet$}};
		\node [style=none] (48) at (5.75, -7.75) {\color{red}{$~\bullet$}};
		\node [style=none] (49) at (-6, -8.75) {$\entsurf^1$};
		\node [style=none] (50) at (-3.5, -8.75) {$\entsurf^2$};
		\node [style=none] (51) at (1.75, -8.75) {$\entsurf^2$};
		\node [style=none] (52) at (5.25, -8.75) {$\entsurf^1$};
		\node [style=none] (53) at (-4.25, -7.75) {$\rA$};
		\node [style=none] (53) at (-3.5, -3.5) {$\mcs_\rA$};
		\node [style=none] (53) at (2, -3.5) {$\mcs_{\rAc}$};
		\node [style=none] (53) at (-11,-5) {$\mcs_{{\bf T}^2}$};
		\node [style=none] (54) at (3.25, -7.75) {$\rAc$};
		\node [style=none] (55) at (-7.25, -7.75) {$p_1$};
		\node [style=none] (56) at (-2, -7.75) {$p_2$};
		\node [style=none] (57) at (0.5, -7.75) {$\widetilde{p}_2$};
		\node [style=none] (58) at (6.5, -7.75) {$\widetilde{p}_1$};
	\end{pgfonlayer}
	\begin{pgfonlayer}{edgelayer}
		\draw [very thick,bend right=105, looseness=1.25] (0.center) to (1.center);
		\draw [very thick,bend left=75] (0.center) to (1.center);
		\draw [very thick,bend left] (2.center) to (3.center);
		\draw [very thick,bend right=45] (4.center) to (5.center);
		\draw [very thick, color=red, in=-90, out=-75, looseness=0.75] (6.center) to (7.center);
		\draw [very thick, color=red, bend left=90] (6.center) to (7.center);
		\draw [very thick, color=gray, bend right=75, looseness=1.50] (0.center) to (2.center);
		\draw [very thick,  color=gray,bend left=90, looseness=1.50] (0.center) to (4.center);
		\draw [very thick,  color=gray,bend right=90, looseness=1.75] (5.center) to (1.center);
		\draw [very thick,  color=gray,bend left=75, looseness=1.50] (5.center) to (1.center);
		\draw [very thick, ->] (11.center) to (12.center);
		\draw [very thick,in=120, out=60, looseness=0.75] (14.center) to (16.center);
		\draw [very thick, red, in=120, out=60] (15.center) to (17.center);
		\draw [very thick,bend left=75] (13.center) to (18.center);
		\draw [very thick, color=gray, bend right=90, looseness=1.25] (13.center) to (14.center);
		\draw [very thick, color=gray, bend left=75, looseness=1.25] (13.center) to (14.center);
		\draw [very thick,  color=gray,bend right=90, looseness=1.25] (16.center) to (18.center);
		\draw [very thick,  color=gray,bend left=75, looseness=1.25] (16.center) to (18.center);
		\draw [very thick,bend left=75, looseness=0.75] (25.center) to (26.center);
		\draw [very thick, red, bend left=75] (24.center) to (27.center);
		\draw [very thick,bend left=75] (23.center) to (28.center);
		\draw [very thick, color=gray, bend right=90, looseness=1.25] (23.center) to (25.center);
		\draw [very thick,  color=gray,bend left=75, looseness=1.25] (23.center) to (25.center);
		\draw [very thick,  color=gray,bend right=90, looseness=1.25] (26.center) to (28.center);
		\draw [very thick,  color=gray,bend left=75, looseness=1.25] (26.center) to (28.center);
		\draw [very thick, ->] (32.center) to (33.center);
		\draw [very thick, ->] (34.center) to (35.center);
		\draw [very thick](36.center) to (39.center);
		\draw [very thick](37.center) to (38.center);
		\draw [very thick](40.center) to (42.center);
		\draw [very thick](41.center) to (43.center);
		\draw [very thick, color=gray, in=180, out=-180, looseness=1.25] (36.center) to (37.center);
		\draw [very thick,  color=gray,bend right=75] (39.center) to (38.center);
		\draw [very thick,  color=gray,bend left=75] (40.center) to (41.center);
		\draw [very thick, color=gray, bend right=90, looseness=1.25] (42.center) to (43.center);
		\draw [very thick,  color=gray,bend left=75, looseness=1.25] (36.center) to (37.center);
		\draw [very thick, color=gray, bend left=90, looseness=1.50] (39.center) to (38.center);
		\draw [very thick, color=gray, bend left=270, looseness=1.25] (40.center) to (41.center);
		\draw [very thick, color=gray, bend left=75, looseness=1.25] (42.center) to (43.center);
		\draw [very thick,bend left=270, looseness=3.00] (36.center) to (37.center);
		\draw [very thick,bend left=90, looseness=3.25] (39.center) to (38.center);
		\draw [very thick,bend left=270, looseness=2.50] (40.center) to (41.center);
		\draw [very thick,bend left=90, looseness=2.75] (42.center) to (43.center);
	\end{pgfonlayer}
\end{tikzpicture}
\end{center}
\caption{Cutting the solid torus $\mcs_{{\bf T}^2}$ having a Wilson loop in representation $R_i$ produces two handlebodies $\mcs_\rA$ with boundary $\rA\cup \mathcal{D}_{\rA}^1 \cup \mathcal{D}_{\rA}^2$ and $\mcs_{\rAc}$ with boundary $\rAc\cup \mathcal{D}_{\rAc}^1 \cup \mathcal{D}_{\rAc}^2$. 
The Wilson line induces marked point $p_1$ carrying representation $R_i$ on $\mathcal{D}_{\rA}^1$ and $p_2$ carrying  the dual representation of $R_i$ on $\mathcal{D}_{\rA}^2$. 
It likewise induces marked point $\widetilde{p}_2$ carrying representation $R_i$ on $\mathcal{D}_{\rAc}^2$ and $\widetilde{p}_1$ carrying  the dual representation of $R_i$ on $\mathcal{D}_{\rAc}^1$. The boundary condition on $\entsurf^1$ can be identified with the marked point $p_1$ and the representation $R_i$. Similar statements can be made for the other boundaries  
$\entsurf^2$ (and we can at the end of the day conflate the boundaries of $\rA$ and $\rAc$).}
\label{fig:torusstatesplit}
\end{figure}

These boundaries  will cut through the Wilson lines passing through them. Thus, for instance, $\mathcal{D}_{\rA}$ will have marked points with associated representations, i.e.,  
there will be operators inserted  at different marked points in $\mathcal{D}_{\rA}$.  According to the state-operator correspondence, the boundary conditions imposed at 
the boundaries of  $\mathcal{D}_{\rA}$  are in one-to-one correspondence with the sets of operators inserted in them.  Note that the boundaries of  $\mathcal{D}_{\rA}$ make up the 
boundary of region $\rA$, which of course forms the entangling surface of interest. Therefore, the allowed boundary conditions for region $\rA$ can be identified with the marked points with associated representations in $\mathcal{D}_{\rA}$. The superselection sectors $\mathcal{H}_{\rA}^{[\alpha]}$ can be characterized by specifying the  number of marked points 
and the associated representations in $\mathcal{D}_{\rA}$. Similar statements hold for the complementary region $\rAc$ and surfaces associated thereto.
We have pictorially depicted this decomposition in Fig.~\ref{fig:torusstatesplit} for $\Sigma={\bf T}^2$.
 
It is perhaps worth noting that owing to the introduction of boundary degrees of freedom, the superselection sectors $\mathcal{H}_{\rA}^{[\alpha]}$ (respectively $\mathcal{H}_{\rAc}^{[\alpha]}$) can have dimensionality greater than one (even when $\mathcal{H}_\rA$ is unidimensional).  Furthermore, the states of $\mathcal{H}_{\rA}^{[\alpha]}$ (respectively $\mathcal{H}_{\rAc}^{[\alpha]}$) can be organized into representations of the symmetry group arising from large gauge transformations. That is to say, the presence of the boundary makes physical precisely those 
gauge transformations which respect the chosen boundary conditions and  are non-vanishing at the boundaries of  the regions $\rA$ ($\rAc$)  \cite{Elitzur:1989nr}. Notice that the Gauss law constraints \eqref{eq:gausslawwilson} make sure that any state obtained by combining the states from $\mathcal{H}_{\rA}$ and $\mathcal{H}_{\rAc}$  can be interpreted as a state produced on $\Sigma$ by performing path integral over the handlebody $\mcs_{\Sigma}$ with appropriate Wilson loops placed along its non-contractible cycles, as required. 
    
\subsection{Entanglement on a Riemann sphere}
\label{sec:eeSp} 
     
Armed with the construction described above, we are now in a position to study the entanglement structure of Chern-Simons theory on $\Sp^3$. Consider an $\Sp^2$ slice of $\Sp^3$ using the decomposition depicted in Fig.~\ref{fig:S3cut}. There is only one independent state in the Hilbert space $\mathcal{H}_{\Sp^2}$, and we denote it by $\psiSk$. 
Let us bipartition the  $\Sp^2$ into two connected regions $\rA$ and $\rAc$. Both the Hilbert space $\mathcal{H}_\rA$ restricted to region $\rA$ and that associated to the  
complement $\rAc$ have dimensionality greater than one. Furthermore, they also do not contain any non-trivial superselection sectors within them. 

In fact, the states in $\mathcal{H}_\rA$  ($\mathcal{H}_{\rAc}$) are in a representation of the loop group associated with the gauge group $G$ of the Chern-Simons theory \cite{Elitzur:1989nr}. Let us denote the independent vectors in $\mathcal{H}_\rA$ by $|\Psi^{\mu}_{_\rA} \rangle, \ \mu=1,\cdots, d_\rA$ and  the independent vectors in $\mathcal{H}_{\rAc}$ by $|\Psi^{\nu}_{_{\rAc}}\rangle,~\nu=1,\cdots,d_\rA$, respectively, where we defined 
$ d_\rA = \text{dim}\left(\mathcal{H}_\rA\right)= \text{dim}\left(\mathcal{H}_{\rAc}\right)$.
Therefore, we can decompose the state $\psiSk$ as follows
\begin{equation}
\psiSk=\sum_{\mu,\nu} \, \mathfrak{c}_{\mu\nu}|\Psi^{\mu}_{_\rA} \rangle \otimes |\Psi^{\nu}_{_{\rAc}}\rangle,
\label{eq:s2decompose}
\end{equation}
where $\mathfrak{c}_{\mu\nu}$ are complex numbers.

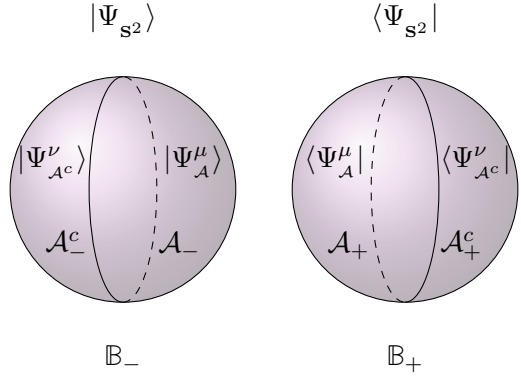
\begin{figure}[htbp]
\begin{center}
\usetikzlibrary{backgrounds}
	\begin{tikzpicture}[scale=.75]
	\shade[ball color = violet!40, opacity = 0.4] (0,0) circle (2cm);
	\draw (0,0) circle (2cm);
  	\draw [dashed]  (0,-2) arc (-90:90:0.6 and 2);
  	\draw (0,2) arc (90:270:0.6 and 2);
    \shade[ball color = violet!40, opacity = 0.4] (5,0) circle (2cm);
 	\draw (5,0) circle (2cm);
  	\draw   (5,-2) arc (-90:90:0.6 and 2);
 	\draw [dashed] (5,2) arc (90:270:0.6 and 2);
 	\node at (1,-1) {$\rAk$};
    \node at (-1,-1) {$\rAck$};
    \node at (6,-1) {$\rAcb$};
  	\node at (4,-1) {$\rAb$};
    \node at (0,-3) {$\bket$};
   	\node at (5,-3) {$\bbra$};
    \node at (0,3) {$\psiSk$};
    \node at (5,3) {$\psiSb$};
    \node at (1.25,.5) {$|\Psi_{_\rA}^{\mu}\rangle$};
    \node at (-1.25,.5) {$|\Psi_{_{\rAc}}^{\nu}\rangle$};
    \node at (6.25,.5) {$\langle\Psi^{\nu}_{_{\rAc}}|$};
   \node at (3.75,.5) {$\langle\Psi^{\mu}_{_\rA}|$};
\end{tikzpicture}
\end{center}
\caption{Starting with the decomposition of the three sphere $\Sp^3$ into two balls $\bket$ and $\bbra$, we can identify the state $\psiSk$  ($\psiSb$)  as being associated with the ball $\bket$ ($\bbra$). These states are obtained by canonical quantization as described earlier and are meant to live on $\Sp^2 = \partial \bket$ 
(respectively $\partial\bbra$). This configuration of solid spheres carrying Chern-Simons path integral without any identification between $\bket$ and $\bbra$ represents total density matrix 
$\rho=\psiSk \psiSb$. If we now further decompose the boundary of the balls into $\partial \bket = \rA \cup \rAc$, and 
$\partial \bbra = \rAb \cup \rAcb$, respectively, then we obtain reduced states on subregions of interest which we have indicated above.}
\label{fig:rdensitym}
\end{figure}

In order to study the entanglement structure, we must construct the relevant density matrices, and this requires identifying the geometric configurations 
associated with the states $\psiSk$, $|\Psi_{_\rA}\rangle$, and $|\Psi_{_{\rAc}}\rangle$. To obtain this, we recall the picture of radial canonical quantization described in \S\ref{sec:H01summary} where we realized that $\Sp^3$ can be obtained by gluing two solid 3-balls $\bket$ and $\bbra$ by identifying their respective boundaries, see Fig.~\ref{fig:S3cut}. 
The boundaries of both $\bket$ and $\bbra$ are 2-spheres. Let us pick the $\Sp^2$ which is the boundary of $\bket$ and define a state $\psiSk$ on it. 
Then the state on the other $\Sp^2$ which is the boundary of $\bbra$ can be identified as the conjugate state  $ \psiSb$. This is illustrated in Fig.~\ref{fig:rdensitym}.  This identification makes complete sense because the operation of taking the inner product between these two states $\psibk$ can be understood as the gluing of the Chern-Simons path integral on $\bket$ and $\bbra$. Consequently, the value of this inner product  is given by
\begin{equation}
\psibk= \zcs(\Sp^3),
\label{eq:innprodzs3}
\end{equation}
where $\zcs(\Sp^3)$ the partition function of Chern-Simons theory on $\Sp^3$. All we are doing here is slicing open the path integral to extract the state, and are aided by the ability to decompose the $\Sp^3$ topologically as indicated. 

Our next task is to identify the states on subregions $\rA$ and $\rAc$ and their corresponding conjugate states. To  do so, we start by bipartitioning $\partial \bket =\Sp^2 
= \rAk \cup\rAck$ and construct the states by canonical quantization  as explained above. An analogous operation on $\partial \bbra$ produces the conjugate states. 
We make the choice to orthogonalize the reduced states, by demanding  
\begin{equation}
\langle \Psi^{\mu_1}_{_\rA}|\Psi^{\mu_2}_{_\rA} \rangle=\delta_{\mu_1,\mu_2} \qquad\qquad 
\langle \Psi^{\nu_1}_{_{\rAc}} |\Psi^{\nu_2}_{_{\rAc}} \rangle=\delta_{\nu_1,\nu_2}.
\label{eq:normalize}
\end{equation}
Combining the equations \eqref{eq:s2decompose} and  \eqref{eq:innprodzs3} and accounting for our normalization choice we learn that 
\begin{equation}
\sum_{\mu,\nu}| \mathfrak{c}_{\mu\nu}|^2= \zcs(\Sp^3).
\label{eq:s2de1}
\end{equation}
Therefore, the reduced density matrix $\rhoA$ takes the form
\begin{equation}
\rhoA=\sum_{\nu,\mu_1,\mu_2}\, \mathfrak{c}_{\mu_1\nu}\, \mathfrak{c}^*_{\mu_2\nu}\,  |\Psi_{_\rA}^{\mu_1}\rangle  \langle \Psi_{_\rA}^{\mu_2}| \,.
\label{eq:rdensitym1}
\end{equation}
We will leave this reduced density matrix unnormalized and account for the normalizations when we compute traces of its powers separately. 

\begin{figure}
\begin{center}
\usetikzlibrary{backgrounds}
\begin{tikzpicture}[scale=.75]
 \shade[ball color = violet!40, opacity = 0.4] (0,0) circle (2cm);
 \draw (0,0) circle (2cm);
 \draw [dashed]  (0,-2) arc (-90:90:0.6 and 2);
 \draw (0,2) arc (90:270:0.6 and 2);
\shade[ball color = violet!40, opacity = 0.4] (5,0) circle (2cm);
 \draw (5,0) circle (2cm);
 \draw   (5,-2) arc (-90:90:0.6 and 2);
 \draw [dashed] (5,2) arc (90:270:0.6 and 2);
 \node at (1,-1) {$\rAk$};
\node at (-1,-1) {$\rAck$};
 \node at (6,-1) {$\rAcb$};
 \node at (4,-1) {$\rAb$};
 \node at (0,-3) {$\bket$};
 \node at (5,-3) {$\bbra$};
 \node at (0,3) {$\psiSk$};
 \node at (5,3) {$\psiSb$};
 \node at (1.25,.5) {$|\Psi_{_\rA}^{\mu}\rangle$};
 \node at (-1.25,.5) {$|\Psi_{_{\rAc}}^{\nu}\rangle$};
 \node at (6.25,.5) {$\langle\Psi^{\nu}_{_{\rAc}}|$};
 \node at (3.75,.5) {$\langle\Psi^{\mu}_{_\rA}|$};  
 \shade[ball color = violet!40, opacity = 0.4] (10,0) circle (2cm);
 \draw (10,0) circle (2cm);
 \draw   (10,-2) arc (-90:90:0.6 and 2);
 \draw [dashed] (10,2) arc (90:270:0.6 and 2);
 \node at (11,-1) {$\rA$};
 \node at (9,-1) {$\rAc$};
 \node at (10,-3) {$\ball$};
 \node at (10,3) {$\psiSk$};
 \node at (11.25,.5) {$|\Psi^{\nu_1}_{_{\rA}}\rangle$};
 \node at (8.75,.5) {$|\Psi^{\mu_1}_{_{\rAc}}\rangle$};
\end{tikzpicture}
\end{center}
\caption{The action of $\rhoA$ on the state $\psiSk$ can be captured by performing the path integral over three $3$-balls, $\bket$, $\bbra$,  and $\ball$ with the following identifications:  region $\rAck$ identified with $\rAcb$, region $\rAb$ identified with $\rA$. The net result is a path integral over a 
single $3$-ball whose boundary consists of two complementary regions $\rAk$ and $\rAc$. }
\label{fig:rdensitymonpsi}
\end{figure}
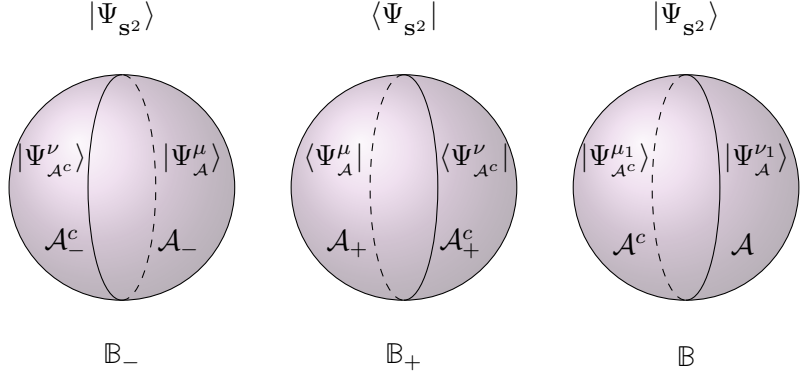

From Fig.~\ref{fig:rdensitym} it is clear that the reduced state $\rhoA$ can be obtained by performing the Chern-Simons path integral over $\bket$ and $\bbra$ with region 
$\rAck$ and $\rAcb$ identified. This is entirely analogous to the usual functional integral definition of the reduced density matrix where we open up the path integral around the region 
$\rA$ of interest and introduce regulatory surfaces at $t = 0^\pm$ to prescribe suitable boundary conditions in order to extract the matrix elements of $\rhoA$ (see eg., 
\cite{Rangamani:2016dms}). This can be exploited to obtain a useful property of the reduced density matrix.

Consider acting  with $\rhoA$ on the state $\psiSk$ which can be done pictorially as illustrated in Fig.~\ref{fig:rdensitymonpsi}. It is clear that this operation involves the path integral over 
three independent $3$-balls, $\bket$, $\bbra$,  and $\ball$ which make up the reduced density matrix and the state $\psiSk$, respectively. The action involves making 
identifications of subregions of these three 3-balls. We identify region $\rAck$ identified with $\rAcb$ to make $\rhoA$ as described earlier. In addition, region 
$\rAb$  has to be identified with $\rA$ to implement the operation of $\rhoA$ acting on $\psiSk$. Effectively we are left with a path integral over a 
single 3-ball whose boundary consists of two regions $\rAk$ and $\rAc$. This statement is to be understood topologically, which is essentially all that we need for the purposes of
Chern-Simons computation. The final result after identifications is however just our original path integral definition for constructing the state $\psiSk$! 
Thus, we have the following relation
\begin{equation}
\rhoA\psiSk=\psiSk.
\label{eq:rhoaonpsi}
\end{equation}
This translates into the following relation between different coefficients $\mathfrak{c}_{\mu\nu}$
\begin{equation}
\sum_{\mu_1,\nu_1} \, \mathfrak{c}_{\mu\nu_1}\, \mathfrak{c}_{\mu_1\nu_1}^* \, \mathfrak{c}_{\mu_1\nu}= \mathfrak{c}_{\mu\nu}.
\label{eq:aconstarint}
\end{equation}
Using the relation \eqref{eq:s2de1} and \eqref{eq:aconstarint}, we obtain a simple result for traces of arbitrary integral powers of the reduced density matrix:
\begin{equation}
\text{Tr}_\rA\left(\rhoA^{\!q} \right)=\zcs(\Sp^3) \,.
\label{eq:trrhon}
\end{equation}
This result was obtained in \cite{Dong:2008ft} by a similar argument. Their observation was that computing $\Tr{\rhoA^{\!q}}$ involves the branched cover manifold $\mcs^{(q)}_3$ which is the $q$-fold branching of an $\Sp^3$ over the equatorial $\Sp^1$ bounding $\rA$ and $\rAc$, respectively. This $3$-manifold is topologically an $\Sp^3$ and thus the computation of the trace 
brings us back to a known evaluation. We have simply chosen to reinterpret this in terms of a state space picture to enable us make statements later for the closed topological string.
Finally, note that entanglement entropy of the state $\psiSk$ for bipartitioning of  $\Sp^2$ is given by:
\begin{equation}
S_\rA= \lim_{q\to1}\frac{1}{1-q} \bigg[ \log \text{Tr}_\rA\left(\rhoA^{\!q} \right) -q\; \log \text{Tr}_\rA\left(\rhoA\right) \bigg] =
\log\zcs(\Sp^3).
\label{eq:entropys2}
\end{equation}

Note that for this choice of bipartitioning we get a universal answer involving the $\Sp^3$ partition function of the Chern-Simons theory, both for the von Neumann and for the 
R\'enyi entropies. It is easy to check $S_\rA^{(q)} = \log \zcs(\Sp^3)$. Such a flat entanglement spectrum is indicative of the topological nature of the underlying theory. What it encodes is the lack of penalty due to physical interactions, illustrated by our ability to freely glue different three-manifolds together as in Fig.~\ref{fig:rdensitymonpsi}. This is somewhat reminiscent of tensor network toy models of holography \cite{Pastawski:2015qua,Hayden:2016cfa} which likewise exhibit a flat entanglement spectrum, as they are unaware of the dynamics of gravitational interactions. We will return to this issue in \S\ref{sec:discuss}.
We would also like to reiterate a point made in \S\ref{sec:intro}: the replica computation described only captures only the topological part of the entanglement. To actually see the entire contribution including the linearly divergent term we would have to go back to the edge mode discussion in \S\ref{sec:rhoCS}.

\begin{figure}[ht]
\begin{center}
\usetikzlibrary{backgrounds}
\begin{tikzpicture}[scale=.45]
\filldraw[fill=blue!20] (-4.2,3) ellipse (.6 and .6);
	\filldraw[fill=green!20] (-6,3) ellipse (.6 and .6);
	\node at (-6,3)  {$\scriptstyle{\rAk^1}$};
	\node at (-4.2,3){$\scriptstyle{\rAk^2}$};
	\filldraw[fill=blue!20] (1.2,3) ellipse (.6 and .6);
	\filldraw[fill=green!20] (-1,3) ellipse (.6 and .6);
	\node at (-1,3)  {$\scriptstyle{\rAb^1}$};
	\node at (1.2,3){$\scriptstyle{\rAb^2}$};
	\node at (-5,2)  {$\scriptstyle{\rAck}$};
	\node at (0,2){$\scriptstyle{\rAcb}$};
	\shade[ball color = violet!40, opacity = 0.4] (0,3) circle (2cm);
	\draw (0,3) circle (2cm);
	\shade[ball color = violet!40, opacity = 0.4] (-5,3) circle (2cm);
	\draw (-5,3) circle (2cm);
	\draw[very thick,->](3,3)--(6,3);
	  \node at (8.5,3)  {${\frac{1}{\zcs(\Sp^3)}}$};
	\shade[ball color = blue!40, opacity = 0.4](17,3) circle (2cm);
	  \draw (17,3) circle (2cm);
	  \shade[ball color = green!40, opacity = 0.4] (12,3) circle (2cm);
	  \draw (12,3) circle (2cm);	  
 \draw [dashed]  (12,1) arc (-90:90:0.4 and 2);
  \draw (12,5) arc (90:270:0.4 and 2);
   \draw [dashed]  (17,1) arc (-90:90:0.4 and 2);
  \draw (17,5) arc (90:270:0.4 and 2);
  \node at (11,3)  {$\scriptstyle{\rAk^1}$};
	\node at (13,3){$\scriptstyle{\rAb^1}$};
	 \node at (16,3)  {$\scriptstyle{\rAk^2}$};
	\node at (18,3){$\scriptstyle{\rAb^2}$};
\end{tikzpicture}
\end{center}
\caption{The $\rhoA$ for Riemann sphere with two regions $\rA=\rA^1\cup\rA^2$ and $\rAc$ with two interfaces  is obtained by gluing the path integrals over the 3-balls by 
identifying the boundary regions $\rAck$ and $\rAcb$ on the boundaries.
The resultant path integral can be understood as a path integral over two 3-balls without any identification (whose boundaries comprise the components of $\rA$), divided by $\zcs(\Sp^3)$. This figure also makes it clear that action of 
$\rhoA$ on the state $\psiSk$ gives $\frac{1}{\zcs(\Sp^3)}\psiSk$.}
\label{fig:rdensitymtsp2intf}
\end{figure}
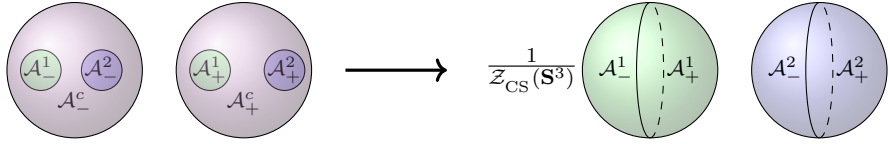

We have thus far considered situations where the entangling surface is a single connected surface. We can easily generalize to situations with of the Cauchy surface $\Sp^2$
partitioned into a set of $M$ disconnected regions $\rA = \cup_i\, \rA^i$, say by cutting along different latitudes, or by picking sub-domains inside the 3-balls. We illustrate the construction of the corresponding reduced density matrix in Fig.~\ref{fig:rdensitymtsp2intf}.

Let us repeat the path integral  construction for $\rhoA\psiSk$: we again can start with  three 3-balls $\ball_i$, $i=1,2,3$, each with $M+1$ 
components corresponding to our partitioning of the system into multi-component subregions $\rA$ and $\rAc$. 
The action $\rhoA\psiSk$ now is achieved by first identifying $\rAck$ in $\bket$ with $\rAcb$ in $\bbra$ as before to make up $\rhoA$, while  the action of the density matrix is achieved by identifying $\rAb$ with $\rA$. We again get back a the functional integral over a single 3-ball, and thus the state $\psiSk$. The only thing we have to be careful 
about is normalization. Each gluing we perform in contracting the components of $\rAb$ and $\rA$, locally produces a copy of the $\Sp^3$ partition function. 
We perform $M$ such gluings, only one of which is used to set the normalization of $\psiSk$. The result should therefore be accounted for as an overall amplitude in the action, resulting therefore in:
\begin{equation}
\rhoA\psiSk=\frac{1}{\big[\zcs(\Sp^3)\big]^{M-1}} \; \psiSk.
\label{eq:rhoaonpsiM}
\end{equation}
Then the R\'enyi and von Neumann entropies are readily computed be computed and differ from the previous result \eqref{eq:entropys2} by a factor of $M$ \cite{Dong:2008ft}. For example,
\begin{equation}
\begin{split}
S_\rA &=  \lim_{q\to1}\frac{1}{1-q} \bigg[ \log \text{Tr}_\rA\left(\rhoA^{\!q}\right) -q\; \log  \text{Tr}_\rA\left(\rhoA\right)\bigg]  \\
&= \lim_{q\to1}\frac{1}{1-q}\bigg[-(q\, M-q -M)\, \log~\zcs(\Sp^3)-q\, \log \zcs(\Sp^3) \bigg] \\
&=M\; \log\zcs(\Sp^3).
\end{split}
\label{eq:entropysM}
\end{equation}

Thus, we see that for the basic state $\psiSk$ we have a universal result that only depends on the number of disconnected components making up $\entsurf$ up to an overall factor, the $\Sp^3$ partition function of Chern-Simons theory. As presaged above, we can extract more interesting information by considering states that have Wilson lines. We will do so once we have established the basic dictionary for mapping the spatially ordered entanglement of the state $\psiSk$ onto the closed string side.

 \section{Entanglement in closed topological string theory}
\label{sec:entctop}

We now turn to figuring out how to define reduced density matrices, and entanglement entropy thereof, in the closed topological string theory. We will give a prescription for defining a state space of the quantum closed topological string, constructing the reduced density matrix therefrom, and finally computing the von Neumann entropy. The key point in our discussion will be that it is possible to follow the replica construction through the geometric transition, quite explicitly.  

For purposes of defining reduced density matrix in the closed topological string theory, we are essentially going to mimic the usual field theory replica construction. One can view this as an attempt to uplift the arguments of \cite{Lewkowycz:2013nqa} who utilize the replica construction of a physical QFT, to closed topological string theory. For such a construction to make sense,  the theory ought to have a consistent path integral formulation in terms of a spacetime action. Moreover, it must be possible to characterize an constant time slice in the target space  (the latter for us will be the resolved conifold) completely by specifying its topology in a physically meaningful manner within the theory. Fortunately,  we happen to be  in a favourable situation. The formalism of closed string field theory which applies equally to the topological setting allows one to construct a path integral formulation in terms of a spacetime action.\footnote{  See \cite{Zwiebach:1992ie,Moosavian:2017qsp,Moosavian:2017sev} for discussions of  the physical closed string field theory.}  We will not need the details of this construction in what follows. As we shall see below, making a choice for a Cauchy slice can be done in a sensible manner only with specification of topological data within the resolved conifold. At the end of the day, all of this follows from the independence of the theory on geometric data, in particular, the absence of non-trivial gravitational dynamics, but it will nevertheless be reassuring that one can indeed carry out the constructions to their logical end simply by following our nose.

In  framing the replica construction, we need several pieces of data. Let us quickly recall some of the basic elements (we have already implicitly used this in our Chern-Simons discussion in 
\S\ref{sec:csent}). Given a field theory on a manifold $\xf$, we define states on a codimension-1 Cauchy slice $\tsb$ (which w.l.o.g.\ we think of as the $t=0$ slice). As in our discussion of Chern-Simons entanglement, consider a bipartition $\tsb = \rxA \cup \rxAc$. In order to define matrix elements of the reduced density matrix $\rhoAx$, we slice open 
$\xf$ across $t=0^\pm$ along $\rxA$ to obtain a manifold with a cut along $\rxA$. Then the path integral with suitable boundary conditions $\Phi_\pm$ at $t=0^\pm$ for fields in 
$\rxA$ leads to matrix elements $\left( \rhoAx\right)_{-+}$.  The next step is to obtain matrix elements of $\rhoAx^{q}$ by taking $q$-fold copies of the path integral computing 
$\rhoAx$ and making the appropriate identifications (respecting the ${\mathbb Z}_q$ replica symmetry). As alluded to earlier, this 
construction involves $q$ functional integrals cut open along the region $\rxA$ with a cyclic gluing condition. We are instructed to  identify the $t=0^+$ configuration 
$j^{\text{th}}$ copy with the $t=0^-$ data on the  $(j+1)^{\text{st}}$ copy. These identifications  of the $q$ copies of the manifold $\xf$ construct our `branched cover' replica 
manifold $\xf^{(q)}$. The R\'enyi entropies of the reduced state are then computed from the functional integrals as:
\begin{equation}
S_\rA^{(q)}=\frac{1}{1-q}\, \bigg(\log   \mathcal{Z}[\xf^{(q)}]  - q\, \log   \mathcal{Z}[\xf]  \bigg).
\label{eq:rentropypathint}
\end{equation}

\begin{figure}[ht]
\begin{center}
\usetikzlibrary{backgrounds}
\begin{tikzpicture}[scale=.65]
	\begin{pgfonlayer}{nodelayer}
		\node [style=none] (11) at (-2, 0) {};
		\node [style=none] (12) at (-2, -4.5) {};
		\node [style=none] (17) at (-5, 1) {};
		\node [style=none] (18) at (-5, -5) {};
		\node [style=none] (16) at (-8, 0) {};
		\node [style=none] (13) at (-8, -4.5) {};
		\node [style=none] (14) at (0.1, 0) {${r=\infty}$};
		\node [style=none] (15) at (0, -4.5) {${r=0}$};
		\node [style=none] at (-2.5, 1.5) {$\mathbf{\mathbf{S^2}}$};
		\node [style=none] at (-4, 1.5) {$\bbra$};
		\node [style=none] at (-7.5, 1.5) {$\mathbf{S^2}$};
		\node [style=none] at (-6, 1.5) {$\bket$};
	\end{pgfonlayer}
	\begin{pgfonlayer}{edgelayer}
 	   \shade[ball color = olive!40, opacity = 0.4] (-7.9,0) circle (.6cm); 
		\shade[ball color = olive!40, opacity = 0.4] (-7.9,-1.65) circle (.5cm);
 		\shade[ball color = olive!40, opacity = 0.4] (-7.9,-2.75) circle (.3cm);
  		\shade[ball color = olive!40, opacity = 0.4] (-7.9,-3.75) circle (.2cm);
 		\shade[ball color = olive!40, opacity = 0.4] (-7.9,-4.5) circle (.05cm);
		\draw [ thick, <-, in=90, out=-90] (-1,0) -- (-1,-4.5);
   		\shade[ball color = olive!40, opacity = 0.4] (-2.2,0) circle (.6cm); 
		\shade[ball color = olive!40, opacity = 0.4] (-2.2,-1.65) circle (.5cm);
		\shade[ball color = olive!40, opacity = 0.4] (-2.2,-2.75) circle (.3cm);
		\shade[ball color = olive!40, opacity = 0.4] (-2.2,-3.75) circle (.2cm);
		\shade[ball color = olive!40, opacity = 0.4] (-2.2,-4.5) circle (.05cm);
		\shade[ball color = violet!120, opacity = 0.4] (-4.65,-4.5) circle (.3cm); 
		\shade[ball color = violet!120, opacity = 0.4] (-4.65,-3.75) circle (.3cm);
		\shade[ball color = violet!120, opacity = 0.4] (-4.5,-2.75) circle (.45cm);
		\shade[ball color = violet!120, opacity = 0.4] (-4.35,-1.65) circle (.6cm);
		\shade[ball color = violet!120, opacity = 0.4] (-4,0) circle (.95cm);
		\shade[ball color = violet!120, opacity = 0.4] (-5.35,-4.5) circle (.3cm); 
		\shade[ball color = violet!120, opacity = 0.4] (-5.35,-3.75) circle (.3cm);
		\shade[ball color = violet!120, opacity = 0.4] (-5.5,-2.75) circle (.45cm);
		\shade[ball color = violet!120, opacity = 0.4] (-5.65,-1.65) circle (.6cm);
		\shade[ball color = violet!120, opacity = 0.4] (-6,-0) circle (.95cm);
		\draw [  color=black,dashed]   (-5.05,0) arc (0:180:0.95 and .2);
		\draw [  color=black] (-5.05,0) arc (360:180:0.95 and .2);
   		\draw [  color=black,dashed]   (-3.05,0) arc (0:180:0.95 and .2);
  		\draw [  color=black] (-3.05,0) arc (360:180:0.95 and .2);
     	\draw [  color=black,dashed]   (-5.05,-1.65) arc (0:180:0.6 and .17);
 		\draw [  color=black] (-5.05,-1.65) arc (360:180:0.6 and .17);
   		\draw  [ color=black,dashed]  (-3.75,-1.65) arc (0:180:0.6 and  .17);
  		\draw [ color=black] (-3.75,-1.65) arc (360:180:0.6 and .17);
     	\draw [  color=black,dashed]   (-5.05,-2.75) arc (0:180:0.45 and .12);
 	    \draw [ color=black] (-5.05,-2.75) arc (360:180:0.45 and .12);
   		\draw  [color=black,dashed]  (-4.05,-2.75) arc (0:180:0.45 and .12);
  		\draw [ color=black] (-4.05,-2.75) arc (360:180:0.45 and .12);
    	\draw [  color=black,dashed]   (-5.05,-3.75) arc (0:180:0.3 and .06);
  		\draw [  color=black] (-5.05,-3.75) arc (360:180:0.3 and .06);
   		\draw [ color=black,dashed]   (-4.35,-3.75) arc (0:180:0.3 and .06);
 		\draw [  color=black] (-4.35,-3.75) arc (360:180:0.3 and .06);
    	\draw [  color=black,dashed]   (-5.05,-4.5) arc (0:180:0.3 and .06);
  		\draw [  color=black] (-5.05,-4.5) arc (360:180:0.3 and .06);
   		\draw [ color=black,dashed]   (-4.35,-4.5) arc (0:180:0.3 and .06);
 		\draw [  color=black] (-4.35,-4.5) arc (360:180:0.3 and .06);
  		 \draw   [dashed]  (-7.3,0) arc (0:180:0.6 and .15);
 		 \draw (-7.3,0) arc (360:180:0.6 and .1);
		  \draw   [dashed]  (-1.6,0) arc (0:180:0.6 and .15);
 		 \draw (-1.6,0) arc (360:180:0.6 and .1);
    	 \draw   [dashed]  (-7.4,-1.65) arc (0:180:0.5 and .125);
  		\draw (-7.4,-1.65) arc (360:180:0.5 and .125);
			 \draw   [dashed]  (-1.7,-1.65) arc (0:180:0.5 and .125);
  		\draw (-1.7,-1.65) arc (360:180:0.5 and .125);
    	 \draw   [dashed]  (-7.6,-2.75) arc (0:180:0.3 and .1);
 		 \draw (-7.6,-2.75) arc (360:180:0.3 and .1);
		  \draw   [dashed]  (-1.9,-2.75) arc (0:180:0.3 and .1);
 		 \draw (-1.9,-2.75) arc (360:180:0.3 and .1);
        \draw   [dashed]  (-7.7,-3.75) arc (0:180:0.2 and .05);
 		 \draw (-7.7,-3.75) arc (360:180:0.2 and .05);
		      \draw   [dashed]  (-2.0,-3.75) arc (0:180:0.2 and .05);
 		 \draw (-2.0,-3.75) arc (360:180:0.2 and .05);
   		 \draw   [dashed]  (-7.85,-4.5) arc (0:180:0.05 and .02);
 		 \draw (-7.85,-4.5) arc (360:180:0.05 and .02);
		  \draw   [dashed]  (-2.15,-4.5) arc (0:180:0.05 and .02);
 		 \draw (-2.15,-4.5) arc (360:180:0.05 and .02);
		\draw[very thick,->,color = orange] (-.5,-2)--(4,-2);
		\node at (1.75,-1.5) {$\text{\tiny{Geometric transition}}$};
	\end{pgfonlayer}
\end{tikzpicture}
\begin{tikzpicture}[scale=.7]
	\begin{pgfonlayer}{nodelayer}
		\node [style=none] (11) at (-2, 0) {};
		\node [style=none] (12) at (-2, -4.5) {};
		\node [style=none] (17) at (-5, 1) {};
		\node [style=none] (18) at (-5, -5) {};
		\node [style=none] (16) at (-8, 0) {};
		\node [style=none] (13) at (-8, -4.5) {};
		\node [style=none] (14) at (.1, 0) {${r=\infty}$};
		\node [style=none] (15) at (0, -4.5) {${r=0}$};
		\node [style=none] at (-2.5, 1.5) {$\mathbf{\mathbf{S^2}}$};
		\node [style=none] at (-4, 1.5) {$\bbra$};
		\node [style=none] at (-7.5, 1.5) {$\mathbf{S^2}$};
		\node [style=none] at (-6, 1.5) {$\bket$};
	\end{pgfonlayer}
	\begin{pgfonlayer}{edgelayer}
		\shade[ball color = violet!120, opacity = 0.4] (-4.9,-4.5) circle (.05cm); 
		\shade[ball color = violet!120, opacity = 0.4] (-4.75,-3.75) circle (.2cm);
		\shade[ball color = violet!120, opacity = 0.4] (-4.5,-2.75) circle (.45cm);
		\shade[ball color = violet!120, opacity = 0.4] (-4.35,-1.65) circle (.6cm);
		\shade[ball color = violet!120, opacity = 0.4] (-4,0) circle (.95cm);
		\shade[ball color = violet!120, opacity = 0.4] (-5.1,-4.5) circle (.05cm); 
		\shade[ball color = violet!120, opacity = 0.4] (-5.25,-3.75) circle (.2cm);
		\shade[ball color = violet!120, opacity = 0.4] (-5.5,-2.75) circle (.45cm);
		\shade[ball color = violet!120, opacity = 0.4] (-5.65,-1.65) circle (.6cm);
		\shade[ball color = violet!120, opacity = 0.4] (-6,-0) circle (.95cm);
		\shade[ball color = olive!40, opacity = 0.4] (-7.9,0) circle (.6cm); 
		\shade[ball color = olive!40, opacity = 0.4] (-7.9,-1.65) circle (.5cm);
 		\shade[ball color = olive!40, opacity = 0.4] (-7.9,-2.75) circle (.3cm);
  		\shade[ball color = olive!40, opacity = 0.4] (-7.9,-3.75) circle (.2cm);
 		\shade[ball color = olive!40, opacity = 0.4] (-7.9,-4.5) circle (.2cm);
		\draw [thick, <-, in=90, out=-90] (-1,0) -- (-1,-4.5);
   		\shade[ball color = olive!40, opacity = 0.4] (-2.2,0) circle (.6cm); 
		\shade[ball color = olive!40, opacity = 0.4] (-2.2,-1.65) circle (.5cm);
		\shade[ball color = olive!40, opacity = 0.4] (-2.2,-2.75) circle (.3cm);
		\shade[ball color = olive!40, opacity = 0.4] (-2.2,-3.75) circle (.2cm);
		\shade[ball color = olive!40, opacity = 0.4] (-2.2,-4.5) circle (.2cm);
		\draw [  color=black,dashed]   (-5.05,0) arc (0:180:0.95 and .2);
		\draw [  color=black] (-5.05,0) arc (360:180:0.95 and .2);
   		\draw [  color=black,dashed]   (-3.05,0) arc (0:180:0.95 and .2);
  		\draw [  color=black] (-3.05,0) arc (360:180:0.95 and .2);
     	\draw [  color=black,dashed]   (-5.05,-1.65) arc (0:180:0.6 and .17);
 		\draw [  color=black] (-5.05,-1.65) arc (360:180:0.6 and .17);
   		\draw  [ color=black,dashed]  (-3.75,-1.65) arc (0:180:0.6 and  .17);
  		\draw [ color=black] (-3.75,-1.65) arc (360:180:0.6 and .17);
  		\draw [  color=black,dashed]   (-5.05,-2.75) arc (0:180:0.45 and .12);
 	    \draw [ color=black] (-5.05,-2.75) arc (360:180:0.45 and .12);
   		\draw  [color=black,dashed]  (-4.05,-2.75) arc (0:180:0.45 and .12);
  		\draw [ color=black] (-4.05,-2.75) arc (360:180:0.45 and .12);
        \draw [  color=black,dashed]   (-5.05,-3.75) arc (0:180:0.2 and .06);
 	    \draw [ color=black] (-5.05,-3.75) arc (360:180:0.2 and .06);
   	    \draw [ color=black,dashed]   (-4.55,-3.75) arc (0:180:0.2 and .06);
  	    \draw [ color=black] (-4.55,-3.75) arc (360:180:0.2 and .06);
    	\draw [  color=black,dashed]   (-5.05,-4.5) arc (0:180:0.05 and .02);
  		\draw [  color=black] (-5.05,-4.5) arc (360:180:0.05 and .02);
   		\draw [ color=black,dashed]   (-4.85,-4.5) arc (0:180:0.05 and .02);
 		\draw [  color=black] (-4.85,-4.5) arc (360:180:0.05 and .02);
		 \draw   [dashed]  (-7.3,0) arc (0:180:0.6 and .15);
 		 \draw (-7.3,0) arc (360:180:0.6 and .1);
		  \draw   [dashed]  (-1.6,0) arc (0:180:0.6 and .15);
 		 \draw (-1.6,0) arc (360:180:0.6 and .1);
    	 \draw   [dashed]  (-7.4,-1.65) arc (0:180:0.5 and .125);
  		\draw (-7.4,-1.65) arc (360:180:0.5 and .125);
			 \draw   [dashed]  (-1.7,-1.65) arc (0:180:0.5 and .125);
  		\draw (-1.7,-1.65) arc (360:180:0.5 and .125);
    	 \draw   [dashed]  (-7.6,-2.75) arc (0:180:0.3 and .1);
 		 \draw (-7.6,-2.75) arc (360:180:0.3 and .1);
		  \draw   [dashed]  (-1.9,-2.75) arc (0:180:0.3 and .1);
 		 \draw (-1.9,-2.75) arc (360:180:0.3 and .1);
        \draw   [dashed]  (-7.7,-3.75) arc (0:180:0.2 and .05);
 		 \draw (-7.7,-3.75) arc (360:180:0.2 and .05);
		      \draw   [dashed]  (-2.0,-3.75) arc (0:180:0.2 and .05);
 		 \draw (-2.0,-3.75) arc (360:180:0.2 and .05);
   		 \draw   [dashed]  (-7.7,-4.5) arc (0:180:0.2 and .05);
 		 \draw (-7.7,-4.5) arc (360:180:0.2 and .05);
		  \draw   [dashed]  (-2,-4.5) arc (0:180:0.2 and .05);
 		 \draw (-2.0,-4.5) arc (360:180:0.2 and .05);
	    \end{pgfonlayer}
\end{tikzpicture}
\end{center}
\caption{The geometric transition between the deformed and resolved conifolds seen from  the perspective 3-manifold surgery. We identify  on the left the uplift of the surgical decomposition of $\Sp^3 \subset T^*\Sp^3$. Following this through the geometric transition brings us to the right figure which will be our starting point for defining the closed string Hilbert space. 
While the $\Sp^2$ is non-trivially fibered over the $\Sp^3$ in the deformed and resolved conifolds, for ease of illustration and discussion, we will indicate it as if it were a direct product; see footnote \ref{fn:fibration}. Note that we have swapped the direction of the radial coordinate of the cone relative to Fig.~\ref{fig:conifold} for ease of illustration.}
\label{fig:rdensitydefcontorescon}
\end{figure}
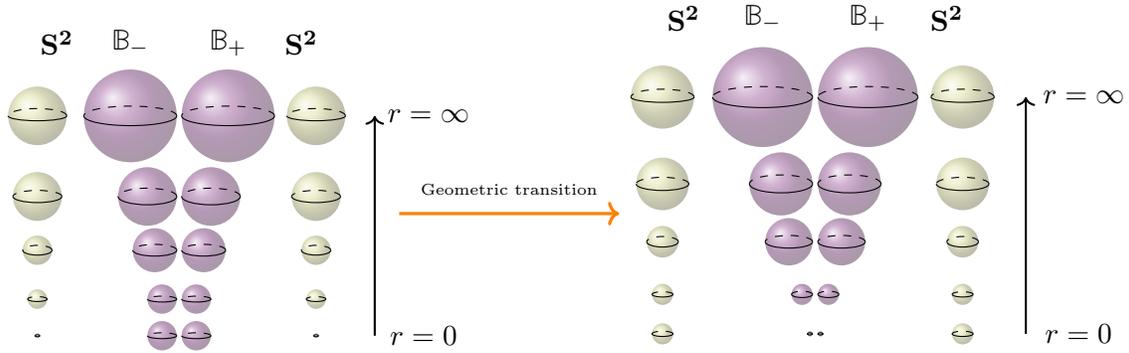

Our basic premise is to start with the picture we  have described for the Chern-Simons theory in \S\ref{sec:csent} and view this taking place on the $\Sp^3 \subset T^*\Sp^3$ in the  A-model  open+closed string theory. Heuristically, this amounts to viewing the Cauchy slices and their bipartitioning as occurring far away from the tip of the cone. However, given a surgical decomposition of $\Sp^3$, we can consider uplifting it directly into $T^*\Sp^3$. This would motivate a splitting  along a Cauchy slice in target space, which will extend all the way close to the tip.  We can then dial the complex structure parameter of the deformed conifold, so that we can pass through the geometric transition onto the resolved conifold side, where the branes and associated open string degrees of freedom disappear and we are left with closed topological strings. In this process we rely on the fact that the geometric transition is localized near the tip of the cone, and can thus motivate a  construction of a state space for the closed string on the resolved conifold. We have attempted to illustrate this in Fig.~\ref{fig:rdensitydefcontorescon}. 

Once we get to this point, we can furthermore take inspiration from \cite{Lewkowycz:2013nqa} for the closed string analysis. We extend our bipartitioning into the bulk of the resolved conifold 
by picking an ansatz for the location of a `topological cosmic brane' which we refer to as the \emph{entangling brane} for reasons described in \S\ref{sec:intro}.
Requiring that the decomposition be compatible with the dynamics of the closed topological string, we learn of the topological constraints on the construction of the reduced density matrix
and replicas thereof. Once we have fixed the replica target space by imposing these, we can immediately compute the closed topological string partition function and check that the answer is compatible with that expected from Chern-Simons theory. 

In what follows we will attempt to motivate a construction in the topological string theory that mimics the replica construction in Chern-Simons theory. As discussed in \S\ref{sec:intro} this construction only captures the topological contribution to entanglement. Focusing on this has the advantage that in the closed string side one does not a-priori have to worry about  `closed string edge modes' and associated UV divergent contributions to the entanglement entropy. It might a-priori seem strange that the construction is tuned to extracting only part of the contribution. Nevertheless since the topological contribution is the part is extractable from a path integral framework and the open/closed string duality is best understood directly as a map between the open and closed string partition functions, it seems reasonable to proceed and learn the lessons we can from the replica construction.

\subsection{Density matrices in topological string theory}
\label{sec:ctsreplica}

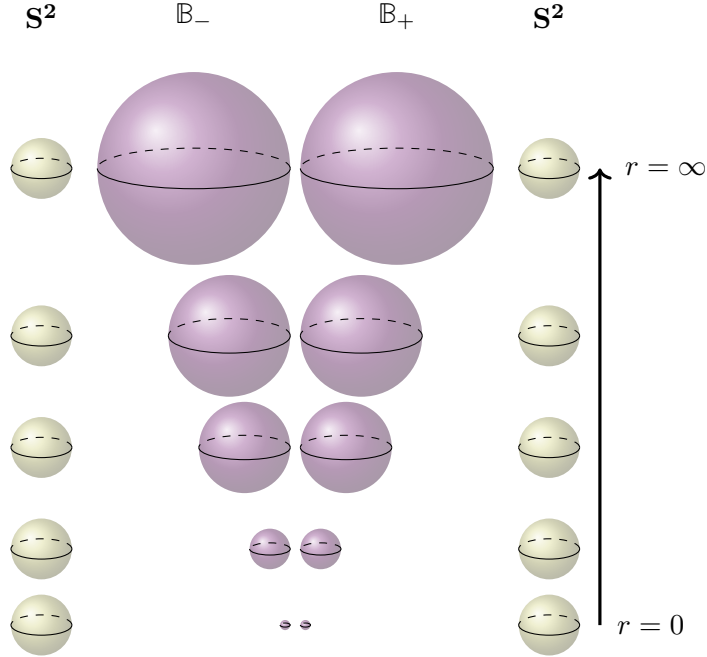
\begin{figure}[ht]
\begin{center}
\usetikzlibrary{backgrounds}
\begin{tikzpicture}[scale=1.35]
	\begin{pgfonlayer}{nodelayer}
		\node [style=none] (11) at (-2, 0) {};
		\node [style=none] (12) at (-2, -4.5) {};
		\node [style=none] (17) at (-5, 1) {};
		\node [style=none] (18) at (-5, -5) {};
		\node [style=none] (16) at (-8, 0) {};
		\node [style=none] (13) at (-8, -4.5) {};
		\node [style=none] (14) at (-1.35, 0) {${r=\infty}$};
		\node [style=none] (15) at (-1.5, -4.5) {${r=0}$};
		\node [style=none] at (-2.5, 1.5) {$\mathbf{\mathbf{S^2}}$};
		\node [style=none] at (-4, 1.5) {$\bbra$};
		\node [style=none] at (-7.5, 1.5) {$\mathbf{S^2}$};
		\node [style=none] at (-6, 1.5) {$\bket$};
	\end{pgfonlayer}
	\begin{pgfonlayer}{edgelayer}
 	   \shade[ball color = olive!40, opacity = 0.4] (-7.5,0) circle (.3cm); 
		\shade[ball color = olive!40, opacity = 0.4] (-7.5,-1.65) circle (.3cm);
 		\shade[ball color = olive!40, opacity = 0.4] (-7.5,-2.75) circle (.3cm);
  		\shade[ball color = olive!40, opacity = 0.4] (-7.5,-3.75) circle (.3cm);
 		\shade[ball color = olive!40, opacity = 0.4] (-7.5,-4.5) circle (.3cm);
		\draw [very thick, <-, in=90, out=-90] (11.center) to (12.center);
   		\shade[ball color = olive!40, opacity = 0.4] (-2.5,0) circle (.3cm); 
		\shade[ball color = olive!40, opacity = 0.4] (-2.5,-1.65) circle (.3cm);
		\shade[ball color = olive!40, opacity = 0.4] (-2.5,-2.75) circle (.3cm);
		\shade[ball color = olive!40, opacity = 0.4] (-2.5,-3.75) circle (.3cm);
		\shade[ball color = olive!40, opacity = 0.4] (-2.5,-4.5) circle (.3cm);
		\shade[ball color = violet!120, opacity = 0.4] (-4.9,-4.5) circle (.05cm); 
		\shade[ball color = violet!120, opacity = 0.4] (-4.75,-3.75) circle (.2cm);
		\shade[ball color = violet!120, opacity = 0.4] (-4.5,-2.75) circle (.45cm);
		\shade[ball color = violet!120, opacity = 0.4] (-4.35,-1.65) circle (.6cm);
		\shade[ball color = violet!120, opacity = 0.4] (-4,0) circle (.95cm);
		\shade[ball color = violet!120, opacity = 0.4] (-5.1,-4.5) circle (.05cm); 
		\shade[ball color = violet!120, opacity = 0.4] (-5.25,-3.75) circle (.2cm);
		\shade[ball color = violet!120, opacity = 0.4] (-5.5,-2.75) circle (.45cm);
		\shade[ball color = violet!120, opacity = 0.4] (-5.65,-1.65) circle (.6cm);
		\shade[ball color = violet!120, opacity = 0.4] (-6,-0) circle (.95cm);
		\draw [  color=black,dashed]   (-5.05,0) arc (0:180:0.95 and .2);
		\draw [  color=black] (-5.05,0) arc (360:180:0.95 and .2);
   		\draw [  color=black,dashed]   (-3.05,0) arc (0:180:0.95 and .2);
  		\draw [  color=black] (-3.05,0) arc (360:180:0.95 and .2);
     	\draw [  color=black,dashed]   (-5.05,-1.65) arc (0:180:0.6 and .17);
 		\draw [  color=black] (-5.05,-1.65) arc (360:180:0.6 and .17);
   		\draw  [ color=black,dashed]  (-3.75,-1.65) arc (0:180:0.6 and  .17);
  		\draw [ color=black] (-3.75,-1.65) arc (360:180:0.6 and .17);
     	\draw [  color=black,dashed]   (-5.05,-2.75) arc (0:180:0.45 and .12);
 	    \draw [ color=black] (-5.05,-2.75) arc (360:180:0.45 and .12);
   		\draw  [color=black,dashed]  (-4.05,-2.75) arc (0:180:0.45 and .12);
  		\draw [ color=black] (-4.05,-2.75) arc (360:180:0.45 and .12);
        \draw [  color=black,dashed]   (-5.05,-3.75) arc (0:180:0.2 and .06);
 	    \draw [ color=black] (-5.05,-3.75) arc (360:180:0.2 and .06);
   	    \draw [ color=black,dashed]   (-4.55,-3.75) arc (0:180:0.2 and .06);
  	    \draw [ color=black] (-4.55,-3.75) arc (360:180:0.2 and .06);
    	\draw [  color=black,dashed]   (-5.05,-4.5) arc (0:180:0.05 and .02);
  		\draw [  color=black] (-5.05,-4.5) arc (360:180:0.05 and .02);
   		\draw [ color=black,dashed]   (-4.85,-4.5) arc (0:180:0.05 and .02);
 		\draw [  color=black] (-4.85,-4.5) arc (360:180:0.05 and .02);
  		 \draw   [dashed]  (-7.2,0) arc (0:180:0.3 and .1);
 		 \draw (-7.2,0) arc (360:180:0.3 and .1);
  		 \draw   [dashed]  (-2.2,0) arc (0:180:0.3 and .1);
  		\draw (-2.2,0) arc (360:180:0.3 and .1);
    	 \draw   [dashed]  (-7.2,-1.65) arc (0:180:0.3 and .1);
  		\draw (-7.2,-1.65) arc (360:180:0.3 and .1);
  		 \draw  [dashed]   (-2.2,-1.65) arc (0:180:0.3 and  .1);
  		\draw (-2.2,-1.65) arc (360:180:0.3 and .1);
    	 \draw   [dashed]  (-7.2,-2.75) arc (0:180:0.3 and .1);
 		 \draw (-7.2,-2.75) arc (360:180:0.3 and .1);
   		\draw    [dashed] (-2.2,-2.75) arc (0:180:0.3 and .1);
 		 \draw (-2.2,-2.75) arc (360:180:0.3 and .1);
        \draw   [dashed]  (-7.2,-3.75) arc (0:180:0.3 and .1);
 		 \draw (-7.2,-3.75) arc (360:180:0.3 and .1);
   		\draw    [dashed] (-2.2,-3.75) arc (0:180:0.3 and .1);
  		\draw  (-2.2,-3.75) arc (360:180:0.3 and .1);
   		 \draw   [dashed]  (-7.2,-4.5) arc (0:180:0.3 and .1);
 		 \draw (-7.2,-4.5) arc (360:180:0.3 and .1);
  		 \draw    [dashed] (-2.2,-4.5) arc (0:180:0.3 and .1);
  		\draw  (-2.2,-4.5) arc (360:180:0.3 and .1);
	\end{pgfonlayer}
\end{tikzpicture}
\end{center}
\caption{The resolved conifold is described as a collection of  $\Sp^3$s placed along an infinite ray with the coordinate $r\in [0,\infty)$. At each point along the ray we have a five-dimensional space $\Sp^3 \wtimes \Sp^2$, with the  $\Sp^3$ having radius $r$ and the $\Sp^2$  being of unit radius. In our illustration, we have decomposed the $\Sp^3$  topologically into  two 3-balls $\bket$ and $\bbra$ (which are assumed to be identified along their boundaries). The purpose of the $\Sp^2$'s drawn alongside each  3-ball is to remind us that each point in the $\Sp^3$ carries with it a  $\Sp^2$ of unit radius, and as explained in footnote \ref{fn:fibration} we are leaving  implicit the fibration structure.}
\label{fig:rdensityrescon}
\end{figure}

Let us now construct the reduced density matrix in closed topological string theory which we will think of in terms of a string field theory on the resolved conifold 
$\rco = \mathcal{O}(-1)\oplus\mathcal{O}(-1)\to \mathbb{P}^1$.  We must first specify a Cauchy surface in this resolved conifold and then specify a splitting of this 
codimension-1 surface into two complementary regions $\rxA$ and $\rxAc$. We want to achieve all of this topologically, so we should remind ourselves of some of the key features of the topology in question. Further details can be found in Appendix~\ref{sec:conifold}.

The resolved conifold  can be constructed in  two steps. The first step is to solve the defining equation:
\begin{equation}
|z_1|^2+|z_2|^2-|z_3|^2-|z_4|^2= t \,.
\label{eq:conifoldeq}
\end{equation}
Here $t$ is the K\"ahler parameter of the resolved conifold. The second step is to quotient the solution obtained by solving \eqref{eq:conifoldeq} with the following $U(1)$  action 
\begin{equation}
\left(z_1,z_2,z_3,z_4\right)\to \left(e^{i\, \theta}z_1, e^{i\, \theta}z_2,e^{-i\, \theta}z_3,e^{-i\, \theta}z_4\right).
\label{eq:u1action}
\end{equation} 

The solution to the equation \eqref{eq:conifoldeq} can be parameterized  by introducing a real variable $r$, so that 
\begin{equation}
|z_1|^2+|z_2|^2=r^2+t \,,\qquad \qquad |z_3|^2+|z_4|^2=r^2\,,\qquad\qquad 0\leq r<\infty.
\label{eq:s3s}
\end{equation}
A-priori we have obtained two $\Sp^3$s defined by the two sets of equations above.  Note that at $r=0$ one of the $\Sp^3$s  shrinks to a point, while  other $\Sp^3$ with coordinates 
$(z_1,z_2)$  is always of finite radius (bounded below by $t$). Therefore, we are allowed to quotient the solution \eqref{eq:s3s} with the  $U(1)$ action \eqref{eq:u1action}. This can be done easily by freezing the phase of either $z_1$ or $z_2$. However, freezing the phase of one of the components in the coordinate duple  $(z_3, z_4)$ will not suffice. This is due to the fact that at 
$r=0$ both $z_3$ and $z_4$ vanish. Let us therefore use this freedom to fix the phase $\theta$. The freezing of the phase of both $z_1$ or $z_2$ then reduces the non-contractible surface from an $\Sp^3$ into an $\Sp^2$. We will find it convenient to redefine coordinates to make this resulting  $\Sp^2$  have unit radius. To wit, 
\begin{equation}
\widetilde{z}_i=\frac{z_i}{\sqrt{|z_3|^2+|z_4|^2+t}}\qquad i=1,2.
\label{eq:rescal}
\end{equation}

We can therefore describe the topology of the resolved conifold $\mathcal{O}(-1)\oplus\mathcal{O}(-1)\to \mathbb{P}^1$ as a collection of  $\Sp^3$s placed along an infinite ray parameterized by the coordinate $r \in [0, \infty)$. At each point in the ray, specified by the value of the coordinate $r$, there is an $\Sp^3$ having radius $r$ as well as a  
unit radius $\Sp^2$. We depict this perspective in Fig.~\ref{fig:rdensityrescon}.  As indicated in the caption of Fig.~\ref{fig:rdensitydefcontorescon} we are ignoring the fact that the $\Sp^2$ is non-trivially fibered over the $\Sp^3$ for ease of discussion. 

This realizes quite explicitly the description of the resolved conifold explained in \S\ref{sec:topduality}; we have a cone over a base $\Sp^2 \wtimes \Sp^3$, with the radial direction of the cone parameterized by $r$, which is valued in  $\mathbb{R}_+\cup \{0\}$. The base is a five dimensional space, which itself may be viewed as a fibre bundle whose base space is an $\Sp^3$ having radius $r$, while the fibre direction is a fixed size $\Sp^2$.
 
 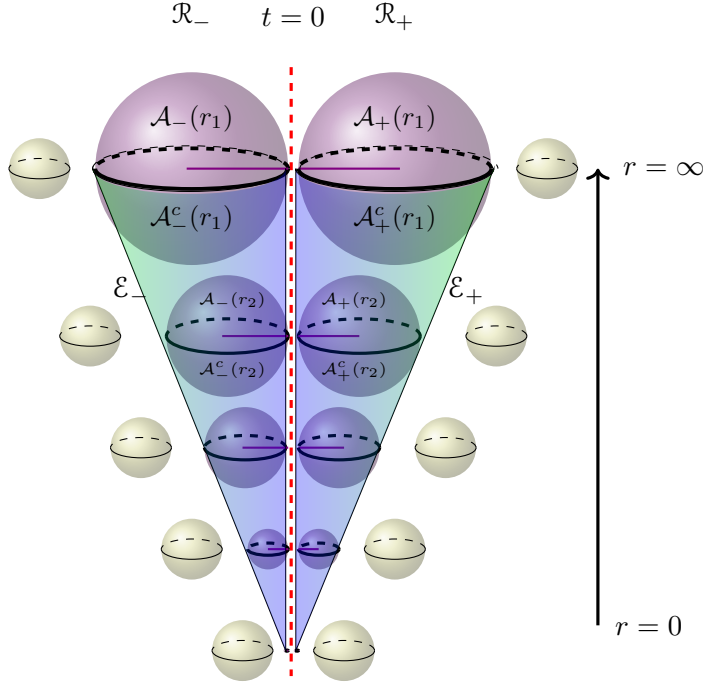
\begin{figure}[ht]
    \begin{center}
\usetikzlibrary{backgrounds}
\begin{tikzpicture}[scale=1.35]
\begin{pgfonlayer}{nodelayer}
		\node [style=none] (11) at (-2, 0) {};
		\node [style=none] (12) at (-2, -4.5) {};
			\node [style=none] (17) at (-5.02, 1) {};
		\node [style=none] (18) at (-5.02, -5) {};
				\node [style=none] (16) at (-8, 0) {};
		\node [style=none] (13) at (-8, -4.5) {};
		\node [style=none] (14) at (-1.35, 0) {$ r=\infty $};
		\node [style=none] (15) at (-1.5, -4.5) {$ r=0 $};
		\node [style=none] at (-4, 1.5) {$ \rco_+$};
		\node [style=none] at (-6, 1.5) {$\rco_-$};
		\node [style=none] at (-5, 1.5) {$ t=0$};
		\node [style=none] at (-6, .5) {\small{$\rAk(r_1)$}};
		\node [style=none] at (-4, .5) {\small{$\rAb(r_1)$}};
		\node [style=none] at (-5.6, -1.3) {\tiny{$\rAk(r_2)$}};
		\node [style=none] at (-4.4, -1.3) {\tiny{$\rAb(r_2)$}};
		\node [style=none] at (-6, -.5) {\small{$\rAck(r_1)$}};
		\node [style=none] at (-4, -.5) {\small{$\rAcb(r_1)$}};
		\node [style=none] at (-5.6, -2) {\tiny{$\rAck(r_2)$}};
		\node [style=none] at (-4.4, -2) {\tiny{$\rAcb(r_2)$}};
		\node [style=none] at (-6.6, -1.2) {$\entsurfx_-$};
		\node [style=none] at (-3.3, -1.2) {$\entsurfx_+$};
	\end{pgfonlayer}
	\begin{pgfonlayer}{edgelayer}
	\draw [very thick, style=dashed,color=red, in=90, out=-90] (17.center) to (18.center);
    \shade[ball color = olive!40, opacity = 0.4] (-7.5,0) circle (.3cm); 
 \shade[ball color = olive!40, opacity = 0.4] (-7,-1.65) circle (.3cm);
 \shade[ball color = olive!40, opacity = 0.4] (-6.5,-2.75) circle (.3cm);
  \shade[ball color = olive!40, opacity = 0.4] (-6,-3.75) circle (.3cm);
 \shade[ball color = olive!40, opacity = 0.4] (-5.5,-4.75) circle (.3cm);
 \draw [very thick, <-, in=90, out=-90] (11.center) to (12.center);
 \shade[ball color = olive!40, opacity = 0.4] (-2.5,0) circle (.3cm); 
 \shade[ball color = olive!40, opacity = 0.4] (-3,-1.65) circle (.3cm);
 \shade[ball color = olive!40, opacity = 0.4] (-3.5,-2.75) circle (.3cm);
  \shade[ball color = olive!40, opacity = 0.4] (-4,-3.75) circle (.3cm);
 \shade[ball color = olive!40, opacity = 0.4] (-4.5,-4.75) circle (.3cm); 
  \shade[ball color = violet!120, opacity = 0.4] (-4.975,-4.75) circle (.02cm); 
 \shade[ball color = violet!120, opacity = 0.4] (-4.75,-3.75) circle (.2cm);
 \shade[ball color = violet!120, opacity = 0.4] (-4.55,-2.75) circle (.4cm);
  \shade[ball color = violet!120, opacity = 0.4] (-4.35,-1.65) circle (.6cm);
 \shade[ball color = violet!120, opacity = 0.4] (-4,0) circle (.95cm);
  \shade[ball color = violet!120, opacity = 0.4] (-5.075,-4.75) circle (.02cm); 
 \shade[ball color = violet!120, opacity = 0.4] (-5.25,-3.75) circle (.2cm);
 \shade[ball color = violet!120, opacity = 0.4] (-5.475,-2.75) circle (.4cm);
  \shade[ball color = violet!120, opacity = 0.4] (-5.65,-1.65) circle (.6cm);
 \shade[ball color = violet!120, opacity = 0.4] (-6,-0) circle (.95cm);
   \draw [line width=0.5mm,dashed]   (-5.05,0) arc (0:180:0.95 and .2);
  \draw [line width=0.5mm] (-5.05,0) arc (360:180:0.95 and .2);
  \draw[color=violet,thick] (-4.95,0) to (-3.95,0);
   \draw [line width=0.5mm,dashed]   (-3.05,0) arc (0:180:0.95 and .2);
  \draw [line width=0.5mm] (-3.05,0) arc (360:180:0.95 and .2);
   \draw[color=violet,thick] (-5.05,0) to (-6.05,0);
     \draw [very thick,dashed]   (-5.05,-1.65) arc (0:180:0.6 and .17);
  \draw [very thick] (-5.05,-1.65) arc (360:180:0.6 and .17);
    \draw[color=violet,thick] (-5.05,-1.65) to (-5.7,-1.65);
   \draw  [very thick,dashed]  (-3.75,-1.65) arc (0:180:0.6 and  .17);
  \draw [very thick] (-3.75,-1.65) arc (360:180:0.6 and .17);
     \draw[color=violet,thick] (-4.95,-1.65) to (-4.35,-1.65);
     \draw [very thick,dashed]   (-5.075,-2.75) arc (0:180:0.4 and .12);
  \draw [very thick] (-5.075,-2.75) arc (360:180:0.4 and .12);
   \draw[color=violet,thick] (-4.95,-2.75) to (-4.5,-2.75);
   \draw  [very thick,dashed]  (-4.15,-2.75) arc (0:180:0.4 and .12);
  \draw [very thick] (-4.15,-2.75) arc (360:180:0.4 and .12);
     \draw[color=violet,thick] (-5.05,-2.75) to (-5.5,-2.75);
       \draw [very thick,dashed]   (-5.05,-3.75) arc (0:180:0.2 and .06);
  \draw [very thick] (-5.05,-3.75) arc (360:180:0.2 and .06);
  \draw[color=violet,thick] (-5.05,-3.75) to (-5.25,-3.75);
   \draw [very thick,dashed]   (-4.55,-3.75) arc (0:180:0.2 and .06);
  \draw [very thick] (-4.55,-3.75) arc (360:180:0.2 and .06);
   \draw[color=violet,thick] (-4.95,-3.75) to (-4.75,-3.75);
     \draw [very thick,dashed]   (-5.05,-4.75) arc (0:180:0.01 and .005);
  \draw [very thick] (-5.05,-4.75) arc (360:180:0.01 and .005);
   \draw[color=violet,thick] (-4.95,-4.75) to (-4.94,-4.75);
   \draw [very thick,dashed]   (-4.95,-4.75) arc (0:180:0.01 and .005);
  \draw [very thick] (-4.95,-4.75) arc (360:180:0.01 and .005);
    \draw[color=violet,thick] (-5.05,-4.75) to (-5.06,-4.75);
     \draw   [dashed]  (-7.2,0) arc (0:180:0.3 and .1);
  \draw (-7.2,0) arc (360:180:0.3 and .1);
   \draw   [dashed]  (-2.2,0) arc (0:180:0.3 and .1);
  \draw (-2.2,0) arc (360:180:0.3 and .1);
     \draw   [dashed]  (-6.7,-1.65) arc (0:180:0.3 and .1);
  \draw (-6.7,-1.65) arc (360:180:0.3 and .1);
   \draw  [dashed]   (-2.7,-1.65) arc (0:180:0.3 and  .1);
  \draw (-2.7,-1.65) arc (360:180:0.3 and .1);
     \draw   [dashed]  (-6.2,-2.75) arc (0:180:0.3 and .1);
  \draw (-6.2,-2.75) arc (360:180:0.3 and .1);
   \draw    [dashed] (-3.2,-2.75) arc (0:180:0.3 and .1);
  \draw (-3.2,-2.75) arc (360:180:0.3 and .1);
       \draw   [dashed]  (-5.7,-3.75) arc (0:180:0.3 and .1);
  \draw (-5.7,-3.75) arc (360:180:0.3 and .1);
   \draw    [dashed] (-3.7,-3.75) arc (0:180:0.3 and .1);
  \draw  (-3.7,-3.75) arc (360:180:0.3 and .1);
     \draw   [dashed]  (-5.2,-4.75) arc (0:180:0.3 and .1);
  \draw (-5.2,-4.75) arc (360:180:0.3 and .1);
   \draw    [dashed] (-4.2,-4.75) arc (0:180:0.3 and .1);
  \draw  (-4.2,-4.75) arc (360:180:0.3 and .1);
  \draw (-4.975,0) arc (180:360:.975cm and 0.225cm) -- (-4.975,-4.75) -- cycle;
    \draw[dashed] (-4.935,0) arc (180:0:.975cm and 0.225cm);
    \shade[left color=blue!5!blue,right color=green!120!white,opacity=0.3] (-4.975,0) arc (180:360:.975cm and 0.25cm) -- (-4.975,-4.75) -- cycle; 
    \draw (-6.975,0) arc (180:360:.95cm and 0.225cm) -- (-5.075,-4.75) -- cycle;
    \draw[dashed] (-6.975,0) arc (180:0:.95cm and 0.225cm);
    \shade[left color=blue!5!green,right color=blue!120!white,opacity=0.3] (-6.975,0) arc (180:360:.95cm and 0.25cm) -- (-5.075,-4.75) -- cycle;
	\end{pgfonlayer}
\end{tikzpicture}
    \end{center}
\caption{The radial direction of the 3-balls in the pair $(\bket,\bbra)$,  depicted by the violet lines in the figure, are taken to be the  time direction. Splitting the resolved conifold at $t=0$ involves decomposing it about the red dashed line that pass through the middle of the pairs of 3-balls, resulting in two geometries $\rco_\pm$ as indicated. 
They are bounded by  codimension-1 constant time surfaces $\tsb_\pm$ (not shown explicitly) which comprise of an $\Sp^2$ of radius $r$ (boundaries of $\ball_\pm$, respectively), over which is fibered a unit-radius $\Sp^2$. The entangling branes $\entsurfx_\pm$ have topology $\Sp^2(1) \wtimes \Sp^1 \wtimes (\mathbb{R}_+ \cup \{0\})$ and will be discussed later in the text.
}
\label{fig:rdensityrescon1}
\end{figure}

Our next step is to specify a Cauchy surface in the resolved conifold. For this, we must identify  one of five directions comprising the base of the cone as the time direction. Happily, we can mimic what we did in the Chern-Simons discussion, i.e.,  we shall identify the radial direction of the pair of 3-balls $(\bket,\bbra)$ which are being identified along their boundaries to obtain the $\Sp^3$, as the time direction. Therefore, choosing a Heegaard splitting of all the $\Sp^3$s that are placed along the radial direction of the resolved conifold  into a 
pair of 3-balls can be understood as the splitting of the resolved conifold along a Cauchy surface $\tsb$. This choice divides the resolved conifold into two pieces $\rco_+$ and $\rco_-$, respectively. Let us for convenience resolve the Cauchy slice by opening it up about the constant time slice. We denote the  boundary of the first piece ($\rco_-$, which lies to the left  of  $\tsb$) as $\tsb_-$, and the boundary of the second  piece ($\mathscr{R_+}$, which lies to the right) as $\tsb_+$, respectively as illustrated in Fig.~\ref{fig:rdensityrescon1}. 

We should now argue that the decomposition of the resolved conifold into $\rco_\pm$ is meaningful from the closed topological string perspective. To do so, let us start by noting a remarkable property of the constant time surfaces $\tsb_\pm$. Neither of them cut through any of the non-contractible $\Sp^2$s in the resolved conifold -- including,  in particular,  the $\Sp^2$ at the origin $r=0$, which is the only non-trivial two-cycle in the resolved conifold. This has the following important consequence:  it allows us to conclude that the splitting of the resolved conifold as described is a consistent operation which doesn't slice  through any closed string configuration. 

To see this we invoke the toric picture of the resolved conifold (see Appendix~\ref{sec:conifold}). The fact that  $\tsb_\pm$ do not slice any of the homology two-cycle, can be interpreted in the toric picture as saying that they do not cut through any of the edges of the toric graph (where a certain fibre degenerates). Since using toric actions, all the topological closed string configurations can be  made to pass  through the edges of the toric diagram \cite{Kontsevich:1994na}, it follows that the surfaces $\tsb_\pm$ cut none of the closed strings. Hence we conclude that the splitting of the resolved conifold along the constant time surface $\tsb$ is a consistent operation to do from the viewpoint of topological closed string theory. In other words, we can consistently formulate topological closed string theory  in $\rco_\pm$, the spacetimes  obtained by cutting the resolved conifold $\rco$ along the constant time Cauchy surface $\tsb$.   

Since $\rco_\pm$ are manifolds with boundaries $\tsb_\pm$, respectively, the closed topological string field theory path integral over them does not compute a number but rather evaluates to a vector. This produces states in the Hilbert space obtained by quantizing the topological closed string theory on the Cauchy surface $\tsb$. Assume that the path integral over 
$\rco_-$ (without  any brane) produces the state $|\PhiR\rangle$ on  $\tsb_-$. The path integral over $\rco_+$ (without any brane) then produces the dual state 
$\langle\PhiR|$ on $\tsb_+$. Consequently, the inner product between these two states is given by fusing the two functional integrals together; i.e., it is given by the partition function 
$\zc(\rco)$ of the closed topological string theory on the resolved conifold:
\begin{equation}
\langle\PhiR|\PhiR\rangle= \zc(\rco).
\label{eq:inprodpartfu}
\end{equation}

Having constructed a state in the closed topological string Hilbert space, we can now proceed to ask how to construct a reduced density matrix on a subregion $\rxA \subset \tsb$. 
Recall that  $\tsb$ has the topology of a cone with base $\Sp^2(r) \wtimes \Sp^2(1)$ where we have now indicated the radii of the spheres for clarity. The $\Sp^2(r)$ is the boundary of the ball  $\ball$.  Away from tip of the cone $r=0$, it can be easily bipartitioned into two across the equator by picking the northern and southern hemispheres, $\rA(r)$, and $\rAc(r)$,  respectively. Each such $\rA(r)$ has the topology of a disc and we still have the $\Sp^2(1)$ fibred over this disk at each point along the ray parameterized by $r$.  At the tip of the cone, the ball $\ball$ has zero size, so there is nothing to bipartition.  Carrying this out for different radii, we get the desired decomposition of $\tsb$. We can then declare the region $\rxA$ of interest to be:  
\begin{equation}
\rxA = \bigg(\bigcup_{r\geq 0}  \, \rA(r)\bigg) \wtimes \Sp^2(1) \,.
\label{eq:rcAdef}
\end{equation}	
The codimension-2 \emph{entangling brane} at the boundary of $\rxA$ and $\rxAc$ is denoted as  $\entsurfx$. It has the topology of a cone, except that the tip at $r=0$ is now removed since we are not bipartitioning the shrunken 3-ball at that locus. The base of the cone is $\Sp^1(r) \times \Sp^2(1)$ and thus 
\begin{equation}
\entsurfx =  \bigg(\bigcup_{r> 0} \, \Sp^1(r)\bigg) \wtimes \Sp^2(1) \,. 
\label{eq:entrcA}
\end{equation}	

We  implement this  construction on both $\tsb_\pm$ and denote the decompositions thus obtained as $\tsb_- = \rxAk \cup \rxAck$ and $\tsb_+ =
\rxAb \cup \rxAcb$ and denote the associated entangling branes as $\entsurfx_\pm$, respectively. Now we are in a position to construct the topological closed string density matrix  
$\rho=|\PhiR\rangle\langle \PhiR|$, and the associated reduced density matrix $\rhoAx$  with respect to the chosen bipartition of $\tsb$.
The density matrix $\rho$ can be identified with the outer product of the topological closed string field theory path integral over  
$\rco_\mp$. Since we are taking the outer product we are not to identify the boundaries $\tsb_\pm$. 
The reduced density matrix $\rhoAx$ can be likewise be identified as the closed topological string field theory path integral over $\rco_\mp$ with the proviso that we identify these geometries along the complement i.e., we identify $\rxAck$ on $\tsb_-$ with   $\rxAcb$ on 
$\tsb_+$.\footnote{ We again remind the reader that we are only interested in extracting the topological contribution to entanglement. This allows us to avoid some further complications with defining what the nature of the entangling brane is, especially in relation to the closed string edge modes (see \S\ref{sec:discuss}).}

\subsection{Replica and R\'enyi entropies}
\label{sec:ctsrep}

Having at hand a functional integral definition of the reduced density matrix, we can now proceed to implement the replica construction. The computation of the R\'enyi entropy $S^{(q)}_{\rxA}$ requires  taking  $q$ copies of the path integral computing the reduced density matrix $\rhoAx$, and cyclically gluing them to produce the `branched cover' target spacetime $\rco_q$.  Recall that the gluing proceeds by identifying the  region $\rxAb$ from the $j^{\text{th}}$  copy  with the region $\rxAk$ from the $(j+1)^{\text{st}}$  copy. At the end of the day  the computation of $\text{Tr}_{\rxA}\left( \rhoAx^{q}\right)$  reduces  to computing the closed topological string theory partition function on the Calabi-Yau threefold $\rco_q$. 

We have however not yet normalized the density matrix $\rhoAx$. Therefore in performing the computation we actually need to compare the partition function on the target 
$\rco_q$ with $q$ copies of the result on the resolved conifold, cf., \eqref{eq:rentropypathint}. The key point to keep in mind is that the time interval on the `branched cover' replica geometry $\rco_q$ is  $q$ times that of the resolved conifold $\rco$. To facilitate direct comparison, it is useful to mimic the discussion in the physical gauge/gravity context \cite{Lewkowycz:2013nqa}. Let us therefore reinterpret $q$-fold product of the resolved conifold partition function $\zc(\rco)^q$, as the partition of closed topological string theory with target $\rco_{\oplus q}$ whose temporal extent is $q$ times that of the resolved conifold $\rco$.

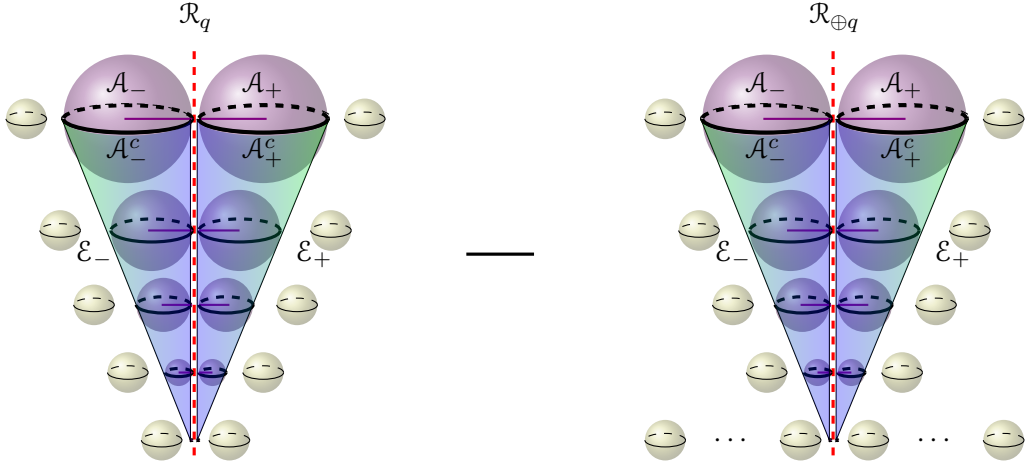
\begin{figure}
\begin{center}
\usetikzlibrary{backgrounds}
\begin{tikzpicture}[scale=.9]
\begin{pgfonlayer}{nodelayer}
		\node [style=none] (11) at (-2, 0) {};
		\node [style=none] (12) at (-2, -4.5) {};
		\node [style=none] (17) at (-5.02, 1) {};
		\node [style=none] (18) at (-5.02, -5) {};
		\node [style=none] (16) at (-8, 0) {};
		\node [style=none] (13) at (-8, -4.5) {};
		\node [style=none] at (-5, 1.5) {$\rco_{q}$};
		\node [style=none] at (-6, .5) {$\rxAk$};
		\node [style=none] at (-4, .5) {$\rxAb$};
		\node [style=none] at (-6, -.5) {$\rxAck$};
		\node [style=none] at (-4, -.5) {$\rxAcb$};
		\node [style=none] at (-6.5, -2) {${\entsurfx_-}$};
		\node [style=none] at (-3.25, -2) {${\entsurfx_+}$};
\end{pgfonlayer}
\begin{pgfonlayer}{edgelayer}
		\draw [very thick, style=dashed,color=red, in=90, out=-90] (17.center) to (18.center);
	    \shade[ball color = olive!40, opacity = 0.4] (-7.5,0) circle (.3cm); 
 		\shade[ball color = olive!40, opacity = 0.4] (-7,-1.65) circle (.3cm);
 		\shade[ball color = olive!40, opacity = 0.4] (-6.5,-2.75) circle (.3cm);
  		\shade[ball color = olive!40, opacity = 0.4] (-6,-3.75) circle (.3cm);
 		\shade[ball color = olive!40, opacity = 0.4] (-5.5,-4.75) circle (.3cm);		
	    \shade[ball color = olive!40, opacity = 0.4] (-2.5,0) circle (.3cm); 
 		\shade[ball color = olive!40, opacity = 0.4] (-3,-1.65) circle (.3cm);
 		\shade[ball color = olive!40, opacity = 0.4] (-3.5,-2.75) circle (.3cm);
 		\shade[ball color = olive!40, opacity = 0.4] (-4,-3.75) circle (.3cm);
 		\shade[ball color = olive!40, opacity = 0.4] (-4.5,-4.75) circle (.3cm);
	   \shade[ball color = violet!120, opacity = 0.4] (-4.975,-4.75) circle (.02cm); 
       \shade[ball color = violet!120, opacity = 0.4] (-4.75,-3.75) circle (.2cm);
	   \shade[ball color = violet!120, opacity = 0.4] (-4.55,-2.75) circle (.4cm);
 	    \shade[ball color = violet!120, opacity = 0.4] (-4.35,-1.65) circle (.6cm);
 \shade[ball color = violet!120, opacity = 0.4] (-4,0) circle (.95cm);
  \shade[ball color = violet!120, opacity = 0.4] (-5.075,-4.75) circle (.02cm); 
 \shade[ball color = violet!120, opacity = 0.4] (-5.25,-3.75) circle (.2cm);
 \shade[ball color = violet!120, opacity = 0.4] (-5.475,-2.75) circle (.4cm);
  \shade[ball color = violet!120, opacity = 0.4] (-5.65,-1.65) circle (.6cm);
 \shade[ball color = violet!120, opacity = 0.4] (-6,-0) circle (.95cm);
   \draw [line width=0.5mm,dashed]   (-5.05,0) arc (0:180:0.95 and .2);
  \draw [line width=0.5mm] (-5.05,0) arc (360:180:0.95 and .2);
  \draw[color=violet,thick] (-4.95,0) to (-3.95,0);
   \draw [line width=0.5mm,dashed]   (-3.05,0) arc (0:180:0.95 and .2);
  \draw [line width=0.5mm] (-3.05,0) arc (360:180:0.95 and .2);
   \draw[color=violet,thick] (-5.05,0) to (-6.05,0);
     \draw [very thick,dashed]   (-5.05,-1.65) arc (0:180:0.6 and .17);
  \draw [very thick] (-5.05,-1.65) arc (360:180:0.6 and .17);
    \draw[color=violet,thick] (-5.05,-1.65) to (-5.7,-1.65);
   \draw  [very thick,dashed]  (-3.75,-1.65) arc (0:180:0.6 and  .17);
  \draw [very thick] (-3.75,-1.65) arc (360:180:0.6 and .17);
     \draw[color=violet,thick] (-4.95,-1.65) to (-4.35,-1.65);
     \draw [very thick,dashed]   (-5.075,-2.75) arc (0:180:0.4 and .12);
  \draw [very thick] (-5.075,-2.75) arc (360:180:0.4 and .12);
   \draw[color=violet,thick] (-4.95,-2.75) to (-4.5,-2.75);
   \draw  [very thick,dashed]  (-4.15,-2.75) arc (0:180:0.4 and .12);
  \draw [very thick] (-4.15,-2.75) arc (360:180:0.4 and .12);
     \draw[color=violet,thick] (-5.05,-2.75) to (-5.5,-2.75);
       \draw [very thick,dashed]   (-5.05,-3.75) arc (0:180:0.2 and .06);
  \draw [very thick] (-5.05,-3.75) arc (360:180:0.2 and .06);
  \draw[color=violet,thick] (-5.05,-3.75) to (-5.25,-3.75);
   \draw [very thick,dashed]   (-4.55,-3.75) arc (0:180:0.2 and .06);
  \draw [very thick] (-4.55,-3.75) arc (360:180:0.2 and .06);
   \draw[color=violet,thick] (-4.95,-3.75) to (-4.75,-3.75);
     \draw [very thick,dashed]   (-5.05,-4.75) arc (0:180:0.01 and .005);
  \draw [very thick] (-5.05,-4.75) arc (360:180:0.01 and .005);
   \draw[color=violet,thick] (-4.95,-4.75) to (-4.94,-4.75);
   \draw [very thick,dashed]   (-4.95,-4.75) arc (0:180:0.01 and .005);
  \draw [very thick] (-4.95,-4.75) arc (360:180:0.01 and .005);
    \draw[color=violet,thick] (-5.05,-4.75) to (-5.06,-4.75);
     \draw   [dashed]  (-7.2,0) arc (0:180:0.3 and .1);
  \draw (-7.2,0) arc (360:180:0.3 and .1);
   \draw   [dashed]  (-2.2,0) arc (0:180:0.3 and .1);
  \draw (-2.2,0) arc (360:180:0.3 and .1);
     \draw   [dashed]  (-6.7,-1.65) arc (0:180:0.3 and .1);
  \draw (-6.7,-1.65) arc (360:180:0.3 and .1);
   \draw  [dashed]   (-2.7,-1.65) arc (0:180:0.3 and  .1);
  \draw (-2.7,-1.65) arc (360:180:0.3 and .1);
     \draw   [dashed]  (-6.2,-2.75) arc (0:180:0.3 and .1);
  \draw (-6.2,-2.75) arc (360:180:0.3 and .1);
   \draw    [dashed] (-3.2,-2.75) arc (0:180:0.3 and .1);
  \draw (-3.2,-2.75) arc (360:180:0.3 and .1);
       \draw   [dashed]  (-5.7,-3.75) arc (0:180:0.3 and .1);
  \draw (-5.7,-3.75) arc (360:180:0.3 and .1);
   \draw    [dashed] (-3.7,-3.75) arc (0:180:0.3 and .1);
  \draw  (-3.7,-3.75) arc (360:180:0.3 and .1);
     \draw   [dashed]  (-5.2,-4.75) arc (0:180:0.3 and .1);
  \draw (-5.2,-4.75) arc (360:180:0.3 and .1);
   \draw    [dashed] (-4.2,-4.75) arc (0:180:0.3 and .1);
  \draw  (-4.2,-4.75) arc (360:180:0.3 and .1);
  \draw (-4.975,0) arc (180:360:.975cm and 0.225cm) -- (-4.975,-4.75) -- cycle;
    \draw[dashed] (-4.935,0) arc (180:0:.975cm and 0.225cm);
    \shade[left color=blue!5!blue,right color=green!120!white,opacity=0.3] (-4.975,0) arc (180:360:.975cm and 0.25cm) -- (-4.975,-4.75) -- cycle;
    \draw (-6.975,0) arc (180:360:.95cm and 0.225cm) -- (-5.075,-4.75) -- cycle;
    \draw[dashed] (-6.975,0) arc (180:0:.95cm and 0.225cm);
    \shade[left color=blue!5!green,right color=blue!120!white,opacity=0.3] (-6.975,0) arc (180:360:.95cm and 0.25cm) -- (-5.075,-4.75) -- cycle;
    \draw[very thick] (-1,-2) -- (0,-2);
	\end{pgfonlayer}
\end{tikzpicture}
\hspace{1cm}
\begin{tikzpicture}[scale=.9]
\begin{pgfonlayer}{nodelayer}
		\node [style=none] (11) at (-2, 0) {};
		\node [style=none] (12) at (-2, -4.5) {};
			\node [style=none] (17) at (-5.02, 1) {};
		\node [style=none] (18) at (-5.02, -5) {};
				\node [style=none] (16) at (-8, 0) {};
		\node [style=none] (13) at (-8, -4.5) {};
		\node [style=none] at (-5, 1.5) {$\rco_{\oplus q}$};
		\node [style=none] at (-6.5, -4.75) {$\cdots$};
		\node [style=none] at (-3.5, -4.75) {$\cdots$};
		\node [style=none] at (-6, .5) {$\rxAk$};
		\node [style=none] at (-4, .5) {$\rxAb$};
		\node [style=none] at (-6, -.5) {$\rxAck$};
		\node [style=none] at (-4, -.5) {$\rxAcb$};
		\node [style=none] at (-6.5, -2) {${\entsurfx_-}$};
		\node [style=none] at (-3.25, -2) {${\entsurfx_+}$};
	\end{pgfonlayer}
	\begin{pgfonlayer}{edgelayer}
	\draw [very thick, style=dashed,color=red, in=90, out=-90] (17.center) to (18.center);
\shade[ball color = olive!40, opacity = 0.4] (-7.5,0) circle (.3cm); 
 \shade[ball color = olive!40, opacity = 0.4] (-7,-1.65) circle (.3cm);
 \shade[ball color = olive!40, opacity = 0.4] (-6.5,-2.75) circle (.3cm);
  \shade[ball color = olive!40, opacity = 0.4] (-6,-3.75) circle (.3cm);
 \shade[ball color = olive!40, opacity = 0.4] (-5.5,-4.75) circle (.3cm);
  \shade[ball color = olive!40, opacity = 0.4] (-7.5,-4.75) circle (.3cm);
 \shade[ball color = olive!40, opacity = 0.4] (-2.5,0) circle (.3cm); 
 \shade[ball color = olive!40, opacity = 0.4] (-3,-1.65) circle (.3cm);
 \shade[ball color = olive!40, opacity = 0.4] (-3.5,-2.75) circle (.3cm);
  \shade[ball color = olive!40, opacity = 0.4] (-4,-3.75) circle (.3cm);
 \shade[ball color = olive!40, opacity = 0.4] (-4.5,-4.75) circle (.3cm);
   \shade[ball color = olive!40, opacity = 0.4] (-2.5,-4.75) circle (.3cm);
  \shade[ball color = violet!120, opacity = 0.4] (-4.975,-4.75) circle (.02cm); 
 \shade[ball color = violet!120, opacity = 0.4] (-4.75,-3.75) circle (.2cm);
 \shade[ball color = violet!120, opacity = 0.4] (-4.55,-2.75) circle (.4cm);
  \shade[ball color = violet!120, opacity = 0.4] (-4.35,-1.65) circle (.6cm);
 \shade[ball color = violet!120, opacity = 0.4] (-4,0) circle (.95cm);
  \shade[ball color = violet!120, opacity = 0.4] (-5.075,-4.75) circle (.02cm); 
 \shade[ball color = violet!120, opacity = 0.4] (-5.25,-3.75) circle (.2cm);
 \shade[ball color = violet!120, opacity = 0.4] (-5.475,-2.75) circle (.4cm);
  \shade[ball color = violet!120, opacity = 0.4] (-5.65,-1.65) circle (.6cm);
 \shade[ball color = violet!120, opacity = 0.4] (-6,-0) circle (.95cm);
   \draw [line width=0.5mm,dashed]   (-5.05,0) arc (0:180:0.95 and .2);
  \draw [line width=0.5mm] (-5.05,0) arc (360:180:0.95 and .2);
  \draw[color=violet,thick] (-4.95,0) to (-3.95,0);
   \draw [line width=0.5mm,dashed]   (-3.05,0) arc (0:180:0.95 and .2);
  \draw [line width=0.5mm] (-3.05,0) arc (360:180:0.95 and .2);
   \draw[color=violet,thick] (-5.05,0) to (-6.05,0);
     \draw [very thick,dashed]   (-5.05,-1.65) arc (0:180:0.6 and .17);
  \draw [very thick] (-5.05,-1.65) arc (360:180:0.6 and .17);
    \draw[color=violet,thick] (-5.05,-1.65) to (-5.7,-1.65);
   \draw  [very thick,dashed]  (-3.75,-1.65) arc (0:180:0.6 and  .17);
  \draw [very thick] (-3.75,-1.65) arc (360:180:0.6 and .17);
     \draw[color=violet,thick] (-4.95,-1.65) to (-4.35,-1.65);
     \draw [very thick,dashed]   (-5.075,-2.75) arc (0:180:0.4 and .12);
  \draw [very thick] (-5.075,-2.75) arc (360:180:0.4 and .12);
   \draw[color=violet,thick] (-4.95,-2.75) to (-4.5,-2.75);
   \draw  [very thick,dashed]  (-4.15,-2.75) arc (0:180:0.4 and .12);
  \draw [very thick] (-4.15,-2.75) arc (360:180:0.4 and .12);
     \draw[color=violet,thick] (-5.05,-2.75) to (-5.5,-2.75);
       \draw [very thick,dashed]   (-5.05,-3.75) arc (0:180:0.2 and .06);
  \draw [very thick] (-5.05,-3.75) arc (360:180:0.2 and .06);
  \draw[color=violet,thick] (-5.05,-3.75) to (-5.25,-3.75);
   \draw [very thick,dashed]   (-4.55,-3.75) arc (0:180:0.2 and .06);
  \draw [very thick] (-4.55,-3.75) arc (360:180:0.2 and .06);
   \draw[color=violet,thick] (-4.95,-3.75) to (-4.75,-3.75);
     \draw [very thick,dashed]   (-5.05,-4.75) arc (0:180:0.01 and .005);
  \draw [very thick] (-5.05,-4.75) arc (360:180:0.01 and .005);
   \draw[color=violet,thick] (-4.95,-4.75) to (-4.94,-4.75);
   \draw [very thick,dashed]   (-4.95,-4.75) arc (0:180:0.01 and .005);
  \draw [very thick] (-4.95,-4.75) arc (360:180:0.01 and .005);
    \draw[color=violet,thick] (-5.05,-4.75) to (-5.06,-4.75);
     \draw   [dashed]  (-7.2,0) arc (0:180:0.3 and .1);
  \draw (-7.2,0) arc (360:180:0.3 and .1);
   \draw   [dashed]  (-2.2,0) arc (0:180:0.3 and .1);
  \draw (-2.2,0) arc (360:180:0.3 and .1);
     \draw   [dashed]  (-6.7,-1.65) arc (0:180:0.3 and .1);
  \draw (-6.7,-1.65) arc (360:180:0.3 and .1);
   \draw  [dashed]   (-2.7,-1.65) arc (0:180:0.3 and  .1);
  \draw (-2.7,-1.65) arc (360:180:0.3 and .1);
     \draw   [dashed]  (-6.2,-2.75) arc (0:180:0.3 and .1);
  \draw (-6.2,-2.75) arc (360:180:0.3 and .1);
   \draw    [dashed] (-3.2,-2.75) arc (0:180:0.3 and .1);
  \draw (-3.2,-2.75) arc (360:180:0.3 and .1);
       \draw   [dashed]  (-5.7,-3.75) arc (0:180:0.3 and .1);
  \draw (-5.7,-3.75) arc (360:180:0.3 and .1);
   \draw    [dashed] (-3.7,-3.75) arc (0:180:0.3 and .1);
  \draw  (-3.7,-3.75) arc (360:180:0.3 and .1);
     \draw   [dashed]  (-5.2,-4.75) arc (0:180:0.3 and .1);
  \draw (-5.2,-4.75) arc (360:180:0.3 and .1);
    \draw   [dashed]  (-7.2,-4.75) arc (0:180:0.3 and .1);
  \draw (-7.2,-4.75) arc (360:180:0.3 and .1);
   \draw    [dashed] (-4.2,-4.75) arc (0:180:0.3 and .1);
  \draw  (-4.2,-4.75) arc (360:180:0.3 and .1);
   \draw    [dashed] (-2.2,-4.75) arc (0:180:0.3 and .1);
  \draw  (-2.2,-4.75) arc (360:180:0.3 and .1); 
  \draw (-4.975,0) arc (180:360:.975cm and 0.225cm) -- (-4.975,-4.75) -- cycle;
    \draw[dashed] (-4.935,0) arc (180:0:.975cm and 0.225cm);
    \shade[left color=blue!5!blue,right color=green!120!white,opacity=0.3] (-4.975,0) arc (180:360:.975cm and 0.25cm) -- (-4.975,-4.75) -- cycle;
    \draw (-6.975,0) arc (180:360:.95cm and 0.225cm) -- (-5.075,-4.75) -- cycle;
    \draw[dashed] (-6.975,0) arc (180:0:.95cm and 0.225cm);
    \shade[left color=blue!5!green,right color=blue!120!white,opacity=0.3] (-6.975,0) arc (180:360:.95cm and 0.25cm) -- (-5.075,-4.75) -- cycle;
	\end{pgfonlayer}
\end{tikzpicture}
\end{center}
\caption{The difference between the manifolds $\rco_q$ and $\rco_{\oplus q}$ from the viewpoint of closed topological string theory is localized at the tip of the 
codimension-two surface $\entsurfx$ in them, obtained by identifying the surfaces $\entsurfx_-$ and $\entsurfx_+$. At the tip, where radial coordinate $r=0$, $\rco_q$ has only one $\Sp^2$, but $\rco_{\oplus q}$ has  $q$-number of $\Sp^2$s. }
\label{fig:rdensityrescon2}
\end{figure}

We are now left with determining the new target geometry $\rco_{\oplus q}$. We claim that  the manifold $\rco_{\oplus q}$ can be obtained by taking $q$ copies of 
$\rco$ and cyclically gluing them as follows. The $j^{\text{th}}$  copy of $\rco$ is glued to the $(j+1)^\text{st}$ copy by identifying the codimension-2 regions associated with the entangling branes.  We identify $\entsurfx_+$ (which is the boundary of $\rxAb$) from the $j^{\text{th}}$ copy  with $\entsurfx_-$ (which is the boundary of the region 
$\rxAk$) from the $(j+1)^{\text{st}}$  copy.  See Fig.~\ref{fig:rdensityrescon2} for an illustration of $\rco_q$ and $\rco_{\oplus q}$.

Let us verify the claim by demonstrating that we indeed get back the correct answer for the partition function, viz., 
$\zc(\rco_{\oplus q}) = \zc(\rco)^q$.   The striking feature of   topological string theory is that the physical quantities  are sensitive only to the topology of the target space. 
More precisely, the partition function depends only on the non-trivial cycles and the Chern classes of  the target. We elaborate on this and provide some details in Appendix~\ref{sec:GWtheory}.

Let us therefore identify the non-trivial cycles in $\rco_{\oplus q}$ which should aid our computation of $\zc(\rco_{\oplus q})$.
The geometry  $\rco_{\oplus q}$ has $q$ pairs of  codimension-2 entangling branes $(\entsurfx_-,\entsurfx_+)$ that are glued together as follows. 
Within each pair, $\entsurfx_-$ is identified with $\entsurfx_+$ so that we make up the resolved conifold. In addition, the cyclic gluing of $\entsurfx_+$ in the  $j^{\text{th}}$  pair   with 
$\entsurfx_-$ in  the $(j+1)^{\rm st}$ pair is necessary to grow the time direction by a factor of $q$.  

Recall that the only non-trivial cycle in $\rco$ is the $\Sp^2$ at the locus $r=0$, which corresponds to the tip of the cone in the resolved conifold. The codimension-2 entangling surfaces $\entsurfx_{\pm}$ have the topology of an infinite cone with the point at the tip removed,  each point carrying an $\Sp^2$ of unit radius.  Since the surfaces $\entsurfx_\pm$ do not intersect with any of the non-trivial cycles in $\rco$, the gluing does not introduce any new compact cycles in  $\rco_{\oplus q}$, on which the closed topological string worldsheets can wrap. However, each $\rco$ has a non-trivial two-cycle, which is the $\Sp^2(1)$ at the origin. As a result  $\rco_{\oplus q}$, which is obtained gluing $q$ copies  of  $\rco$, has $q$ number of  non-trivial two-cycles, all having the same K\"ahler parameter $t$ as in the resolved conifold.

Armed with this information, we can analyze the Chern class of $\rco_{\oplus q}$. The manifold $\rco_{\oplus q}$ is a cyclic $q$-sheeted covering of the resolved conifold
 $\rco$. Interestingly, it is possible to express the Chern class $\widetilde{\mathbf{c}}$ of the covering space $\rco_{\oplus q}$  in terms of the Chern class of each sheet. Let us denote the Chern class of the $j^{\text{th}}$ sheet as $\mathbf{c}^j$. Then the Chern class of $\rco_{\oplus q}$ is given by the product of the classes on the sheets, up to a residue term: 
\begin{equation}
\widetilde{\mathbf{c}}=\mathbf{c}^1\times \cdots \times \mathbf{c}^q+\mathbf{e},
\label{eq:chern}
\end{equation}
The residue term $\mathbf{e}$ has the property that it vanishes if the first Chern class $c_1$ of $\rco$ vanishes \cite{Evens:1978che,Fulton:1987cha}. 
Since, $\rco$ is a Calabi-Yau manifold, its first Chern class vanishes. Hence, we have 
\begin{equation}
\widetilde{\mathbf{c}}=\mathbf{c}^1\times \cdots \times \mathbf{c}^q.
\label{eq:chernrq}
\end{equation}

Therefore, from the viewpoint of topological closed string theory, $\rco_{\oplus q}$ effectively behaves as $q$ copies of $\rco$ that are not identified. 
As a result, the partition function of the closed topological string theory with target space  $\rco_{\oplus q}$ is $q^{\text{th}}$ power of the partition function  of the closed topological string theory with target space  $\rco$.

\begin{figure}[ht]
\begin{center}
\usetikzlibrary{backgrounds}
\begin{tikzpicture}[scale=.6]
\begin{pgfonlayer}{nodelayer}
		\node [style=none] (0) at (-7, 5) {};
		\node [style=none] (1) at (-6, 4) {};
		\node [style=none] (2) at (-7, 3) {};
		\node [style=none] (3) at (-4, 4) {};
		\node [style=none] (4) at (-3, 5) {};
		\node [style=none] (5) at (-3, 3) {};
		\node [style=none] (6) at (-2, 2) {};
		\node [style=none] (7) at (-1, 1) {};
		\node [style=none] (8) at (0, 2) {};
		\node [style=none] (9) at (-1, -1) {};
		\node [style=none] (10) at (-2, -2) {};
		\node [style=none] (11) at (0, -2) {};
		\node [style=none] (12) at (-3, -3) {};
		\node [style=none] (13) at (-4, -4) {};
		\node [style=none] (14) at (-3, -5) {};
		\node [style=none] (15) at (-6, -4) {};
		\node [style=none] (16) at (-7, -5) {};
		\node [style=none] (17) at (-7, -3) {};
		\node [style=none] (18) at (-8, -2) {};
		\node [style=none] (19) at (-9, -1) {};
		\node [style=none] (20) at (-10, -2) {};
		\node [style=none] (21) at (-9, 1) {};
		\node [style=none] (22) at (-8, 2) {};
		\node [style=none] (23) at (-10, 2) {};
		\node [style=none] (24) at (-5.25, 4.5) {};
		\node [style=none] (25) at (-5.25, 4.5) {};
		\node [style=none] (26) at (-5.25, 4.5) {$\Sp^2(t)$};
		\node [style=none] (27) at (-1.75, 3) {$\Sp^2(\infty)$};
		\node [style=none] (29) at (-0.25, 0) {$\Sp^2(t)$};
		\node [style=none] (31) at (-1.75, -3) {$\Sp^2(\infty)$};
		\node [style=none] (32) at (-5, -4.5) {$\Sp^2(t)$};
		\node [style=none] (33) at (-8.5, -2.75) {$\Sp^2(\infty)$};
		\node [style=none] (34) at (-8.5, 3) {$\Sp^2(\infty)$};
		\node [style=none] (35) at (-9.75, 0) {$\Sp^2(t)$};
	\end{pgfonlayer}
	\begin{pgfonlayer}{edgelayer}
		\draw [thick] (0.center) to (1.center);
		\draw [thick] (2.center) to (1.center);
		\draw [thick] (1.center) to (3.center);
		\draw [thick] (3.center) to (4.center);
		\draw [thick] (3.center) to (5.center);
		\draw [thick, style=dashed] (5.center) to (6.center);
		\draw [thick] (6.center) to (7.center);
		\draw [thick] (7.center) to (8.center);
		\draw [thick] (7.center) to (9.center);
		\draw [thick] (9.center) to (10.center);
		\draw [thick] (9.center) to (11.center);
		\draw [thick, style=dashed] (12.center) to (10.center);
		\draw [thick] (17.center) to (15.center);
		\draw [thick] (15.center) to (16.center);
		\draw [thick] (15.center) to (13.center);
		\draw [thick] (13.center) to (12.center);
		\draw [thick] (13.center) to (14.center);
		\draw [thick, style=dashed] (22.center) to (2.center);
		\draw [thick, style=dashed] (18.center) to (17.center);
		\draw [thick] (21.center) to (22.center);
		\draw [thick] (21.center) to (23.center);
		\draw [thick] (21.center) to (19.center);
		\draw [thick] (19.center) to (18.center);
		\draw [thick] (19.center) to (20.center);
	\end{pgfonlayer}
\end{tikzpicture}
\end{center}
\caption{Toric diagram for $\rco_{\oplus q}$ obtained by connecting the toric diagrams of resolved conifolds, each with an $\Sp^2$ having K\"ahler parameter $t$, by connecting their vertices using infinitely long lines. The dashed lines are meant to represent infinity long two-cycles, i.e.,  $\Sp^2$s having $t=\infty$. Partition function of closed topological A-model string theory on the Calabi-Yau threefold represented by this toric diagram is given by $ \zc(\rco)^q$. We have illustrated the $q=4$ example above. }
\label{fig:rdensityrescontoric}
\end{figure}
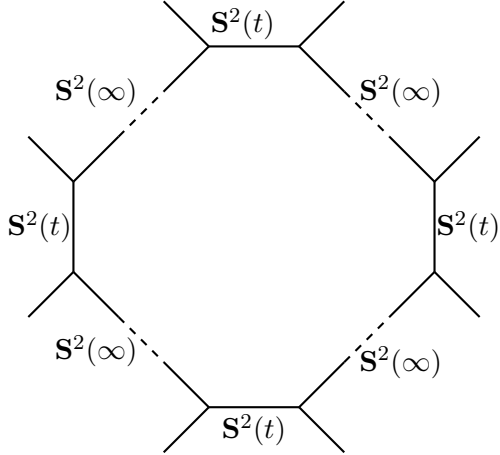

This result matches with the topological vertex rules. It is known that if the length of an edge in a toric diagram representing a Calabi-Yau threefold $\xf$ is infinite then the partition function of the topological closed string theory on $\xf$ factorizes into the partition functions of topological closed string theories on the manifolds represented by the toric diagrams obtained by cutting across  infinitely long edge.
This suggests that the toric diagram of  $\rco_{\oplus q}$ can be obtained by cyclically connecting $q$ copies of a toric diagram representing $\rco$ via infinitely long edges, see Fig.~\ref{fig:rdensityrescontoric}.  The infinitely long edges in the toric diagram represents $q$-number of $\Sp^2$s having infinite K\"ahler parameters in $\rco_{\oplus q}$. Let us examine these $\Sp^2$s in $\rco_{\oplus q}$. We constructed $\rco_{\oplus q}$ by gluing $q$-number of $\rco$s  along the entangling surfaces $\entsurfx_{\pm}$s. The entangling surface $\entsurfx_{\pm}$ has the topology $\mathscr{K}_{\pm}\times \Sp^2$ where $\mathscr{K}_{\pm}$ is an infinitely long  cone with the point at the tip removed. Then the $\Sp^2$s  having infinite K\"ahler parameters can be identified with the two-cycles obtained by considering the union of  $\mathscr{K}_{-}$ and $\mathscr{K}_{+}$  of adjacent $\rco$s. Therefore, 
\begin{equation}
\zc(\rco_{\oplus q}) = \zc(\rco)^q
\label{eq:qfold}
\end{equation}	

We are finally ready to compute the R\'enyi entropy $S_\rxA^{(q)}$, by finding the difference between the (logarithms of) the partition functions of the closed topological string theory on 
$\rco_q$ and $\rco_{\oplus q}$ (and then normalizing the answer with $1-q$). The difference can be computed by directly comparing the topology of the two spacetimes. Both the spacetimes are obtained by cyclically gluing $q$ copies of $\rco$. In $\rco_q$ the identification is along the region $\rxA$. This region has the topology of solid cone with an $\Sp^2(1)$ at each point of the solid cone. The tip of the solid cone carries this $\Sp^2(1)$, which is the non-trivial two-cycle in the resolved conifold. 
As a result, when we glue $q$ copies of $\rco$ to obtain  $\rco_q$, we identify all the non-trivial $\Sp^2$s from each copy of the resolved conifold. 
Therefore,  $\rco_q$ has only one non-trivial two-cycle. Moreover, the Chern class of $\rco_q$ is same as that of $\rco$ using \eqref{eq:chernrq}. 
Hence, from the viewpoint of closed topological string theory, $\rco_q$ is same as the resolved conifold $\rco$. This is clear for the constant map from the worldsheet. However, for non-constant maps we have not explicitly verified that there are no new Gromov-Witten invariants from non-trivial wrappings.\footnote{ We thank Xi Yin for raising this point.} At the very least our construction works at large $N$ with the aforementioned subtlety contributing  at most to  $\alpha'$ corrections to the relation $\zc(\rco_q) = \zc(\rco)$.

The conclusion is that the difference between the manifolds $\rco_q$ and $\rco_{\oplus q}$ from the viewpoint of closed topological string theory is localized at the tip of the codimension-two surface $\entsurfx$ in them, obtained by identifying the surfaces $\entsurfx_-$ and $\entsurfx_+$. At the tip, where radial coordinate $r=0$, $\rco_q$ has only one 
$\Sp^2$, but $\rco_{\oplus q}$ has  $q$-number of $\Sp^2$s. We have attempted to illustrate this in Fig.~\ref{fig:rdensityrescon2} and also provided a toric perspective of the geometry in Fig.~\ref{fig:rdensityrescontoric}. This implies that 
\begin{equation}
\begin{split}
S_\rxA^{(q)}  
&= \frac{1}{1-q} \log \left[\zc(\rco_q) - \log \zc(\rco_{\oplus q}) \right] \\
&= \frac{1}{1-q} \log \left[\zc(\rco) - q \,  \log \zc(\rco) \right] \\
& = \log \zc(\rco)
\end{split}
\label{eq:Renyibulk}
\end{equation}
This is of course what we expect from the point of view of the open/closed topological string duality as we know that $\zc(\rco) = \zcs(\Sp^3)$.

\section{Generalizations to other states}
\label{sec:genPsi}

Our discussion thus far has been confined to entanglement properties of the state $\psiSk$ which we uplifted onto the state $\ket{\PhiR}$ of the closed string theory on the resolved conifold.
While we focused on spatial bipartitions of this state, it should be clear that the closed string story for decomposing the spatial domain into multiple disconnected regions parallels the Chern-Simons discussion.  Similarly, the perspective that surgery commutes through the geometric transition, implies that we can consider other states of the Chern-Simons theory and determine the analogous picture in the resolved conifold. We will briefly discuss some of these generalizations below, explaining in detail the surgery on the Chern-Simons side, and indicating the necessary changes for the closed string on the resolved conifold.

\subsection{Entanglement on a Riemann surface}
\label{sec:eeRs}  

In \S\ref{sec:csent} we understood the situation when we decompose the $\Sp^3$ into two $3$-balls to define $\psiSk$. Suppose instead that we are now interested in $\Sigma_g$, a  codimension-1 Cauchy slice in $\Sp^3$ with a non-trivial topology, viz.,  a Riemann surface with $g$ handles. Unlike $\mathcal{H}_{\Sp^2}$, the Hilbert space $\mathcal{H}_{\Sigma_g}$ for $\Sigma_g$ contains more than one independent element. As described in \S\ref{sec:canq} these states are obtained by performing the Chern-Simons path integral on the handlebody  
$\mcs_{\Sigma_g}$ bounded by $\Sigma_g$ with Wilson loops placed along the non-contractible cycles of $\mcs_{\Sigma_g}$.

\begin{figure}[h]
\begin{center}
\usetikzlibrary{backgrounds}
\begin{tikzpicture}[scale=.75]
	\begin{pgfonlayer}{nodelayer}
		\node [style=none] (0) at (-11.5, 3.25) {};
		\node [style=none] (1) at (-5.75, 3.25) {};
		\node [style=none] (2) at (-10, 3.25) {};
		\node [style=none] (3) at (-7.25, 3.25) {};
		\node [style=none] (9) at (-5.25, 3.25) {};
		\node [style=none] (11) at (-9.5, 2.75) {};
		\node [style=none] (12) at (-8, 2.75) {};
		\node [style=none] (13) at (-4, 2.5) {};
		\node [style=none] (14) at (-9.5, 4) {};
		\node [style=none] (15) at (-7.75, 4) {};
		\node [style=none] (16) at (-6, 4.25) {};
		\node [style=none] (17) at (-8.25, 5.5) {};
		\node [style=none] (18) at (-7.5, 1.25) {};
		\node [style=none] (19) at (-9, 1) {};
		\node [style=none] (20) at (-5.25, 2.5) {};
	\end{pgfonlayer}
	\begin{pgfonlayer}{edgelayer}
		\draw [very thick,bend right=90, looseness=1.25] (0.center) to (1.center);
		\draw [very thick,in=-120, out=-75] (2.center) to (3.center);
		\draw [very thick,bend left=90, looseness=2.00] (11.center) to (13.center);
		\draw [very thick,bend left=90, looseness=1.75] (12.center) to (9.center);
		\draw [very thick,in=75, out=-15] (15.center) to (3.center);
		\draw [very thick,bend left=15] (16.center) to (1.center);
		\draw [very thick,in=90, out=180] (17.center) to (0.center);
		\draw [very thick,in=105, out=-165, looseness=0.75] (14.center) to (2.center);
		\draw [very thick,in=-75, out=-45, looseness=1.25] (18.center) to (9.center);
		\draw [very thick,in=-90, out=-60, looseness=1.50] (19.center) to (13.center);
		\draw [very thick,style=dashed, bend left=45, looseness=0.75] (20.center) to (13.center);
		\draw [very thick,bend right=75, looseness=0.75] (20.center) to (13.center);
		\draw [very thick,bend right=105, looseness=0.50] (11.center) to (12.center);
		\draw [very thick,style=dashed, bend left, looseness=0.75] (11.center) to (12.center);
		\draw [very thick,style=dashed, bend right=90, looseness=1.25] (0.center) to (2.center);
		\draw [very thick,bend left=75] (0.center) to (2.center);
		\draw [very thick,style=dashed, bend right=90, looseness=1.25] (3.center) to (1.center);
		\draw [very thick,bend left=75] (3.center) to (1.center);
	\end{pgfonlayer}
	\filldraw[fill=violet!20] (-5.2,1) ellipse (.6 and .5);
	\filldraw[fill=violet!20] (-7.5,2) ellipse (.6 and .5);
	\node at (-7.5,2)  {$\rAk$};
	\node at (-5.2,1){$\rAb$};
	\node at (-9.5,2)  {$\rAck$};
	\node at (-6.5,0.25){$\rAcb$};
	\shade[ball color = violet!40, opacity = 0.4] (2,3) circle (2cm);
  \draw (2,3) circle (2cm);
  \draw [dashed]  (2,1) arc (-90:90:0.6 and 2);
  \draw (2,5) arc (90:270:0.6 and 2);
  \node at (1,3)  {$\rAk$};
	\node at (2.95,3){$\rAb$};
	\draw[very thick,->](-3,3)--(-1,3);
\end{tikzpicture}
\end{center}
\caption{The $\rhoA$ for torus is obtained by gluing the path integrals over the two solid tori by identifying the boundary regions $\rAck$ and $\rAcb$ on the boundary tori. The resulting path integral is same as the reduced density matrix $\rhoA$ for  Riemann sphere. }
\label{fig:rdensitymtorus}
\end{figure}
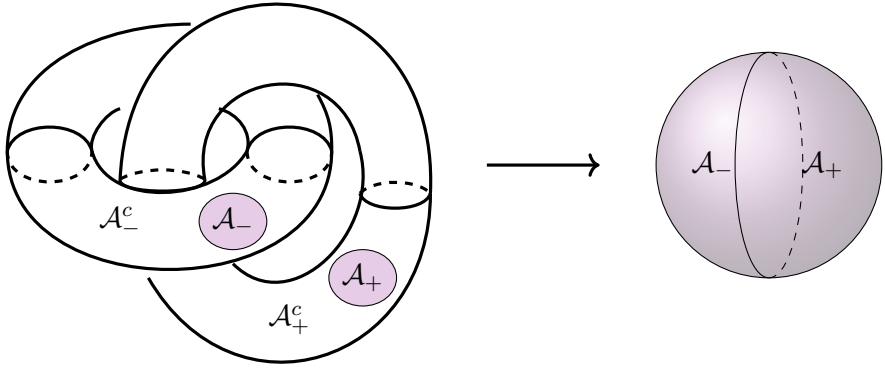

Consider the state $|\Psi_{g}\rangle \in \mathcal{H}_{\Sigma_g}$ that corresponds to performing the path integral over $\mcs_{\Sigma_g}$ with no Wilson loops placed along its non-contractible cycles. Let us compute the entanglement entropy of $|\Psi_{g}\rangle $ for the bipartition of $\Sigma_g$ into two regions $\rA$ and $\rAc$ with $M$ number of disconnected interfaces between them. We illustrate in Fig.~\ref{fig:rdensitymtorus} the state prepared on a solid torus obtained by slicing open, in an alternate manner, the $\Sp^3$ partition function. 

Assume that the states 
\begin{equation}
\begin{split}
&|\phi^{[\alpha]}_\mu \rangle, \;  \mu=1,\cdots, \text{dim}\left(\mathcal{H}_{A}^{[\alpha]}\right)  \\
& |\chi^{[\alpha]}_\nu \rangle,\;  \nu=1,\cdots, \text{dim}\left(\mathcal{H}_{\rAc}^{[\alpha]}\right)
\end{split}
\label{eq:ssbasis}
\end{equation}
 form a basis for the $\alpha^{\text{th}}$ superselection sectors
of the Hilbert spaces  $\mathcal{H}_{\rA}$ and $\mathcal{H}_{\rAc}$, respectively.  The set $\left\{|\phi^{[\alpha]}_\mu \rangle \otimes |\chi^{[\alpha]}_\nu \rangle\right\}$ 
satisfies the Gauss law constraints and form a basis for $\mathcal{H}_{\Sigma_g}$. We can therefore express the state $|\Psi_g\rangle$ as
\begin{equation}
|\Psi_g\rangle=\sum_{\alpha} \, \sum_{\mu,\nu}\; \mathfrak{c}^{[\alpha]}_{\mu\nu} \  |\phi^{[\alpha]}_\mu \rangle \otimes | \chi^{[\alpha]}_\nu\rangle \,.
\end{equation} 
If the state $|\Psi_g\rangle$ corresponds to one where no Wilson loops are inserted in $\mcs_{\Sigma_g}$, then the r.h.s.\ of above expression receives contributions only from the trivial superselection sector, i.e., $\mathfrak{c}^{[\alpha]}_{\mu\nu} =0$ for $\alpha\neq 0$.
 
The state $\langle\Psi_g|$  conjugate to the state $|\Psi_g\rangle$ has also a  nice path integral description. We again invoke the Heegard splitting theorem for 3-manifolds. 
Applied to  $\Sp^3$ it asserts that an $\Sp^3$ can be obtained gluing two handlebodies $\mcs^+_{_{\Sigma_g}}$ and 
$\mcs^-_{_{\Sigma_g }}$ by identifying their boundaries which have the same topology as that of $\Sigma_g$  \cite{Prasolov:1997kno}. We have already encountered the simpler versions of this statement in \S\ref{sec:H01summary}.   If we identify the state $|\Psi_g\rangle$ with the Chern-Simons path integral in the handlebody  $\mcs^-_{_{\Sigma_g}}$, then we identify the dual state $\langle \Psi_g|$ with the path integral in  $\mcs^+_{_{\Sigma_g}}$. As a result, we have by a similar reasoning that led to \eqref{eq:innprodzs3}, 
\begin{equation}
\langle\Psi_g|\Psi_g\rangle=\zcs(\Sp^3).
\label{eq:psig}
\end{equation}
This translates into the following normalization condition for the variables $\mathfrak{c}^{[0]}_{\mu\nu}$
\begin{equation}
\sum_{\mu,\nu} \big| \mathfrak{c}^{[0]}_{\mu\nu} \big|^2=\zcs(\Sp^3).
\label{eq:norm}
\end{equation}
We can therefore define the  normalized reduced density matrix $\rhoAh$  by
\begin{equation}
\rhoAh=\frac{\rho^{[0]}_\rA}{\zcs(\Sp^3)}=
\frac{1}{\zcs(\Sp^3)} \; \sum_{\nu} \, \langle \chi^{[0]}_\nu |\Psi_g\rangle \; \langle \Psi_g| \chi^{[0]}_\nu \rangle.
\end{equation}   

Armed with this information we are in position to compute the $q^{\text{th}}$ R\'enyi entropy of the state $|\Psi_g\rangle$ for our bipartition  $\Sigma_g=\rA \cup \rAc$. We obtain
\begin{equation}
\begin{split}
S_\rA^{(q)} &= \frac{1}{1-q}\, \log \text{Tr}_{\rA}\left(\rhoAh^{\,q} \right) \\
&= \frac{1}{1-q} \log \left[\frac{1}{\zcs(\Sp^3)^{q}} \; \text{Tr}_{\rA}\left(\sum_{\nu,\mu_1,\mu_2}  \mathfrak{c}^{[0]}_{\mu_1\nu}( \mathfrak{c}^{[0]}_{\mu_2\nu})^*
|\phi^{[0]}_{\mu_1} \rangle\; \langle \phi^{[0]}_{\mu_2}| \right)^q \; \right] \\
&=M\; \log \zcs(\Sp^3) \,.
\end{split}
\label{eq:renSigma}
\end{equation}
Here we used the normalization \eqref{eq:norm} and the following property of the $ \mathfrak{c}^{[0]}_{\mu\nu}$s:
\begin{equation}
\sum_{\mu_1,\nu_1} \mathfrak{c}^{[0]}_{\mu \nu_1}( \mathfrak{c}^{[0]}_{\mu_1\nu_1})^* \mathfrak{c}^{[0]}_{\mu_1\nu}=\frac{ \mathfrak{c}^{[0]}_{\mu\nu}}{\left[ \zcs(\Sp^3)\right]^{M-1}}\,.
\end{equation}
These relations are induced directly from the action of the reduced density matrix on the state $|\Psi_g\rangle$
\begin{equation}
\rhoA|\Psi_g\rangle=\frac{1}{\left[ \zcs(\Sp^3)\right]^{M-1}}|\Psi_g\rangle.
\end{equation}
 which is a direct consequence of the Heegard splitting of $\Sp^3$ into two handle bodies. The fact that it fails to depend on the genus of the Riemann surface (see Eq.~\eqref{eq:rhoaonpsiM} for a comparison), reflects again the underlying topological nature. We once again find that the  $q^{\text{th}}$ Reyni entropy $S_\rA^{(q)}$ is independent of $q$, and equals the von Neumann entropy.

To understand this construction in the resolved conifold, we invoke the following algorithm:
\begin{itemize}
\item Given a Heegard splitting of $\Sp^3$, we uplift this to a topological decomposition of $T^*\Sp^3$. This decomposition leaves untouched the $\Sp^2$ and gives rise to Cauchy surfaces 
of the bulk geometry which have topology $\Sigma_g^\pm \times \Sp^2 \times (\mathbb{R}_+ \cup {0})$.  
\item We follow this decomposition of $\Sp^3$ through the geometric transition, obtaining a similar topological decomposition of the resolved conifold into two components $\rco_\pm$. This again can be used to define the closed string states $\ket{\!\Phi_g}$ which live on $\Gamma_\pm$ whose topology is described above. The only difference is that for the resolved conifold there is no $\Sp^3$ at the tip of the cone to decompose. The reader is invited to visualize Fig.~\ref{fig:rdensitydefcontorescon} with the $\ball_\pm$ replaced by the interlocked three manifolds $\mcs^\pm_{_{\Sigma_g}}$.
\end{itemize}

Once this is done, we are in a position to  bipartition $\Gamma_\pm$ into $\rxA_\pm \cup \rxAc_\pm$ and carry through the replica analysis. We would construct the analog the manifolds $\rco_q$ and $\rco_{\oplus q}$ and compute the partition functions thereupon as before \eqref{eq:Renyibulk}. The closed string partition functions will still only care about the homology two-cycles 
in the resulting geometry, and we recover \eqref{eq:renSigma}.

\subsection{Entanglement in the presence of Wilson lines }
\label{sec:eeWs}

All states of Chern-Simons on $\Sigma_g$ without Wilson lines end up giving a simple answer for the entanglement entropy, which only depends on the number of disconnected components of the entangling surface $\entsurf$ and the three-sphere partition function. We can consider other states, which contain Wilson lines along non-contractible cycles of $\mcs_{_{\Sigma_g}}$. It will turn out that the answer for entanglement measures will depend on whether or not our bipartitioning  slices through the Wilson lines. This makes physical sense for reasons explained in \S\ref{sec:rhoCS}, for we have to sum over the appropriate set of superselection sectors, which depends on the presence/absence of charged states on the boundary. 

Let us first review the construction of the reduced density matrix in Chern-Simons theory before turning again to the closed string side of the story. Say we want to understand the entanglement in the state $\psiTk$ defined in Fig.~\ref{fig:torusstate}. We prepare the state as explained there by performing the path integral over the solid torus with a Wilson line in representation $R_i$. We need to pick the conjugate state to prepare the density operator. A natural choice would be picked by the state that sets the normalization to unity, but this will  not be convenient for our purposes.\footnote{ We understand how to map Chern-Simons theory on $\Sp^3$ onto the topological closed string on the resolved conifold, but do not have an analogous statement for Chern-Simons theory on $\Sp^2 \times \Sp^1$ (see \S\ref{sec:discuss}). The computations described in \cite{Dong:2008ft} for topological entanglement entropy exploit the latter.}
We will instead pick the state $\psiTb$ to be obtained by performing the functional integral over another solid torus, also with a Wilson line in the same representation $R_i$ inserted along its non-contractible cycle. However, we will take the non-contractible cycles to be related by an S-transformation (say it is the $b$-cycle for the ket solid torus $\mcs^-_{_{{\bf T}^2}}$, and the $a$-cycle for the bra solid torus $\mcs^+_{_{{\bf T}^2}}$), so that when we take the inner product after identifying the boundaries, we end up with a Chern-Simons partition function on a three-sphere with a link $L$, linking two unknots, each carrying the representation $R_i$.
\begin{equation}\label{eq:wTnorm}
\psiTkb = \zcs(\Sp^3, L;R_i) \,.
\end{equation}	
We will also have occasion to revisit states prepared on the three-ball, but for now we will proceed with the state described above.

We can consider bipartitioning the ${\bf T}^2$ into $\rA \cup \rAc$, but we have a choice whether or not the Wilson line is cut in the process of bipartitioning. There are two basic cases to consider:
\begin{itemize}
\item The bipartition is achieved without interfering with any Wilson line. Technically, this requires that the region $\rA$ we identify on the boundary of the handlebody, can be separated from the handlebody by scooping out a domain that does not contain the Wilson lines. For example, for $\rA$ having the topology of a disc, we scoop out a half-ball which is not intersecting the Wilson line placed inside the handlebody.
\item Alternately, the bipartitioning slices through the Wilson line insertion, either because we cannot separate $\rA$ from $\rAc$ without cutting through the Wilson line, or because an open Wilson line ends in the interior of $\rA$ (or $\rAc$). 
\end{itemize}

\subsubsection{Bipartitions avoiding Wilson lines}
\label{sec:wbipart1}

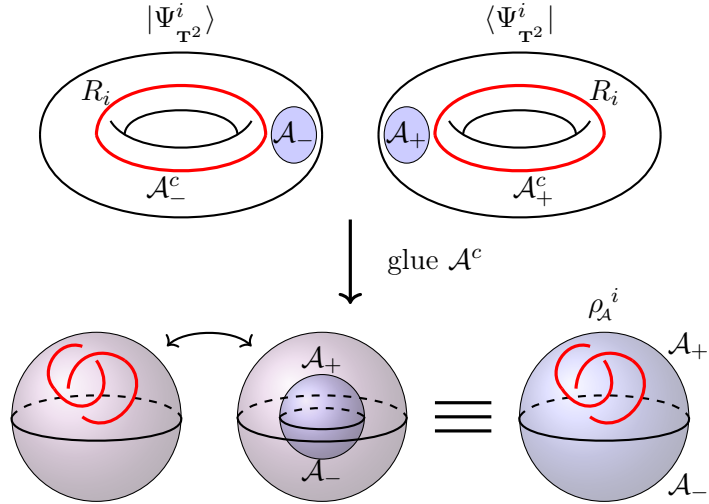
\begin{figure}[h]
\begin{center}
\usetikzlibrary{backgrounds}
\begin{tikzpicture}[scale=.75]
  \begin{pgfonlayer}{nodelayer}
    \node [style=none] (0) at (-12, 4) {};
    \node [style=none] (1) at (-7, 4) {};
    \node [style=none] (2) at (-10.5, 4) {};
    \node [style=none] (3) at (-8.5, 4) {};
    \node [style=none] (4) at (-10.75, 4.25) {};
    \node [style=none] (5) at (-8.25, 4.25) {};
    \node [style=none] (6) at (-11, 4) {};
    \node [style=none] (7) at (-8, 4) {};
    \node [style=none] (8) at (-6, 4) {};
    \node [style=none] (9) at (-1, 4) {};
    \node [style=none] (10) at (-4.5, 4) {};
    \node [style=none] (11) at (-2.5, 4) {};
    \node [style=none] (12) at (-4.75, 4.25) {};
    \node [style=none] (13) at (-2.25, 4.25) {};
    \node [style=none] (14) at (-5, 4) {};
    \node [style=none] (15) at (-2, 4) {};
    \node [style=none] (16) at (-5.75, 4) {};
    \node [style=none] (17) at (-7.5, 4) {};
    \node [style=none] (20) at (-3.25, 3) {$\rAc_+$};
    \node [style=none] (21) at (-9.75, 3) {$\rAc_-$};
    \node [style=none] (22) at (-7.5, 4) {$\rA_-$};
    \node [style=none] (23) at (-5.5, 4) {$\rA_+$};
    \node [style=none] (24) at (-9.5, 6) {$\psiTk$};
    \node [style=none] (25) at (-3.5, 6) {$\psiTb$};
    \node [style=none] (26) at (-11, 4.75) {$R_i$};
    \node [style=none] (27) at (-2, 4.75) {$R_i$};
    \node [style=none] (28) at (-6.5, 2.5) {};
    \node [style=none] (29) at (-6.5, 1) {};
    \node [style=none] (30) at (-5, 1.75) {glue $\rAc$};
    \node [style=none] (31) at (-12.5, -1) {};
    \node [style=none] (32) at (-9.5, -1) {};
    \node [style=none] (33) at (-8.5, -1) {};
    \node [style=none] (34) at (-5.5, -1) {};
    \node [style=none] (35) at (-5, -0.75) {};
    \node [style=none] (36) at (-5, -1) {};
    \node [style=none] (37) at (-5, -1.25) {};
    \node [style=none] (38) at (-4, -1.25) {};
    \node [style=none] (39) at (-4, -1) {};
    \node [style=none] (40) at (-4, -0.75) {};
    \node [style=none] (41) at (-7.75, -1) {};
    \node [style=none] (42) at (-6.25, -1) {};
    \node [style=none] (43) at (-7, 0) {$\rA_+$};
    \node [style=none] (44) at (-7, -2) {$\rA_-$};
    \node [style=none] (45) at (-9.75, 0.25) {};
    \node [style=none] (46) at (-8.25, 0.25) {};
    \node [style=none] (47) at (-11.25, 0.25) {};
    \node [style=none] (48) at (-10.5, 0) {};
    \node [style=none] (49) at (-11.5, -0.5) {};
    \node [style=none] (50) at (-11.25, -1) {};
    \node [style=none] (51) at (-11, 0) {};
    \node [style=none] (52) at (-11.75, -0.5) {};
    \node [style=none] (53) at (-3.5, -1) {};
    \node [style=none] (54) at (-0.5, -1) {};
    \node [style=none] (56) at (-2.25, 0.25) {};
    \node [style=none] (57) at (-1.5, 0) {};
    \node [style=none] (58) at (-2.5, -0.5) {};
    \node [style=none] (59) at (-2.25, -1) {};
    \node [style=none] (60) at (-2, 0) {};
    \node [style=none] (61) at (-2.75, -0.5) {};
    \node [style=none] (62) at (-2, 1) {$\rhoA^{\! i}$};
    \node [style=none] (62) at (-.5, .25) {$\rA_+$};
    \node [style=none] (63) at (-.5, -2.25) {$\rA_-$};
  \end{pgfonlayer}
  \begin{pgfonlayer}{edgelayer}
  \filldraw[fill=blue!20] (-7.5, 4) ellipse (.4 and .5);
  \filldraw[fill=blue!20] (-5.5, 4) ellipse (.4 and .5);
    \shade[ball color = violet!40, opacity = 0.4] (-11, -1) circle (1.5cm);
    \shade[ball color = violet!40, opacity = 0.4] (-7, -1) circle (1.5cm);
     \shade[ball color = blue!40, opacity = 0.4] (-2, -1) circle (1.5cm);
  \shade[ball color = blue!40, opacity = 0.4] (-7,-1) circle (.75cm);
        \draw (-11, -1) circle (1.5cm);
    \draw (-7, -1) circle (1.5cm);
  \draw (-2, -1) circle (1.5cm);
  \draw (-7,-1) circle (.75cm);
    \draw [thick, bend right=90] (0.center) to (1.center);
    \draw [thick, bend left=90] (0.center) to (1.center);
    \draw [thick, bend left=90, looseness=0.75] (2.center) to (3.center);
    \draw [thick, bend right=60, looseness=0.75] (4.center) to (5.center);
    \draw [very thick, red, bend left=90] (6.center) to (7.center);
    \draw [very thick, red, bend right=75, looseness=0.75] (6.center) to (7.center);
    \draw [thick, bend right=90] (8.center) to (9.center);
    \draw [thick, bend left=90] (8.center) to (9.center);
    \draw [thick, bend left=90, looseness=0.75] (10.center) to (11.center);
    \draw [thick, bend right=60, looseness=0.75] (12.center) to (13.center);
    \draw [very thick, red, bend left=90] (14.center) to (15.center);
    \draw [very thick, red, bend right=75, looseness=0.75] (14.center) to (15.center);
    \draw [very thick, ->] (28.center) to (29.center);
    \draw [very thick] (35.center) to (40.center);
    \draw [very thick] (36.center) to (39.center);
    \draw [very thick] (37.center) to (38.center);
    \draw [thick, bend right=105, looseness=0.50] (31.center) to (32.center);
    \draw [thick, dashed, bend left=60, looseness=0.50] (31.center) to (32.center);
    \draw [thick, bend right=90, looseness=0.50] (33.center) to (34.center);
    \draw [thick, dashed, bend left=60, looseness=0.50] (33.center) to (34.center);
    \draw [thick, bend right=90, looseness=0.50] (41.center) to (42.center);
    \draw [thick, dashed, bend left=45, looseness=0.50] (41.center) to (42.center);
    \draw [thick, <->, bend left=45, looseness=0.75] (45.center) to (46.center);
    \draw [very thick, red, bend right=75, looseness=1.25] (47.center) to (52.center);
    \draw [very thick, red, bend right=90, looseness=2.25] (52.center) to (51.center);
    \draw [very thick, red, bend left=60, looseness=1.25] (49.center) to (48.center);
    \draw [very thick, red, bend left=90, looseness=1.50] (48.center) to (50.center);
    \draw [thick, bend right=105, looseness=0.50] (53.center) to (54.center);
    \draw [thick, dashed, bend left=60, looseness=0.50] (53.center) to (54.center);
    \draw [very thick, red, bend right=75, looseness=1.25] (56.center) to (61.center);
    \draw [very thick, red, bend right=90, looseness=2.25] (61.center) to (60.center);
    \draw [very thick, red, bend left=60, looseness=1.25] (58.center) to (57.center);
    \draw [very thick, red, bend left=90, looseness=1.50] (57.center) to (59.center);
  \end{pgfonlayer}
\end{tikzpicture}
\end{center}
\caption{The construction of the density matrix $\rhoA^{\! i}$ from the state $\psiTk$ and $\psiTb$ for regions $\rA$ that avoid interfering with the Wilson line inside the solid torus. }
\label{fig:rdensityWilsonTSingle}
\end{figure}

We first consider the case where the Wilson line is entirely contained in $\rAc$. 
 In this case, we can glue the functional integral on the solid torii along regions $\rAc_\pm$. The result can be represented as the path integral over an $\Sp^3$ with the link $(L;R_i)$ and a three-ball $\ball_{\rA}$ scooped out. The $\Sp^2$ boundary of the 3-ball is the union of the two regions $\rA_\pm$. Identifying  the two regions clearly gives back the normalization \eqref{eq:wTnorm}. We can represent the density operator more simply by inverting out the scooped ball, so that we have a three-ball, whose boundary is split into $\rA_\pm$, but now there is a link $(L;R_i)$ inside the ball; see Fig.~\ref{fig:rdensityWilsonTSingle}.
\begin{figure}[h]
\begin{center}
\usetikzlibrary{backgrounds}
\begin{tikzpicture}[scale=.65]
  \begin{pgfonlayer}{nodelayer}
    \node [style=none] (0) at (-5.5, 3.5) {};
    \node [style=none] (1) at (-0.5, 3.5) {};
    \node [style=none] (2) at (-4, 3.5) {};
    \node [style=none] (3) at (-2, 3.5) {};
    \node [style=none] (4) at (-4.25, 3.75) {};
    \node [style=none] (5) at (-1.75, 3.75) {};
    \node [style=none] (6) at (-4.5, 3.5) {};
    \node [style=none] (7) at (-1.5, 3.5) {};
    \node [style=none] (17) at (-1, 3.5) {};
    \node [style=none] (19) at (-3.25, 2.5) {$\rAc$};
    \node [style=none] (20) at (-1.1, 3.7) {$\scriptstyle{\rA}$};
    \node [style=none] (22) at (-3, 5.5) {$\psiTk$};
    \node [style=none] (24) at (-4.55, 4.25) {$R_i$};
    \node [style=none] (26) at (0, 3.5) {};
    \node [style=none] (27) at (4.75, 3.5) {};
    \node [style=none] (28) at (2.5, 4.5) {glue $\rA$ and $\rA_+$};
    \node [style=none] (51) at (-9.75, 3.75) {};
    \node [style=none] (52) at (-6.75, 3.75) {};
    \node [style=none] (53) at (-8.5, 5) {};
    \node [style=none] (54) at (-7.75, 4.75) {};
    \node [style=none] (55) at (-8.75, 4.25) {};
    \node [style=none] (56) at (-8.5, 3.75) {};
    \node [style=none] (57) at (-8.25, 4.75) {};
    \node [style=none] (58) at (-9, 4.25) {};
    \node [style=none] (59) at (-10.25, 4.75) {$\rA_+$};
    \node [style=none] (60) at (-10.25, 2.75) {$\rA_-$};
    \node [style=none] (61) at (-8.25, 6) {$\rhoA^{\! i}$};
    \node [style=none] (62) at (5.25, 3.5) {};
    \node [style=none] (63) at (12.5, 3.5) {};
    \node [style=none] (64) at (7.75, 3.5) {};
    \node [style=none] (65) at (9.75, 3.5) {};
    \node [style=none] (66) at (7.5, 3.75) {};
    \node [style=none] (67) at (10, 3.75) {};
    \node [style=none] (68) at (7.25, 3.5) {};
    \node [style=none] (69) at (10.5, 3.5) {};
    \node [style=none] (71) at (9, 2.25) {$\rAc$};
    \node [style=none] (72) at (11.5, 3.7) {$\scriptstyle{\rA_-}$};
    \node [style=none] (74) at (7.25, 4.25) {$R_i$};
    \node [style=none] (75) at (6, 4) {};
    \node [style=none] (76) at (6.75, 3.75) {};
    \node [style=none] (77) at (5.75, 3.25) {};
    \node [style=none] (78) at (6, 2.75) {};
    \node [style=none] (79) at (6.25, 3.75) {};
    \node [style=none] (80) at (5.5, 3.25) {};
  \end{pgfonlayer}
  \begin{pgfonlayer}{edgelayer}
  \filldraw[fill=blue!20] (-1.02, 3.6) ellipse (.4 and .6);
  \filldraw[fill=blue!20] (11.5, 3.75) ellipse (.5 and .6);
  \shade[ball color = violet!40, opacity = 0.4]  (-8.25, 3.75) circle (1.5cm);
  \draw (-8.25, 3.75) circle (1.5cm);
    \draw [thick, bend right=90] (0.center) to (1.center);
    \draw [thick, bend left=90] (0.center) to (1.center);
    \draw [thick, bend left=90, looseness=0.75] (2.center) to (3.center);
    \draw [thick, bend right=60, looseness=0.75] (4.center) to (5.center);
    \draw [very thick, red, bend left=90] (6.center) to (7.center);
    \draw [very thick, red, bend right=75, looseness=0.75] (6.center) to (7.center);
    \draw [very thick, ->] (26.center) to (27.center);
    \draw [thick, bend right=105, looseness=0.50] (51.center) to (52.center);
    \draw [thick, dashed, bend left=60, looseness=0.50] (51.center) to (52.center);
    \draw [very thick, red, bend right=75, looseness=1.25] (53.center) to (58.center);
    \draw [very thick, red, bend right=90, looseness=2.25] (58.center) to (57.center);
    \draw [very thick, red, bend left=60, looseness=1.25] (55.center) to (54.center);
    \draw [very thick, red, bend left=90, looseness=1.50] (54.center) to (56.center);
    \draw [thick, bend right=90] (62.center) to (63.center);
    \draw [thick, bend left=90] (62.center) to (63.center);
    \draw [thick, bend left=90, looseness=0.75] (64.center) to (65.center);
    \draw [thick, bend right=60, looseness=0.75] (66.center) to (67.center);
    \draw [very thick, red, bend left=90] (68.center) to (69.center);
    \draw [very thick, red, bend right=75, looseness=0.75] (68.center) to (69.center);
    \draw [very thick, red, bend right=75, looseness=1.25] (75.center) to (80.center);
    \draw [very thick, red, bend right=90, looseness=2.25] (80.center) to (79.center);
    \draw [very thick, red, bend left=60, looseness=1.25] (77.center) to (76.center);
    \draw [very thick, red, bend left=90, looseness=1.50] (76.center) to (78.center);
  \end{pgfonlayer}
\end{tikzpicture}
\begin{tikzpicture}[scale=.55]
  \begin{pgfonlayer}{nodelayer}
    \node [style=none] (13) at (13, 3.5) {};
    \node [style=none] (14) at (12, 3.5) {};
    \node [style=none] (16) at (14, 5) {};
    \node [style=none] (17) at (17, 5) {};
    \node [style=none] (18) at (15.25, 6.25) {};
    \node [style=none] (19) at (16, 6) {};
    \node [style=none] (20) at (15, 5.5) {};
    \node [style=none] (21) at (15.25, 5) {};
    \node [style=none] (22) at (15.5, 6) {};
    \node [style=none] (23) at (14.75, 5.5) {};
    \node [style=none] (27) at (4.25, 3.5) {};
    \node [style=none] (28) at (11.5, 3.5) {};
    \node [style=none] (29) at (6.75, 3.5) {};
    \node [style=none] (30) at (8.75, 3.5) {};
    \node [style=none] (31) at (6.5, 3.75) {};
    \node [style=none] (32) at (9, 3.75) {};
    \node [style=none] (33) at (6.25, 3.5) {};
    \node [style=none] (34) at (9.5, 3.5) {};
    \node [style=none] (37) at (6.25, 4.5) {$R_i$};
    \node [style=none] (38) at (5, 4) {};
    \node [style=none] (39) at (5.75, 3.75) {};
    \node [style=none] (40) at (4.75, 3.25) {};
    \node [style=none] (41) at (5, 2.75) {};
    \node [style=none] (42) at (5.25, 3.75) {};
    \node [style=none] (43) at (4.5, 3.25) {};
    \node [style=none] (44) at (17.25, 3.5) {};
    \node [style=none] (45) at (25, 3.5) {};
    \node [style=none] (46) at (20.25, 3.5) {};
    \node [style=none] (47) at (22.25, 3.5) {};
    \node [style=none] (48) at (20, 3.75) {};
    \node [style=none] (49) at (22.5, 3.75) {};
    \node [style=none] (50) at (19.25, 3.5) {};
    \node [style=none] (51) at (23, 3.5) {};
    \node [style=none] (53) at (24, 3.7) {$\scriptstyle{\rA_-}$};
    \node [style=none] (54) at (21.25, 5) {$R_i$};
    \node [style=none] (55) at (14.25, 0.75) {};
    \node [style=none] (56) at (17.25, 0.75) {};
    \node [style=none] (66) at (0.5, 5) {};
    \node [style=none] (67) at (3.25, 5) {};
    \node [style=none] (77) at (0.5, 1.25) {};
    \node [style=none] (78) at (3.5, 1.25) {};
    \node [style=none] (88) at (13, 3.25) {};
    \node [style=none] (89) at (12, 3.25) {};
    \node [style=none] (90) at (13, 3) {};
    \node [style=none] (91) at (12, 3) {};
    \node [style=none] (92) at (10.5, 3.7) {$\scriptstyle{\rA_-}$};
    \node [style=none] (93) at (16.25, 4.75) {$R_i$};
  \end{pgfonlayer}
  \begin{pgfonlayer}{edgelayer}
    \filldraw[fill=blue!20] (10.5, 3.65) ellipse (.5 and .6);
  \filldraw[fill=blue!20] (24, 3.75) ellipse (.5 and .6);
    \shade[ball color = violet!40, opacity = 0.4]  (1.9, 5) circle (1.4cm);
  \draw (1.9, 5) circle (1.4cm);
    \shade[ball color = violet!40, opacity = 0.4]  (2, 1.25) circle (1.5cm);
  \draw (2, 1.25) circle (1.5cm);
      \shade[ball color = violet!40, opacity = 0.4]  (15.5, 5) circle (1.5cm);
  \draw (15.5, 5) circle (1.5cm);
    \shade[ball color = violet!40, opacity = 0.4]  (15.75, 0.75) circle (1.5cm);
  \draw (15.75, 0.75) circle (1.5cm);
    \draw [very thick] (13.center) to (14.center);
    \draw [thick, bend right=105, looseness=0.50] (16.center) to (17.center);
    \draw [thick, dashed, bend left=60, looseness=0.50] (16.center) to (17.center);
    \draw [very thick, red, bend right=75, looseness=1.25] (18.center) to (23.center);
    \draw [very thick, red, bend right=90, looseness=2.25] (23.center) to (22.center);
    \draw [very thick, red, bend left=60, looseness=1.25] (20.center) to (19.center);
    \draw [very thick, red, bend left=90, looseness=1.50] (19.center) to (21.center);
    \draw [thick, bend right=90] (27.center) to (28.center);
    \draw [thick, bend left=90] (27.center) to (28.center);
    \draw [thick, bend left=90, looseness=0.75] (29.center) to (30.center);
    \draw [thick, bend right=60, looseness=0.75] (31.center) to (32.center);
    \draw [very thick, red, bend left=90] (33.center) to (34.center);
    \draw [very thick, red, bend right=75, looseness=0.75] (33.center) to (34.center);
    \draw [very thick, red, bend right=75, looseness=1.25] (38.center) to (43.center);
    \draw [very thick, red, bend right=90, looseness=2.25] (43.center) to (42.center);
    \draw [very thick, red, bend left=60, looseness=1.25] (40.center) to (39.center);
    \draw [very thick, red, bend left=90, looseness=1.50] (39.center) to (41.center);
    \draw [thick, bend right=90] (44.center) to (45.center);
    \draw [thick, bend left=90] (44.center) to (45.center);
    \draw [thick, bend left=90, looseness=0.75] (46.center) to (47.center);
    \draw [thick, bend right=60, looseness=0.75] (48.center) to (49.center);
    \draw [very thick, red, bend left=90] (50.center) to (51.center);
    \draw [very thick, red, bend right=75, looseness=0.75] (50.center) to (51.center);
    \draw [thick, bend right=105, looseness=0.50] (55.center) to (56.center);
    \draw [thick, dashed, bend left=60, looseness=0.50] (55.center) to (56.center);
    \draw [thick, bend right=105, looseness=0.50] (66.center) to (67.center);
    \draw [thick, dashed, bend left=60, looseness=0.50] (66.center) to (67.center);
    \draw [thick, bend right=105, looseness=0.50] (77.center) to (78.center);
    \draw [thick, dashed, bend left=60, looseness=0.50] (77.center) to (78.center);
    \draw [very thick] (88.center) to (89.center);
    \draw [very thick] (90.center) to (91.center);
  \end{pgfonlayer}
\end{tikzpicture}
\end{center}
\caption{The action of the reduced density matrix on $\psiTk$ is obtained by gluing the region $\rA_+$ from $\rhoA^{\! i}$ into $\rA$ on the ${\bf T}^2$. The resulting action can be simplified to obtain the representation given in \eqref{eq:rhoTpsi} by surgery as indicated on the second row.}
\label{fig:rdensityWilsonTSingle1}
\end{figure}
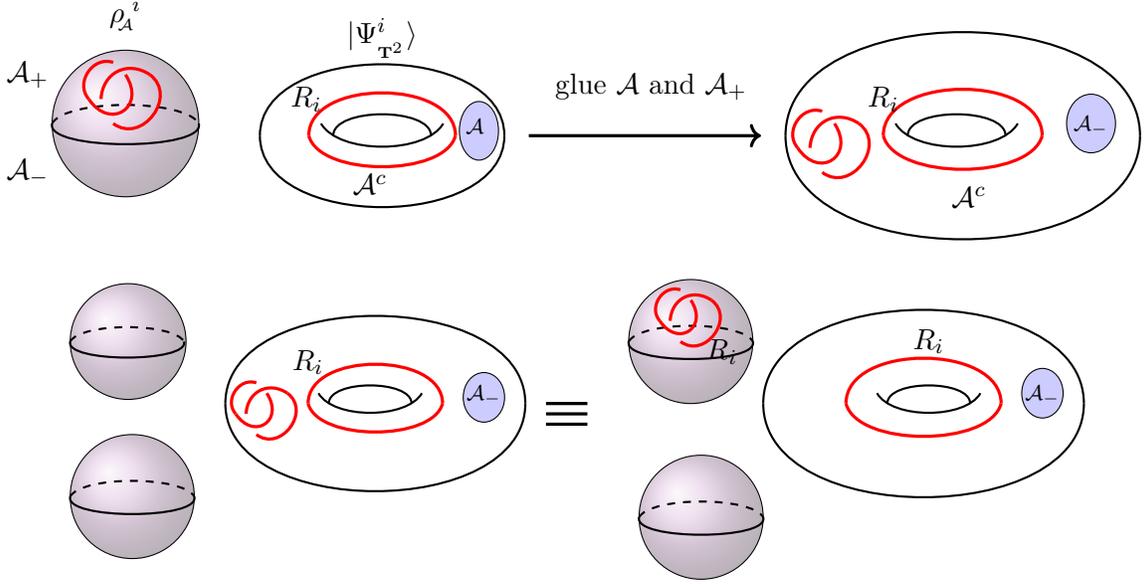

Given this description, we can consider again the action $\rhoA^{\! i} \psiTk$. It is not hard to see that 
\begin{equation}\label{eq:rhoTpsi}
\rhoA^{\! i} \, \psiTk  = \frac{\zcs(\Sp^3, L; R_i)}{\zcs(\Sp^3)} \, \psiTk
\end{equation}	
To obtain this, it is useful to multiply through by $\zcs(\Sp^3)$ and use its decomposition into three-balls. Further performing surgery to transplant the link from the solid torus to one of the balls, as described in Fig.~\ref{fig:rdensityWilsonTSingle1}, results in \eqref{eq:rhoTpsi}.
From here it is easy to see that
\begin{equation}\label{eq:RenT2Ri}
S_\rA^{(q)}(\rhoA^{\! i}) = \frac{1}{1-q} \,\left[ \log  \frac{\zcs(\Sp^3, L; R_i)^q}{\zcs(\Sp^3)^{q-1}} - q\, \log \zcs(\Sp^3, L; R_i) \right] 
= \log \zcs(\Sp^3) \,.
\end{equation}	

\begin{figure}[h]
\begin{center}
\usetikzlibrary{backgrounds}
\begin{tikzpicture}[scale=.7]
  \begin{pgfonlayer}{nodelayer}
\node [style=none] (0) at (-12, 2.5) {};
    \node [style=none] (1) at (-12.25, 0.75) {};
    \node [style=none] (2) at (-10.5, 0) {};
    \node [style=none] (3) at (-10.5, 2.75) {};
    \node [style=none] (6) at (-3, 1.75) {};
    \node [style=none] (7) at (-3.25, 0.5) {};
    \node [style=none] (8) at (-11.5, 1.5) {};
    \node [style=none] (9) at (-11.25, 2) {};
    \node [style=none] (10) at (-8.5, 1.5) {};
    \node [style=none] (11) at (-6.75, 1.5) {};
    \node [style=none] (12) at (-5.5, 1.5) {};
    \node [style=none] (13) at (-3.75, 1.5) {};
    \node [style=none] (14) at (-8.25, 0.25) {};
    \node [style=none] (15) at (-6.5, 0.5) {};
    \node [style=none] (16) at (-5.5, 0.5) {};
    \node [style=none] (17) at (-9.75, 1.75) {};
    \node [style=none] (20) at (-8.5, 1.25) {};
    \node [style=none] (21) at (-7, 1.25) {};
    \node [style=none] (22) at (-5.25, 1.25) {};
    \node [style=none] (23) at (-5.5, 2.75) {};
    \node [style=none] (24) at (-4, 1.25) {};
    \node [style=none] (25) at (-2.5, 1.25) {};
    \node [style=none] (26) at (-1.5, 0.5) {};
    \node [style=none] (27) at (-8.5, 2.75) {};
    \node [style=none] (28) at (-10, 1.5) {};
    \node [style=none] (29) at (-11.25, 1.25) {};
    \node [style=none] (30) at (-10.75, 2.5) {};
    \node [style=none] (31) at (-8.75, 0.25) {};
    \node [style=none] (32) at (-6.75, 0.25) {};
    \node [style=none] (33) at (-5.5, 0.5) {};
    \node [style=none] (34) at (-3.75, 0.25) {};
    \node [style=none] (35) at (-3.5, -0.75) {};
    \node [style=none] (36) at (-6.5, -0.75) {};
    \node [style=none] (37) at (-9, -1) {};
    \node [style=none] (38) at (-11, 0) {};
    \node [style=none] (39) at (-7.25, 2.5) {};
    \node [style=none] (40) at (-5.25, -0.5) {};
    \node [style=none] (41) at (-4, 2.5) {};
    \node [style=none] (42) at (-7.5, -0.25) {};
    \node [style=none] (43) at (-9.75, 0) {};
    \node [style=none] (48) at (-10.5, 0) {};
    \node [style=none] (49) at (-10.75, 0.75) {};
    \node [style=none] (50) at (-9.25, 1.75) {};
    \node [style=none] (51) at (-8, 1.25) {};
    \node [style=none] (52) at (-10.5, 1.25) {};
    \node [style=none] (57) at (4.75, 1.25) {};
    \node [style=none] (62) at (3.75, 1.5) {};
    \node [style=none] (66) at (3.75, 0.75) {};
    \node [style=none] (67) at (3.5, 0.75) {};
    \node [style=none] (68) at (4, 1.25) {};
    \node [style=none] (69) at (4.75, 1.25) {};
    \node [style=none] (70) at (3.75, 1) {};
  \end{pgfonlayer}
  \begin{pgfonlayer}{edgelayer}
    \draw [thick, in=120, out=-135] (0.center) to (1.center);
    \draw [thick, bend right, looseness=0.75] (1.center) to (2.center);
    \draw [thick, in=45, out=165, looseness=0.75] (3.center) to (0.center);
    \draw [thick, bend left=45, looseness=1.25] (6.center) to (7.center);
    \draw [thick, bend left=45] (8.center) to (9.center);
    \draw [thick, bend right=120] (10.center) to (11.center);
    \draw [thick, bend right=90, looseness=0.75] (12.center) to (13.center);
    \draw [thick, in=150, out=30, looseness=0.75] (15.center) to (16.center);
    \draw [thick, bend right] (16.center) to (7.center);
    \draw [thick, bend left=75, looseness=1.50] (21.center) to (22.center);
    \draw [thick, bend left=75, looseness=1.25] (24.center) to (25.center);
    \draw [thick, bend left=60] (29.center) to (27.center);
    \draw [thick, in=75, out=75, looseness=1.25] (28.center) to (20.center);
    \draw [thick, bend right=60, looseness=1.25] (34.center) to (25.center);
    \draw [thick, in=-105, out=0] (35.center) to (26.center);
    \draw [thick, bend right=60] (32.center) to (33.center);
    \draw [thick, in=-165, out=-75, looseness=0.75] (38.center) to (37.center);
    \draw [thick, in=-60, out=-90] (8.center) to (17.center);
    \draw [thick, in=225, out=-30, looseness=1.25] (27.center) to (39.center);
    \draw [thick, bend left] (39.center) to (23.center);
    \draw [thick, in=-165, out=-15, looseness=1.25] (36.center) to (40.center);
    \draw [thick, in=180, out=15, looseness=1.25] (40.center) to (35.center);
    \draw [thick, in=210, out=-30, looseness=1.25] (23.center) to (41.center);
    \draw [thick, bend left=75, looseness=1.25] (41.center) to (26.center);
    \draw [thick, in=165, out=15] (37.center) to (42.center);
    \draw [thick, in=180, out=-15, looseness=0.75] (42.center) to (36.center);
    \draw [thick, in=-120, out=-45] (43.center) to (31.center);
    \draw [thick, bend right=45, looseness=0.75] (14.center) to (15.center);
    \draw [thick, in=150, out=0] (48.center) to (14.center);
    \draw [very thick, red, in=180, out=-180, looseness=2.25] (30.center) to (49.center);
    \draw [very thick, red, in=90, out=90, looseness=1.50] (52.center) to (51.center);
    \draw [very thick, red, in=255, out=-60, looseness=1.25] (48.center) to (14.center);
    \draw [very thick, red, in=-75, out=0] (49.center) to (50.center);
    \draw [very thick, red, in=180, out=150, looseness=2.00] (62.center) to (67.center);
    \draw [very thick, red, in=105, out=90, looseness=1.50] (70.center) to (69.center);
    \draw [very thick, red, in=-75, out=-60, looseness=1.25] (66.center) to (57.center);
    \draw [very thick, red, in=-75, out=0] (67.center) to (68.center);
    \filldraw[fill=violet!20] (-2,1) ellipse (.3 and .5);
          \filldraw[fill=violet!20] (-3.5,1) ellipse (.5 and .3);
          \filldraw[fill=blue!20] (-6,1.25) ellipse (.4 and .4);
          \filldraw[fill=blue!20] (-6,2.35) ellipse (.5 and .3);
    \node [style=none]  at (-10.55, 1.5) {$R_i$};
    \node [style=none]  at (-12.15, 1.5) {$R_i$};
    \node [style=none]  at (-2.0, 1) {$\scriptstyle{\rA^2_+}$};
    \node [style=none]  at (-3.5, 1) {$\scriptstyle{\rA^2_-}$};
    \node [style=none]  at (-6, 2.35) {$\scriptstyle{\rA^1_+}$};
    \node [style=none]  at (-6, 1.25) {$\scriptstyle{\rA^1_-}$};
    \node [style=none]  at (-7.25, 2) {$\rAc_+$};
    \node [style=none]  at (-7.25, .5) {$\rAc_-$};
          \draw [very thick, ->] (-1,1)--(.5,1);
          \shade[ball color = blue!60, opacity = 0.4] (4,1) circle (1.5cm);
          \draw (4,1) circle (1.5cm);
          \shade[ball color = violet!60, opacity = 0.4] (8,1) circle (1.5cm);
          \draw (8,1) circle (1.5cm);
          \draw [dashed]  (4,-.5) arc (-90:90:0.4 and 1.5);
                \draw (4,2.5) arc (90:270:0.4 and 1.5);
         \draw [dashed]  (8,-.5) arc (-90:90:0.4 and 1.5);
               \draw (8,2.5) arc (90:270:0.4 and 1.5);
               \node [style=none]  at (8.75, 1) {$\rA^2_+$};
    \node [style=none]  at (7.25, 1) {$\rA^2_-$};
    \node [style=none]  at (4.75, 1) {$\rA^1_+$};
    \node [style=none]  at (3.25, 1) {$\rA^1_-$};
    \node at (1.5,1)  {$ \frac{1}{\zcs(\Sp^3)}$};
    \node at (-5,3) {$\mcs^+_{\Sigma_3}$};
    \node at (-11,3.25) {$\mcs^-_{\Sigma_3}$};
  \end{pgfonlayer}
\end{tikzpicture}
\end{center}
\caption{The state $|\Psi_3^i\rangle$ on a genus three Riemann surface $\Sigma_3$ can be constructed by performing the  path integral over the handle body $\mcs^-_{\Sigma_3}$,  a 3-manifold obtained by filling  all the three independent $a$-cycles of $\Sigma_3$, with a Wilson loop along one of its non-contractible $b$-cycle in representation $R_i$. 
The dual state $\langle\Psi^i_3|$  can be constructed by performing the path integral over another handle body $\mcs^+_{\Sigma_3}$  obtained by filling the $b$-cycles of $\Sigma_3$, with a Wilson loop along its non-contractible $a$-cycle in representation  $R_i$.   The path integral representation of  ${\rhoA}$  of $|\Psi_{i}\rangle$  for the bipartition of $\Sigma_3$ into two regions $\rA=\rA^1\cup\rA^2$ and $\rAc$ can be obtained by gluing the path integrals over  $\mcs^+_{\Sigma_3}$ and $\mcs^-_{\Sigma_3}$  by identifying the  regions $\rAck$ and $\rAcb$ on their boundaries.  The resultant path integral can be understood as a path integral over two  3-balls, one without any Wilson lines and the other with a Wilson line placed along a knot in it, normalized by $\zcs(\Sp^3)$. The boundary of one of them is given by the union of the complementary regions $\rA^1_-$ and $\rA^1_+$, and the boundary of the other is given by the union of the complementary regions $\rA^2_-$ and $\rA^2_+$. }
\label{fig:nrdensitymg3RS}
\end{figure}
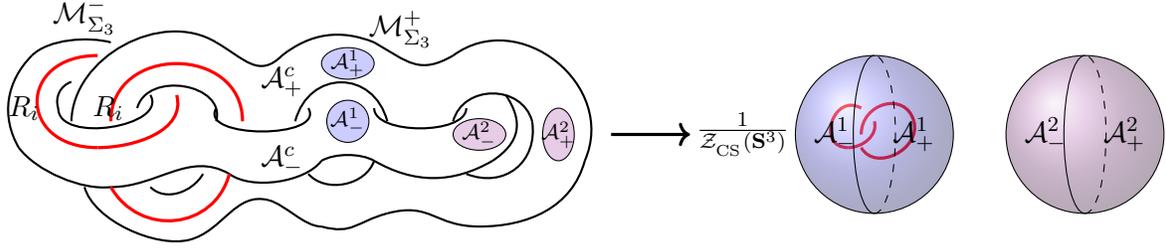

The final answer is the same as that for the bipartitioning of the state $\psiSk$ and in general only depends on the number of components of the entangling surface. For example, the reader can check immediately that for a state $|\Psi_g^i\rangle$ on a Riemann surface with $M$ components for $\rA$, and with a Wilson line in representation $R_i$ contained entirely in $\rAc$, the entropies evaluate to $M\, \log \zcs(\Sp^3)$. We illustrate the construction of the density matrix $\rhoA$ for $M=2$ in Fig.~\ref{fig:nrdensitymg3RS} for a genus-3 surface.

Let us now consider uplifting this computation to the closed topological string on the resolved conifold. As before, we will start by first identifying the analogous construction in the open+closed string on the deformed conifold, and then proceed to take the resulting answer through the geometric transition. The object we should first uplift into the deformed conifold is the partition function which we are decomposing \eqref{eq:wTnorm} which involves a Wilson loop in a specific representation on a link $L$. 

These Wilson loops uplift to probe topological D-branes wrapping non-compact Lagrangian cycles. They intersect  the $\Sp^3$ on the link $L$ \cite{Ooguri:1999bv}.  The Lagrangian cycles corresponding to an unknot has topology $\Sp^1 \times \mathbb{R}^2$ where the $\Sp^1$ is the unknot and the non-compact directions are along the normals to the $\Sp^3$.  After passing through the geometric transition the Lagrangian cycles, and the probe D-branes wrapping them, remain. The Lagrangian cycle $\lagr_{R_i} \subset \rco$ on the resolved conifold side still has topology $\Sp^1 \times {\mathbb R}^2$. For instance, the Lagrangian cycle corresponding to the unknot, the $\Sp^1$ is heuristically viewed as the knot/link in the $\Sp^3$ at $r=\infty$.  This cycle ends up wrapping a circle in $\Sp^2$, which we take to be the equator. When we have non-trivial knots/links, then the cycles corresponding to links are suitably braided; see \cite{Taubes:2001wk} for a construction of such Lagrangian cycles in $\rco$. It is important for our purposes that all this action happens away from the non-contractible $\Sp^2$ at  the tip of the resolved conifold. To be specific, we can follow the discussion of \cite{Gomis:2006mv,Gomis:2007kz} and use the specifics of the representation  $R_i$ (eg., its Dynkin labels) to find a representative set of Lagrangian cycles  $\lagr_{\rco,R_i}$ on the resolved conifold that end on the non-contractible $\Sp^2$. 

On the deformed conifold side we can use our earlier observation that every $\Sp^3$ along the radial direction can be Heegard split to construct a state space. The split we choose is the one that is informed by the manner in which we performed our Heegard decomposition. So each $\Sp^3(r)$ in the deformed conifold will be split into two solid torii with Wilson lines in representation $R_i$, exactly mimicking the Chern-Simons construction. The Lagrangian cycles corresponding to the knots in the link $L$ are braided within the $\Sp^3(r)$.

We follow this through split through the geometric transition, and continue to decompose each $\Sp^3(r) \in \rco$ in the manner described above. In this manner we generate at each radial position of the resolved conifold a copy of $\mcs^\pm_{_{{\bf T}^2}}$ from $\Sp^3(r)$ along with our unit $\Sp^2(1)$ (the non-contractible cycle). The split brings out the two Lagrangian cycles representing the link $L$, which we collectively refer to as $\lagr_{R_i}$. Having exposed the handlebodies through three-manifold surgery, we bring forth the Cauchy surfaces $\Gamma_\pm$ which define for us closed topological string uplifts of the state $\psiTk$ which we refer to as $\ket{\! \PhiR^i}$, and its conjugate, respectively.   Since the Wilson line along $L$ can be located entirely on one of the handlebodies in the Chern-Simons side, it follows that the Lagrangian $\lagr_{R_i}$ is likewise contained within one of the solid torii obtained by the Heegard splitting. Even more conveniently, the brane can be located w.l.o.g.\ entirely on $\Gamma_-$. 

We now have a construction of the closed string state. It is easy to see that the rest of the story can now proceed as before. Specifically, the Cauchy surfaces $\Gamma_\pm$ are bipartitioned  into $\rxA \cup \rxAc$ and we build the reduced state $\rhoAx$ by gluing the $\rxAc_\pm$ components together. The main novelty is the presence of the probe branes on $\lagr_{\rco,R_i}$, but they pretty much play a spectator role in the computation, being localized on $\Gamma_-$ by our choice. While the cycle $\lagr_{R_i}$ does end on the $\Sp^2$, at that locus we do not have an $\Sp^3$ to bipartition, so as before our bipartitioning leaves untouched the only 2-cycle of the resolved conifold. 
Given the construction of $\rhoAx$ the replica construction proceeds analogously to the earlier computations. 

The replica spacetime $\rco_q\left(\mathcal{L}_{\rco,R_i}\right)$ is obtained by gluing  copies of the resolved conifold and the Lagrangian cycle to each other, subject to the above manipulations, producing for us the manifold $\rco_q\left( \lagr_{R_i}\right)$. All of the above can be succinctly summarized by noting that our entangling brane does not interfere with the Lagrangian cycle.  It is not hard to see that the closed string partition function on this geometry will be equal to the first term in \eqref{eq:RenT2Ri}. The normalization factor on the other hand comes from $\rco_{\oplus q}\left(\lagr_{\rco,R_i}\right)$. The latter ends up contributing $q$ copies of the Wilson loop partition function, so that at the end of the day, we find the desired answer $S^{(q)}_{\rxA} = \log \zc(\rco)$.  

Thus, our prescription for understanding closed topological string entanglement continues to work when there are Lagrangian cycles corresponding to Wilson lines on knots/links, so far, at least for the case where they are both consigned to spectator roles. We finally need to consider situation where the entangling brane interferes in some manner with the Lagrangian cycles corresponding to the probe branes.  

\subsubsection{Bipartitions slicing through Wilson lines}
\label{sec:wbipart2}

Let us now turn to situations where the bipartitioning of our state results in some Wilson line being cut, leaving behind marked points. There are two possibilities to consider:
\begin{itemize}
\item The marked points lie in the regions $\rA$ (or $\rAc$, or both) representing insertion of charged states. 
\item The cut of the Wilson line is in the interior of the three-manifold, with no imprint on the regions themselves as in Fig.~\ref{fig:torusstatesplit}. 
\end{itemize}
The distinction can be viewed in terms of the Wilson lines being `timelike' when they pierce through the Cauchy surface and thus pass through $\rA$ or $\rAc$, or `spacelike' when they do 
not. However unlike the situation in \S\ref{sec:wbipart1}, in the latter case the bipartitioning does not allow us to separate out a domain around $\rA$ without interfering with the Wilson line. 

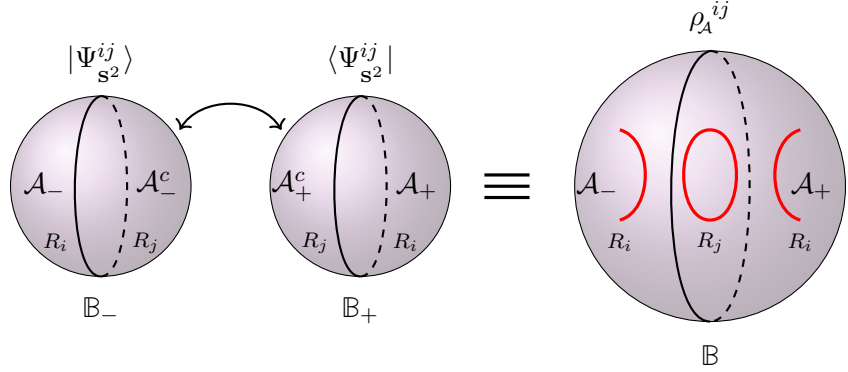
\begin{figure}[ht]
\begin{center}
\usetikzlibrary{backgrounds}
\begin{tikzpicture}[scale=.6]
\begin{pgfonlayer}{nodelayer}
    \node [style=none] (0) at (-11, 5.5) {};
    \node [style=none] (1) at (-11, 1.5) {};
    \node [style=none] (2) at (-9.75, 4.5) {};
    \node [style=none] (3) at (-10, 2.75) {};
    \node [style=none] (4) at (-12, 4.5) {};
    \node [style=none] (5) at (-12, 2.75) {};
    \node [style=none] (26) at (-10.25, 3.5) {};
    \node [style=none] (27) at (-10, 2.25) {$\scriptstyle{R_j}$};
    \node [style=none] (28) at (-12, 2.25) {$\scriptstyle{R_i}$};
    \node [style=none] (33) at (-12.25, 3.5) {$\rA_-$};
    \node [style=none] (34) at (-9.75, 3.5) {$\rAc_-$};
    \node [style=none] (35) at (-6.75, 3.5) {$\rAc_+$};
    \node [style=none] (37) at (-2.5, 3.5) {};
    \node [style=none] (38) at (-2.5, 3.5) {};
    \node [style=none] (39) at (-1.5, 3.5) {};
    \node [style=none] (40) at (-2.5, 3.75) {};
    \node [style=none] (41) at (-1.5, 3.75) {};
    \node [style=none] (42) at (-2.5, 3.25) {};
    \node [style=none] (43) at (-1.5, 3.25) {};
    \node [style=none] (44) at (-5.25, 5.5) {};
    \node [style=none] (45) at (-5.25, 1.5) {};
    \node [style=none] (46) at (-4, 4.5) {};
    \node [style=none] (47) at (-4.25, 2.75) {};
    \node [style=none] (48) at (-6.25, 4.5) {};
    \node [style=none] (49) at (-6.25, 2.75) {};
    \node [style=none] (50) at (-4.5, 3.5) {};
    \node [style=none] (51) at (-6.25, 2.25) {$\scriptstyle{R_j}$};
    \node [style=none] (52) at (-4.25, 2.25) {$\scriptstyle{R_i}$};
    \node [style=none] (55) at (-4, 3.5) {$\rA_+$};
    \node [style=none] (56) at (2.5, 6.5) {};
    \node [style=none] (57) at (2.5, 0.5) {};
    \node [style=none] (58) at (2.5, 4.75) {};
    \node [style=none] (59) at (2.5, 2.75) {};
    \node [style=none] (60) at (0.5, 4.75) {};
    \node [style=none] (61) at (0.5, 2.75) {};
    \node [style=none] (62) at (3, 3.5) {};
    \node [style=none] (63) at (2.5, 2.25) {$\scriptstyle{R_j}$};
    \node [style=none] (64) at (0.5, 2.25) {$\scriptstyle{R_i}$};
    \node [style=none] (65) at (0, 3.5) {$\rA_-$};
    \node [style=none] (66) at (4.75, 3.5) {$\rA_+$};
    \node [style=none] (70) at (4.5, 4.75) {};
    \node [style=none] (71) at (4.5, 2.75) {};
    \node [style=none] (72) at (2.5, 4.75) {};
    \node [style=none] (73) at (2.5, 2.75) {};
    \node [style=none] (76) at (4.5, 2.25) {$\scriptstyle{R_i}$};
    \node [style=none] (78) at (3.25, 4.75) {};
    \node [style=none] (79) at (-11, 0.75) {};
    \node [style=none] (80) at (-11, 0.75) {$\ball_-$};
    \node [style=none] (81) at (-5.25, 0.75) {$\ball_+$};
    \node [style=none] (82) at (-11, 6.25) {$\psiSkij$};
    \node [style=none] (83) at (-5.25, 6.25) {$\psiSbij$};
    \node [style=none] (84) at (2.5, 7.25) {$\rhoA^{\!ij}$};
    \node [style=none] (86) at (2.5, -0.25) {$\ball$};
    \node [style=none] (87) at (-9.25, 4.75) {};
    \node [style=none] (88) at (-7, 4.75) {};
  \end{pgfonlayer}
  \begin{pgfonlayer}{edgelayer}
      \shade[ball color = violet!40, opacity = 0.4]  (-11, 3.5) circle (2cm);
  \draw (-11, 3.5) circle (2cm);
    \shade[ball color = violet!40, opacity = 0.4]  (-5.25, 3.5) circle (2cm);
  \draw (-5.25, 3.5) circle (2cm);
    \shade[ball color = violet!40, opacity = 0.4]  (2.5, 3.5) circle (3cm);
  \draw (2.5, 3.5) circle (3cm);
    \draw [thick, in=180, out=-165, looseness=0.50] (0.center) to (1.center);
    \draw [thick, style=dashed, in=0, out=0, looseness=0.50] (0.center) to (1.center);
    \draw [very thick] (40.center) to (41.center);
    \draw [very thick] (38.center) to (39.center);
    \draw [very thick] (42.center) to (43.center);
    \draw [thick, in=180, out=-165, looseness=0.50] (44.center) to (45.center);
    \draw [thick, style=dashed, in=0, out=0, looseness=0.50] (44.center) to (45.center);
    \draw [thick, in=180, out=-165, looseness=0.50] (56.center) to (57.center);
    \draw [thick, style=dashed, in=0, out=0, looseness=0.50] (56.center) to (57.center);
    \draw [very thick, color=red, in=-180, out=180] (58.center) to (59.center);
    \draw [very thick, color=red, in=15, out=-15] (60.center) to (61.center);
    \draw [very thick, color=red, in=165, out=-165] (70.center) to (71.center);
    \draw [very thick, color=red, in=0, out=0] (72.center) to (73.center);
    \draw [thick, <->, bend left=60] (87.center) to (88.center);
  \end{pgfonlayer}
\end{tikzpicture}
 \end{center}
 \caption{The state $\psiSkij$ is obtained by slicing open $\zcs(\Sp^3, R_i,R_j)$ to expose a three-ball $\ball_-$ with the two Wilson lines piercing through the $\Sp^2$. The conjugate state is prepared on the other ball $\ball_+$. Bipartitioning the $\Sp^2$ so that the Wilson line $R_i$ pierces through $\rA$ while $R_j$ pierces through $\rAc$ results in a density matrix $\rhoA^{\! ij}$
which can be seen to be representable via a functional integral on a single three-ball $\ball$ whose boundary is $\rA_-\cup \rA_+$ each of which has the two marked points with representation label $R_i$. The interior of the ball contains a Wilson line on an unknot with  the other representation $R_j$. 
 }
\label{fig:nrdensitymS2W}
\end{figure}

\paragraph{1. Marked points in $\rA, \rAc$:} Let us first consider the first situation, with marked points inside our region $\rA$. A simple example is to consider the state $\psiSkij$ on $\Sp^2$ prepared by a path integral over a three-ball with two open Wilson lines carrying representations $R_i$ and $R_j$, respectively, which end on four marked points on the boundary $\Sp^2$. We depict this construction in 
Fig.~\ref{fig:nrdensitymS2W}. We choose $\rA$ to contain the marked points associated with the two ends of the Wilson line in representation $R_i$ and $\rAc$ to contain those with representation $R_j$. We have also engineered the situation so that the Wilson lines are on unknots and are not linked inside the three-ball. If we were agnostic about the interior of the three-ball, we would have to sum over two configurations, with and without linking. This is discussed in \cite{Dong:2008ft}, but the simpler situation we are discussing will suffice for our purposes. We pick the dual state $\psiSbij$ so that the inner product produces $\zcs(\Sp^3, R_i,R_j)$. Identifying the regions $\rAc_\pm$ from the balls $\ball_\pm$, we end up with a simple picture of having to perform a functional integral over a three-ball with boundary $\rA_- \cup \rA_+$ with the associated $R_i$ marked points, and a Wilson loop in representation $R_j$ sitting in its interior. This is our desired representation for $\rhoA^{\! ij}$.

\begin{figure}[h]
 \begin{center}
\begin{tikzpicture}[scale=.35]
\begin{pgfonlayer}{nodelayer}
    \node [style=none] (127) at (26.5, 3.75) {$\cdots\cdots$};
    \node [style=none] (128) at (36.75, 3.5) {};
    \node [style=none] (129) at (37.75, 3.5) {};
    \node [style=none] (155) at (40, 2.25) {};
    \node [style=none] (156) at (53, 4) {};
    \node [style=none] (167) at (45, 3.75) {};
    \node [style=none] (168) at (45, 3.75) {$\cdots\cdots$};
    \node [style=none] (169) at (46.5, 10) {};
    \node [style=none] (170) at (46.5, -3.75) {};
    \node [style=none] (171) at (41.25, 3.5) {$\scriptstyle{\rA_-}$};
    \node [style=none] (172) at (51.75, 3.5) {$\scriptstyle{\rA_+}$};
    \node [style=none] (173) at (42, 0.5) {};
    \node [style=none] (174) at (42.75, 0) {};
    \node [style=none] (175) at (45.25, 0) {};
    \node [style=none] (176) at (46, -0.25) {};
    \node [style=none] (177) at (46.5, 0) {};
    \node [style=none] (178) at (50, 0) {};
    \node [style=none] (179) at (51, 0.5) {};
    \node [style=none] (180) at (45.75, -0.75) {};
    \node [style=none] (181) at (45.75, -0.75) {$q$-pairs};
    \node [style=none] (182) at (42.25, 9) {};
    \node [style=none] (184) at (50.5, 9) {};
    \node [style=none] (185) at (12, -0.75) {};
    \node [style=none] (186) at (12.75, -1.25) {};
    \node [style=none] (187) at (22, -1.25) {};
    \node [style=none] (188) at (22.75, -1.75) {};
    \node [style=none] (189) at (23.5, -1.25) {};
    \node [style=none] (190) at (31.5, -1.25) {};
    \node [style=none] (191) at (32, -0.75) {};
    \node [style=none] (192) at (22.75, -2.25) {$q$ copies};
    \node [style=none] (193) at (10.75, 6.5) {};
    \node [style=none] (194) at (10.25, 7.5) {};
    \node [style=none] (195) at (13.75, 9) {};
    \node [style=none] (196) at (32, 9) {};
    \node [style=none] (197) at (34, 7.75) {};
    \node [style=none] (198) at (33.75, 6.5) {};
    \node [style=none] (199) at (15.25, 6.75) {};
    \node [style=none] (200) at (18.75, 6.75) {};
    \node [style=none] (201) at (36.75, 3.25) {};
    \node [style=none] (202) at (37.75, 3.25) {};
    \node [style=none] (203) at (36.75, 3) {};
    \node [style=none] (204) at (37.75, 3) {};
    \node [style=none] (205) at (12.75, 6.75) {};
    \node [style=none] (206) at (12.75, 0.75) {};
    \node [style=none] (207) at (12.75, 5) {};
    \node [style=none] (208) at (12.75, 3) {};
    \node [style=none] (209) at (10.75, 5) {};
    \node [style=none] (210) at (10.75, 3) {};
    \node [style=none] (211) at (13.25, 3.75) {};
    \node [style=none] (212) at (12.75, 2.5) {$\scriptstyle{R_j}$};
    \node [style=none] (213) at (10.75, 2.5) {$\scriptstyle{R_i}$};
    \node [style=none] (214) at (10.5, 3.75) {$\scriptstyle{\rA_-}$};
    \node [style=none] (215) at (15, 3.75) {$\scriptstyle{\rA_+}$};
    \node [style=none] (216) at (14.75, 5) {};
    \node [style=none] (217) at (14.75, 3) {};
    \node [style=none] (218) at (12.75, 5) {};
    \node [style=none] (219) at (12.75, 3) {};
    \node [style=none] (220) at (14.75, 2.5) {$\scriptstyle{R_i}$};
    \node [style=none] (221) at (13.5, 5) {};
    \node [style=none] (222) at (12.75, 7.5) {$\rhoA^{\! ij}$};
    \node [style=none] (223) at (12.75, 0) {$\ball$};
    \node [style=none] (224) at (21, 6.75) {};
    \node [style=none] (225) at (21, 0.75) {};
    \node [style=none] (226) at (21, 5) {};
    \node [style=none] (227) at (21, 3) {};
    \node [style=none] (228) at (19, 5) {};
    \node [style=none] (229) at (19, 3) {};
    \node [style=none] (230) at (21.5, 3.75) {};
    \node [style=none] (231) at (21, 2.5) {$\scriptstyle{R_j}$};
    \node [style=none] (232) at (19, 2.5) {$\scriptstyle{R_i}$};
    \node [style=none] (233) at (18.75, 3.75) {$\scriptstyle{\rA_-}$};
    \node [style=none] (234) at (23.25, 3.75) {$\scriptstyle{\rA_+}$};
    \node [style=none] (235) at (23, 5) {};
    \node [style=none] (236) at (23, 3) {};
    \node [style=none] (237) at (21, 5) {};
    \node [style=none] (238) at (21, 3) {};
    \node [style=none] (239) at (23, 2.5) {$\scriptstyle{R_i}$};
    \node [style=none] (240) at (21.75, 5) {};
    \node [style=none] (241) at (21, 7.5) {$\rhoA^{\! ij}$};
    \node [style=none] (242) at (21, 0) {$\ball$};
    \node [style=none] (243) at (31.75, 6.75) {};
    \node [style=none] (244) at (31.75, 0.75) {};
    \node [style=none] (245) at (31.75, 5) {};
    \node [style=none] (246) at (31.75, 3) {};
    \node [style=none] (247) at (29.75, 5) {};
    \node [style=none] (248) at (29.75, 3) {};
    \node [style=none] (249) at (32.25, 3.75) {};
    \node [style=none] (250) at (31.75, 2.5) {$\scriptstyle{R_j}$};
    \node [style=none] (251) at (29.75, 2.5) {$\scriptstyle{R_i}$};
    \node [style=none] (252) at (29.5, 3.75) {$\scriptstyle{\rA_-}$};
    \node [style=none] (253) at (34, 3.75) {$\scriptstyle{\rA_+}$};
    \node [style=none] (254) at (33.75, 5) {};
    \node [style=none] (255) at (33.75, 3) {};
    \node [style=none] (256) at (31.75, 5) {};
    \node [style=none] (257) at (31.75, 3) {};
    \node [style=none] (258) at (33.75, 2.5) {$\scriptstyle{R_i}$};
    \node [style=none] (259) at (32.5, 5) {};
    \node [style=none] (260) at (31.75, 7.5) {$\rhoA^{\! ij}$};
    \node [style=none] (261) at (31.75, 0) {$\ball$};
    \node [style=none] (263) at (42.75, 6.25) {};
    \node [style=none] (264) at (42.75, 4.25) {};
    \node [style=none] (265) at (43.25, 5) {};
    \node [style=none] (266) at (42.75, 6.75) {$\scriptstyle{R_i}$};
    \node [style=none] (267) at (42.75, 6.25) {};
    \node [style=none] (268) at (42.75, 4.25) {};
    \node [style=none] (269) at (42.75, 3.25) {};
    \node [style=none] (270) at (42.75, 1.25) {};
    \node [style=none] (271) at (43.25, 2) {};
    \node [style=none] (272) at (42.75, 0.75) {$\scriptstyle{R_j}$};
    \node [style=none] (273) at (42.75, 3.25) {};
    \node [style=none] (274) at (42.75, 1.25) {};
    \node [style=none] (275) at (50.25, 3.25) {};
    \node [style=none] (276) at (50.25, 1.25) {};
    \node [style=none] (277) at (50.75, 2) {};
    \node [style=none] (279) at (50.25, 3.25) {};
    \node [style=none] (280) at (50.25, 1.25) {};
    \node [style=none] (281) at (50.25, 6.25) {};
    \node [style=none] (282) at (50.25, 4.25) {};
    \node [style=none] (283) at (50.75, 5) {};
    \node [style=none] (285) at (50.25, 6.25) {};
    \node [style=none] (286) at (50.25, 4.25) {};
    \node [style=none] (287) at (36, 9.75) {};
    \node [style=none] (288) at (50.25, 6.75) {$\scriptstyle{R_i}$};
    \node [style=none] (289) at (50.25, 0.75) {$\scriptstyle{R_j}$};
    \node [style=none] (290) at (46.5, 12.5) {$\zcs\left(\mathbf{S}^3,\left(R_i,R_j,\right)^q \right)$};
    \node [style=none] (291) at (34, 16.75) {};
  \end{pgfonlayer}
  \begin{pgfonlayer}{edgelayer}
      \shade[ball color = violet!40, opacity = 0.4]  (12.75, 3.75) circle (3cm);
  \draw (12.75, 3.75) circle (3cm);
    \shade[ball color = violet!40, opacity = 0.4]  (21, 3.75) circle (3cm);
  \draw (21, 3.75) circle (3cm);
    \shade[ball color = violet!40, opacity = 0.4]  (31.75, 3.75) circle (3cm);
  \draw (31.75, 3.75) circle (3cm);
     \shade[ball color = violet!40, opacity = 0.4]  (46.5, 3.1) circle (6.85cm);
  \draw (46.5, 3.1) circle (6.85cm);
    \draw [thick] (128.center) to (129.center);
    \draw [thick, in=-180, out=-165, looseness=0.25] (169.center) to (170.center);
    \draw [thick, style=dashed, in=-15, out=0, looseness=0.25] (169.center) to (170.center);
    \draw [thick, bend right, looseness=0.75] (173.center) to (174.center);
    \draw [thick] (174.center) to (175.center);
    \draw [thick, bend left, looseness=0.75] (175.center) to (176.center);
    \draw [thick, bend left, looseness=0.75] (176.center) to (177.center);
    \draw [thick] (177.center) to (178.center);
    \draw [thick, bend right] (178.center) to (179.center);
    \draw [thick, <->, bend left=75, looseness=1.25] (182.center) to (184.center);
    \draw [thick, bend right, looseness=0.75] (185.center) to (186.center);
    \draw [thick] (186.center) to (187.center);
    \draw [thick, bend left] (187.center) to (188.center);
    \draw [thick, bend left] (188.center) to (189.center);
    \draw [thick] (189.center) to (190.center);
    \draw [thick, bend right, looseness=1.25] (190.center) to (191.center);
    \draw [thick, <-, in=-90, out=135] (193.center) to (194.center);
    \draw [thick, in=-180, out=90] (194.center) to (195.center);
    \draw [thick] (195.center) to (196.center);
    \draw [thick, in=105, out=0] (196.center) to (197.center);
    \draw [thick, ->, in=45, out=-75, looseness=0.50] (197.center) to (198.center);
    \draw [thick, <->, bend left=45] (199.center) to (200.center);
    \draw [thick] (201.center) to (202.center);
    \draw [thick] (203.center) to (204.center);
    \draw [thick, in=180, out=-165, looseness=0.50] (205.center) to (206.center);
    \draw [thick, style=dashed, in=0, out=0, looseness=0.50] (205.center) to (206.center);
    \draw [very thick, color=red, in=-180, out=180] (207.center) to (208.center);
    \draw [very thick, color=red, in=15, out=-15] (209.center) to (210.center);
    \draw [very thick, color=red, in=165, out=-165] (216.center) to (217.center);
    \draw [very thick, color=red, in=0, out=0] (218.center) to (219.center);
    \draw [thick, in=180, out=-165, looseness=0.50] (224.center) to (225.center);
    \draw [thick, style=dashed, in=0, out=0, looseness=0.50] (224.center) to (225.center);
    \draw [very thick, color=red, in=-180, out=180] (226.center) to (227.center);
    \draw [very thick, color=red, in=15, out=-15] (228.center) to (229.center);
    \draw [very thick, color=red, in=165, out=-165] (235.center) to (236.center);
    \draw [very thick, color=red, in=0, out=0] (237.center) to (238.center);
    \draw [thick, in=180, out=-165, looseness=0.50] (243.center) to (244.center);
    \draw [thick, style=dashed, in=0, out=0, looseness=0.50] (243.center) to (244.center);
    \draw [very thick, color=red, in=-180, out=180] (245.center) to (246.center);
    \draw [very thick, color=red, in=15, out=-15] (247.center) to (248.center);
    \draw [very thick, color=red, in=165, out=-165] (254.center) to (255.center);
    \draw [very thick, color=red, in=0, out=0] (256.center) to (257.center);
    \draw [very thick, color=red, in=-180, out=180] (263.center) to (264.center);
    \draw [very thick, color=red, in=0, out=0] (267.center) to (268.center);
    \draw [very thick, color=red, in=-180, out=180] (269.center) to (270.center);
    \draw [very thick, color=red, in=0, out=0] (273.center) to (274.center);
    \draw [very thick, color=red, in=-180, out=180] (275.center) to (276.center);
    \draw [very thick, color=red, in=0, out=0] (279.center) to (280.center);
    \draw [very thick, color=red, in=-180, out=180] (281.center) to (282.center);
    \draw [very thick, color=red, in=0, out=0] (285.center) to (286.center);
  \end{pgfonlayer}
\end{tikzpicture}
\end{center}
 \caption{The replica computation for the density matrix $\rhoA^{\!ij}$ constructed in Fig.~\ref{fig:nrdensitymS2W} glues $q$ balls $\ball$ whose boundary is the union of $\rA_\pm$  for the chosen bipartition. As a result one ends up with a functional integral which computes $\zcs(\Sp^3, (R_i, R_j)^q) \equiv \zcs(\Sp^3, R_i,R_j, \cdots, R_i, R_j)$, the $\Sp^3$ partition function with 
 $q$-pairs of Wilson lines along unknots in representations $R_i$ and $R_j$, respectively. 
 }
\label{fig:nrdensitymS2Wq}
\end{figure}
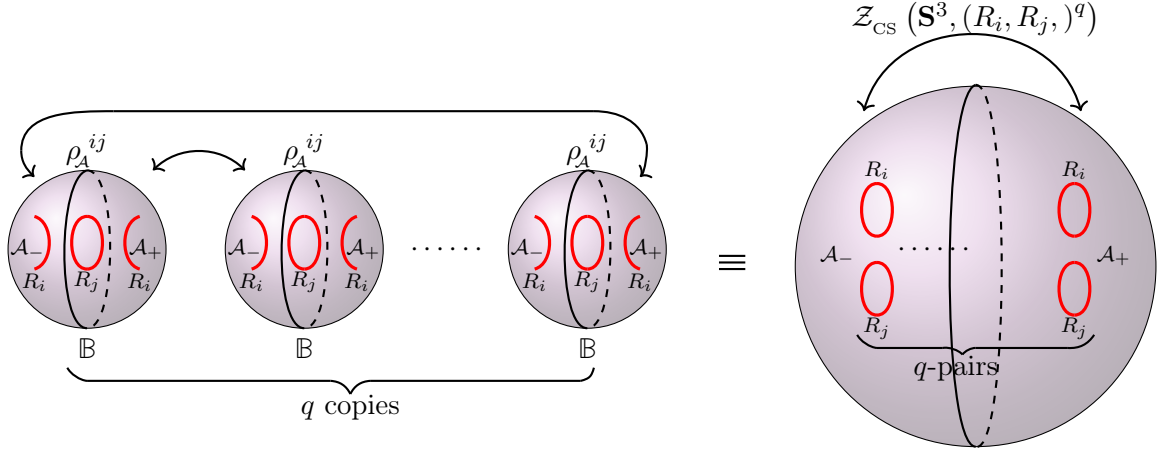

With this picture for the density matrix, the replica computation is straightforward and is explained in Fig.~\ref{fig:nrdensitymS2Wq}. We find:
\begin{equation}\label{eq:SAij}
S_\rA^{(q)}(\rhoA^{\! ij}) = \frac{1}{1-q} \log\left[ \frac{\zcs(\Sp^3, (R_i, R_j)^q) }{\left(\zcs(\Sp^3, R_i,R_j)\right)^q}\right] = \log \zcs(\Sp^3)\,,
\end{equation}	
where the numerator in the argument of the first log is to be understood as the partition function with $q$ pairs of unlinked Wilson lines carrying representations $R_i$ and $R_j$. To derive the second equality we are making use of  three-manifold surgery to show that $\zcs(\Sp^3, (R_i, R_j)^q) =\zcs(\Sp^3, R_i,R_j)^q\, \zcs(\Sp^3)^{q-1} $. 
The argument is as follows: we supply  $q-1$ copies of the three-sphere partition function, each decomposed into a pair of three-balls. We scoop out three-balls containing the 
pair $(R_i,R_j)$  from the decomposition of the partition function $\zcs(\Sp^3, (R_i, R_j)^q)$ and replace the resulting hole with one the three-balls we have supplied. It is clear that we can then extract $(q-1)$ pairs of $(R_i,R_j)$, leaving behind one-pair in the original $\Sp^3$. Piecing all of this together, we see that we get the  answer given in \eqref{eq:SAij}.

The uplift of this construction to the deformed conifold is straightforward. We start with the representation of $\zcs(\Sp^3, R_i,R_j)$ in terms of having two sets of probe branes on 
Lagrangian cycles. Passing through the geometric transition, we land up with the resolved conifold $\rco$ also with these Lagrangian cycles, which we can label as $\lagr_{R_i}$
and $\lagr_{R_j}$.  To build the state $\ket{\! \PhiR^{ij}}$ dual to $\psiSkij$ on the Chern-Simons side, we again employ the Heegard decomposition of the $\Sp^3 \in \rco$ and open it up into three-balls $\ball_\pm$. The new feature now is that the Lagrangian cycles are sliced through by the decomposition.

The Lagrangians $\lagr_{R_i}$  and  $\lagr_{R_j}$ in the resolved conifold, as mentioned above, have topology $\Sp^1 \times \mathbb{R}^2$, where the $\Sp^1$ is to be identified with the unknot. Since we are cutting through the unknot, the result will be to split the $\lagr_{\rco} = \lagr_{\rco}^\pm$ each with topology $\mathcal{I} \times \mathbb{R}^2$. $\mathcal{I}$ here is an interval which is obtained from the open Wilson line. Note that the marked points have been uplifted to marked planes, the ends of the interval, extending along the non-compact direction of the Lagrangian cycles. Despite the fact that our construction cuts open the Wilson lines, it is worth keeping in mind that they do not pass through the entangling brane. So while the global state preparation on the Cauchy surfaces $\Gamma_\pm$ involves  a more complicated construction, opening up the probe D-branes, not much is happening at the level of the degrees of freedom being bipartitioned by the entangling brane. This already suggests the answer we find in \eqref{eq:SAij}, and we can recover this in a similar manner from the closed string replica.

\begin{figure}[ht]
    \begin{center}
\usetikzlibrary{backgrounds}
\begin{tikzpicture}[scale=1.35]
\begin{pgfonlayer}{nodelayer}
		\node [style=none] (11) at (-2, 0) {};
		\node [style=none] (12) at (-2, -4.5) {};
		\node [style=none] (17) at (-5.02, 1) {};
		\node [style=none] (18) at (-5.02, -5) {};
		\node [style=none] (16) at (-8, 0) {};
		\node [style=none] (13) at (-8, -4.5) {};
		\node [style=none] (14) at (-1.35, 0) {$ r=\infty $};
		\node [style=none] (15) at (-1.5, -4.5) {$ r=0 $};
		\node [style=none] at (-4, 1.5) {$ \rco_+$};
		\node [style=none] at (-6, 1.5) {$\rco_-$};
		\node [style=none] at (-5, 1.5) {$ t=0$};
		\node [style=none] at (-6, .5) {\small{$\rxAk(r_1)$}};
		\node [style=none] at (-4, .5) {\small{$\rxAb(r_1)$}};
		\node [style=none] at (-5.6, -1.3) {\tiny{$\rxAk(r_2)$}};
		\node [style=none] at (-4.4, -1.3) {\tiny{$\rxAb(r_2)$}};
		\node [style=none] at (-6, -.5) {\small{$\rxAck(r_1)$}};
		\node [style=none] at (-4, -.5) {\small{$\rxAcb(r_1)$}};
		\node [style=none] at (-5.6, -2) {\tiny{$\rxAck(r_2)$}};
		\node [style=none] at (-4.4, -2) {\tiny{$\rxAcb(r_2)$}};
		\node [style=none] at (-6.5, -2.2) {$\entsurfx_-$};
		\node [style=none] at (-3.4, -2.2) {$\entsurfx_+$};
	\end{pgfonlayer}
	\begin{pgfonlayer}{edgelayer}
	\draw [very thick, style=dashed,color=red, in=90, out=-90] (17.center) to (18.center);
    \shade[ball color = olive!40, opacity = 0.4] (-7.5,0) circle (.3cm); 
 \shade[ball color = olive!40, opacity = 0.4] (-7,-1.65) circle (.3cm);
 \shade[ball color = olive!40, opacity = 0.4] (-6.5,-2.75) circle (.3cm);
  \shade[ball color = olive!40, opacity = 0.4] (-6,-3.75) circle (.3cm);
 \shade[ball color = olive!40, opacity = 0.4] (-5.5,-4.75) circle (.3cm);
 \draw [very thick, <-, in=90, out=-90] (11.center) to (12.center);
 \shade[ball color = olive!40, opacity = 0.4] (-2.5,0) circle (.3cm); 
 \shade[ball color = olive!40, opacity = 0.4] (-3,-1.65) circle (.3cm);
 \shade[ball color = olive!40, opacity = 0.4] (-3.5,-2.75) circle (.3cm);
  \shade[ball color = olive!40, opacity = 0.4] (-4,-3.75) circle (.3cm);
 \shade[ball color = olive!40, opacity = 0.4] (-4.5,-4.75) circle (.3cm); 
  \shade[ball color = violet!120, opacity = 0.4] (-4.975,-4.75) circle (.02cm); 
 \shade[ball color = violet!120, opacity = 0.4] (-4.75,-3.75) circle (.2cm);
 \shade[ball color = violet!120, opacity = 0.4] (-4.55,-2.75) circle (.4cm);
  \shade[ball color = violet!120, opacity = 0.4] (-4.35,-1.65) circle (.6cm);
 \shade[ball color = violet!120, opacity = 0.4] (-4,0) circle (.95cm);
  \shade[ball color = violet!120, opacity = 0.4] (-5.075,-4.75) circle (.02cm); 
 \shade[ball color = violet!120, opacity = 0.4] (-5.25,-3.75) circle (.2cm);
 \shade[ball color = violet!120, opacity = 0.4] (-5.475,-2.75) circle (.4cm);
  \shade[ball color = violet!120, opacity = 0.4] (-5.65,-1.65) circle (.6cm);
 \shade[ball color = violet!120, opacity = 0.4] (-6,-0) circle (.95cm);
   \draw [line width=0.5mm,dashed]   (-5.05,0) arc (0:180:0.95 and .2);
  \draw [line width=0.5mm] (-5.05,0) arc (360:180:0.95 and .2);
  \draw[color=violet,thick] (-4.95,0) to (-3.95,0);
   \draw [line width=0.5mm,dashed]   (-3.05,0) arc (0:180:0.95 and .2);
  \draw [line width=0.5mm] (-3.05,0) arc (360:180:0.95 and .2);
   \draw[color=violet,thick] (-5.05,0) to (-6.05,0);
     \draw [very thick,dashed]   (-5.05,-1.65) arc (0:180:0.6 and .17);
  \draw [very thick] (-5.05,-1.65) arc (360:180:0.6 and .17);
    \draw[color=violet,thick] (-5.05,-1.65) to (-5.7,-1.65);
   \draw  [very thick,dashed]  (-3.75,-1.65) arc (0:180:0.6 and  .17);
  \draw [very thick] (-3.75,-1.65) arc (360:180:0.6 and .17);
     \draw[color=violet,thick] (-4.95,-1.65) to (-4.35,-1.65);
     \draw [very thick,dashed]   (-5.075,-2.75) arc (0:180:0.4 and .12);
  \draw [very thick] (-5.075,-2.75) arc (360:180:0.4 and .12);
   \draw[color=violet,thick] (-4.95,-2.75) to (-4.5,-2.75);
   \draw  [very thick,dashed]  (-4.15,-2.75) arc (0:180:0.4 and .12);
  \draw [very thick] (-4.15,-2.75) arc (360:180:0.4 and .12);
     \draw[color=violet,thick] (-5.05,-2.75) to (-5.5,-2.75);
       \draw [very thick,dashed]   (-5.05,-3.75) arc (0:180:0.2 and .06);
  \draw [very thick] (-5.05,-3.75) arc (360:180:0.2 and .06);
  \draw[color=violet,thick] (-5.05,-3.75) to (-5.25,-3.75);
   \draw [very thick,dashed]   (-4.55,-3.75) arc (0:180:0.2 and .06);
  \draw [very thick] (-4.55,-3.75) arc (360:180:0.2 and .06);
   \draw[color=violet,thick] (-4.95,-3.75) to (-4.75,-3.75);
     \draw [very thick,dashed]   (-5.05,-4.75) arc (0:180:0.01 and .005);
  \draw [very thick] (-5.05,-4.75) arc (360:180:0.01 and .005);
   \draw[color=violet,thick] (-4.95,-4.75) to (-4.94,-4.75);
   \draw [very thick,dashed]   (-4.95,-4.75) arc (0:180:0.01 and .005);
  \draw [very thick] (-4.95,-4.75) arc (360:180:0.01 and .005);
    \draw[color=violet,thick] (-5.05,-4.75) to (-5.06,-4.75);
     \draw   [red,dashed]  (-7.2,0) arc (0:180:0.3 and .1);
  \draw [red](-7.2,0) arc (360:180:0.3 and .1);
   \draw   [red,dashed]  (-2.2,0) arc (0:180:0.3 and .1);
  \draw [red](-2.2,0) arc (360:180:0.3 and .1);
     \draw   [red,dashed]  (-6.7,-1.65) arc (0:180:0.3 and .1);
  \draw [red](-6.7,-1.65) arc (360:180:0.3 and .1);
   \draw  [red,dashed]   (-2.7,-1.65) arc (0:180:0.3 and  .1);
  \draw [red](-2.7,-1.65) arc (360:180:0.3 and .1);
     \draw   [red,dashed]  (-6.2,-2.75) arc (0:180:0.3 and .1);
  \draw [red](-6.2,-2.75) arc (360:180:0.3 and .1);
   \draw    [red,dashed] (-3.2,-2.75) arc (0:180:0.3 and .1);
  \draw [red](-3.2,-2.75) arc (360:180:0.3 and .1);
       \draw   [red,dashed]  (-5.7,-3.75) arc (0:180:0.3 and .1);
  \draw [red](-5.7,-3.75) arc (360:180:0.3 and .1);
   \draw    [red,dashed] (-3.7,-3.75) arc (0:180:0.3 and .1);
  \draw  [red](-3.7,-3.75) arc (360:180:0.3 and .1);
     \draw   [red,dashed]  (-5.2,-4.75) arc (0:180:0.3 and .1);
  \draw [red](-5.2,-4.75) arc (360:180:0.3 and .1);
   \draw    [red,dashed] (-4.2,-4.75) arc (0:180:0.3 and .1);
  \draw  [red](-4.2,-4.75) arc (360:180:0.3 and .1);
  \draw (-4.975,0) arc (180:360:.975cm and 0.225cm) -- (-4.975,-4.75) -- cycle;
    \shade[left color=blue!5!blue,right color=green!120!white,opacity=0.3] (-4.975,0) arc (180:360:.975cm and 0.25cm) -- (-4.975,-4.75) -- cycle; 
    \draw (-6.975,0) arc (180:360:.95cm and 0.225cm) -- (-5.075,-4.75) -- cycle;
    \shade[left color=blue!5!green,right color=blue!120!white,opacity=0.3] (-6.975,0) arc (180:360:.95cm and 0.25cm) -- (-5.075,-4.75) -- cycle;
        \shade[left color=blue!5!blue,right color=green!120!white,opacity=0.3] (-4.975,0) arc (180:360:.975cm and 0.25cm) -- (-4.975,-4.75) -- cycle; 
    \draw (-5.475,0.2) arc (90:270:.45cm and 0.2cm) -- (-5.075,-4.75) -- cycle;
    \shade[left color=black!75!violet,right color=red!120!white,opacity=0.7] (-5.475,0.2) arc (90:270:.45cm and 0.2cm) -- (-5.075,-4.75) -- cycle;
       \draw (-5.925,0) arc (180:270:.45cm and 0.2cm) -- (-5.075,-4.75) -- cycle;
    \shade[left color=red!125!violet,right color=red!120!white,opacity=0.3] (-5.925,0) arc (180:270:.45cm and 0.2cm) -- (-5.075,-4.75) -- cycle;
    \draw (-3.475,0.2) arc (90:255:.45cm and 0.2cm) -- (-4.975,-4.75) -- cycle;
    \shade[left color=red!125!black,right color=black!75!white,opacity=0.7] (-3.475,0.2) arc (90:255:.45cm and 0.2cm) -- (-4.975,-4.75) -- cycle;
       \draw (-3.925,0) arc (180:255:.45cm and 0.2cm) -- (-4.975,-4.75) -- cycle;
    \shade[left color=red!125!violet,right color=red!120!white,opacity=0.3] (-3.925,0) arc (180:255:.45cm and 0.2cm) -- (-4.975,-4.75) -- cycle;
    \draw (-4.575,0.2) arc (90:-90:.45cm and 0.2cm) -- (-4.975,-4.75) -- cycle;
    \shade[left color=red!125!black,right color=black!75!white,opacity=0.7] (-4.575,0.2) arc (90:-90:.45cm and 0.2cm) -- (-4.975,-4.75) -- cycle;
       \draw (-4.125,0) arc (0:-90:.45cm and 0.2cm) -- (-4.975,-4.75) -- cycle;
    \shade[left color=red!125!violet,right color=red!120!white,opacity=0.3] (-4.125,0) arc (0:-90:.45cm and 0.2cm) -- (-4.975,-4.75) -- cycle;
    \draw (-6.575,0.2) arc (90:-75:.45cm and 0.2cm) -- (-5.075,-4.75) -- cycle;
    \shade[left color=black!75!violet,right color=red!120!white,opacity=0.7] (-6.575,0.2) arc (90:-75:.45cm and 0.2cm) -- (-5.075,-4.75) -- cycle;
       \draw (-6.125,0) arc (0:-75:.45cm and 0.2cm) -- (-5.075,-4.75) -- cycle;
    \shade[left color=red!75!violet,right color=red!120!white,opacity=0.3] (-6.125,0) arc (0:-75:.45cm and 0.2cm) -- (-5.075,-4.75) -- cycle;
	\end{pgfonlayer}
\end{tikzpicture}
    \end{center}
\caption{The surgical split of the resolved conifold $\rco = \rco_- \cup \rco_+$ when the configuration of interest contains Wilson lines. The Wilson lines uplift to Lagrangian cycles (indicated by the red surfaces) $\lagr_{R_i}$ and $\lagr_{R_j}$ if we consider the closed string dual of the density matrix prepared in Fig.~\ref{fig:nrdensitymS2W}. These cycles form an unknot in the $\Sp^3$, are stretched out along the radial direction of the cone, and wrap the equatorial $\Sp^1 \subset \Sp^2(1)$. 
 When we decompose the $\Sp^3$ into $\ball_\pm$, these Lagrangian cycles are decomposed into 
 $ \lagr_{R} = \left(\mathcal{I}_\pm \times {\mathbb R}^2\right)_{R}$ and split endpoints lie on $\Gamma_\pm$ as we have tried to illustrate.
 }
\label{fig:rdensityLagconif}
\end{figure}
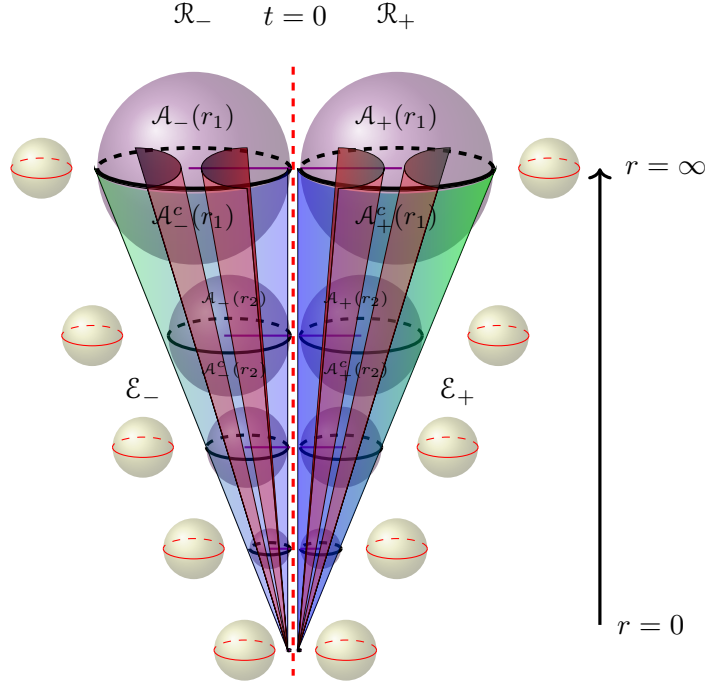

To be explicit, we start with the data $(\rco, \lagr_{R_i}, \lagr_{R_j})$ which we decompose into the corresponding bra and ket pieces $(\rco_\pm, \lagr^\pm_{R_i}, 
\lagr^\pm_{R_j})$ and then proceed to bipartition $\rco_\pm$ across $\entsurfx$. We have attempted to illustrate this construction in Fig.~\ref{fig:rdensityLagconif}. 
We then construct the branched cover geometry for $\rhoAx$ as $\left(\rco_q, \lagr^\pm_{R_i}, 
(\lagr_{R_i}, \lagr_{R_j})^{q-1} , \lagr_{R_j} \right)$, where we have already acknowledged the fact that we glue back  the whole cycle with the $R_j$ representation in each realization of $\rhoAx$, and obtain $(q-1)$ copies of the whole cycle in the $R_i$ representation from the inner product.  The closed string partition function on this geometry gives us 
$\zc\left(\rco, (\lagr_{R_i}, \lagr_{R_j} )^q\right) = \zc(\rco, \lagr_{R_i}, \lagr_{R_j})^q \, \zc(\rco)^{q-1}$. We prove this relation in a manner analogous to the one employed above for Chern-Simons. We can compute the normalization factor for the density matrix $\rhoAx$ straightforwardly, and conclude that $S_\rxA = \log \zc(\rco)$, as expected.

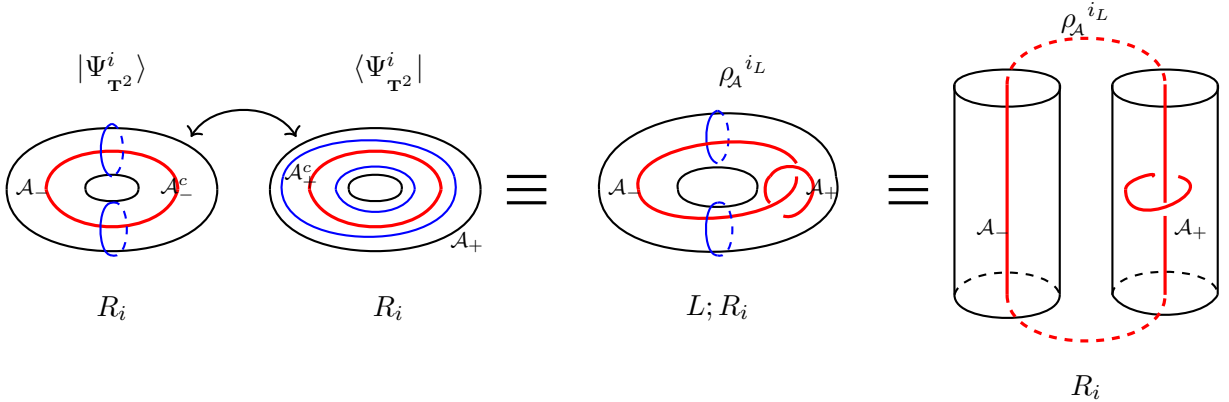
\begin{figure}[h]
\begin{center}
\usetikzlibrary{backgrounds}
\begin{tikzpicture}[scale=.7]
\begin{pgfonlayer}{nodelayer}
    \node [style=none] (0) at (-13, 2) {};
    \node [style=none] (1) at (-9, 2) {};
    \node [style=none] (2) at (-11.5, 2) {};
    \node [style=none] (3) at (-10.5, 2) {};
    \node [style=none] (4) at (-12.25, 2) {};
    \node [style=none] (5) at (-9.75, 2) {};
    \node [style=none] (6) at (-11, 3.25) {};
    \node [style=none] (7) at (-11, 2.25) {};
    \node [style=none] (8) at (-11, 1.75) {};
    \node [style=none] (9) at (-11, 0.75) {};
    \node [style=none] (10) at (-12.5, 2) {$\scriptstyle{\rA_-}$};
    \node [style=none] (12) at (-9.75, 2) {$\scriptstyle{\rAc_-}$};
    \node [style=none] (13) at (-8, 2) {};
    \node [style=none] (14) at (-4, 2) {};
    \node [style=none] (15) at (-6.5, 2) {};
    \node [style=none] (16) at (-5.5, 2) {};
    \node [style=none] (17) at (-7.25, 2) {};
    \node [style=none] (18) at (-4.75, 2) {};
    \node [style=none] (19) at (-5.25, 2) {};
    \node [style=none] (20) at (-6.75, 2) {};
    \node [style=none] (21) at (-4.5, 2.25) {};
    \node [style=none] (22) at (-7.75, 2.25) {};
    \node [style=none] (23) at (-4.25, 1) {$\scriptstyle{\rA_+}$};
    \node [style=none] (24) at (-7.4, 2.25) {$\scriptstyle{\rAc_+}$};
    \node [style=none] (26) at (-11, -.25) {$R_i$};
    \node [style=none]  at (-11, 4.25) {$\psiTk$};
    \node [style=none] (27) at (-5.75, -.25) {$R_i$};
    \node [style=none]  at (-5.75, 4.25) {$\psiTb$};
    \node [style=none] (28) at (-9.5, 3) {};
    \node [style=none] (29) at (-7.5, 3) {};
    \node [style=none] (30) at (-3.5, 2.25) {};
    \node [style=none] (31) at (-3.5, 2) {};
    \node [style=none] (32) at (-3.5, 1.75) {};
    \node [style=none] (33) at (-2.75, 1.75) {};
    \node [style=none] (34) at (-2.75, 2) {};
    \node [style=none] (35) at (-2.75, 2.25) {};
    \node [style=none] (47) at (-1.75, 2) {};
    \node [style=none] (48) at (2.75, 2.25) {};
    \node [style=none] (49) at (-0.25, 2) {};
    \node [style=none] (50) at (1.25, 2) {};
    \node [style=none] (51) at (-1, 2) {};
    \node [style=none] (52) at (2, 2.25) {};
    \node [style=none] (53) at (0.5, 3.5) {};
    \node [style=none] (54) at (0.5, 2.5) {};
    \node [style=none] (55) at (0.5, 1.75) {};
    \node [style=none] (56) at (0.5, 0.75) {};
    \node [style=none] (57) at (-1.25, 2) {$\scriptstyle{\rA_-}$};
    \node [style=none] (58) at (2.5, 2) {$\scriptstyle{\rA_+}$};
    \node [style=none] (59) at (0.5, -.25) {$L;R_i$};
    \node [style=none]  at (7.5, 5.25) {$\rhoA^{i_L}$};
    \node [style=none] (60) at (1.5, 2.25) {};
    \node [style=none] (61) at (2.25, 2.25) {};
    \node [style=none] (62) at (1.5, 1.75) {};
    \node [style=none] (63) at (1.75, 1.5) {};
    \node [style=none] (64) at (2, 2.5) {};
    \node [style=none] (65) at (3.75, 2.25) {};
    \node [style=none] (66) at (3.75, 2) {};
    \node [style=none] (67) at (3.75, 1.75) {};
    \node [style=none] (68) at (4.5, 1.75) {};
    \node [style=none] (69) at (4.5, 2) {};
    \node [style=none] (70) at (4.5, 2.25) {};
    \node [style=none] (71) at (5, 4) {};
    \node [style=none] (72) at (5, 0) {};
    \node [style=none] (73) at (7, 0) {};
    \node [style=none] (74) at (7, 4) {};
    \node [style=none] (75) at (8, 4) {};
    \node [style=none] (76) at (8, 0) {};
    \node [style=none] (77) at (10, 0) {};
    \node [style=none] (78) at (10, 4) {};
    \node [style=none] (79) at (6, 4) {};
    \node [style=none] (80) at (6, 0) {};
    \node [style=none] (81) at (9, 4) {};
    \node [style=none] (82) at (9, 1.75) {};
    \node [style=none] (83) at (8.75, 2.25) {};
    \node [style=none] (84) at (9.25, 2.25) {};
    \node [style=none] (85) at (8.25, 2) {};
    \node [style=none] (86) at (9, 1.5) {};
    \node [style=none] (87) at (9, 0) {};
    \node [style=none] (88) at (5.75, 1.25) {$\scriptstyle{\rA_-}$};
    \node [style=none] (89) at (2.25, 2.25) {};
    \node [style=none] (90) at (2.25, 2.25) {};
    \node [style=none] (91) at (9.5, 1.25) {$\scriptstyle{\rA_+}$};
    \node [style=none] (92) at (7.5, -1.75) {$R_i$};
    \node [style=none] (93) at (7.75, 4) {};
    \node [style=none] (95) at (1, 4.25) {$\rhoA^{i_L}$};
  \end{pgfonlayer}
  \begin{pgfonlayer}{edgelayer}
    \draw [thick, in=-90, out=-90] (0.center) to (1.center);
    \draw [thick, bend left=90] (0.center) to (1.center);
    \draw [thick, bend right=90, looseness=0.75] (2.center) to (3.center);
    \draw [thick, in=75, out=105] (2.center) to (3.center);
    \draw [very thick, color=red, bend right=75] (4.center) to (5.center);
    \draw [very thick, color=red, bend left=90] (4.center) to (5.center);
    \draw [thick, color=blue, bend right=75, looseness=0.75] (6.center) to (7.center);
    \draw [thick, color=blue, style=dashed, bend left=75, looseness=0.75] (6.center) to (7.center);
    \draw [thick, color=blue, bend right=90, looseness=0.75] (8.center) to (9.center);
    \draw [style=dashed, color=blue, thick, bend left=105] (8.center) to (9.center);
    \draw [thick, in=-90, out=-90] (13.center) to (14.center);
    \draw [thick, bend left=90] (13.center) to (14.center);
    \draw [thick, bend right=90, looseness=0.75] (15.center) to (16.center);
    \draw [thick, in=75, out=105] (15.center) to (16.center);
    \draw [very thick, color=red, bend right=75] (17.center) to (18.center);
    \draw [very thick, color=red, bend left=90] (17.center) to (18.center);
    \draw [thick,color=blue,  bend left=270] (19.center) to (20.center);
    \draw [thick, color=blue, bend left=75] (19.center) to (20.center);
    \draw [thick,color=blue, bend right=75, looseness=0.75] (21.center) to (22.center);
    \draw [thick, color=blue,bend left=105, looseness=1.25] (21.center) to (22.center);
    \draw [thick, <->, bend left=60] (28.center) to (29.center);
    \draw [very thick] (30.center) to (35.center);
    \draw [very thick] (31.center) to (34.center);
    \draw [very thick] (32.center) to (33.center);
    \draw [thick, in=-75, out=-90] (47.center) to (48.center);
    \draw [thick, bend left=90] (47.center) to (48.center);
    \draw [thick, bend right=90, looseness=0.75] (49.center) to (50.center);
    \draw [thick, in=75, out=105] (49.center) to (50.center);
    \draw [very thick, color=red, bend right=90, looseness=0.75] (51.center) to (52.center);
    \draw [thick,color=blue, bend right=75, looseness=0.75] (53.center) to (54.center);
    \draw [thick, color=blue,style=dashed, bend left=75, looseness=0.75] (53.center) to (54.center);
    \draw [thick, color=blue,bend right=90, looseness=0.75] (55.center) to (56.center);
    \draw [style=dashed, color=blue,thick, bend left=105] (55.center) to (56.center);
    \draw [very thick, red, bend right=60, looseness=0.75] (60.center) to (62.center);
    \draw [very thick, red, bend right=75] (63.center) to (61.center);
    \draw [very thick, red, bend left=300] (61.center) to (60.center);
    \draw [very thick, red, in=120, out=105, looseness=0.75] (51.center) to (64.center);
    \draw [very thick] (65.center) to (70.center);
    \draw [very thick] (66.center) to (69.center);
    \draw [very thick] (67.center) to (68.center);
    \draw [thick] (71.center) to (72.center);
    \draw [thick] (74.center) to (73.center);
    \draw [thick] (75.center) to (76.center);
    \draw [thick] (78.center) to (77.center);
    \draw [thick, bend right=105, looseness=0.75] (71.center) to (74.center);
    \draw [thick, bend left=75, looseness=0.50] (71.center) to (74.center);
    \draw [thick, style=dashed, bend left=75, looseness=0.75] (72.center) to (73.center);
    \draw [thick, bend right=90, looseness=0.75] (72.center) to (73.center);
    \draw [thick, bend right=105, looseness=0.75] (75.center) to (78.center);
    \draw [thick, bend left=75, looseness=0.50] (75.center) to (78.center);
    \draw [thick, style=dashed, bend left=90, looseness=0.75] (76.center) to (77.center);
    \draw [thick, bend right=60, looseness=0.75] (76.center) to (77.center);
    \draw [very thick, red] (79.center) to (80.center);
    \draw [very thick, red] (81.center) to (82.center);
    \draw [very thick, red, in=30, out=90] (85.center) to (83.center);
    \draw [very thick, red, in=-15, out=-90, looseness=2.25] (85.center) to (84.center);
    \draw [very thick, red] (86.center) to (87.center);
    \draw [very thick, style=dashed, red, bend left=90] (79.center) to (81.center);
    \draw [very thick, style=dashed, red, bend right=90] (80.center) to (87.center);
  \end{pgfonlayer}
  \end{tikzpicture}
\end{center}
 \caption{Density matrix and R\'enyi entropy for the bipartitioning of the state $\psiTk$. The regions $\rA$ and $\rAc$ split the ${\bf T}^2$ into two-connected cylinders, and have a two-component entangling surface. We have constructed the density matrix by gluing the complement region $\rAc_\pm$ from the solid torii $\mcsT^\mp$ used to build the state and its conjugate, and found a convenient representation for it. The density matrix $\rhoA^{i_L}$ can be obtained by performing the Chern-Simons path integral over a solid torus with a Wilson line placed along a link $L$ in it. The boundary of the solid torus is given by the union of the complementary regions $\rA_-$ and $\rA_+$. }
\label{fig:nrdensitymTW}
\end{figure}

\paragraph{2. Slicing Wilson lines in the interior:} We now turn to the case of the Wilson line keeping away from the regions $\rA$ and $\rAc$, but the separation of a domain around $\rA$ necessarily slicing through the Wilson line. As an example we will consider the state  $\psiTk$  introduced in \S\ref{sec:wbipart1}. We take $\rA$ to be a single-connected region having a two-component entangling surface with $\rAc$ as illustrated in Fig.~\ref{fig:nrdensitymTW}. 
Given our construction of the state $\psiTk$ on a solid torus $\mcsT^-$ with a contractible $a$-cycle, it follows that the entangling surface is a pair of circles along the $a$-cycle of the ket ${\bf T}^2_-$. The disks which fill in the entangling surfaces are pierced through by the Wilson line.  In the dual state, $\psiTb$, the entangling surfaces are also two circles which still lie along the 
$a$-cycle. We should however bear in mind that the $a$-cycle is non-contractible in the dual $\mcsT^+$. It is useful to visualize the cross-section of the cut along the entangling surfaces as a filled annulus, with boundaries $\entsurf$. The Wilson line runs along the non-contractible cycle of this annulus.

Gluing the regions $\rAc_\pm$ from these two torii can be seen to lead to a configuration which can be mapped to the path integral over a solid torus with a Wilson line placed along a 
link $L$ in it. The presence of the link can independently be inferred from the fact that we obtained the state by opening up $\zcs(\Sp^3, L;R_i)$. The boundary of the resulting solid torus representing $\rhoA^{i_L}$ is a union of the complementary regions $\rA_-$ and $\rA_+$, see the top part of Fig.~\ref{fig:nrdensitymTW}. 
Further identifying $\rA_\pm$ gives us back a three-sphere with the link inside, so we recover again \eqref{eq:wTnorm}.
 
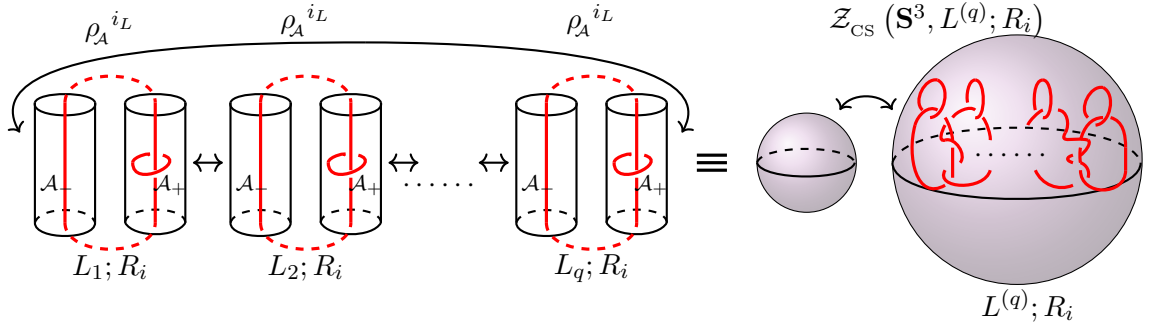
\begin{figure}[t]
\begin{center}
\usetikzlibrary{backgrounds}
\begin{tikzpicture}[scale=.4]
\begin{pgfonlayer}{nodelayer}
    \node [style=none] (0) at (4.5, 3) {};
    \node [style=none] (1) at (16, 6) {};
    \node [style=none] (2) at (26.5, 3) {};
    \node [style=none] (3) at (5, 4) {};
    \node [style=none] (4) at (5, 0) {};
    \node [style=none] (5) at (7, 0) {};
    \node [style=none] (6) at (7, 4) {};
    \node [style=none] (7) at (8, 4) {};
    \node [style=none] (8) at (8, 0) {};
    \node [style=none] (9) at (10, 0) {};
    \node [style=none] (10) at (10, 4) {};
    \node [style=none] (11) at (6, 4) {};
    \node [style=none] (12) at (6, 0) {};
    \node [style=none] (13) at (9, 4) {};
    \node [style=none] (14) at (9, 1.75) {};
    \node [style=none] (15) at (8.75, 2.25) {};
    \node [style=none] (16) at (9.25, 2.25) {};
    \node [style=none] (17) at (8.25, 2) {};
    \node [style=none] (18) at (9, 1.5) {};
    \node [style=none] (19) at (9, 0) {};
    \node [style=none] (20) at (5.75, 1.25) {$\scriptstyle{\rA}_-$};
    \node [style=none] (21) at (9.5, 1.25) {$\scriptstyle{\rA}_+$};
    \node [style=none] (22) at (7.5, 6.5) {$\rhoA^{i_L}$};
    \node [style=none] (23) at (7.75, 4) {};
    \node [style=none] (24) at (16.75, 2) {};
    \node [style=none] (25) at (11.25, 2) {};
    \node [style=none] (26) at (10.25, 2) {};
    \node [style=none] (27) at (11.5, 4) {};
    \node [style=none] (28) at (11.5, 0) {};
    \node [style=none] (29) at (13.5, 0) {};
    \node [style=none] (30) at (13.5, 4) {};
    \node [style=none] (31) at (14.5, 4) {};
    \node [style=none] (32) at (14.5, 0) {};
    \node [style=none] (33) at (16.5, 0) {};
    \node [style=none] (34) at (16.5, 4) {};
    \node [style=none] (35) at (12.5, 4) {};
    \node [style=none] (36) at (12.5, 0) {};
    \node [style=none] (37) at (15.5, 4) {};
    \node [style=none] (38) at (15.5, 1.75) {};
    \node [style=none] (39) at (15.25, 2.25) {};
    \node [style=none] (40) at (15.75, 2.25) {};
    \node [style=none] (41) at (14.75, 2) {};
    \node [style=none] (42) at (15.5, 1.5) {};
    \node [style=none] (43) at (15.5, 0) {};
    \node [style=none] (44) at (12.25, 1.25) {$\scriptstyle{\rA}_-$};
    \node [style=none] (45) at (16, 1.25) {$\scriptstyle{\rA}_+$};
    \node [style=none] (46) at (14, 6.75) {$\rhoA^{i_L}$};
    \node [style=none] (47) at (14.25, 4) {};
    \node [style=none] (49) at (18.5, 1.25) {$\cdots\cdots$};
    \node [style=none] (50) at (20.75, 2) {};
    \node [style=none] (51) at (19.75, 2) {};
    \node [style=none] (52) at (17.75, 2) {};
    \node [style=none] (53) at (21, 4) {};
    \node [style=none] (54) at (21, 0) {};
    \node [style=none] (55) at (23, 0) {};
    \node [style=none] (56) at (23, 4) {};
    \node [style=none] (57) at (24, 4) {};
    \node [style=none] (58) at (24, 0) {};
    \node [style=none] (59) at (26, 0) {};
    \node [style=none] (60) at (26, 4) {};
    \node [style=none] (61) at (22, 4) {};
    \node [style=none] (62) at (22, 0) {};
    \node [style=none] (63) at (25, 4) {};
    \node [style=none] (64) at (25, 1.75) {};
    \node [style=none] (65) at (24.75, 2.25) {};
    \node [style=none] (66) at (25.25, 2.25) {};
    \node [style=none] (67) at (24.25, 2) {};
    \node [style=none] (68) at (25, 1.5) {};
    \node [style=none] (69) at (25, 0) {};
    \node [style=none] (70) at (21.75, 1.25) {$\scriptstyle{\rA}_-$};
    \node [style=none] (71) at (25.5, 1.25) {$\scriptstyle{\rA}_+$};
    \node [style=none] (72) at (23.5, 6.75) {$\rhoA^{i_L}$};
    \node [style=none] (73) at (23.75, 4) {};
    \node [style=none] (74) at (27, 2.25) {};
    \node [style=none] (75) at (27, 2) {};
    \node [style=none] (76) at (27, 1.75) {};
    \node [style=none] (77) at (28, 2.25) {};
    \node [style=none] (78) at (28, 2) {};
    \node [style=none] (79) at (28, 1.75) {};
    \node [style=none] (80) at (29, 2) {};
    \node [style=none] (81) at (32.25, 2) {};
    \node [style=none] (82) at (33.5, 2) {};
    \node [style=none] (83) at (41.75, 2) {};
    \node [style=none] (98) at (35, 6.75) {$\zcs\left(\Sp^3,L^{(q)}; R_i\right)$};
    \node [style=none] (107) at (34.25, 3.25) {};
    \node [style=none] (108) at (35.25, 1.25) {};
    \node [style=none] (109) at (35, 3.75) {};
    \node [style=none] (111) at (34.5, 3.75) {};
    \node [style=none] (112) at (34.75, 3.5) {};
    \node [style=none] (113) at (35, 4.75) {};
    \node [style=none] (117) at (37.5, 2.25) {$\cdots\cdots$};
    \node [style=none] (123) at (31.75, 3.75) {};
    \node [style=none] (124) at (33.5, 3.75) {};
    \node [style=none] (125) at (7.5, -1.5) {$L_1; R_i$};
    \node [style=none] (127) at (38, -2.75) {$L^{(q)}; R_i$};
    \node [style=none] (129) at (35.5, 3.5) {};
    \node [style=none] (130) at (35.25, 2.75) {};
    \node [style=none] (131) at (36.25, 3.75) {};
    \node [style=none] (132) at (35.25, 3.25) {};
    \node [style=none] (133) at (39.5, 2.25) {};
    \node [style=none] (135) at (39.5, 2) {};
    \node [style=none] (138) at (35.25, 3.5) {};
    \node [style=none] (139) at (35.5, 2.75) {};
    \node [style=none] (140) at (35.75, 2) {};
    \node [style=none] (141) at (35.25, 1.75) {};
    \node [style=none] (142) at (36.75, 1.75) {};
    \node [style=none] (143) at (35.5, 1.5) {};
    \node [style=none] (144) at (35.5, 2.5) {};
    \node [style=none] (145) at (36, 4) {};
    \node [style=none] (146) at (36.25, 3.5) {};
    \node [style=none] (147) at (36.25, 4.75) {};
    \node [style=none] (149) at (36.5, 3.5) {};
    \node [style=none] (150) at (36.75, 2.75) {};
    \node [style=none] (151) at (39.75, 2.75) {};
    \node [style=none] (152) at (38.25, 1.75) {};
    \node [style=none] (153) at (40.5, 3.5) {};
    \node [style=none] (154) at (40, 3.5) {};
    \node [style=none] (155) at (40.25, 3.25) {};
    \node [style=none] (156) at (40.5, 4.5) {};
    \node [style=none] (157) at (40, 2.75) {};
    \node [style=none] (158) at (39.25, 3) {};
    \node [style=none] (159) at (39.75, 2.25) {};
    \node [style=none] (160) at (39, 3.75) {};
    \node [style=none] (161) at (40.75, 3.5) {};
    \node [style=none] (162) at (41.25, 2.5) {};
    \node [style=none] (163) at (40, 2.75) {};
    \node [style=none] (164) at (39.75, 2) {};
    \node [style=none] (165) at (39.75, 2.5) {};
    \node [style=none] (166) at (41.25, 1.75) {};
    \node [style=none] (167) at (41.25, 2.5) {};
    \node [style=none] (168) at (38.25, 3.75) {};
    \node [style=none] (169) at (38.5, 3.5) {};
    \node [style=none] (170) at (38.5, 4.75) {};
    \node [style=none] (171) at (38, 3.25) {};
    \node [style=none] (172) at (38.5, 3.75) {};
    \node [style=none] (173) at (39.75, 1.75) {};
    \node [style=none] (174) at (40, 1.5) {};
    \node [style=none] (175) at (39.5, 1.25) {};
    \node [style=none] (176) at (14, -1.5) {};
    \node [style=none] (177) at (14, -1.5) {$L_2; R_i$};
    \node [style=none] (178) at (23.5, -1.5) {};
    \node [style=none] (179) at (23.5, -1.5) {$L_q; R_i$};
    \node [style=none] (180) at (43, 2) {};
    \node [style=none] (181) at (18.25, 0) {};
    \node [style=none] (182) at (60, 2) {};
  \end{pgfonlayer}
  \begin{pgfonlayer}{edgelayer}
     \shade[ball color = violet!40, opacity = 0.4]  (30.65, 2) circle (1.65cm);
  \draw (30.65, 2) circle (1.65cm);
     \shade[ball color = violet!40, opacity = 0.4]  (37.65, 2.1) circle (4.15cm);
  \draw (37.65, 2.1) circle (4.15cm);
    \draw [thick] (3.center) to (4.center);
    \draw [thick] (6.center) to (5.center);
    \draw [thick] (7.center) to (8.center);
    \draw [thick] (10.center) to (9.center);
    \draw [thick, bend right=105, looseness=0.75] (3.center) to (6.center);
    \draw [thick, bend left=75, looseness=0.50] (3.center) to (6.center);
    \draw [thick, style=dashed, bend left=75, looseness=0.75] (4.center) to (5.center);
    \draw [thick, bend right=90, looseness=0.75] (4.center) to (5.center);
    \draw [thick, bend right=105, looseness=0.75] (7.center) to (10.center);
    \draw [thick, bend left=75, looseness=0.50] (7.center) to (10.center);
    \draw [thick, style=dashed, bend left=90, looseness=0.75] (8.center) to (9.center);
    \draw [thick, bend right=60, looseness=0.75] (8.center) to (9.center);
    \draw [very thick, red] (11.center) to (12.center);
    \draw [very thick, red] (13.center) to (14.center);
    \draw [very thick, red, in=30, out=90] (17.center) to (15.center);
    \draw [very thick, red, in=-15, out=-90, looseness=2.25] (17.center) to (16.center);
    \draw [very thick, red] (18.center) to (19.center);
    \draw [very thick, style=dashed, red, bend left=90] (11.center) to (13.center);
    \draw [very thick, style=dashed, red, bend right=90] (12.center) to (19.center);
    \draw [thick] (27.center) to (28.center);
    \draw [thick] (30.center) to (29.center);
    \draw [thick] (31.center) to (32.center);
    \draw [thick] (34.center) to (33.center);
    \draw [thick, bend right=105, looseness=0.75] (27.center) to (30.center);
    \draw [thick, bend left=75, looseness=0.50] (27.center) to (30.center);
    \draw [thick, style=dashed, bend left=75, looseness=0.75] (28.center) to (29.center);
    \draw [thick, bend right=90, looseness=0.75] (28.center) to (29.center);
    \draw [thick, bend right=105, looseness=0.75] (31.center) to (34.center);
    \draw [thick, bend left=75, looseness=0.50] (31.center) to (34.center);
    \draw [thick, style=dashed, bend left=90, looseness=0.75] (32.center) to (33.center);
    \draw [thick, bend right=60, looseness=0.75] (32.center) to (33.center);
    \draw [very thick, red] (35.center) to (36.center);
    \draw [very thick, red] (37.center) to (38.center);
    \draw [very thick, red, in=30, out=90] (41.center) to (39.center);
    \draw [very thick, red, in=-15, out=-90, looseness=2.25] (41.center) to (40.center);
    \draw [very thick, red] (42.center) to (43.center);
    \draw [very thick, style=dashed, red, bend left=90] (35.center) to (37.center);
    \draw [very thick, style=dashed, red, bend right=90] (36.center) to (43.center);
    \draw [thick] (53.center) to (54.center);
    \draw [thick] (56.center) to (55.center);
    \draw [thick] (57.center) to (58.center);
    \draw [thick] (60.center) to (59.center);
    \draw [thick, bend right=105, looseness=0.75] (53.center) to (56.center);
    \draw [thick, bend left=75, looseness=0.50] (53.center) to (56.center);
    \draw [thick, style=dashed, bend left=75, looseness=0.75] (54.center) to (55.center);
    \draw [thick, bend right=90, looseness=0.75] (54.center) to (55.center);
    \draw [thick, bend right=105, looseness=0.75] (57.center) to (60.center);
    \draw [thick, bend left=75, looseness=0.50] (57.center) to (60.center);
    \draw [thick, style=dashed, bend left=90, looseness=0.75] (58.center) to (59.center);
    \draw [thick, bend right=60, looseness=0.75] (58.center) to (59.center);
    \draw [very thick, red] (61.center) to (62.center);
    \draw [very thick, red] (63.center) to (64.center);
    \draw [very thick, red, in=30, out=90] (67.center) to (65.center);
    \draw [very thick, red, in=-15, out=-90, looseness=2.25] (67.center) to (66.center);
    \draw [very thick, red] (68.center) to (69.center);
    \draw [very thick, style=dashed, red, bend left=90] (61.center) to (63.center);
    \draw [very thick, style=dashed, red, bend right=90] (62.center) to (69.center);
    \draw [thick, <->] (26.center) to (25.center);
    \draw [thick, <->] (24.center) to (52.center);
    \draw [thick, <->] (51.center) to (50.center);
    \draw [thick, <-, in=0, out=60, looseness=0.75] (2.center) to (1.center);
    \draw [thick, ->, in=120, out=180, looseness=0.75] (1.center) to (0.center);
    \draw [very thick] (74.center) to (77.center);
    \draw [very thick] (75.center) to (78.center);
    \draw [very thick] (76.center) to (79.center);
    \draw [thick, bend right=75, looseness=0.50] (80.center) to (81.center);
    \draw [thick, style=dashed, bend left=75, looseness=0.50] (80.center) to (81.center);
    \draw [thick, bend right=90, looseness=0.50] (82.center) to (83.center);
    \draw [thick, style=dashed, bend left=75, looseness=0.50] (82.center) to (83.center);
    \draw [very thick, red, in=-150, out=-105, looseness=1.25] (107.center) to (108.center);
    \draw [very thick, red, bend right=75] (113.center) to (111.center);
    \draw [very thick, red, bend left=75] (113.center) to (112.center);
    \draw [very thick, red, in=150, out=75] (107.center) to (109.center);
    \draw [thick, <->, bend left=60] (123.center) to (124.center);
    \draw [very thick, red, in=195, out=165, looseness=2.50] (135.center) to (133.center);
    \draw [very thick, red, in=135, out=60, looseness=0.75] (129.center) to (131.center);
    \draw [very thick, red, in=-165, out=-120, looseness=1.25] (132.center) to (130.center);
    \draw [very thick, red, bend left=15, looseness=0.75] (138.center) to (139.center);
    \draw [very thick, red, bend right=75] (140.center) to (130.center);
    \draw [very thick, red, in=-105, out=-120, looseness=1.25] (141.center) to (142.center);
    \draw [very thick, red, in=105, out=180, looseness=0.75] (147.center) to (145.center);
    \draw [very thick, red, in=-15, out=0, looseness=0.75] (147.center) to (146.center);
    \draw [very thick, red, in=-90, out=90, looseness=0.75] (143.center) to (144.center);
    \draw [very thick, red, bend left, looseness=0.50] (149.center) to (150.center);
    \draw [very thick, red, bend right=75] (156.center) to (154.center);
    \draw [very thick, red, bend left=75] (156.center) to (155.center);
    \draw [very thick, red, in=180, out=75] (151.center) to (153.center);
    \draw [very thick, red, in=30, out=-30, looseness=0.75] (157.center) to (159.center);
    \draw [very thick, red, in=60, out=0] (160.center) to (158.center);
    \draw [very thick, red, in=90, out=-15] (161.center) to (162.center);
    \draw [very thick, red, in=-120, out=165, looseness=1.50] (163.center) to (158.center);
    \draw [very thick, red, in=-105, out=105, looseness=0.50] (164.center) to (165.center);
    \draw [very thick, red, in=105, out=180, looseness=0.75] (170.center) to (168.center);
    \draw [very thick, red, in=-15, out=0, looseness=0.75] (170.center) to (169.center);
    \draw [very thick, red, in=-90, out=90, looseness=0.75] (166.center) to (167.center);
    \draw [very thick, red, bend left, looseness=0.50] (171.center) to (172.center);
    \draw [very thick, red, bend right=90, looseness=1.50] (173.center) to (166.center);
    \draw [very thick, red, bend left=45] (164.center) to (174.center);
    \draw [very thick, red, bend right=45] (152.center) to (175.center);
  \end{pgfonlayer}
  \end{tikzpicture}
\end{center}
 \caption{
To compute the  $q^{\text{th}}$-R\'enyi entropy for the reduced density matrix $\rhoA^{i_L}$ we need to cyclically glue copies of the solid torii with a link $L$ carrying representation $R_i$ on each of its components. As indicated in Fig.~\ref{fig:nrdensitymTW} it is helpful to view this solid torus in terms of gluing together two solid cylinders with a link $L$ contained within them. Gluing these together we see that we are required to compute the path integral over $\Sp^3$ with a Wilson line representation $R_i$ placed along the direct sum of links 
$L^{(q)}  = L_1+\cdots+L_q$. The $q$ links get intertwined in the process of identifying the pieces of the density matrices across replica copies. }
\label{fig:nrdensitymT2Wq}
\end{figure}

The $q^{\text{th}}$-R\'enyi entropy is obtained by cyclically gluing $q$ copies of the path integral representing the reduced density matrix  ${\rhoA^{i_L}}$. To obtain the result, it is useful to open up the solid torus into a pair of solid cylinders, and implement the cyclic gluing. Identification of the regions $\rA_+$ from the $i^{\rm th}$-copy of $\rhoA^{i_L}$ with the $(i+1)^{\rm st}$ one ends up linking the loops $L_i$ and $L_{i+1}$, respectively. The final trace converts the solid torus into an $\Sp^3$, but the link is still present. In fact, each gluing creates an  
overcross and an undercross for neighboring pairs of links. We denote the final result as $L^{(q)}  = L_1+\cdots+L_q$. The computation is depicted in Fig.~\ref{fig:nrdensitymT2Wq}. As a consequence, the $q$-fold replica can be understood as a path integral over $\Sp^3$ with a Wilson line representation $R_i$ placed along a link $L^{(q)}$. We therefore have
\begin{equation}\label{eq:RenT2Ri2comp}
S_\rA^{(q)} (\rhoA^{\!\!iL})= \frac{1}{1-q} \left[ \log \zcs(\Sp^3, L^{(q)}; R_i) - q\, \log \zcs(\Sp^3, L; R_i) \right]  = 2\, \log \zcs(\Sp^3, \bar{R_i}) \,.
\end{equation}	
We have simplified the answer above to one involving the partition function of Chern-Simons on $\Sp^3$ with a Wilson line in the conjugate representation $\bar{R}_i$. This can be explained as follows. Each pair of neighboring links $L_i$ and $L_{i+1}$ in $L^{(q)}$ have two crossings between them, We consider $\zcs(\Sp^3, L^{(q)}; R_i) $ and supply $2(q-1)$ copies of $\zcs(\Sp^3, \bar{R}_i)$. We play the usual game of decomposing the latter factors into three-balls, and use the balls containing $\bar{R}_i$ to replace the local neighbourhood of the overcross and undercross between $L_i$ and $L_{i+1}$. This results in cutting out the links, leaving behind the factor coming from the normalization. The remainder then boils down to the supplied factor which gives the answer quoted. We note that  \cite{Dong:2008ft} derive this more simply by using the normalized state on the torus.
Note that here we  finally obtain an answer for the topological entanglement entropy that is different from  the previous cases, all of which reduced, up to an integral factor counting the number of 
components of the entangling surface, to the three-sphere partition function.
 
At this point, it should be clear how we wish to compute the same result from the closed topological string on the resolved conifold. We can repeat our algorithm of performing the surgery of  $\Sp^3 \subset \rco$ in the presence of the Lagrangian cycles, etc. We assume that the local surgery procedure we are using for the Wilson lines continues to work for the topological branes (we are however unaware of explicit construction of Lagrangian cycles in the resolved conifold for linked knots).  Since supplying the conjugate representation involves topological anti-branes, the simplification of \eqref{eq:RenT2Ri2comp} is a consequence of topological brane anti-brane annihilation.  The point as always is that the $\Sp^2$ at the tip continues to be a spectator, all the action involving the Lagrangian branes happening elsewhere. So unsurprisingly the final answer works out as desired, as the reader can check.

\section{Discussion}
\label{sec:discuss}

Our broad motivation was to understand how we can encode spatially ordered quantum field theory entanglement in a dual closed string description. As is by now well appreciated, the remarkable aspect of the holographic entanglement entropy proposals is that they give a geometric meaning to such entanglement in a particular corner of parameter space where the closed string theory can be truncated to  classical  gravitational dynamics. Ideally, we would like to have a picture for the situation where the closed string description remains classical, but is not simply reducible to classical (super) gravity. Rather than tackle this question head on, we have chosen to examine the problem in the context of topological open/closed string duality. While care should be taken to export lessons from this line of investigation to the physical string theory, it is nevertheless apposite to take stock of the lessons learnt and speculate on  those that might have implications for the physical gauge/gravity dualities.

We have argued that it is possible to give a meaningful definition of topological entanglement  in closed topological string theory. This was done by taking inspiration from surgery techniques of \cite{Witten:1988hf} used to extract the Chern-Simons Hilbert space and performing the selfsame surgery within the closed string target space. Specifically, given a spatial decomposition of the field theory (or open string field theory) across an entangling surface, there is a corresponding decomposition of the closed string target. We have chosen to call the target space separatrix an entangling brane, as it captures the essence of both  being an entangling surface from the closed string perspective, as well as the locus for the cosmic brane that we have been inspired to introduce in the semiclassical gravity limit following \cite{Lewkowycz:2013nqa}. 

The closed topological string theory construction of the reduced density matrix relied not just on the ability of performing 3-manifold surgery within the Calabi-Yau target space, but also the fact that such a surgery can be done without slicing through closed string worldsheets. This relied heavily on the worldsheets in the topological A-model being holomorphic maps which are not cut open during the surgery. We have argued that this is possible in the specific case of the resolved conifold, including situations where we have additional Lagrangian cycles on which probe topological D-branes representing Wilson lines are included. Depending on our choice of spatial bipartitioning, sometimes these Lagrangian cycles can be cut open. The crucial fact is that  surgery leaves untouched the intersection locus of these cycles with the homology 2-cycle of the resolved conifold (the $\Sp^2$ at the tip). 
Moreover, much of the analysis only involved specifying the topological characteristics of the entangling brane or even the Cauchy surface where the state space is defined. This is of course reasonable given that the theory depends only on certain key topological properties, viz., two-cycles in the target and K\"ahler structure. In part owing to this simplification, we got away without having to understand the detailed aspects of the closed string field theory Hilbert space, of which we have, as yet, limited understanding. 

Performing a similar exercise in physical string theory seems a lot more formidable, and we remain agnostic as to whether we can port lessons from our investigations to that context. At a conceptual level our construct of the entangling brane is similar to the idea of entanglement brane introduced earlier in \cite{Donnelly:2016jet,Donnelly:2018ppr} in the context of 
topological field theories, especially two-dimensional Yang-Mills theory. They define such entangling brane by demanding consistent factorization of the topological theory across this separatrix, which is again made possible by the gluing rules for topological partition functions. In theories with physical degrees of freedom one must however confront issues relating to the set of allowed boundary conditions, and more generally in string theory, define a notion of what it means to specify a subregion in spacetime. 

Perhaps it is opportune to ask what is it about topological entanglement that enables such simplification. Given a state in a field theory, partitioning it into reduced states, across some spatial domain, requires that we impose certain consistency conditions on the set of allowed boundary conditions. While this is familiar in other contexts, in the topological context, we believe these splitting conditions are more tractable owing to their rigidity, and can be axiomatized succinctly \cite{Donnelly:2018ppr}. Conversely, given reduced states of a topological theory, the presence of the topological entanglement is what allows us to glue back the pieces to recover a state that only cares about the topological data, and is incognizant of the geometric features. A key to this is the factorization of topological partition functions under manifold surgery \cite{Atiyah:1989vu}, which we have used extensively in our discussion, both in the context of surgery in Chern-Simons and in its uplift to the topological closed string. In the Chern-Simons description the surgery formula of \cite{Witten:1988hf} notes that for a 3-manifold $\mcs$ which is the connected sum of two others, $\mcs_1$ and $\mcs_2$ glued together with an $\Sp^3$, one has $\zcs(\mcs) \, \zcs(\Sp^3)  = \zcs(\mcs_1) \, \zcs(\mcs_2)$. Our proposal amounts to noting in specific cases that this has an analogous closed topological string uplift; heuristically: $\zc(\xf) \, \zc(\rco) \sim \zc(\xf_1) \, \zc(\xf_2) $ for situations where the target space $\xf$ can be viewed as a connected sum of $\xf_{1,2}$ glued together by a resolved conifold $\rco$. Generalizations with open string degrees of freedom ought to be possible, but we are not attempting here to give a general prescription. 

A natural corollary of the above observation is a simple explanation for the flat entanglement spectrum and its rather weak dependence on the state of the theory we are bipartitioning. As we have seen, in all but one example discussed, the R\'enyi and von Neumann entropies were given by $M\, \log \zcs(\Sp^3) $, with $M$ being the number of entangling interfaces bipartitioning the state. Viewing topological entanglement as the glue that binds these pieces together without prescribing additional structure implies that we should only be sensitive to the number of pieces being glued and an overall factor providing a measure of the number of topological degrees of freedom being pieced together. The latter is captured by the quantum dimension 
which is related to the $\Sp^3$  partition function. This is highly suggestive that  the closed string partition function on the resolved conifold is a measure of the topological closed string degrees of freedom, viz., a string quantum dimension. 

Moreover, it also suggests that our replica construction on the closed string side is penalty free -- the location of the entangling brane is not constrained by the dynamics. This is in contrast to the  physical context, where as explained in \cite{Headrick:2010zt,Lewkowycz:2013nqa}, placing a cosmic brane induces non-trivial backreaction. In a certain sense, the topological string is a natural home for the original attempt of \cite{Fursaev:2006ih} to prove the RT prescription.\footnote{ We thank Matt Headrick for this suggestion and useful discussions on this issue.} From this vantage point, it is natural to argue that the tensor network models of holography \cite{Pastawski:2015qua,Hayden:2016cfa}  which give flat entanglement spectra are at best capturing some topological features of the AdS/CFT correspondence. Thus they can be useful in encoding some information theoretic properties of the holographic map 
\cite{Harlow:2016vwg} but may not provide a deep rationale for the emergence of gravitational dynamics, which would constitute and essential limitation (a complementary viewpoint is articulated in \cite{Dong:2018seb,Akers:2018fow}). It would be interesting to test this hypothesis  (see below for a suggestion).

Of course, all of the above statements are predicated upon the fact that the replica construction in the topological setting is insensitive to the UV divergent terms in the entanglement entropy. While it is unfortunate that the construction we have described does not capture the entire spatially partitioned entanglement, the fact that we can actually relate the open and closed string replica constructions provides sufficient rationale for the analysis. One disadvantage of this approach is the fact that the topological contribution we extracted is by itself is not necessarily positive in Chern-Simons theory \cite{Dong:2008ft} (the topological entanglement defined in \cite{Kitaev:2005dm} is negative of what we compute). The positivity of the von Neumann entropy requires keeping track of the UV divergent terms as well. As we noted in the main text, this is possible to do in the Chern-Simons context as in \S\ref{sec:rhoCS}, but the analogous statement in closed string side is not clear to us at present. Closing this loophole requires understanding how one can partition the closed string Hilbert space, a question that we have assiduously worked to avoid by working with the functional integral. 

 While our justifications for the closed string bipartitioning, and interpretations this affords for topological closed string entanglement,  were made on physical grounds, based in part on requiring consistency of the GV duality, it would useful to make these statements more precise.  There are several  questions that we feel are worth exploring in the context of the topological string theory itself, which could shed further light into our analysis. For one, the arguments presented here have mostly relied on various consistency checks, and not invoked any of the known explicit formulae for Chern-Simons observables. We outline some of these questions below.

\paragraph{Replica and Wilson lines:} We have seen that the replica computation in Chern-Simons theory gives a simple  answer for the trace of powers of the reduced density matrix \eqref{eq:trrhon}. As is clear from the discussion  around Fig.~\ref{fig:rdensitymtsp2intf}, the $q^{\rm th}$ power of the normalized density matrix $\rhoAh$ can be viewed in terms of prescribing boundary conditions on half-balls from which $q$ 3-balls have been scooped out. This is equivalently understood in terms of realizing that ${\rm Tr}(\rhoAh)$ is computed by $\zcs(\Sp^2 \times \Sp^1)$, 
as we explain in Appendix~\ref{sec:nrdensitym}. 

One can use this picture to motivate another interpretation of the traces of powers of the reduced density matrix.  Employing surgery to scoop out $q$ 3-balls from each of the balls obtained in the Heegard decomposition of the $\Sp^3$, the result can be expressed in terms of Wilson line expectation values by generalizing \eqref{eq:s2xs1part}. One finds a decomposition in terms of Wilson line expectation values,
\begin{equation}
\Tr{\rhoA^{\!q}} = 
	\frac{1}{\zcs(\Sp^3)^{q-1}} = \sum_{i_1, i_2, \cdots i_q} \left(\prod_{j=1}^q  \zcs(\Sp^3; R_{i_j})\right) 
	\zcs(\Sp^3; R_{i_1},R_{i_2}, \cdots R_{i_q}) \,.
\label{eq:rennth}
\end{equation}	
The closed string encoding of Wilson loop expectation values after the geometric transition involves Lagrangian branes
\cite{Ooguri:1999bv}  which can be viewed as a bubbling Calabi-Yau space \cite{Gomis:2006mv,Gomis:2007kz}. It would be interesting to compare the resulting geometry with the picture we are proposing. The comparison would have been more straightforward if the topological closed string description for Chern-Simons on $\Sp^2 \times \Sp^1$ were available (which appears as the natural intermediate step when computing  $\Tr{\rhoA}$). 

\paragraph{Chern-Simons on other 3-manifolds:} Our focus was on features of Chern-Simons on $\Sp^3$, but we could likewise consider other three-manifolds. For instance it would be interesting to examine a version of our story for Lens spaces. Explicit expressions for the Lens space partition functions are obtained using a matrix model description \cite{Aganagic:2002qg}. These are however obtained in a particular Coulomb branch vacuum, obtained by a choice of background $\mathbb{Z}_q$ flat connection. However, an interesting expression summing over all vacua was derived in 
\cite{Gopakumar:1998vy} assuming that $q|N$ and $q|k$: the result is simply related to the $\Sp^3$ partition function, 
$\zcs(\Sp^3/\mathbb{Z}_q, N, k) = \left( \zcs(\Sp^3,N/q, k/q)\right)^q$. It would be interesting to understand if we can upgrade our story to this case. 

\paragraph{Adding physical degrees of freedom:} While the tractability of Chern-Simons theory provided us with much insight, it is interesting to consider adding physical matter degrees of freedom. Such theories have been widely studied in recent literature and have many interesting features such as Bose-Fermi duality.  Large $N$ models of such matter Chern-Simons theories are furthermore solvable \cite{Aharony:2011jz,Giombi:2011kc}.  With the addition of matter degrees of freedom we expect under spatial bipartitioning the von Neumann entropy to behave as 
$S_\rA = \frac{L_{\entsurf}}{\epsilon} - \gamma_\text{phys} + S_\rA^\text{top}$ where we now pick up a UV divergent piece proportional to the length of the entangling surface. We have split the finite contribution into a piece that is physical and one that has topological origins. The latter is the part we have focused on. For one, it would be an interesting exercise to understand the entanglement properties of this class of theories, and for another, by coupling in physical degrees of freedom we can hope to learn some useful lessons about how the physical dynamics is imprinted into the entanglement structure. As mentioned above, it would be worthwhile to understand how to upgrade the tensor network models of holography to construct realizations with non-flat entanglement spectrum (with the flat part being captured by the non-dynamical sector). As an intermediate step in the analysis it might be worthwhile examining the partially topological Chern-Simons matter theories discussed in \cite{Aganagic:2017tvx,Aharony:2019suq}. 

\paragraph{Tree level entanglement entropy:}  There is an interesting question for which the topological strings provide a useful context to develop intuition. Consider off-shell string theory, which in the low-energy limit reduces to off-shell supergravity. Working with the Euclidean configuration we learn that for asymptotic boundary conditions containing a thermal circle, the path integral localizes around black hole saddle points. In particular, the functional integral evaluates to a non-trivial answer at tree level, giving the Gibbons-Hawking black hole free energy \cite{Gibbons:1976ue}. Naively this contribution is hard to capture in the tree level string theory, although there are strong arguments suggesting that this must indeed be possible \cite{Susskind:1994sm}. 
The primary issue is that from a worldsheet perspective, the sphere diagram with no insertions vanishes owing to the underlying $SL(2,\mathbb{C})$ invariance of worldsheet theory. Per se, this does not imply that such a contribution is not possible off-shell, but we do not, as yet, have the technical tools to extract the same.\footnote{ We thank Raghu Mahajan for a useful discussion on this point.} What we need is a physical principle that allows us to freeze the worldsheet moduli and prevent them from trivializing the computation. 

Note that the topological closed string theory manifestly contains such a tree level contribution, which comes from classical intersection theory (see Appendix~\ref{sec:tsreview}).  The origins of this term in the topological string context owes to the fact that one has additional data. One is not simply looking the worldsheet theory, but also into its embedding in the target space. Given the structure of the topological A-model, we are forced to look at stable holomorphic maps, which in turn end up providing the data necessary to fix the worldsheet Weyl invariance. While the details are specific to the topological string, it is perhaps worth keeping in mind the lesson that one needs to take into account the embedding of the worldsheet into the target space even in the physical context.

\acknowledgments

It is a pleasure to thank Mina Aganagic, Tudor Dimofte, Rajesh Gopakumar, Matthew Headrick, R.~Loganayagam,  P.~Ramadevi, Cumrun Vafa,  Edward Witten, 
and Xi Yin for illuminating discussions. We would also like to thank the anonymous referee for raising questions about the absence of UV divergent terms in the replica formalism, and their implications for the string theory discussion.
VH and MR would like to thank KITP, UCSB for hospitality during the workshop ``Chaos and Order: From strongly correlated systems to black holes'', 
where the research was supported in part by the National Science Foundation under Grant No. NSF PHY17-48958 to the KITP.
The authors were supported by  U.S.\ Department of Energy grant DE-SC0019480 under the HEP-QIS QuantISED program and by funds from the University of California.

\appendix

    \section{The Topology of the Conifold}
\label{sec:conifold}

Let us discuss the topology of the target spacetimes of both the open and the closed string theory description of the $SU(N)$ Chern-Simons theory on $\Sp^3$ a bit further. Some of the basic features have already been sketched out in \S\ref{sec:topduality}.  We recall from that discussion that the smooth target spacetime of open A-model string theory is obtained by considering a deformation of the singular conifold, and the smooth target spacetime of closed A-model string theory is obtained by resolving  the same singular conifold.

\paragraph{The singular conifold:}   The singular conifold of our interest is a hypersurface in $\mathbb{C}^4$ with complex coordinates $(x,y,w,z)$ satisfying the equation \eqref{eq:scfld2}. It is a Calabi-Yau threefold with a conifold singularity at $(x,y,w,z)=(0,0,0,0)$. The coordinate map to go from \eqref{eq:scfld2} to \eqref{eq:scfld} can be easily inferred to be:
\begin{equation}
\begin{split}
x &= \zeta_1 + i\, \zeta_2  \,,\qquad \; y  = \zeta_1 - i\, \zeta_2  \,, \\
w &= i\, (\zeta_3 + i\, \zeta_4 ) \,, \quad z = i\, (\zeta_3 - i\, \zeta_4)  \,, 
\end{split}
\label{eq:varch}
\end{equation}	

The geometry of  singular conifold is however more transparent in the following set of coordinates: 
\begin{equation}
\begin{split}
\vec{u}&=\frac{1}{2}\left(\text{Re}\left(x+y\right), \text{Re}\left(i\, x-i\, y\right),\text{Re}\left(i\, z+i\, w\right),\text{Re}\left(z+w\right)\right)  \\
\vec{v}&=\frac{1}{2}\left(\text{Im}\left(x+y\right), \text{Im}\left(i\, x-i\, y\right),\text{Im}\left(i\, z+i\, w\right),\text{Im}\left(z+w\right)\right) 
\end{split}
\label{eq:scfld2a}
\end{equation}
The complex conifold equation \eqref{eq:scfld2} in this coordinate becomes two real equations:
\begin{equation}
\vec{u}^2-\vec{v}^2=0\qquad\qquad \vec{u}\cdot\vec{v}=0
\label{eq:scfld3}
 \end{equation}
These equations can be solved by setting $\vec{u}^2=r^2$ and $\vec{v}^2=r^2$ such that $\vec{u}\cdot\vec{v}=0$. This implies that $\vec{u}$ lives on $\Sp^3$, and for a fixed $r^2$ and $\vec{u}$, the vector $\vec{v}$ lives on an $\Sp^2$.    Therefore we can identify the topology of the singular conifold as a cone with a base which is topologically $\Sp^2\times \Sp^3$, and as we go from the base of the cone to the tip $r$ varies from $\infty $ to $0$. Notice that at the tip of the cone, where $r=0$, both the $\Sp^3$ and the $\Sp^2$ shrinks to a point.

\paragraph{The Deformed Conifold:} The singular geometry of the conifold can be made smooth by considering the following deformation of the defining equation \eqref{eq:scfld3} 
\begin{equation}
    \vec{u}^2-\vec{v}^2=\mu^2\qquad\qquad \vec{u}\cdot\vec{v}=0
\label{eq:dconifold}    
\end{equation}
which makes manifest the equivalence to our earlier presentation in \eqref{eq:dcfld}. Even after this deformation the geometry can be understood as a cone over $\Sp^3\times \Sp^2$. However, after deformation only the $\Sp^2$ shrinks to a point  when $r=0$. The deformed equation \eqref{eq:dconifold} represents the total space $\dco$. This can be made obvious by rewriting the deformed conifold using the coordinate $\vec{q}=\frac{\vec{u}}{\sqrt{\vec{v}^2+\mu^2}}$ 
\begin{equation}
\vec{q}^{\,2}=1 \qquad\qquad \vec{q}\cdot \vec{v}=0
\end{equation}

\paragraph{The Resolved Conifold:} The resolution is a different way for obtaining a smooth geometry from the singular conifold. The equation for the singular conifold \eqref{eq:scfld2} is equivalent to demanding that the determinant of a matrix vanishes, viz.,
\begin{equation}
\text{det}\left(\begin{array}{cc}x & w \\z & y\end{array}\right)=0 \,.
\end{equation}	
The singular point can be resolved by considering another hypersurface $\rco$ in $\mathbb{C}^4\times \mathbb{P}^1$ with coordinates $(x,y,w,z,\lambda_1,\lambda_2)$
\begin{equation}
\left(\begin{array}{cc}x & w \\z & y\end{array}\right)\left(\begin{array}{c}\lambda_1 \\\lambda_2\end{array}\right)=0
\label{eq:rconifold}
\end{equation}      
The map $(x,y,w,z,\lambda_1,\lambda_2)\to (x,y,w,z)$ is a map from $\rco$ to the singular conifold except at the singular point. The singular point at the origin is  now replaced by $\mathbb{P}^1\equiv \Sp^2$. Therefore, the resolved conifold has an extra element in its second homology class $H_2$. By varying the size of $\mathbb{P}^1$ and letting it shrink, we can recover the singular conifold. 

The process of varying the complex structure from a smooth Calabi-Yau so that a conifold singularity appears, and then resolving that conifold so that a new $\Sp^2$ appears is referred to  as the {\it conifold} or {\it geometric transition}. The deformed conifold has a complex modulus $\mu$ and the resolved conifold has a single K\"ahler modulus for its Calabi-Yau metric, naturally parameterized by the complexified K\"ahler parameter $t$. The real part of the complexified K\"ahler parameter corresponds to the deformation of the Calabi-Yau metric, i.e., 
K\"ahler class $[\omega]$ and the imaginary part comes from the NS-NS $B$ field which is a class in $H^2$. If there is a non-trivial cohomology class $[B]$, we can modify the topological string theory by putting a phase factor $$\text{exp}\left(i\, \int \phi^*B \right)$$ in the path integral. This factor is invariant under a continuous deformation of the map $\phi$ from the worldsheet to the target  space. In particular, it is invariant under the supersymmetry variation and this modification does not break the supersymmetry. Furthermore, the forms of the supercurrent and the supercharges remain the same as above.  This modification of the topological string theory is necessary to make it invariant under the T-duality transformation. \par
      
      The complexified K\"ahler class is given by $$[\omega]-i\, [B].$$
    For $\Sp^2$ we have 
    \begin{equation}
    B=\frac{\theta}{2\pi}\omega^{\text{FS}}\qquad\qquad     \omega=\frac{r}{2\pi}\omega^{\text{FS}},
    \end{equation}
    where $\omega^{\text{FS}}$ is the Fubini-Study metric on $\Sp^2$. Then the complexified K\"ahler parameter is given by $$t=r-i\, \theta.$$

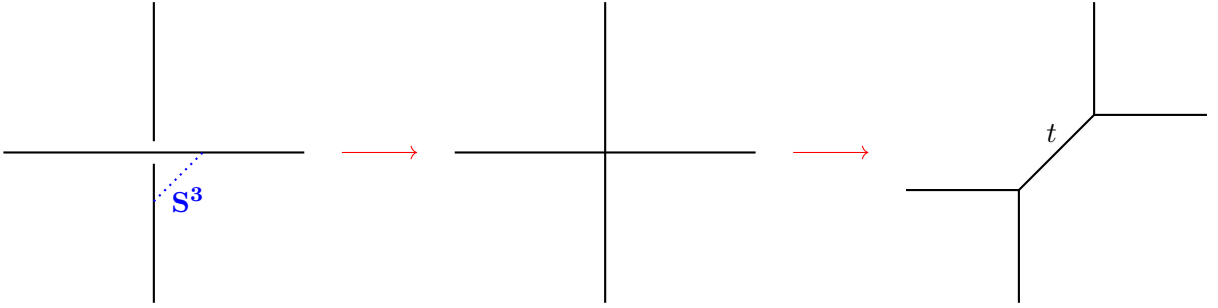
\begin{figure}[h]
\centering
\begin{tikzpicture}
\draw [thick, black] (-8,0) -- (-4,0);
\draw [thick, black] (-6,-2) -- (-6,-0.15);
\draw [thick, black] (-6,0.15) -- (-6,2);
\draw[thick, color = blue, dotted] (-6,-0.65) -- node[below]{$ \ \ {\bf S^3}$} (-5.35,0);
\draw[color = red, ->] (-3.5,0) -- (-2.5,0);
\draw [thick, black] (-2,0) -- (2,0);
\draw [thick, black] (0,-2) -- (0,2);
\draw[color = red, ->] (2.5,0) -- (3.5,0);
\draw [thick, black] (4,-0.5) -- (5.5,-0.5) -- (5.5,-2);
\draw [thick, black] (5.5,-0.5) -- node[above] {$t\ $} (6.5,0.5)  -- (6.5,2);
\draw [thick, black] (6.5,0.5)--(8,0.5);
\end{tikzpicture}
\caption{ The toric structure of the deformed (left), singular (middle), and resolved (right) conifold geometries, displaying the geometric transition. 
We display the two plane $\mathbb{R}^2$ coordinatized by $(r_\alpha, r_\beta)$ and draw thereupon the loci where the cycles of the ${\bf T}^2$ degenerate. The third direction normal the plane of the paper is   $r_\gamma$.
}
\label{fig:toriccfld}
\end{figure}

\paragraph{Toric structure of the conifold:} The resolved conifold geometry admits a toric structure, which refers to the fact that it can be viewed a 
a ${\bf T}^2$ fibration over a base manifold. The rough idea is view these geometries as gluings of local copies of $\mathbb{C}^3$ (for three-folds), which clearly has a ${\bf T}^2$ fibration (it of course has three circles corresponding to rotations in each two-plane, but only two are relevant). 
The fibres may degenerate, and the information about the degeneration loci is encoded in toric diagrams which are usually drawn on a two-plane. We can exhibit the toric structure for the conifold geometry by providing relative phase rotations of our coordinates used to define the singular conifold 
\eqref{eq:scfld2}
\begin{equation}
(x,\,  y ,\, w,\, z) \mapsto (e^{-i\,\alpha}\, x,\, e^{i\,\alpha}\, y , \,
e^{-i\,\beta}\,w, \, e^{-i\,\beta}\,z) 
\label{eq:t2fibres}
\end{equation}	
Translating to the coordinates used to parameterize the resolved conifold \eqref{eq:rcfld} via $x = \xi_1\, \xi_3 \,, \; y = \xi_2 \, \xi_4\,, \;  u = \xi_1 \, \xi_2\,,\; v = \xi_3\,\xi_4$ we infer that 
\begin{equation}
(\xi_1,\,  \xi_2 ,\, \xi_3,\, \xi_4) \mapsto (e^{-i\,(\alpha+\beta)}\, \xi_1,\, e^{i\,\alpha}\, \xi_2 , \, e^{i\,\beta} \,\xi_3,\,  \xi_4) 
\label{eq:zt2fibres}
\end{equation}	
In the resolved conifold, we can pick  a coordinate chart for $\mathbb{R}^3$ with  say with coordinates
$r_\alpha = |\xi_2|^2 - |\xi_1|^2$, $r_\beta = |\xi_3|^2 - |\xi_1|^2$ and $r_\gamma =  {\rm Im}(\xi_1\xi_2 \xi_3 \xi_4)$. These are to be viewed as the base for the torus fibration. The $\alpha$ and $\beta$ circles are fibered over this $\mathbb{R}^3$ along with another fibre which is non-compact. Let us visualize the $(r_\alpha, r_\beta)$ plane (see Fig.~\ref{fig:toriccfld}). The $\alpha$ and $\beta$ circles degenerate at various points along this plane. We have  the $(1,0)$ cycle degenerating along the horizontal, and the $(0,1)$ cycle degenerating along the vertical axes, while inclined lines refer to an appropriate $(a,b)$-cycle's degeneration. Note that in the singular conifold both the cycles degenerate at the origin. Deforming this gives us an $\Sp^3$ which is to be viewed as a ${\bf T}^2$ fibration over an interval (in this case a finite interval in $r_\gamma$). Resolving it gives a geometry where the $(1,-1)$ cycle degenerates at the locus where we have a finite size $\mathbb{P}^1$, which is viewed as a circle fibration over an interval.

\section{Normalized Chern-Simons density matrices}
\label{sec:nrdensitym}

As discussed in \S\ref{sec:H01summary} removing a solid torus from $\Sp^3$ leaves behind another solid torus. Within $\Sp^3$ these two solid tori are interlocked and their boundaries are identified; the non-contractible $a$-cycle of one boundary torus is identified with the contractible $b$-cycle of the other torus, see Fig.~\ref{fig:S3cuttori}.

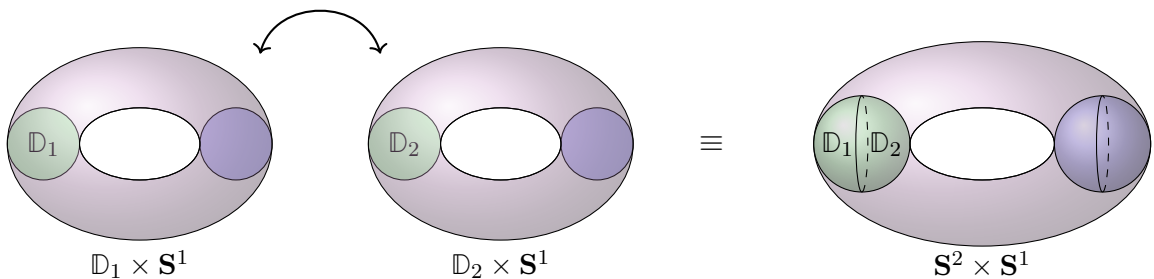
\begin{figure}
\begin{center}
\usetikzlibrary{backgrounds}
\begin{tikzpicture}[scale=.8]

\filldraw[fill=blue!20] (-1.4,3) ellipse (.6 and .6);
  \filldraw[fill=green!20] (-4.6,3) ellipse (.6 and .6);
  \node at (-4.6,3)  {${\disk_1}$};
  \node at (-3,1){${\disk_1\times \Sp^1}$};
  \shade[ball color = violet!40, opacity = 0.4] (-3,3) ellipse (2.2 and 1.6);
  \draw(-3,3) ellipse (2.2 and 1.6);
  \filldraw[fill=white]  (-3,3) ellipse (1 and 0.6);
  \draw (-3,3) ellipse (1 and 0.6);
  \filldraw[fill=blue!20] (4.6,3) ellipse (.6 and .6);
  \filldraw[fill=green!20] (1.4,3) ellipse (.6 and .6);
  \node at (1.4,3)  {${\disk_2}$};
  \node at (3,1){${\disk_2\times \Sp^1}$};  
  \shade[ball color = violet!40, opacity = 0.4] (3,3) ellipse (2.2 and 1.6);
  \draw (3,3) ellipse (2.2 and 1.6);
  \filldraw[fill=white]  (3,3) ellipse (1 and 0.6);
  \draw(3,3) ellipse (1 and 0.6);
   \draw[thick, <->] (-1,4.5)  to [out=75,in=105, looseness=1.25] (1,4.5);
   \node at (6.5,3)  {\large{$\equiv $}};
  \shade[ball color = violet!40, opacity = 0.4] (11,3) ellipse (2.8 and 1.7);
  \draw(11,3) ellipse (2.8 and 1.7);
  \filldraw[fill=white]  (11,3) ellipse (1.2 and 0.6);
  \draw(11,3) ellipse (1.2 and 0.6);
  \shade[ball color = blue!40, opacity = 0.4](13,3) circle (.8cm);
  \draw (13,3) circle (.8cm);
  \shade[ball color = green!40, opacity = 0.4] (9,3) circle (.8cm);
  \draw (9,3) circle (.8cm);
  \draw [dashed]  (9,2.2) arc (-90:90:0.1 and .8);
    \draw (9,3.8) arc (90:270:0.1 and .8);
    \draw [dashed]  (13,2.2) arc (-90:90:0.1 and .8);
    \draw (13,3.8) arc (90:270:0.1 and .8);
    \node at (8.6,3)  {$\disk_1$};
  \node at (9.4,3){${\disk_2}$};
  \node at (11,1){${\Sp^2\times \Sp^1}$};
\end{tikzpicture}
\end{center}
\caption{Gluing two solid tori $\disk_1\times \Sp^1$ and $\disk_2\times \Sp^1$ such that the $a$-cycles on the boundary tori are identified. The resulting 3-manifold is $\Sp^2\times\Sp^1$. }
\label{fig:s2xs1}
\end{figure}

According to the Dehn-Lickorish theorem, by changing way the cycles on the boundary tori are identified we can obtain different 3-manifolds. For instance, consider gluing two solid tori such that the $a$-cycles on the boundary tori are identified. The resulting 3-manifold is $\Sp^2\times\Sp^1$. This claim can be verified without much effort. For this, notice that a solid torus is the space $\disk\times \Sp^1$, where $\disk$ is a disk. Consider   two solid tori $\disk_1\times \Sp^1$ and $\disk_2\times \Sp^1$. Let us glue them along the boundary tori by identifying the same cycles  on each while gluing. First of all this identifies  the two $\Sp^1$. Moreover, the edges of the disks $\disk_1$ and $\disk_2$ at each point on  
the resulting $\Sp^1$ are also being identified. Since identifying the edges $\disk_1$ and $\disk_2$  produces  $\Sp^2$,  the resulting 3-manifold is $\Sp^2\times\Sp^1$, see 
Fig.~\ref{fig:s2xs1}. Therefore,  $\Sp^2\times\Sp^1$ can be obtained by cutting out a solid torus from $\Sp^3$ and pasting it back in, after applying a modular $\mathcal{S}$-transformation on 
its boundary that interchanges the $a$-cycle with the $b$-cycle. 

 The Dehn-Lickorish theorem allows us to express the partition function of Chern-Simons theory as an inner product between states belongs to  the Hilbert space $\mathcal{H}_{\mathbf{T}^2}$ obtained by quantizing Chern-Simons theory on a torus \cite{Witten:1988hf}. Assume that the state $|\Psi^0_{\mathbf{T}^2}\rangle$ is obtained by computing the Chern-Simons path integral over a solid torus having a Wilson line along the non-contractible cycle in the trivial representation. We  define $\langle\widetilde{\Psi}_{\mathbf{T}^2}^0|$ the state  dual to the state 
 $|\Psi^0_{\mathbf{T}^2}\rangle$, by demanding that the inner product between them give us partition  function of Chern-Simons theory in $\Sp^3$, viz., 
\begin{equation}
\label{eq:zs3torus}
\zcs\left(\Sp^3 \right)=\langle \widetilde{\Psi}^0_{\mathbf{T}^2}|\Psi^0_{\mathbf{T}^2}\rangle.
\end{equation}
The diffeomorphism on the boundary  of the second solid torus can be represented as an action on the state $|\Psi^0_{\mathbf{T}^2}\rangle$.  We are interested in a diffeomorphism which 
exchanges the $a$-cycle and the $b$-cycle of the boundary torus obtained via  a $\mathcal{S}$-transform of $PSL(2,\mathbb{Z})$.  Its action on the state $|\Psi^0_{\mathbf{T}^2}\rangle$ 
leads to the following state:
\begin{equation}
\label{eq:stranpsit}
|\widetilde{\Psi}^0_{\mathbf{T}^2}\rangle=\sum_{j}\mathcal{S}^0_j|\, \Psi^j_{\mathbf{T}_2}\rangle \,.
\end{equation}
Here, the state $|\Psi^j_{\mathbf{T}^2}\rangle$ is obtained by computing the Chern-Simons path integral over a solid torus having a Wilson line along the non-contractible cycle in   representation $R_j$. $\mathcal{S}^0_j$ is given by $\zcs(\Sp^3;R_{j}^T)$, the (unnormalized) partition function of Chern-Simons theory in $\Sp^3$ with a Wilson line in the representation $R^T_j$ which is dual to the representation $R_j$. It is equivalent to the modular $\mathcal{S}$ matrix element in the associated two dimensional WZW model. 

Therefore, we conclude that the partition function $\zcs(\Sp^2\times\Sp^1)$ is given by
\begin{equation}
\label{eq:s2xs1part}
\zcs(\Sp^2\times\Sp^1)=\sum_{j}\mathcal{S}^0_j~\zcs(\Sp^3;R_j) 
\end{equation}
The conventional normalization in Chern-Simons theory sets $\zcs(\Sp^2\times\Sp^1)$ to unity owing to the unidimensional Hilbert space on $\Sp^2$ \cite{Witten:1988hf}:
\begin{equation}\label{eq:s2xs1fix}
\zcs(\Sp^2\times\Sp^1) \equiv 1 \,.
\end{equation}  
Note that this assignment is consistent with the following completeness relation for modular $\mathcal{S}$-
matrix elements:
\begin{equation}\label{eq:sunit}
\sum_j \, \mathcal{S}^0_j \, \mathcal{S}^j_0  = 1 \,.
\end{equation}  
It also immediately gives us $\zcs(\Sp^3) = \mathcal{S}_0^{0}$.

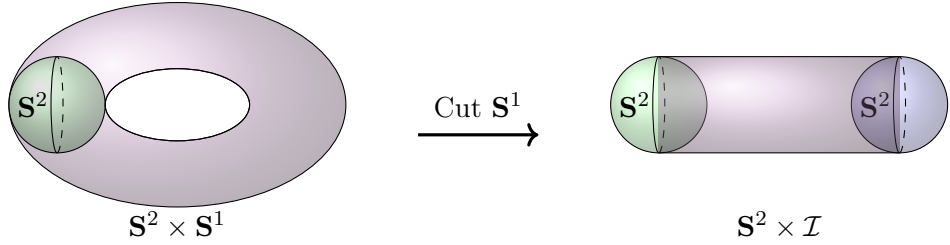
\begin{figure}
\begin{center}
\usetikzlibrary{backgrounds}
\begin{tikzpicture}[scale=.8]
  \shade[ball color = violet!40, opacity = 0.4] (1,3) ellipse (2.8 and 1.7);
    \draw(1,3) ellipse (2.8 and 1.7);
    \filldraw[fill=white]  (1,3) ellipse (1.2 and 0.6);
    \draw(1,3) ellipse (1.2 and 0.6);
    \shade[ball color = green!40, opacity = 0.4] (-1,3) circle (.8cm);
    \draw (-1,3) circle (.8cm);
 \draw [dashed]  (-1,2.2) arc (-90:90:0.1 and .8);
  \draw (-1,3.8) arc (90:270:0.1 and .8);
  \node at (-1.4,3)  {$\Sp^2$};
  \node at (1,1){${\Sp^2\times \Sp^1}$};
   \node at (6,3)  {\text{Cut $\Sp^1$}};
  \draw[very thick,->] (5,2.5)--(7,2.5);
  \shade[ball color = blue!40, opacity = 0.4](13,3) circle (.8cm);
    \draw (13,3) circle (.8cm);
    \shade[ball color = green!40, opacity = 0.4] (9,3) circle (.8cm);
    \draw (9,3) circle (.8cm);
 \draw [dashed]  (9,2.2) arc (-90:90:0.1 and .8);
  \draw (9,3.8) arc (90:270:0.1 and .8);
   \draw [dashed]  (13,2.2) arc (-90:90:0.1 and .8);
  \draw (13,3.8) arc (90:270:0.1 and .8);
  \node at (8.6,3)  {$\Sp^2$};
  \node at (12.6,3){$\Sp^2$};
  \node at (11,1){${\Sp^2\times \mathcal{I}}$};
  \shade[ball color = violet!40, opacity = 0.4]  (9,2.2) rectangle (13,3.8);
  \draw  (9,2.2) rectangle (13,2.2);
  \draw  (9,3.8) rectangle (13,3.8);
\end{tikzpicture}
\end{center}
\caption{ Normalized density matrix $\widehat{\rho}$ is obtained by cutting the Chern-Simons path integral on $\Sp^2\times\Sp^1$ along a fixed Euclidean time slice gives us the path integral over $\Sp^2\times \mathcal{I}$, where $\mathcal{I}$ is an interval obtained by cutting the euclidean time circle.  }
\label{fig:cuts2xs1}
\end{figure}

In \S\ref{sec:eeSp}, we described the path integral construction of reduced density matrix of the state $\psiSk \in \mathcal{H}_{\Sp^2}$ of Chern-Simons theory for the bipartitioning of $\Sp^2$ into two hemispheres $\rA$ and $\rAc$. We shall now describe a direct path integral construction of the \emph{normalized} reduced density matrix $\rhoAh$ for the same bipartition. At first sight, this seems academic. We should simply divide out the normalization factor to obtain  $\rhoAh$. However, there is a nice geometric story behind this construction which makes it interesting in its own right. We will also find this perspective useful when we uplift to the topological string.

Recall that, in \S\ref{sec:eeSp}, we defined the conjugate state $\psiSb$ dual to $\psiSk$ such that their inner product returned 
the partition function $\zcs(\Sp^3)$. Let us introduce another dual state $\psiSbn$ such that its inner product with $\psiSk$ gives the partition function $\zcs(\Sp^2\times\Sp^1)$, i.e., 
\begin{equation}
\label{eq:ndualstate}
\langle \widehat{\Psi}_{_{\Sp^2}}\psiSk=\zcs(\Sp^2\times\Sp^1)=1.
\end{equation}
Then we can define the normalized density matrix for the bipartitioning of the state $\psiSk$ to be
\begin{equation}
\label{eq:nrdensitym}
\widehat{\rho}=\psiSk\psiSbn\,. 
\end{equation}

\begin{figure}
\begin{center}
\usetikzlibrary{backgrounds}
\begin{tikzpicture}[scale=.7]
\shade[ball color = blue!40, opacity = 0.4](3,3) circle (.8cm);
 \draw (3,3) circle (.8cm);
 \shade[ball color = green!40, opacity = 0.4] (-1,3) circle (.8cm);
 \draw (-1,3) circle (.8cm); 
 \draw [dashed]  (-1,2.2) arc (-90:90:0.1 and .8);
 \draw (-1,3.8) arc (90:270:0.1 and .8);
 \draw [dashed]  (3,2.2) arc (-90:90:0.1 and .8);
 \draw (3,3.8) arc (90:270:0.1 and .8);
 \node at (-1.4,3)  {$\scriptstyle{\rAk}$};
 \node at (2.6,3){$\scriptstyle{\rAcb}$};
 \node at (-.5,3)  {$\scriptstyle{\rAck}$};
 \node at (3.5,3){$\scriptstyle{\rAb}$};
 \node at (1,1.5){${\Sp^2\times \mathcal{I}}$};
 \shade[ball color = violet!40, opacity = 0.4]  (-1,2.2) rectangle (3,3.8);
 \draw  (-1,2.2) rectangle (3,2.2);
 \draw  (-1,3.8) rectangle (3,3.8);
\node at (6,3)  {\large{$\equiv$}};
  \filldraw[fill = blue!20](13,5) ellipse (.4 and .8);
    \draw (13,5)ellipse (.4 and .8);
    \filldraw[fill = green!10] (9,5) ellipse (.4 and .8);
    \draw (9,5) ellipse (.4 and .8);
  \node at (9,5)  {$\rAk$};
  \node at (13,5){$\rAb$};
  \node at (11,6.2){${\disk_1\times \mathcal{I}}$};
  \shade[ball color = violet!40, opacity = 0.4]  (9,4.2) rectangle (13,5.8);
  \draw  (9,4.2) rectangle (13,4.2);
  \draw  (9,5.8) rectangle (13,5.8);
\draw[ thick,<->] (11,4)--(11,2);
  \filldraw[fill = blue!20](13,1) ellipse (.4 and .8);
    \draw (13,1)ellipse (.4 and .8);
    \filldraw[fill = green!10] (9,1) ellipse (.4 and .8);
    \draw (9,1) ellipse (.4 and .8);
  \node at (9,1)  {$\rAck$};
  \node at (13,1){$\rAcb$};
  \node at (11,-.2){${\disk_2\times \mathcal{I}}$};
  \shade[ball color = violet!40, opacity = 0.4]  (9,0.2) rectangle (13,1.8);
  \draw  (9,0.2) rectangle (13,0.2);
  \draw  (9,1.8) rectangle (13,1.8);
   \draw[thick, <->] (-0.5,3.5) to [out=70,in=110, looseness=1.25] (2.5,3.5);
   \draw[thick, <->] (8.5,1.5)  to [out=100,in=80, looseness=1.25] (13.5,1.5);  
\end{tikzpicture}
\end{center}
\caption{ The path integral over $\Sp^2\times \mathcal{I}$ can be understood as the path integrals over $\disk_1\times \mathcal{I}$ and $\disk_2\times \mathcal{I}$  glued along the boundaries $\partial\disk_1\times \mathcal{I}$
and $\partial\disk_2\times \mathcal{I}$. The normalized reduced density matrix $\rhoAh$ is obtained by gluing this path integral along the regions $\rAck$ and $\rAcb$. }
\label{fig:nrdensitymatirx}
\end{figure}
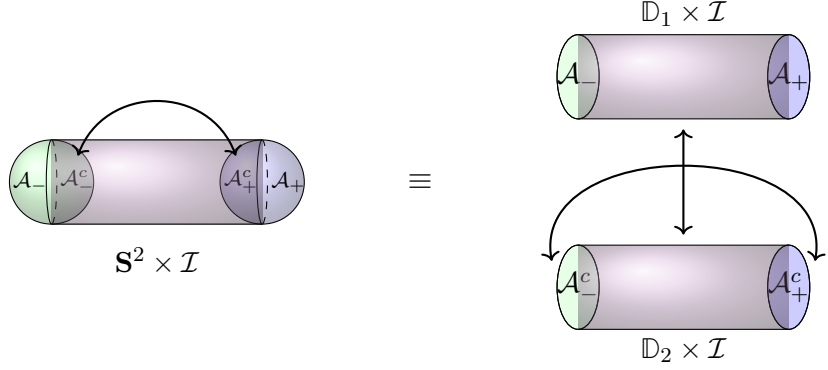

We can provide a path integral construction of $\widehat{\rho}$ by  considering Chern-Simons path integral on $\Sp^2\times\Sp^1$, with $\Sp^1$ treated to be the Euclidean time circle. 
We cut the functional integral along a fixed Euclidean time slice, obtaining thereby the geometry $\Sp^2\times \mathcal{I}$, as depicted in Fig.~\ref{fig:cuts2xs1}. 
We identify the state on $\Sp^2$ at one end of the interval to be $\psiSk$, but pick the state on other  boundary of $\mathcal{I}$  to be the  dual state $\psiSbn$. Gluing them together will give by the partition function on $\Sp^2 \times \Sp^1$ as desired. The path integral over $\Sp^2\times \mathcal{I}$ can be understood as the path integrals over $\disk_1\times \mathcal{I}$ and $\disk_2\times \mathcal{I}$  glued along the boundaries $\partial\disk_1\times \mathcal{I}$
and $\partial\disk_2\times \mathcal{I}$.

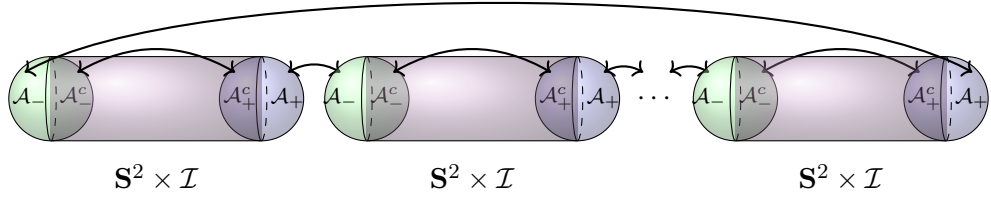
\begin{figure}
\begin{center}
\usetikzlibrary{backgrounds}
\begin{tikzpicture}[scale=.7]
\shade[ball color = blue!40, opacity = 0.4](3,3) circle (.8cm);
    \draw (3,3) circle (.8cm);
    \shade[ball color = green!40, opacity = 0.4] (-1,3) circle (.8cm);
    \draw (-1,3) circle (.8cm);
 \draw [dashed]  (-1,2.2) arc (-90:90:0.1 and .8);
  \draw (-1,3.8) arc (90:270:0.1 and .8);
   \draw [dashed]  (3,2.2) arc (-90:90:0.1 and .8);
  \draw (3,3.8) arc (90:270:0.1 and .8);
  \node at (-1.4,3)  {$\scriptstyle{\rAk}$};
  \node at (2.6,3){$\scriptstyle{\rAcb}$};
    \node at (-.5,3)  {$\scriptstyle{\rAck}$};
  \node at (3.5,3){$\scriptstyle{\rAb}$};
  \node at (1,1.5){${\Sp^2\times \mathcal{I}}$};
  \shade[ball color = violet!40, opacity = 0.4]  (-1,2.2) rectangle (3,3.8);
  \draw  (-1,2.2) rectangle (3,2.2);
  \draw  (-1,3.8) rectangle (3,3.8);
   \draw[thick, <->] (-0.5,3.5) to [out=30,in=150, looseness=1] (2.5,3.5);
   \draw[thick, <->] (3.5,3.5)  to [out=30,in=150, looseness=1] (4.5,3.5);
\shade[ball color = blue!40, opacity = 0.4](9,3) circle (.8cm);
    \draw (9,3) circle (.8cm);
    \shade[ball color = green!40, opacity = 0.4] (5,3) circle (.8cm);
    \draw (5,3) circle (.8cm);
 \draw [dashed]  (5,2.2) arc (-90:90:0.1 and .8);
  \draw (5,3.8) arc (90:270:0.1 and .8);
   \draw [dashed]  (9,2.2) arc (-90:90:0.1 and .8);
  \draw (9,3.8) arc (90:270:0.1 and .8);
  \node at (4.5,3)  {$\scriptstyle{\rAk}$};
  \node at (8.6,3){$\scriptstyle{\rAcb}$};
    \node at (5.4,3)  {$\scriptstyle{\rAck}$};
  \node at (9.5,3){$\scriptstyle{\rAb}$};
  \node at (7,1.5){${\Sp^2\times \mathcal{I}}$};
  \shade[ball color = violet!40, opacity = 0.4]  (5,2.2) rectangle (9,3.8);
  \draw  (5,2.2) rectangle (9,2.2);
  \draw  (5,3.8) rectangle (9,3.8);
   \draw[thick, <->] (5.5,3.5) to [out=30,in=150, looseness=1] (8.5,3.5);
   \draw[thick, <->] (12.5,3.5) to [out=30,in=150, looseness=1] (15.5,3.5);
   \draw[thick, <->] (9.5,3.5)  to [out=30,in=150, looseness=1] (10.25,3.5);
   \node at (10.5,3) {$\cdots$};
   \draw[thick, <->] (10.75,3.5)  to [out=30,in=150, looseness=1] (11.5,3.5);
\shade[ball color = blue!40, opacity = 0.4](16,3) circle (.8cm);
    \draw (16,3) circle (.8cm);
    \shade[ball color = green!40, opacity = 0.4] (12,3) circle (.8cm);
    \draw (12,3) circle (.8cm);   
 \draw [dashed]  (12,2.2) arc (-90:90:0.1 and .8);
  \draw (12,3.8) arc (90:270:0.1 and .8);
   \draw [dashed]  (16,2.2) arc (-90:90:0.1 and .8);
  \draw (16,3.8) arc (90:270:0.1 and .8);
  \node at  (11.5,3) {$\scriptstyle{\rAk}$};
  \node at (15.6,3){$\scriptstyle{\rAcb}$};
    \node at (12.4,3)  {$\scriptstyle{\rAck}$};
  \node at (16.5,3){$\scriptstyle{\rAb}$};
  \node at (14,1.5){${\Sp^2\times \mathcal{I}}$};
  \shade[ball color = violet!40, opacity = 0.4]  (12,2.2) rectangle (16,3.8);
  \draw  (12,2.2) rectangle (16,2.2);
  \draw  (12,3.8) rectangle (16,3.8);
   \draw[thick, <->] (-1.5,3.5) to [out=30,in=150, looseness=.5] (16.5,3.5);  
\end{tikzpicture}
\end{center}
\caption{ The $q^{\text{th}}$ replica of the normalized reduced density matrix $\rhoAh$ is represented by the path integral over the 3-manifold shown in the figure.}
\label{fig:replicacs}
\end{figure}

Consider the bipartitioning of the $\Sp^2$s at the end points of the interval $\mathcal{I}$. Once again let  us denote the hemispheres of the $\Sp^2$ at the left end of the interval 
$\mathcal{I}$ to be $\rAk$ and $ \rAck$, and denote the hemispheres of the $\Sp^2$ at the right end of the interval $\mathcal{I}$ to be $\rAb$ and $ \rAcb$. 
Then the normalized reduced density matrix  $\rhoAh$ is obtained by gluing this path integral along the regions $\rAck$ and $\rAcb$, cf., 
Fig.~\ref{fig:nrdensitymatirx}. Once we have this construction for the reduced density matrix we can easily compute integral powers of it. The $q^{\text{th}}$ replica of the normalized reduced density matrix $\rhoAh$ is represented by the path integral over the 3-manifold $(\Sp^2\times\Sp^1)_q$ shown in Fig.~\ref{fig:replicacs}. 

To obtain the R\'enyi entropies we need to evaluate the partition function of Chern-Simons theory in $(\Sp^2\times\Sp^1)_q$. This can be easily computed by aid of the following trick. 
We first multiply $\zcs\left((\Sp^2\times\Sp^1)_q\right)$ by $\left[\zcs(\Sp^3)\right]^{q-1}$ and show that the result is unity. To do so we represent the latter by decomposing each of 
$\Sp^3$s supplied into 3-balls as before. On the boundary of these 3-balls live copies of our state $\psiSk$ or $\psiSb$. 
To proceed, it is helpful to note that within the replica 3-manifold $(\Sp^2\times\Sp^1)_q$, in  each component that represents a copy of  $\widehat{\rho}_{\rA}$, the regions $\rAk$ and 
$\rAb$ form an $\Sp^2$. 

When computing $\zcs\left((\Sp^2\times\Sp^1)_q\right)$,   the geometry representing $\rhoAh$  is glued to its immediate neighbour, which represents
another copy of $\rhoAh$ along the two hemispheres  (view the two as carrying a state and its dual). When we supply  $\zcs(\Sp^3)$, we effectively provide many more copies of the state $\psiSk$ and $\psiSb$ as noted above. These states themselves can be further partitioned into states restricted to each hemisphere. We then combine the regions 
$\rAk$ and $\rAb$ from each copy of $\rhoAh$ with the state obtained from decomposition of $\zcs(\Sp^3)$. All this amounts to is gluing in a 3-ball into the each replica copy. 
Such a gluing (respecting replica symmetry) results in disconnecting each copy of $\rhoAh$ from its neighbours. We then end up with $q$ disconnected pieces of path 
integrals on $\Sp^2\times \Sp^1$ which we know to be unity.  We therefore conclude through this surgery operation that 
\begin{equation}
\label{zs2xs1q}
\zcs\left((\Sp^2\times\Sp^1)_q\right) = \left[\zcs(\Sp^3)\right]^{1-q}\,,
\end{equation}
and thence, the $q^{\text{th}}$ R\'enyi entrpy is given by
\begin{equation}
\label{nreynis2xs1}
S_q=\frac{1}{1-q}\, \log[\zcs\left((\Sp^2\times\Sp^1)_q\right)] = \log \zcs(\Sp^3) \,,
\end{equation}
as expected. 

\section{Chern-Simons and topological string observables}
\label{sec:}

In our discussion thus far we have confined our attention to using the topological open/closed string duality abstractly to motivate a construction of the closed string density matrices, and show
that it can be derived by following the surgery procedure employed in Chern-Simons theory through a geometric transition. The duality however relies on the fact that the observables on the Chern-Simons side match with those of the topological closed string on a resolved geometry, order by order in a large $N$ expansion. Since explicit results are known for the partition function 
it is worthwhile to examine the answers in some detail. 

\subsection{Chern-Simons partition sums}
\label{sec:cspart}

As noted in the main text the Chern-Simons partition function on $\Sp^3$ is obtained in terms of the matrix elements of the  modular $S$-matrix $\mathcal{S}_i^{j}$: 
\begin{equation}
\zcs(\Sp^3) = \mathcal{S}_0^{0}
\label{eq:zs3}
\end{equation}	
This quantity has an explicit representation of group theoretic data. For $SU(N)_k$ Chern-Simons theory the $\Sp^3$ partition sum is  given by:
\begin{equation}
\begin{split}
\zcs(\Sp^3) 
&=
	\frac{e^{i\,\frac{\pi}{8}\, N(N-1)}}{(k+N)^\frac{N}{2}} \; \sqrt{\frac{k+N}{N}} \  \prod_{r=1}^{N-1}\, 
	\left(2 \sin\left(\frac{r\, \pi}{k+N}\right)\right)^{N-r} 
\end{split}
\label{eq:zs3sun}
\end{equation}	

We can write the final answer after performing the large $N$ expansion in terms of the closed string variables, viz., the string coupling and the K\"ahler data of the resolved conifold using the map \eqref{eq:gstmap} as originally done in  \cite{Gopakumar:1998ki}:
\begin{equation}
\begin{split}
\zcs(\Sp^3) 
&=  	e^{-F(g_s, t)} = 
	\exp\left[-\sum_{g=0}^\infty\; \frac{1}{g_s^{2-2g}}\; F_g(t)
	\right]	\\
&= 	\exp\left[	-\sum_{n=1}^\infty\, \frac{1}{g_s^2\, n^3} \left( \,e^{-n t} -1
	\right) - \frac{1}{g_s^2} \sf{p}_0(t)\right] \\
&= \exp\left[ -\frac{1}{g_s^2} \bigg( {\rm Li}_3(e^{-t}) -\zeta(3) + {\sf p}_0(t) \bigg) + \cdots \right]	
\end{split}
\label{eq:zs3}
\end{equation}	
We have exhibited the tree level answer alone in the last line as it will be useful for some simple checks. 
Explicit expressions for the genus expansion of the free energy can be extracted from the above. For example,  at genus zero one has 
\begin{equation}
\begin{split}
F_0(t) 
&= 
	 -\zeta(3) + {\rm Li}_3(e^{-t}) + {\sf p}_0(t) \\
{\sf p}_0(t) 
&= 
	i \bigg[ \frac {\pi^2}{6} t -   \left( m + \frac{1}{4} \right) \pi\, t^2 +  \frac{1}{12} \,t^3  \bigg] \equiv \frac{(2\pi i)^3}{2\times 3!}\, \mathcal{B}_3(1-\frac{t}{2\pi})
\end{split}
\label{eq:F0}
\end{equation}
where $\mathcal{B}_3$ is a Bernoulli polynomial. 
The corresponding answer for genus $1$ is 
\begin{equation}
\begin{split}
F_1(t) = \frac{1}{24} \, t + \log(1-e^{-t}) 
\end{split}
\label{eq:F1}
\end{equation}
Note that this can also be expressed in terms of the Bernoulli polynomial $\mathcal{B}_1$  which is linear and the polylogarithm ${\rm Li}_1(e^{-t})$.  The higher genus expressions are similar:
\begin{equation}\label{eq:Fn}
F_g(t) = \frac{(-1)^{g-1}}{2g\,(2g-2)} \, B _g \left[ 
	\frac{(-1)^{g-1}}{(2\pi)^{2g-2}} \, 2\,\zeta(2g-2) - \frac{1}{(2g-3)!} \, \sum_{n=1}^\infty \, \frac{1}{n^{3-2g}}\, e^{-nt}
\right]
\end{equation}	
%

\subsection{Wilson loop generating functions}
\label{sec:wlgen}

Having obtained the partition functions we next move on to the computation of the Wilson loop expectation values. For an unknot $\mathcal{K}_0 \subset \Sp^3$ we have from \cite{Witten:1988hf}: 
\begin{equation}
\begin{split}
\vev{W_R(\mathcal{K}_0)}   \equiv 
	\frac{1}{\zcs(\Sp^3)} \, \int\, [\mathcal{D}A]\, e^{-S_{_\text{CS}}[A] } \, W_R(\mathcal{K}_0) 
	=	\frac{\mathcal{S}^0_R}{\mathcal{S}^0_0}		
\end{split}
\label{eq:wRvev}
\end{equation}	
In \cite{Ooguri:1999bv} an interesting generating function for computing $\vev{W_R(\mathcal{K}_0)} $ is derived. This relies on having two sets of topological D-branes 
intersecting on the knot. One set of D-branes are the $N$ topological branes that wrap the $\Sp^3 \subset T^*\Sp^3$ on which our Chern-Simons resides. The other set comprises of $M$ 
probe branes wrapping a Lagrangian cycle which is topologically $\Sp^1 \times \mathbb{R}^2$. At the intersection there is an additional degree of freedom, whose statistics depends on whether we consider branes or anti-branes (see eg., \cite{Gomis:2006mv}).   From a Chern-Simons viewpoint one can simply consider another probe gauge field $\tilde{A}$ which is used to 
capture the generating function.  All in all, one defines the following \emph{Ooguri-Vafa (OV) operator}:\footnote{ We will adhere to a convention where ${\rm tr}$ will refer to the trace taken in the fundamental representation. Traces in other representations will be indicated explicitly as ${\rm Tr}_R$.}
\begin{equation}
\begin{split}
\mathcal{O}(U,V) &= \exp \left(\sum_{n=1}^\infty \, \frac{1}{n} \, {\rm tr}(U^n)\, {\rm tr}( V^{-n})\right) , \\
\end{split}
\label{eq:ovop}
\end{equation}	
where 
\begin{equation}\label{eq:uvdef}
U = \mathcal{P}  \left(e^{i\oint_\gamma\, A}\right) \in SU(N) \,,   \qquad   V = \mathcal{P} \left(e^{i\oint_\gamma\, \tilde{A}} \right)\in SU(M) \,.
\end{equation}	

It will be helpful to consider a one-parameter generalization of this operator along the lines discussed in \cite{Aganagic:2002qg}. 
We  introduce a complex parameter ${\mathfrak r}$ (which maps onto a K\"ahler parameter in the closed string dual) and define 
the \emph{regulated OV operator}:
\begin{equation}
\begin{split}
\mathcal{O}_\mathfrak{r}(U,V) 
&= 
	\exp\left( \sum_{n=1}^\infty \, \frac{1}{n} \, e^{-n\,\mathfrak{r}\,   {\rm tr}(U^n)\, {\rm tr}( V^{-n})} \right) \\
&= 
	\sum_R \, e^{-B\, \mathfrak{r}} \, {\rm Tr}_R (U) \, {\rm Tr}_{R} (V^{-1})\,,
\end{split}
\label{eq:rovop}
\end{equation}	
where we expanded out the exponential, switched to cycle counting notation, and then used Frobenius relation along with the orthogonality of characters to arrive at the final expression. $R$ now is a label over $SU(N)$ irreps and $B$ is the total number of boxes in the Young tableaux associated with the representation.

Thus far the discussion is general, but now let us specialize to the case where the two gauge groups are the same, i.e., $N=M$. We want to evaluate the expectation value of the regulated OV operator. We will assume that the two Wilson lines are on some knots $(\mathcal{K}, \mathcal{K}') \subset \mcs_3 \times \mcs_3$.
 We will furthermore be interested in taking averages with respect to both the gauge fields. To wit, define:
\begin{align}
\vev{\mathcal{O}_\mathfrak{r}(U,V) } 
&= 
	\frac{1}{\left[\zcs(\Sp^3)\right]^2} \, \int_{\mcs_3 \times \mcs_3} [DA] \, [D\tilde A] \, e^{-S_{CS}[A] - S_{CS}[\tilde A]} \mathcal{O}_\mathfrak{r}(U,V)  \nonumber \\
&= 
	\sum_R\, \frac{1}{\left[\zcs(\Sp^3)\right]^2} \, \int_{\mcs_3} [DA] \,  e^{-S_{CS}[A]} {\rm Tr}_R (U) \, e^{-B\, \mathfrak{r}} \, 
	\int_{\mcs_3} [D\tilde A] \, e^{- S_{CS}[\tilde A]}  {\rm Tr}_{R} (V^{-1})  \nonumber \\
&=
	\sum_R \,e^{-B\, \mathfrak{r}} \,  \vev{W_R(\mathcal{K})} \, \vev{W_{\bar{R}} ( \mathcal{K}')} 		\,.
\label{eq:rovvev}
\end{align}
We used the fact that the two sets of gauge fields live on independent 3-manifolds, and the only linking element between them is the exponential box counting weight factor. The evaluations on each copy of $\mcs$ proceeds as usual, and as the 3-manifolds are uncorrelated we end up with the answer quoted.

The general answer can be simplified using \eqref{eq:wRvev} for the case when both the two Wilson lines are on unknots in an $\Sp^3$.  Then we obtain the result:
\begin{equation}
\vev{\mathcal{O}_\mathfrak{r}(U,V) }\bigg|_{\left(\mathcal{K}_0, {\bf S^3}\right),\left(\mathcal{K}'_0, {\bf S^3}\right) }
= 
	\sum_R \, e^{-B\, \mathfrak{r}} \, \frac{\mathcal{S}^0_R}{\mathcal{S}^0_0}\, \frac{\mathcal{S}^R_0}{\mathcal{S}^0_0} \,.
\label{eq:rovuk}
\end{equation}	
In the limit $\mathfrak{r} \to 0$ we can use the orthogonality of the modular S-matrix elements \eqref{eq:sunit} to conclude that 
\begin{equation}
\vev{\mathcal{O}(U,V) }\bigg|_{\left(\mathcal{K}_0, {\bf S^3}\right),\left(\mathcal{K}'_0, {\bf S^3}\right) } = 
\left(\frac{1}{\zcs(\Sp^3)}\right)^2
\label{eq:ovuk}
\end{equation}	
%

\section{Brief Review of  Topological String Theory}
\label{sec:tsreview}
        
We briefly review some of the salient features of topological string theory; more details can be found in the original papers \cite{Witten:1988ze,Witten:1988xj} and the book \cite{Hori:2003ic}. Topological strings are obtained by coupling  two-dimensional topological sigma models, obtained by twisting an Euclidean $\mathcal{N}=2$ supersymmetric theory,  to  worldsheet gravity. One starts with a $\mathcal{N}=(2,2)$ sigma model which has four supercharges $Q_\pm,\overline{Q}_\pm$ and vector and axial $R$-symmetries  $U(1)_V$ and $U(1)_A$,  generated by 
$F_{V,A}$, respectively. The supercharges  transform under the $R$-symmetries and the Euclidean rotation group $U(1)_E$ (generated by $M_E$) as:
\begin{equation}
\begin{split}
&[F_V,Q_{\pm}]=-Q_{\pm}\qquad\qquad [F_V,\overline{Q}_{\pm}]=\overline{Q}_{\pm}\\
&[F_A,Q_{\pm}]=\mp Q_{\pm}\qquad\qquad [F_A,\overline{Q}_{\pm}]=\pm\overline{Q}_{\pm}\\ 
&[M_E,Q_{\pm}]=\mp Q_{\pm}\qquad\qquad [M_E,\overline{Q}_{\pm}]=\mp\overline{Q}_{\pm}.
\end{split}
\end{equation}

The topological twist is performed by replacing $U(1)_E$ by a new rotation group  $U(1)_{E'}$. We can generate $U(1)_{E'}$ by $M_E+F_V$, which defines the {\it A-twist}, or alternately generate it by $M_E+F_A$, which is the {\it B-twist}. On a curved worldsheet we gauge the new rotation group 
by the spin connection, thereby altering the spins of fields. The resulting theory only has bosonic fields constrained by a fermionic symmetry, generated by 
\begin{equation}
\begin{split}
Q_A &=Q_-+\overline{Q}_+ \\
Q_B&=\overline{Q}_-+\overline{Q}_+
\end{split}
\label{eq:brst}
\end{equation}

The topological feature of the twisted theory, viz., independence from metric deformations, results from the stress tensor being a $Q$-commutator. 
The twisted and untwisted theories agree on flat regions of the worldsheet. Since this would be manifest on a flat cylinder worldsheet, one can use the untwisted Hilbert space for the twisted theory. The states of interest in this Hilbert space are the ground states of the untwisted $(2,2)$ theory.
    
We focus on the  A-twist of the sigma model on a K\"ahler threefold $\xf_6$. It is described by three chiral multiplets $\Phi^i,~i=1,2,3$. The bottom components $\phi^i$ represent complex coordinates of the map of the worldsheet $\Sigma_{ws}$, with a fixed complex structure,  to the target space 
$$\phi:\Sigma_{ws}\to \xf_6\,.$$ 
$Q_A$ of this theory can be identified with the de Rham operator $d=\partial+\overline{\partial}$ implying the $Q_A$-cohomology classes are identified with de Rham cohomology classes on
$\xf_6$. If $\upsilon$ is a homology cycle of real codimension-$r$, its Poincar\`e dual $[\Upsilon] \in H^r(\xf_6)$ can be represented as  a delta function $r$-form supported on $\upsilon$. We denote the corresponding operator in the topological sigma model as $\mathcal{O}_{\upsilon}(x)$, $x \in \Sigma_{ws}$. If the map $\phi$ maps $x$ outside the cycle $\upsilon$ then the operator $\mathcal{O}_\upsilon$ vanishes.

Computing physical observables, which are correlation function $\langle \mathcal{O}_1\cdots\mathcal{O}_s \rangle$ of these operators, involves evaluating an integral over all possible maps 
$\phi$ from $\Sigma_{ws}$ to $\xf_6$.  The maps $\phi$ can be classified by the homology class of the image:
\begin{equation}
\beta=\phi_*[\Sigma_{ws}] \in H_2(\xf_6,\mathbb{Z}).
\end{equation}
 Consequently, the correlation function can be decomposed into the sum over these homology classes
\begin{equation}
\langle \mathcal{O}_1\cdots \mathcal{O}_s\rangle=\sum_{\beta \in H_2(X,\mathbb{Z})}\langle \mathcal{O}_1\cdots \mathcal{O}_s \rangle_{\beta}\,.
\end{equation}

Let us discuss the structure  of $\langle \mathcal{O}_1\cdots \mathcal{O}_s\rangle_{\beta}$, whose computation involves maps $\phi$ restricted the homology class 
$\beta$, using the $R$-symmetries and the topological charge $Q_A$. The $R$-symmetries $U(1)_A$ and $U(1)_{V}$ impose certain selection rules on the correlation functions .
If $\mathcal{O}_i$ corresponds to a differential form that has Hodge degree $(p_i,q_i)$.  It has vector $R$-charge $q_V = -p_i + q_i$ and axial $R$-charge $q_A =p_i + q_i$. These charges lead to a selection rule for the correlation function $\langle \mathcal{O}_1\cdots \mathcal{O}_s\rangle_{\beta}$. It is  is non-vanishing only when
\begin{equation}
\sum_{i=1}^s p_i=\sum _{j=1}^s q_j=3(1-g)+\int_{\beta}c_1( \xf_6),
\end{equation}
where $c_1(\xf_6)$ denotes the  first Chern class. The topological charge $Q_A$ reduces the path integral that defines  the correlator 
into a sum over all possible holomorphic maps $\phi$ from $\Sigma_{ws}$ to $\xf_6$. In the presence of a non-trivial $B$-field, the action for a holomorphic map is given by
\begin{equation}
S_b=\int_{\beta}(\omega-i\, B)=(\omega-i\, B)\cdot \beta\,,
\end{equation}
 where $\omega$ is the K\"ahler form. 
 
Let us elaborate a little more on this.  Suppose that $\mathcal{M}\left( \xf_6,\beta\right)$ denotes the moduli space of holomorphic maps that belongs to the class $\beta$.  
The partition function of the $A$-twisted sigma model reduces to a finite dimensional integral over $\mathcal{M}\left( \xf_6,\beta\right)$. Moreover, the determinants in the measure cancel out and the integrand only involves the inserted operators and the weight factor $e^{-(\omega-i\, B)\cdot \beta}$. As a result, the correlation function is given by
\begin{equation}
\langle \mathcal{O}_1\cdots \mathcal{O}_s \rangle_{\beta}=e^{-(\omega-i\, B)\cdot \beta}
\int_{\mathcal{M}_{\Sigma_{ws}}(\xf_6,\beta)} \text{ev}_1^*\Upsilon_1\wedge\cdots\wedge \text{ev}_s^*\Upsilon_s,
\label{eq:crrfbeta}
\end{equation}
where ${\Upsilon}_i$ is the $(p_i,q_i)$ differential form corresponds to operator $\mathcal{O}_i$ inserted at $x_i\in \Sigma$, and $\text{ev}_i$ denotes the evaluation map at $x_i$:
\begin{equation}
\text{ev}_i:\mathcal{M}_{\Sigma_{ws}}\left(\xf_6,\Sigma_{ws}\right)\rightarrow \xf_6 \,.
\label{eq:evi}
\end{equation}

Assume that $[{\Upsilon}_i]$ are the Poincar\'e duals of the cycles $\upsilon_i$ in $\xf_6$ which implies that $\Upsilon_i$ can be chosen to be the delta function form supported on 
$\upsilon_i$. Then the integral in (\ref{eq:crrfbeta}) has an interesting geometrical interpretation. It can be identified as the number,  $n_{\beta,\upsilon_1,\cdots,\upsilon_s}$ of the holomorphic maps belongs to the class $\beta$ which map $x_i$ to $\upsilon_i$ for $i=1,\cdots, s$. Therefore, the total correlation function can be written as
\begin{equation}
\langle \mathcal{O}_1\cdots\mathcal{O}_s\rangle=\sum_{\beta\in H_2(X,\mathbb{Z})}e^{-(\omega-i\, B)\cdot \beta}\; n_{\beta,\upsilon_1,\cdots,\upsilon_s}\,.
\label{eq:corrsigmamodel}
\end{equation}
Since the K\"ahler form $\omega$ restricted to holomorphic curve $\phi(\Sigma_{ws})$ is positive semi-definite, $\omega\cdot\beta\geq 0$, the correlation function is dominated by 
holomorphic maps with $\beta =0$. Such maps map the worldsheet to a point and the associated moduli space is simply $\xf_6$. Since 
$\mathcal{M}_{\Sigma_{ws}}(\xf_6,0)\cong \xf_6$, the evaluation map $\text{ev}_i$ is the identity map, and hence we have:
\begin{equation}
\langle\mathcal{O}_1\cdots\mathcal{O}_s\rangle_0=\int_{\xf_6} \, \Upsilon_1\wedge\cdots\wedge\Upsilon_s.
\label{eq:beta0}
\end{equation}
This computes the intersection number of the dual cycles  $\upsilon_1,\cdots, \upsilon_s$.

It is possible to construct the topological string theory on a Calabi-Yau threefold by first coupling the $A$-twisted topological sigma model to topological gravity, and then by integrating over all complex structures of the worldsheet $\Sigma_{ws}$. The measure on the moduli space $\mathcal{M}_g$ of a genus $g$ worldsheet is defined by
\begin{equation}\label{measureg}        
\left\langle \prod_{i=1}^{3g-3}G_{++}({\mu}_i)\prod_{i=1}^{3g-3}G_{--}({\overline{\mu}}_{\overline{i}})\right\rangle.
\end{equation}
Here
\begin{equation}
G_{++}(\mu_i)\equiv \int G_{zz}\, \mu_{\overline{z}}^z \,d^2z
\qquad\qquad 
G_{--}(\mu_i)\equiv \int G_{\overline{z}\overline{z}}\, \mu^{\overline{z}}_z \, d^2z.
\end{equation}
where $\mu_i,~i=1,\cdots,3g-3$, are the Beltrami differentials, and $G_{\mu\nu}$ is related to the stress tensor $T_{\mu\nu}$ for the twisted sigma model as follows
\begin{equation}
T_{\mu\nu}=\left\{ Q,G_{\mu\nu}\right\}.
\end{equation}
Notice that, since the $G$'s each have axial charge $-1$, the measure has no net axial charge, and it can be nonzero. Then the genus $g$ topological string amplitude is given by
\begin{equation}\label{Fg}
F_g=\int_{\mathcal{M}_g}\prod_{i=1}^{3g-3} \, dm_i d\overline{m}_i \, \left\langle \prod_{i=1}^{3g-3}G_{++}({\mu}_i)\prod_{i=1}^{3g-3}G_{--}({\overline{\mu}}_{\overline{i}})\right\rangle,
\end{equation}
where $dm_i$ are the dual one-forms to the $\mu_i$.

\paragraph{Gromov-Witten theory and topological strings:}
\label{sec:GWtheory}

Consider a non-singular, projective, algebraic threefold $\xf$. Let $T_0,\cdots, T_m$ be a basis of $H^*(X,\mathbb{Z})$. 
Moreover, for simplicity assume that the classes $T_i$ are all even. We denote the fundamental class by $T_0$, the real degree 2 classes by 
$\{T_1,\cdots,T_p\}$, and the higher degree classes by $T_{p+1},\cdots, T_m$. The Gromov-Witten potential of the threefold $\xf$ can be written as:
\begin{equation}
F(t,g_s)=F_{\beta=0}^0+F_{\beta=0}^1+\sum_{g\geq 2}F^g_{\beta=0}+\sum_{g\geq 0}\sum_{\beta,\int_{\beta\neq 0}c_1(X)=0}F^g_{\beta}+\sum_{g\geq 0}\sum_{\beta,\int_{\beta\neq 0}c_1(X)>0}F^g_{\beta}.
\label{eq:GWpotential}
\end{equation}
The genus-0 constant contribution comes from the classical intersection theory of $\xf$, and is given by
\begin{equation}
F_{\beta=0}^0=\frac{1}{g_s^2}\sum_{0\geq i_1,i_2,i_3\leq 3}\frac{t_{i_1}t_{i_2}t_{i_3}}{3}\int_{\xf}\, T_{i_1}\cup T_{i_2}\cup T_{i_3}.
\label{eq:g0constatnt}
\end{equation} 
The genus-1 constant contribution is as follows
\begin{equation}
F^1_{\beta=0}=\sum_{i=1}^pt_i\langle T_i\rangle_{g=1,\beta=0}=-\sum_{i=1}^p\frac{t_i}{24}\int_{\xf}\, T_i\cup c_2(\xf).
\label{eq:g1constant}
\end{equation}
The genus $g\geq 2$ contributions are 
\begin{equation}
F_{\beta=0}^g=(-1)^g\; \frac{g_s^{2g-2}}{2} \, \int_\xf\bigg(c_3(\xf) -c_1(\xf)\cup c_2(\xf)\bigg)\cdot \int_{\overline{\mathcal{M}}_g}\lambda_{g-1}^3\,.
\label{eq:ggeq2constant}
\end{equation}
Finally, the non-constant contribution is given by
\begin{equation}
\begin{split}
F^g_{\beta}(t,\lambda)&=
       g_s^{2g-2}\, q^{\beta}\,\sum_{n\geq 0}\frac{1}{n!}\sum_{p+1\leq i_1,\cdots,i_n\leq m}t_{i_1}\cdots t_{i_n}\left\langle T_{i_1}\cdots T_{i_n}\right\rangle_{g,\beta}\,,\\
q^\beta &\equiv \prod_{i=1}^p\,  \left(q_i^{\int_\beta\, T_i} \right)\,, \qquad q_i = e^{t_i} \,.
\end{split}
\end{equation}
$\left\langle T_{i_1}\cdots T_{i_n}\right\rangle_{g,\beta}$ are the Gromov-Witten invariants. The $c_i(\xf)$ denotes the $i^{\text{th}}$ Chern class of  $\xf$, and 
$\lambda_{g-1}$ denotes the characteristic class in the moduli space $\overline{\mathcal{M}}_g$.\par

There is a beautiful connection between the Gromov-Witten theory and topological string theory. When $\xf$ is a Calabi-Yau threefold we can define the A-model closed topological string theory on it as described above The closed string free energy can be identified with the Gromov-Witten prepotential via
\begin{equation}
F_{cl}(X,t)=\sum_{g\geq 0}^{\infty}g_s^{2g-2}F_g(t)\,, \qquad F_{g}(t)=\sum_{\beta\in H_2(X,\mathbb{Z})}N_{\beta}^ge^{\omega.\beta}\,.
\end{equation}
where $\omega.\beta=\left(n_1t_1,\cdots,n_{h^{1,1}}t_{h^{1,1}} \right)$, $[t]=\left(t_1,\cdots, t_{h^{1,1}} \right)$ is the K\"ahler parameter and $g_s$ is the string coupling constant. The quantities $N_{\beta}^g$ are the genus $g$ Gromov-Witten invariants of $\xf$, corresponding to the class $\beta$.


\begin{thebibliography}{10}

\bibitem{Gopakumar:1998ki}
R.~Gopakumar and C.~Vafa, {\it {On the gauge theory / geometry
  correspondence}},  {\em Adv. Theor. Math. Phys.} {\bf 3} (1999) 1415--1443,
  [\href{http://arxiv.org/abs/hep-th/9811131}{{\tt hep-th/9811131}}]. [AMS/IP
  Stud. Adv. Math.23,45(2001)].

\bibitem{Ooguri:1999bv}
H.~Ooguri and C.~Vafa, {\it {Knot invariants and topological strings}},  {\em
  Nucl. Phys.} {\bf B577} (2000) 419--438,
  [\href{http://arxiv.org/abs/hep-th/9912123}{{\tt hep-th/9912123}}].

\bibitem{Aganagic:2002qg}
M.~Aganagic, M.~Marino, and C.~Vafa, {\it {All loop topological string
  amplitudes from Chern-Simons theory}},  {\em Commun. Math. Phys.} {\bf 247}
  (2004) 467--512, [\href{http://arxiv.org/abs/hep-th/0206164}{{\tt
  hep-th/0206164}}].

\bibitem{Aganagic:2003db}
M.~Aganagic, A.~Klemm, M.~Marino, and C.~Vafa, {\it {The Topological vertex}},
  {\em Commun. Math. Phys.} {\bf 254} (2005) 425--478,
  [\href{http://arxiv.org/abs/hep-th/0305132}{{\tt hep-th/0305132}}].

\bibitem{Maldacena:1997re}
J.~M. Maldacena, {\it {The Large N limit of superconformal field theories and
  supergravity}},  {\em Int. J. Theor. Phys.} {\bf 38} (1999) 1113--1133,
  [\href{http://arxiv.org/abs/hep-th/9711200}{{\tt hep-th/9711200}}]. [Adv.
  Theor. Math. Phys.2,231(1998)].

\bibitem{Gubser:1998bc}
S.~S. Gubser, I.~R. Klebanov, and A.~M. Polyakov, {\it {Gauge theory
  correlators from noncritical string theory}},  {\em Phys. Lett.} {\bf B428}
  (1998) 105--114, [\href{http://arxiv.org/abs/hep-th/9802109}{{\tt
  hep-th/9802109}}].

\bibitem{Witten:1998qj}
E.~Witten, {\it {Anti-de Sitter space and holography}},  {\em Adv. Theor. Math.
  Phys.} {\bf 2} (1998) 253--291,
  [\href{http://arxiv.org/abs/hep-th/9802150}{{\tt hep-th/9802150}}].

\bibitem{Ryu:2006bv}
S.~Ryu and T.~Takayanagi, {\it {Holographic derivation of entanglement entropy
  from AdS/CFT}},  {\em Phys. Rev. Lett.} {\bf 96} (2006) 181602,
  [\href{http://arxiv.org/abs/hep-th/0603001}{{\tt hep-th/0603001}}].

\bibitem{Hubeny:2007xt}
V.~E. Hubeny, M.~Rangamani, and T.~Takayanagi, {\it {A Covariant holographic
  entanglement entropy proposal}},  {\em JHEP} {\bf 07} (2007) 062,
  [\href{http://arxiv.org/abs/0705.0016}{{\tt arXiv:0705.0016}}].

\bibitem{VanRaamsdonk:2010pw}
M.~Van~Raamsdonk, {\it {Building up spacetime with quantum entanglement}},
  {\em Gen. Rel. Grav.} {\bf 42} (2010) 2323--2329,
  [\href{http://arxiv.org/abs/1005.3035}{{\tt arXiv:1005.3035}}]. [Int. J. Mod.
  Phys.D19,2429(2010)].

\bibitem{Maldacena:2013xja}
J.~Maldacena and L.~Susskind, {\it {Cool horizons for entangled black holes}},
  {\em Fortsch. Phys.} {\bf 61} (2013) 781--811,
  [\href{http://arxiv.org/abs/1306.0533}{{\tt arXiv:1306.0533}}].

\bibitem{Rangamani:2016dms}
M.~Rangamani and T.~Takayanagi, {\it {Holographic Entanglement Entropy}},  {\em
  Lect. Notes Phys.} {\bf 931} (2017) pp.1--246,
  [\href{http://arxiv.org/abs/1609.01287}{{\tt arXiv:1609.01287}}].

\bibitem{Dong:2013qoa}
X.~Dong, {\it {Holographic Entanglement Entropy for General Higher Derivative
  Gravity}},  {\em JHEP} {\bf 01} (2014) 044,
  [\href{http://arxiv.org/abs/1310.5713}{{\tt arXiv:1310.5713}}].

\bibitem{Camps:2013zua}
J.~Camps, {\it {Generalized entropy and higher derivative Gravity}},  {\em
  JHEP} {\bf 03} (2014) 070, [\href{http://arxiv.org/abs/1310.6659}{{\tt
  arXiv:1310.6659}}].

\bibitem{Faulkner:2013ana}
T.~Faulkner, A.~Lewkowycz, and J.~Maldacena, {\it {Quantum corrections to
  holographic entanglement entropy}},  {\em JHEP} {\bf 11} (2013) 074,
  [\href{http://arxiv.org/abs/1307.2892}{{\tt arXiv:1307.2892}}].

\bibitem{Witten:2018xfj}
E.~Witten, {\it {Open Strings On The Rindler Horizon}},  {\em JHEP} {\bf 01}
  (2019) 126, [\href{http://arxiv.org/abs/1810.11912}{{\tt arXiv:1810.11912}}].

\bibitem{He:2014gva}
S.~He, T.~Numasawa, T.~Takayanagi, and K.~Watanabe, {\it {Notes on Entanglement
  Entropy in String Theory}},  {\em JHEP} {\bf 05} (2015) 106,
  [\href{http://arxiv.org/abs/1412.5606}{{\tt arXiv:1412.5606}}].

\bibitem{Balasubramanian:2018axm}
V.~Balasubramanian and O.~Parrikar, {\it {Remarks on entanglement entropy in
  string theory}},  {\em Phys. Rev.} {\bf D97} (2018), no.~6 066025,
  [\href{http://arxiv.org/abs/1801.03517}{{\tt arXiv:1801.03517}}].

\bibitem{Lewkowycz:2013nqa}
A.~Lewkowycz and J.~Maldacena, {\it {Generalized gravitational entropy}},  {\em
  JHEP} {\bf 08} (2013) 090, [\href{http://arxiv.org/abs/1304.4926}{{\tt
  arXiv:1304.4926}}].

\bibitem{Dong:2016hjy}
X.~Dong, A.~Lewkowycz, and M.~Rangamani, {\it {Deriving covariant holographic
  entanglement}},  {\em JHEP} {\bf 11} (2016) 028,
  [\href{http://arxiv.org/abs/1607.07506}{{\tt arXiv:1607.07506}}].

\bibitem{Headrick:2010zt}
M.~Headrick, {\it {Entanglement Renyi entropies in holographic theories}},
  {\em Phys. Rev.} {\bf D82} (2010) 126010,
  [\href{http://arxiv.org/abs/1006.0047}{{\tt arXiv:1006.0047}}].

\bibitem{Faulkner:2013yia}
T.~Faulkner, {\it {The Entanglement Renyi Entropies of Disjoint Intervals in
  AdS/CFT}},  \href{http://arxiv.org/abs/1303.7221}{{\tt arXiv:1303.7221}}.

\bibitem{Jafferis:2015del}
D.~L. Jafferis, A.~Lewkowycz, J.~Maldacena, and S.~J. Suh, {\it {Relative
  entropy equals bulk relative entropy}},  {\em JHEP} {\bf 06} (2016) 004,
  [\href{http://arxiv.org/abs/1512.06431}{{\tt arXiv:1512.06431}}].

\bibitem{Headrick:2014cta}
M.~Headrick, V.~E. Hubeny, A.~Lawrence, and M.~Rangamani, {\it {Causality \&
  holographic entanglement entropy}},  {\em JHEP} {\bf 12} (2014) 162,
  [\href{http://arxiv.org/abs/1408.6300}{{\tt arXiv:1408.6300}}].

\bibitem{Czech:2012bh}
B.~Czech, J.~L. Karczmarek, F.~Nogueira, and M.~Van~Raamsdonk, {\it {The
  Gravity Dual of a Density Matrix}},  {\em Class. Quant. Grav.} {\bf 29}
  (2012) 155009, [\href{http://arxiv.org/abs/1204.1330}{{\tt
  arXiv:1204.1330}}].

\bibitem{Wall:2012uf}
A.~C. Wall, {\it {Maximin Surfaces, and the Strong Subadditivity of the
  Covariant Holographic Entanglement Entropy}},  {\em Class. Quant. Grav.} {\bf
  31} (2014), no.~22 225007, [\href{http://arxiv.org/abs/1211.3494}{{\tt
  arXiv:1211.3494}}].

\bibitem{Dong:2016eik}
X.~Dong, D.~Harlow, and A.~C. Wall, {\it {Reconstruction of Bulk Operators
  within the Entanglement Wedge in Gauge-Gravity Duality}},  {\em Phys. Rev.
  Lett.} {\bf 117} (2016), no.~2 021601,
  [\href{http://arxiv.org/abs/1601.05416}{{\tt arXiv:1601.05416}}].

\bibitem{Cotler:2017erl}
J.~Cotler, P.~Hayden, G.~Penington, G.~Salton, B.~Swingle, and M.~Walter, {\it
  {Entanglement Wedge Reconstruction via Universal Recovery Channels}},
  \href{http://arxiv.org/abs/1704.05839}{{\tt arXiv:1704.05839}}.

\bibitem{Engelhardt:2014gca}
N.~Engelhardt and A.~C. Wall, {\it {Quantum Extremal Surfaces: Holographic
  Entanglement Entropy beyond the Classical Regime}},  {\em JHEP} {\bf 01}
  (2015) 073, [\href{http://arxiv.org/abs/1408.3203}{{\tt arXiv:1408.3203}}].

\bibitem{Dong:2017xht}
X.~Dong and A.~Lewkowycz, {\it {Entropy, Extremality, Euclidean Variations, and
  the Equations of Motion}},  {\em JHEP} {\bf 01} (2018) 081,
  [\href{http://arxiv.org/abs/1705.08453}{{\tt arXiv:1705.08453}}].

\bibitem{Donnelly:2016jet}
W.~Donnelly and G.~Wong, {\it {Entanglement branes in a two-dimensional string
  theory}},  {\em JHEP} {\bf 09} (2017) 097,
  [\href{http://arxiv.org/abs/1610.01719}{{\tt arXiv:1610.01719}}].

\bibitem{Donnelly:2018ppr}
W.~Donnelly and G.~Wong, {\it {Entanglement branes, modular flow, and extended
  topological quantum field theory}},
  \href{http://arxiv.org/abs/1811.10785}{{\tt arXiv:1811.10785}}.

\bibitem{Witten:1992fb}
E.~Witten, {\it {Chern-Simons gauge theory as a string theory}},  {\em Prog.
  Math.} {\bf 133} (1995) 637--678,
  [\href{http://arxiv.org/abs/hep-th/9207094}{{\tt hep-th/9207094}}].

\bibitem{Gomis:2006mv}
J.~Gomis and T.~Okuda, {\it {Wilson loops, geometric transitions and bubbling
  Calabi-Yau's}},  {\em JHEP} {\bf 02} (2007) 083,
  [\href{http://arxiv.org/abs/hep-th/0612190}{{\tt hep-th/0612190}}].

\bibitem{Gomis:2007kz}
J.~Gomis and T.~Okuda, {\it {D-branes as a Bubbling Calabi-Yau}},  {\em JHEP}
  {\bf 07} (2007) 005, [\href{http://arxiv.org/abs/0704.3080}{{\tt
  arXiv:0704.3080}}].

\bibitem{Ooguri:2002gx}
H.~Ooguri and C.~Vafa, {\it {World sheet derivation of a large N duality}},
  {\em Nucl. Phys.} {\bf B641} (2002) 3--34,
  [\href{http://arxiv.org/abs/hep-th/0205297}{{\tt hep-th/0205297}}].

\bibitem{Kitaev:2005dm}
A.~Kitaev and J.~Preskill, {\it {Topological entanglement entropy}},  {\em
  Phys. Rev. Lett.} {\bf 96} (2006) 110404,
  [\href{http://arxiv.org/abs/hep-th/0510092}{{\tt hep-th/0510092}}].

\bibitem{Levin:2006zz}
M.~Levin and X.-G. Wen, {\it {Detecting Topological Order in a Ground State
  Wave Function}},  {\em Phys. Rev. Lett.} {\bf 96} (2006) 110405,
  [\href{http://arxiv.org/abs/cond-mat/0510613}{{\tt cond-mat/0510613}}].

\bibitem{Pakman:2008ui}
A.~Pakman and A.~Parnachev, {\it {Topological Entanglement Entropy and
  Holography}},  {\em JHEP} {\bf 07} (2008) 097,
  [\href{http://arxiv.org/abs/0805.1891}{{\tt arXiv:0805.1891}}].

\bibitem{Witten:1988hf}
E.~Witten, {\it {Quantum Field Theory and the Jones Polynomial}},  {\em Commun.
  Math. Phys.} {\bf 121} (1989) 351--399. [,233(1988)].

\bibitem{Elitzur:1989nr}
S.~Elitzur, G.~W. Moore, A.~Schwimmer, and N.~Seiberg, {\it {Remarks on the
  Canonical Quantization of the Chern-Simons-Witten Theory}},  {\em Nucl.
  Phys.} {\bf B326} (1989) 108--134.

\bibitem{Dong:2008ft}
S.~Dong, E.~Fradkin, R.~G. Leigh, and S.~Nowling, {\it {Topological
  Entanglement Entropy in Chern-Simons Theories and Quantum Hall Fluids}},
  {\em JHEP} {\bf 05} (2008) 016, [\href{http://arxiv.org/abs/0802.3231}{{\tt
  arXiv:0802.3231}}].

\bibitem{Balasubramanian:2016sro}
V.~Balasubramanian, J.~R. Fliss, R.~G. Leigh, and O.~Parrikar, {\it
  {Multi-Boundary Entanglement in Chern-Simons Theory and Link Invariants}},
  {\em JHEP} {\bf 04} (2017) 061, [\href{http://arxiv.org/abs/1611.05460}{{\tt
  arXiv:1611.05460}}].

\bibitem{Balasubramanian:2018por}
V.~Balasubramanian, M.~DeCross, J.~Fliss, A.~Kar, R.~G. Leigh, and O.~Parrikar,
  {\it {Entanglement Entropy and the Colored Jones Polynomial}},  {\em JHEP}
  {\bf 05} (2018) 038, [\href{http://arxiv.org/abs/1801.01131}{{\tt
  arXiv:1801.01131}}].

\bibitem{McGough:2013gka}
L.~McGough and H.~Verlinde, {\it {Bekenstein-Hawking Entropy as Topological
  Entanglement Entropy}},  {\em JHEP} {\bf 11} (2013) 208,
  [\href{http://arxiv.org/abs/1308.2342}{{\tt arXiv:1308.2342}}].

\bibitem{Witten:1988xj}
E.~Witten, {\it {Topological Sigma Models}},  {\em Commun. Math. Phys.} {\bf
  118} (1988) 411.

\bibitem{Marino:2004uf}
M.~Marino, {\it {Chern-Simons theory and topological strings}},  {\em Rev. Mod.
  Phys.} {\bf 77} (2005) 675--720,
  [\href{http://arxiv.org/abs/hep-th/0406005}{{\tt hep-th/0406005}}].

\bibitem{Marino:2005sj}
M.~Marino, {\it {Chern-Simons theory, matrix models, and topological strings}},
   {\em Int. Ser. Monogr. Phys.} {\bf 131} (2005) 1--197.

\bibitem{Lickorish:1962rep}
W.~R. Lickorish, {\it A representation of orientable combinatorial
  3-manifolds},  {\em Ann. Math} {\bf 76} (1962), no.~2 531--540.

\bibitem{Prasolov:1997kno}
V.~V. Prasolov and A.~B. Sossinsky, {\em Knots, links, braids, and 3-manifolds:
  an introduction to the new invariants in low-dimensional topology}.
\newblock No.~154. American Mathematical Soc., 1997.

\bibitem{Wen:2016snr}
X.~Wen, S.~Matsuura, and S.~Ryu, {\it {Edge theory approach to topological
  entanglement entropy, mutual information and entanglement negativity in
  Chern-Simons theories}},  {\em Phys. Rev.} {\bf B93} (2016), no.~24 245140,
  [\href{http://arxiv.org/abs/1603.08534}{{\tt arXiv:1603.08534}}].

\bibitem{Das:2015oha}
D.~Das and S.~Datta, {\it {Universal features of left-right entanglement
  entropy}},  {\em Phys. Rev. Lett.} {\bf 115} (2015), no.~13 131602,
  [\href{http://arxiv.org/abs/1504.02475}{{\tt arXiv:1504.02475}}].

\bibitem{Wen:2016bla}
X.~Wen, P.-Y. Chang, and S.~Ryu, {\it {Topological entanglement negativity in
  Chern-Simons theories}},  {\em JHEP} {\bf 09} (2016) 012,
  [\href{http://arxiv.org/abs/1606.04118}{{\tt arXiv:1606.04118}}].

\bibitem{Wong:2017pdm}
G.~Wong, {\it {A note on entanglement edge modes in Chern Simons theory}},
  {\em JHEP} {\bf 08} (2018) 020, [\href{http://arxiv.org/abs/1706.04666}{{\tt
  arXiv:1706.04666}}].

\bibitem{Buividovich:2008gq}
P.~V. Buividovich and M.~I. Polikarpov, {\it {Entanglement entropy in gauge
  theories and the holographic principle for electric strings}},  {\em Phys.
  Lett.} {\bf B670} (2008) 141--145,
  [\href{http://arxiv.org/abs/0806.3376}{{\tt arXiv:0806.3376}}].

\bibitem{Casini:2013rba}
H.~Casini, M.~Huerta, and J.~A. Rosabal, {\it {Remarks on entanglement entropy
  for gauge fields}},  {\em Phys. Rev.} {\bf D89} (2014), no.~8 085012,
  [\href{http://arxiv.org/abs/1312.1183}{{\tt arXiv:1312.1183}}].

\bibitem{Donnelly:2014fua}
W.~Donnelly and A.~C. Wall, {\it {Entanglement entropy of electromagnetic edge
  modes}},  {\em Phys. Rev. Lett.} {\bf 114} (2015), no.~11 111603,
  [\href{http://arxiv.org/abs/1412.1895}{{\tt arXiv:1412.1895}}].

\bibitem{Soni:2015yga}
R.~M. Soni and S.~P. Trivedi, {\it {Aspects of Entanglement Entropy for Gauge
  Theories}},  {\em JHEP} {\bf 01} (2016) 136,
  [\href{http://arxiv.org/abs/1510.07455}{{\tt arXiv:1510.07455}}].

\bibitem{Ghosh:2015iwa}
S.~Ghosh, R.~M. Soni, and S.~P. Trivedi, {\it {On The Entanglement Entropy For
  Gauge Theories}},  {\em JHEP} {\bf 09} (2015) 069,
  [\href{http://arxiv.org/abs/1501.02593}{{\tt arXiv:1501.02593}}].

\bibitem{Pastawski:2015qua}
F.~Pastawski, B.~Yoshida, D.~Harlow, and J.~Preskill, {\it {Holographic quantum
  error-correcting codes: Toy models for the bulk/boundary correspondence}},
  {\em JHEP} {\bf 06} (2015) 149, [\href{http://arxiv.org/abs/1503.06237}{{\tt
  arXiv:1503.06237}}].

\bibitem{Hayden:2016cfa}
P.~Hayden, S.~Nezami, X.-L. Qi, N.~Thomas, M.~Walter, and Z.~Yang, {\it
  {Holographic duality from random tensor networks}},  {\em JHEP} {\bf 11}
  (2016) 009, [\href{http://arxiv.org/abs/1601.01694}{{\tt arXiv:1601.01694}}].

\bibitem{Zwiebach:1992ie}
B.~Zwiebach, {\it {Closed string field theory: Quantum action and the B-V
  master equation}},  {\em Nucl. Phys.} {\bf B390} (1993) 33--152,
  [\href{http://arxiv.org/abs/hep-th/9206084}{{\tt hep-th/9206084}}].

\bibitem{Moosavian:2017qsp}
S.~F. Moosavian and R.~Pius, {\it {Hyperbolic Geometry and Closed Bosonic
  String Field Theory I: The String Vertices Via Hyperbolic Riemann Surfaces}},
   \href{http://arxiv.org/abs/1706.07366}{{\tt arXiv:1706.07366}}.

\bibitem{Moosavian:2017sev}
S.~F. Moosavian and R.~Pius, {\it {Hyperbolic Geometry and Closed Bosonic
  String Field Theory II: The Rules for Evaluating the Quantum BV Master
  Action}},  \href{http://arxiv.org/abs/1708.04977}{{\tt arXiv:1708.04977}}.

\bibitem{Kontsevich:1994na}
M.~Kontsevich, {\it {Enumeration of rational curves via Torus actions}},
  \href{http://arxiv.org/abs/hep-th/9405035}{{\tt hep-th/9405035}}.

\bibitem{Evens:1978che}
L.~Evens and D.~S. Kahn, {\it Chern classes of certain representations of
  symmetric groups},  {\em Transactions of the American Mathematical Society}
  {\bf 245} (1978) 309--330.

\bibitem{Fulton:1987cha}
W.~Fulton and R.~MacPherson, {\it Characteristic classes of direct image
  bundles for covering maps},  {\em Ann. of Math} {\bf 125} (1987) 1--92.

\bibitem{Taubes:2001wk}
C.~H. Taubes, {\it {Lagrangians for the Gopakumar-Vafa conjecture}},  {\em Adv.
  Theor. Math. Phys.} {\bf 5} (2001) 139--163,
  [\href{http://arxiv.org/abs/math/0201219}{{\tt math/0201219}}]. [Geom. Topol.
  Monographs8,73(2006)].

\bibitem{Atiyah:1989vu}
M.~Atiyah, {\it {Topological quantum field theories}},  {\em Inst. Hautes
  Etudes Sci. Publ. Math.} {\bf 68} (1989) 175--186.

\bibitem{Fursaev:2006ih}
D.~V. Fursaev, {\it {Proof of the holographic formula for entanglement
  entropy}},  {\em JHEP} {\bf 09} (2006) 018,
  [\href{http://arxiv.org/abs/hep-th/0606184}{{\tt hep-th/0606184}}].

\bibitem{Harlow:2016vwg}
D.~Harlow, {\it {The Ryu–Takayanagi Formula from Quantum Error Correction}},
  {\em Commun. Math. Phys.} {\bf 354} (2017), no.~3 865--912,
  [\href{http://arxiv.org/abs/1607.03901}{{\tt arXiv:1607.03901}}].

\bibitem{Dong:2018seb}
X.~Dong, D.~Harlow, and D.~Marolf, {\it {Flat entanglement spectra in
  fixed-area states of quantum gravity}},
  \href{http://arxiv.org/abs/1811.05382}{{\tt arXiv:1811.05382}}.

\bibitem{Akers:2018fow}
C.~Akers and P.~Rath, {\it {Holographic Renyi Entropy from Quantum Error
  Correction}},  {\em JHEP} {\bf 05} (2019) 052,
  [\href{http://arxiv.org/abs/1811.05171}{{\tt arXiv:1811.05171}}].

\bibitem{Gopakumar:1998vy}
R.~Gopakumar and C.~Vafa, {\it {Topological gravity as large N topological
  gauge theory}},  {\em Adv. Theor. Math. Phys.} {\bf 2} (1998) 413--442,
  [\href{http://arxiv.org/abs/hep-th/9802016}{{\tt hep-th/9802016}}].

\bibitem{Aharony:2011jz}
O.~Aharony, G.~Gur-Ari, and R.~Yacoby, {\it {d=3 Bosonic Vector Models Coupled
  to Chern-Simons Gauge Theories}},  {\em JHEP} {\bf 03} (2012) 037,
  [\href{http://arxiv.org/abs/1110.4382}{{\tt arXiv:1110.4382}}].

\bibitem{Giombi:2011kc}
S.~Giombi, S.~Minwalla, S.~Prakash, S.~P. Trivedi, S.~R. Wadia, and X.~Yin,
  {\it {Chern-Simons Theory with Vector Fermion Matter}},  {\em Eur. Phys. J.}
  {\bf C72} (2012) 2112, [\href{http://arxiv.org/abs/1110.4386}{{\tt
  arXiv:1110.4386}}].

\bibitem{Aganagic:2017tvx}
M.~Aganagic, K.~Costello, J.~McNamara, and C.~Vafa, {\it {Topological
  Chern-Simons/Matter Theories}},  \href{http://arxiv.org/abs/1706.09977}{{\tt
  arXiv:1706.09977}}.

\bibitem{Aharony:2019suq}
O.~Aharony, A.~Feldman, and M.~Honda, {\it {A String Dual for Partially
  Topological Chern-Simons-Matter Theories}},
  \href{http://arxiv.org/abs/1903.06433}{{\tt arXiv:1903.06433}}.

\bibitem{Gibbons:1976ue}
G.~W. Gibbons and S.~W. Hawking, {\it {Action Integrals and Partition Functions
  in Quantum Gravity}},  {\em Phys. Rev.} {\bf D15} (1977) 2752--2756.

\bibitem{Susskind:1994sm}
L.~Susskind and J.~Uglum, {\it {Black hole entropy in canonical quantum gravity
  and superstring theory}},  {\em Phys. Rev.} {\bf D50} (1994) 2700--2711,
  [\href{http://arxiv.org/abs/hep-th/9401070}{{\tt hep-th/9401070}}].

\bibitem{Witten:1988ze}
E.~Witten, {\it {Topological Quantum Field Theory}},  {\em Commun. Math. Phys.}
  {\bf 117} (1988) 353.

\bibitem{Hori:2003ic}
K.~Hori, S.~Katz, A.~Klemm, R.~Pandharipande, R.~Thomas, C.~Vafa, R.~Vakil, and
  E.~Zaslow, {\em {Mirror symmetry}}, vol.~1 of {\em Clay mathematics
  monographs}.
\newblock AMS, Providence, USA, 2003.

\end{thebibliography}
\providecommand{\href}[2]{#2}\begingroup\raggedright\endgroup

\end{document}